\pdfoutput=1

\documentclass[12pt,openright]{report} 

\newcommand{\thesistitle}{Dark Energy, Anthropic Selection Effects, Entropy and Life}

\newcommand{\fullname}{Chas Astro Egan}
\newcommand{\shortname}{C.\ A.\ Egan}
\newcommand{\thesissubjects}{cosmology}

\usepackage[british]{babel}

\usepackage{geometry}
\usepackage[pdftex]{graphicx}
\DeclareGraphicsExtensions{.pdf}
\usepackage[twoside]{rotating}
\usepackage[T1]{fontenc}
\usepackage{textcomp}
\usepackage{amsmath,amssymb}
\usepackage{pxfonts}
\usepackage{tgheros,tgpagella}

\usepackage{microtype}
\usepackage[round]{natbib}
\usepackage{longtable,booktabs,multirow,lscape}
\usepackage[]{caption,subfig}
\usepackage[unicode,pageanchor,colorlinks,pdftex]{hyperref}
\hypersetup{pdftitle=\thesistitle\ - \shortname,%
  pdfauthor=\fullname,%
  pdfsubject={\thesissubjects},%
  citecolor=Firebrick1,%
  linkcolor=DarkOrchid1,%
  anchorcolor=Cyan4}
 \hypersetup{pdftitle=\thesistitle\ - \shortname,%
   pdfauthor=\fullname,%
   pdfsubject={\thesissubjects},%
   citecolor=black,%
   filecolor=black,%
   linkcolor=black,%
   urlcolor=black%
 }
\usepackage[dvips]{dropping}
\usepackage[pdftex,hyperref,x11names]{xcolor}
\usepackage[section]{placeins}
\usepackage[subfigure,titles]{tocloft}
\usepackage[toc,page,title,titletoc]{appendix}
\usepackage[l2tabu, orthodox]{nag}
\usepackage{ifthen}

\newcommand{\titleDS}{\begingroup
\vspace*{0.1\textheight}
\centering
{\LARGE\usefont{OT1}{qbk}{m}{n}\selectfont\thesistitle}\par
\vspace{2.0\baselineskip}
{\Large \fullname}\\[1.0\baselineskip]
{A thesis submitted for the degree of\\[0.5\baselineskip]
Doctor of Philosophy\\[0.5\baselineskip]
of The University of New South Wales}\par
\centering
\vspace{1\baselineskip}
\begin{figure}[!h]
  \centering
  \includegraphics[width=0.5\textwidth]{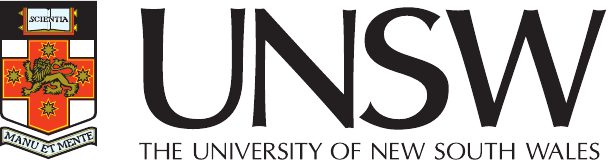}
\end{figure}
School of Physics \\
The University of New South Wales \\
Sydney NSW \\
Australia \\[\baselineskip]
{28$^{\textrm{th}}$ of August, 2009}\par
\vfill
\endgroup}

\usepackage{fncychap}
\makeatletter
\ChNameUpperCase
\ChTitleUpperCase  
\ChNameVar{\centering\Huge\usefont{OT1}{qbk}{m}{n}\selectfont}
\ChNumVar{\Huge}
\ChTitleVar{\centering\Huge\usefont{OT1}{qbk}{m}{n}\selectfont}
\ChRuleWidth{2pt}
\renewcommand{\DOCH}{%
  \CNV\FmN{\@chapapp}\space \CNoV\thechapter
  \par\nobreak
  \vskip -0.5\baselineskip
}
\renewcommand{\DOTI}[1]{%
  \mghrulefill{\RW}\par\nobreak
  \CTV\FmTi{#1}\par\nobreak
  \vskip 60\p@
}
\renewcommand{\DOTIS}[1]{%
  \mghrulefill{\RW}\par\nobreak
  \CTV\FmTi{#1}\par\nobreak
  \vskip 60\p@
}

\usepackage{fancyhdr}
\makeatletter
\def\cleardoublepage{\clearpage\if@twoside \ifodd\c@page\else
	\hbox{}
	\vspace*{\fill}
	\thispagestyle{empty}
	\newpage
	\if@twocolumn\hbox{}\newpage\fi\fi\fi}
\makeatother

\usepackage[calcwidth]{titlesec}
\titlelabel{\thetitle.\quad}

\usepackage{amsmath}
\usepackage{amsfonts}
\usepackage{amssymb}
\usepackage{wasysym}

\newcommand {\lsim}{\mbox{$\:\stackrel{<}{_{\sim}}\:$} }
\newcommand {\gsim}{\mbox{$\:\stackrel{>}{_{\sim}}\:$} }

\def\be{\begin{equation}}
\def\ee{\end{equation}}
\def\bea{\begin{eqnarray}}
\def\eea{\end{eqnarray}}

\def\deg {^\circ\!}

\def\lcdm{$\Lambda$CDM{ }}

\def\rhor{\rho_{r}}
\def\rhoro{\rho_{r_{o}}}

\def\rhom{\rho_{m}}
\def\rhomo{\rho_{m_{o}}}
\def\rhol{\rho_{\Lambda}}
\def\rhoc{\rho_{crit}}

\def\ol{\Omega_{\Lambda}}
\def\om{\Omega_{m}}

\def\orad{\Omega_{r}}

\def\olo{\Omega_{\Lambda_{o}}}
\def\omo{\Omega_{m_{o}}}
\def\orado{\Omega_{r_{o}}}

\def\msol{M_{\odot}}

\def\tp{t_{Planck}}
\def\ap{a_{Planck}}

\def\sol{\phantom{}_{\odot}}

\def\xt{\times 10^}

\long\def\symbolfootnote[#1]#2{\begingroup%
\def\thefootnote{\fnsymbol{footnote}}\footnote[#1]{#2}\endgroup}
\let\oldsqrt\sqrt
\def\sqrt{\mathpalette\DHLhksqrt}
\def\DHLhksqrt#1#2{%
\setbox0=\hbox{$#1\oldsqrt{#2\,}$}\dimen0=\ht0
\advance\dimen0-0.2\ht0
\setbox2=\hbox{\vrule height\ht0 depth -\dimen0}%
{\box0\lower0.4pt\box2}}

\newcommand\FigDiff[1]{\hyperref[#1]{Figure~\ref*{#1}} on Page~\pageref*{#1}}
\newcommand\FigSame[1]{\hyperref[#1]{Figure~\ref*{#1}}}
\newcommand\Figref[1]{\ifthenelse{\value{page}=\pageref{#1}}
                     {\FigSame{#1}}{\FigDiff{#1}}}
\newcommand\TabDiff[1]{\hyperref[#1]{Table~\ref*{#1}} on Page~\pageref*{#1}}
\newcommand\TabSame[1]{\hyperref[#1]{Table~\ref*{#1}}}
\newcommand\Tabref[1]{\ifthenelse{\value{page}=\pageref{#1}}
                     {\TabSame{#1}}{\TabDiff{#1}}}

\let\footnote=\endnote

\newcommand{\jnlref}[1]{\textrm{#1}}

\newcommand{\apj}{\jnlref{ApJ}}

\begin{document}

%
%
\pagestyle{empty}
\titleDS\clearpage
\cleardoublepage

%
%

\vspace*{0.3\textheight}
\begin{center}
\begin{it}
To Tonga, may your coconuts grow.
\end{it}
\end{center}

\cleardoublepage

\section*{Statement of Originality}

I hereby declare that this submission is my own work and to the best of my 
knowledge it contains no materials previously published or written by another 
person, or substantial proportions of material which have been accepted for the 
award of any other degree or diploma at UNSW or any other educational 
institution, except where due acknowledgement is made in the thesis. Any 
contribution made to the research by others, with whom I have worked at 
UNSW or elsewhere, is explicitly acknowledged in the thesis. I also declare that 
the intellectual content of this thesis is the product of my own work, except to 
the extent that assistance from others in the project's design and conception or 
in style, presentation and linguistic expression is acknowledged.

\vspace{2cm}
\hfill (Signed)\dotfill

\vspace{1cm}
\hfill (Date)\dotfill

\cleardoublepage

%
%
\fancypagestyle{plain}{									
\fancyfoot[LE]{\usefont{OT1}{qbk}{m}{n}\selectfont \thepage}		
\fancyfoot[RO]{\usefont{OT1}{qbk}{m}{n}\selectfont \thepage}		
\renewcommand{\headrulewidth}{0pt}						
\renewcommand{\footrulewidth}{0.5pt}						
}
%
\pagestyle{plain}
\pagenumbering{roman}
\setlength{\parindent}{0pt}
\setlength{\parskip}{1.8ex}

\pagestyle{plain}
\tableofcontents
\cleardoublepage

\pagestyle{plain}
\listoffigures
\cleardoublepage

\pagestyle{plain}
\listoftables
\cleardoublepage

\phantomsection\addcontentsline{toc}{section}{Abstract}
\section*{Abstract}


According to the standard $\Lambda$CDM model, the matter and dark energy densities 
($\rho_{m}$ and $\rho_{DE}$) are only comparable for a brief time. Using the temporal distribution of 
terrestrial planets inferred from the cosmic star formation history, we show that the 
observation $\rho_m \sim \rho_{DE}$ is expected for terrestrial-planet-bound observers 
under $\Lambda$CDM, or under any model of dark energy consistent with 
observational constraints. Thus we remove the coincidence problem as a factor motivating dark 
energy models.


We compare the Sun to representative stellar samples in $11$ properties 
plausibly related to life. We find the Sun to be most anomalous in mass and galactic
orbital eccentricity. When the $11$ properties are considered together we find that the 
probability of randomly selecting a star more typical than the Sun is only $29 \pm 11 \%$.
Thus the observed ``anomalies'' are consistent with statistical noise. This contrasts
with previous work suggesting anthropic explanations for the Sun's high mass.

The long-term future of dissipative processes (such as life) depends on the continued availability 
of free energy to dissipate thereby increasing entropy. The entropy budget of the present observable 
Universe is dominated by supermassive black holes in galactic cores. Previous estimates of 
the total entropy in the observable Universe were between $\sim 10^{101}\ k$ and 
$\sim 10^{103}\ k$. Using recent measurements of the supermassive black hole mass function 
we find the total entropy in the observable Universe to be $S_{obs} = 3.1^{+3.0}_{-1.7}\xt{104}\ k$, 
at least an order of magnitude higher than previous estimates. We compute the entropy in $3$ 
new subdominant components and report a new entropy budget of the Universe with 
quantified uncertainties. 
We evaluate upper bounds on the entropy of a comoving volume (normalized to the present 
observable Universe). Under the assumption that energy in matter is constant in a comoving 
volume, the availability of free energy is found to be finite and the future entropy in the volume 
is limited to a constant of order $10^{123} k$. Through this work we uncover a number of 
unresolved questions with implications for the ultimate fate of the Universe.


\cleardoublepage

\phantomsection\addcontentsline{toc}{section}{Acknowledgments}
\section*{Acknowledgments}

I want to thank my unofficial supervisor, Charles Lineweaver, for the years of 
inspiration and guidance. Charley has an enthusiasm for getting to the bottom
of fundamental scientific issues which is not dissuaded by dogma or conventional
disciplinary borders. 

Secondly, to the Research School of Astronomy and Astrophysics at the Australian 
National University, thank you for generously hosting me for the majority of my 
candidature. Mount Stromlo is steeped in academic heritage. It is a wonderful and 
inspiring place to work. The drive up the mountain is always a delight.

I would like to thank my official supervisor, John Webb and others in the School 
of Physics at the University of New South Wales for making it possible for me to
carry out my candidature remotely. 
 
A  warm gracias my most excellent academic brother Jose Robles for turning me into
a mac user. It was a pleasure coding with you. 
To Shane, my office-mate and pepsi lover, and to Josh,
the gentoo guru, to Leith my mediocre table tennis partner, to Brad for organizing 
all the great events, and to Dan and Grant for those mellow days at the beach,
thanks for all the good times.
 
I am indebted to my big family at the Woden Fish Market for always making me 
feel close to the ocean.
 
Mum, Dad and Hayley, thank you for 27 years of love and support.
 
Finally and most importantly, Anna, my soulmate, you have given me more than it 
is fair to ask of anyone. Every day with you is grand no matter where we are. 
Thanks for supporting this dream of mine. I love you forever.
\cleardoublepage

\phantomsection\addcontentsline{toc}{section}{Preface}
\section*{Preface}

The contents of this thesis are based on research articles that I have published during 
the course of my PhD candidature. Although I use the first person plural throughout, all of the work 
presented in this thesis is my own unless explicitly stated here, or in section \ref{C1disclamour}
of the Introduction.


\begin{description}
\item{$\RHD\ $} {\bf C.~H.~Lineweaver \& C.~A.~Egan, 2007},
``The Cosmic Coincidence as a Temporal Selection Effect Produced by the Age Distribution of Terrestrial Planets 
in the Universe'',
\apj, v671, 853--860.
\end{description}

\begin{description}
\item{$\RHD\ $} {\bf C.~A.~Egan \& C.~H.~Lineweaver, 2008}, 
``Dark Energy Dynamics Required to Solve the Cosmic Coincidence'',
Phys.\ Rev.\ D., v78(8), 083528.
\end{description}

\begin{description}
\item{$\RHD\ $} {\bf J.~A.~Robles, C.~H.~Lineweaver, D.~Grether, C.~Flynn, C.~A.~Egan, M.~Pracy, J.~Holmberg \& E.~Gardner, 2008}, 
``A comprehensive comparison of the Sun to other stars: searching for self-selection effects'', 
\apj, v684, 691--706.
\end{description}

\begin{description}
\item{$\RHD\ $} {\bf J.~A.~Robles,  C.~A.~Egan \& C.~H.~Lineweaver, 2009},
``Statistical Analysis of Solar and Stellar Properties'', 
Australian Space Science Conference Series: 8th Conference Proceedings NSSA 
Full Referreed Proceedings CD, (ed) National Space Society of Australia Ltd, edt. 
W. Short, conference held in Canberra, Australia September 23--25, 2008, 
ISBN 13: 978-0-9775740-2-5.
\end{description}

\begin{description}
\item{$\RHD\ $} {\bf C.A. Egan \& C.H. Lineweaver, 2010},
``A Larger Estimate of the Entropy of the Universe'', 
\apj, v710, 1825-1834.
\end{description}

\begin{description}
\item{$\RHD\ $} {\bf C.A. Egan \& C.H. Lineweaver, 2010},
``The Cosmological Heat Death'',
in preparation.
\end{description}

%

I have attached as appendix \ref{appendix1} a publication that is predominantly the work of 
my supervisor, Charles. H. Lineweaver, but to which I made minor contributions.

\begin{description}
\item{$\bullet\ $} {\bf C.H. Lineweaver \& C.A. Egan, 2008}, 
``Life, gravity and the second law of thermodynamics.'' 
Physics of Life Reviews, v5, 225--242.
\end{description}

\cleardoublepage

%

%
%
\fancyfoot[LE]{\usefont{OT1}{qbk}{m}{n}\selectfont \thepage}		
\fancyfoot[RO]{\usefont{OT1}{qbk}{m}{n}\selectfont \thepage}		
\renewcommand{\footrulewidth}{0.5pt}						
\renewcommand{\headrulewidth}{0.5pt}						
\fancypagestyle{plain}{									
\fancyfoot[LE]{\usefont{OT1}{qbk}{m}{n}\selectfont \thepage}		
\fancyfoot[RO]{\usefont{OT1}{qbk}{m}{n}\selectfont \thepage}		
\renewcommand{\headrulewidth}{0pt}
\renewcommand{\footrulewidth}{0.5pt}						
}
%
%
\pagenumbering{arabic}
%
\pagestyle{fancy}
\chapter{Introduction}

\begin{center}
\emph{
I may be reckless, may be a fool, \\
but I get excited when I get confused. 
}
\end{center}
\begin{flushright}
- Fischerspooner, ``The Best Revenge'' \hspace*{2cm}
\end{flushright}
\vspace{1cm}

\section{Copernicanism and Anthropic Selection}

The Copernican idea, that we perceive the Universe from an entirely mediocre vantage point, is deeply embedded
in the modern scientific world view. Before the influences of Copernicus, Galileo and Newton in the 16th and 17th 
centuries the prevalent world view was anthropocentric: we and the Earth were at the center of the Universe, and 
the heavenly bodies lived on spherical planes around us. The paradigm shift to a Copernican world view was
ferociously resisted by theologians and philosophers, but was eventually adopted because of its ability to explain
mounting physical and astronomical observations.

It is with great esteem that we remember these pioneers of modern science, who taught us that observational 
evidence trumps philosophical aesthetics. However, upon pedantic inspection, the Copernican idea leads to untrue 
predictions. For example, if we did occupy a mediocre vantage point then the density of our immediate environment
would be $\sim 10^{-30}\ g\ cm^{-3}$. However the density of our actual environment is $\sim 1\ g\ cm^{-3}$. A napkin 
calculation considering the density and size of collapsed objects suggest the chance of us living in an environment 
as dense or denser by pure chance is around $1$ in $10^{30}$ $-$ a significant signal.

There are selection effects connected with being an observer. They determine, to some degree, 
where and when we observe the Universe. At the cost of strict Copernicanism we must make considerations for 
anthropic selection as a class of observational selection effect \citep{Dicke1961,Carter1974,Barrow1986,Bostrom2002}, 
and we \emph{must} take the appropriate steps to remove anthropic selection effects from our data.

\section{The Cosmic Coincidence Problem}

Recent cosmological observations including observations of the cosmic microwave background temperature
fluctuations, the luminosity-redshift relation from supernova light-curves and the matter power spectrum measured 
in the large scale structure and Lyman-Alpha forests of quasar spectra, have converged on a cosmological 
model which is expanding, and whose energy density is dominated by a mysterious component referred to 
generally as dark energy ($\sim 73\%$) but contains a comparable amount of matter ($\sim 27\%$) and some 
radiation ($\sim 5\xt{-5}\%$). See e.g.\ \citep{Seljak2006} and references therein.

The energy in these components drives the expansion of the Universe via the Friedmann equation, and in turn
responds to the expansion via their equations of state: radiation dilutes as $a^{-4}$, matter dilutes as $a^{-3}$ 
the dark energy density remains constant (assuming that dark energy is Einstein's cosmological constant) where 
$a$ is the scalefactor of the Universe \citep{Carroll2004book}.

Since matter and dark energy dilute at different rates during cosmic expansion, these two components only have 
comparable densities for a brief interval during cosmic history. Thus we are faced with the ``cosmic coincidence 
problem'': Why, just now, do the matter and dark energy densities happen to be of the same order \citep{Weinberg1989,Carroll2001a}?
Ad-hoc dynamic dark energy (DDE) models have been designed to solve the cosmic coincidence problem by 
arranging that the dark energy density is similar to the matter density for significant fractions of the age of the 
Universe. 

Whether or not there is a coincidence problem depends on the range of times during which the Universe may
be observed. In Chapter \ref{chap2}, we quantify the severity of the coincidence problem under $\Lambda$CDM 
by using the temporal distribution of terrestrial planets as a basis for the probable times of observation.

In Chapter \ref{chap3} we generalize this approach to quantify the severity of the coincidence problem for all 
models of dark energy (using a standard parameterization). The two possible outcomes of this line of investigation 
are both valuable. One the one hand finding a significant coincidence problem for otherwise observationally allowed 
dark energy models would rule them out, complementing observational constraints on dark energy. On the other 
hand, finding that the coincidence problem vanishes for all observationally allowed models would remove the 
cosmic coincidence problem as a factor motivating dark energy models.

\section{Searching for Life Tracers Amongst the Solar Properties}

If the origin and evolution of terrestrial-planet-bound observers depend on anomalous properties of the planet's 
host star, then the stars that host such observers (including the Sun) are anthropically selected to have those 
properties. 

\citet{Gonzalez1999a,Gonzalez1999b} found that the Sun was more massive than $\sim 91\%$ of stars, and
suggested that this may be explained if observers may develop preferentially around very massive stars. 
A star's mass determines, in large part, its lifetime, luminosity, temperature and the location of the terrestrial 
habitable zone, all of which may influence the probability of that star hosting observers. 
But the statistical significance of this ``anomalous'' mass depends on the number of other solar properties, 
also plausibly related to life, from which mass was selected.
Thus while Gonzalez's proposition is plausible, it is unclear how strongly it is supported by the data.

In Chapter \ref{chap4} of this thesis we compare the Sun to representative samples of stars in $11$ 
independent parameters plausibly related to life (including mass), with the aim of quantifying the overall
typicality of the Sun and potentially identifying statistically significant anomalous properties - potential tracers 
of life in the Universe.

\section{The Entropy of the Present and Future Universe}

One feature that we can count on as being important to all life in the Universe is the availability of free 
energy. Indeed we can only be sure of this because \emph{all} irreversible processes in the Universe
consume free energy and contribute to the increasing total entropy of the Universe. 

The current entropy of the observable Universe was estimated by \citet{Frampton2008} to be 
$\sim 10^{102}\ k$ of a maximum possible value of $\sim 10^{123}\ k$. The current entropy of the
observable Universe is dominated by the entropy in supermassive black holes at the centers of 
galaxies, followed distantly by the cosmic microwave background, neutrino background and other
components. 

If the entropy of the Universe reaches a value from which it could not be further increased, then 
all dissipative processes would cease. The idea that the future of the universe could end in such a 
state of thermodynamic equilibrium (a so-called heat death) was written about by \citet{Thomson1852},
and later revived within the context of an expanding Universe by \citet{Eddington1931}.
Scientific and popular science literature over the past three decades is ambiguous about whether or not 
there will be a heat death, and if so, in what form.

In Chapter \ref{chap5} we present an improved budget of the entropy of the observable Universe using
new measurements of the supermassive black hole mass function. In Chapter \ref{chap6} we compare 
the growing entropy of the Universe to upper bounds that have been proposed, and draw conclusions
about the future heat death.

\section{About the Papers Presented in this Thesis} \label{C1disclamour}

Chapter \ref{chap2} was published as \citet{Lineweaver2007}. The text was co-written with Charles 
Lineweaver, who is also to be credited for the original idea. However, the work presented in the paper is 
predominantly mine: details of the method, quantitative analyses, the preparation of all figures. For these 
reasons, and with Dr.\ Lineweaver's endorsement, it has been included here verbatim.

Chapter \ref{chap3} is my own and was published as \citet{EganLineweaver2008}.

In Chapter \ref{chap4} I describe work published in \citet{Robles2008a}, \citet{Robles2008c} and the erratum, \citet{Robles2008b}. 
I was a co-author of this work, which was lead by Jose Robles, and I contributed in part to the collection of data 
(age; see figure \ref{fig:C4josef2}), data analysis (advice on, and implementation of statistical, methods, as well as
coding other parts of the analysis pipeline), interpretation and presentation of the results (contributing to figures 
and published articles). 
This chapter summarizes the main results paper, \citet{Robles2008a}, in words that are my own. The figures are 
taken, with permission, from \citet{Robles2008a}.

Chapter \ref{chap5} is my own and has been submitted for publication to \apj\ as \citet{EganLineweaver2009}.

Chapter \ref{chap6} is my own and will contribute towards an article currently in preparation, which we refer to 
as \citet{EganLineweaver2009b}.

Appendix \ref{appendix1} has been published as \citet{LineweaverEgan2008}. The text, and most of the work
presented in that paper is that of my supervisor. 
My contributions include the contribution of the preparation of Figure \ref{fig:A1rhoT}. 
The paper is included in the appendix of this thesis as it is referred to several times, and motivates the work 
presented in Chapters \ref{chap5} and \ref{chap6}.

%
\pagestyle{fancy}
\chapter{The Cosmic Coincidence as a Temporal Selection Effect Produced by the Age Distribution of Terrestrial Planets in the Universe} 
\label{chap2}

\begin{center}
\emph{
Late at night, stars shining bright \\
on me, down by the sea. \\
And when I see them in the sky \\
constantly I'm asking why \\
I was stranded here. \\
I wish I could be out in space.
}
\end{center}
\begin{flushright}
- S.P.O.C.K, ``Out in Space'' \hspace*{2cm}
\end{flushright}
\vspace{1cm}

\section{Is the Cosmic Coincidence Remarkable or Insignificant?}
\label{sec:C2intro}

\subsection{Dicke's argument}

\citet{Dirac1937} pointed out the near equality of several large fundamental dimensionless numbers of the order $10^{40}$. 
One of these large numbers varied with time since it depended on the age of the Universe. 
Thus there was a limited time during which this near equality would hold.  
Under the assumption that observers could exist at any time during the history of the Universe, 
this large number coincidence could not be explained in the standard cosmology.  This problem
motivated \citet{Dirac1938} and \citet{Jordan1955} to construct an ad hoc new cosmology.
Alternatively, \citet{Dicke1961} proposed that our observations of the Universe could only be 
made during a time interval after carbon had been produced in the Universe and before the 
last stars stop shining.  Dicke concluded that this temporal observational selection 
effect -- even one so loosely delimited -- could explain Dirac's large number coincidence 
without invoking a new cosmology.

\begin{figure}[!p]
\includegraphics[width=\linewidth]{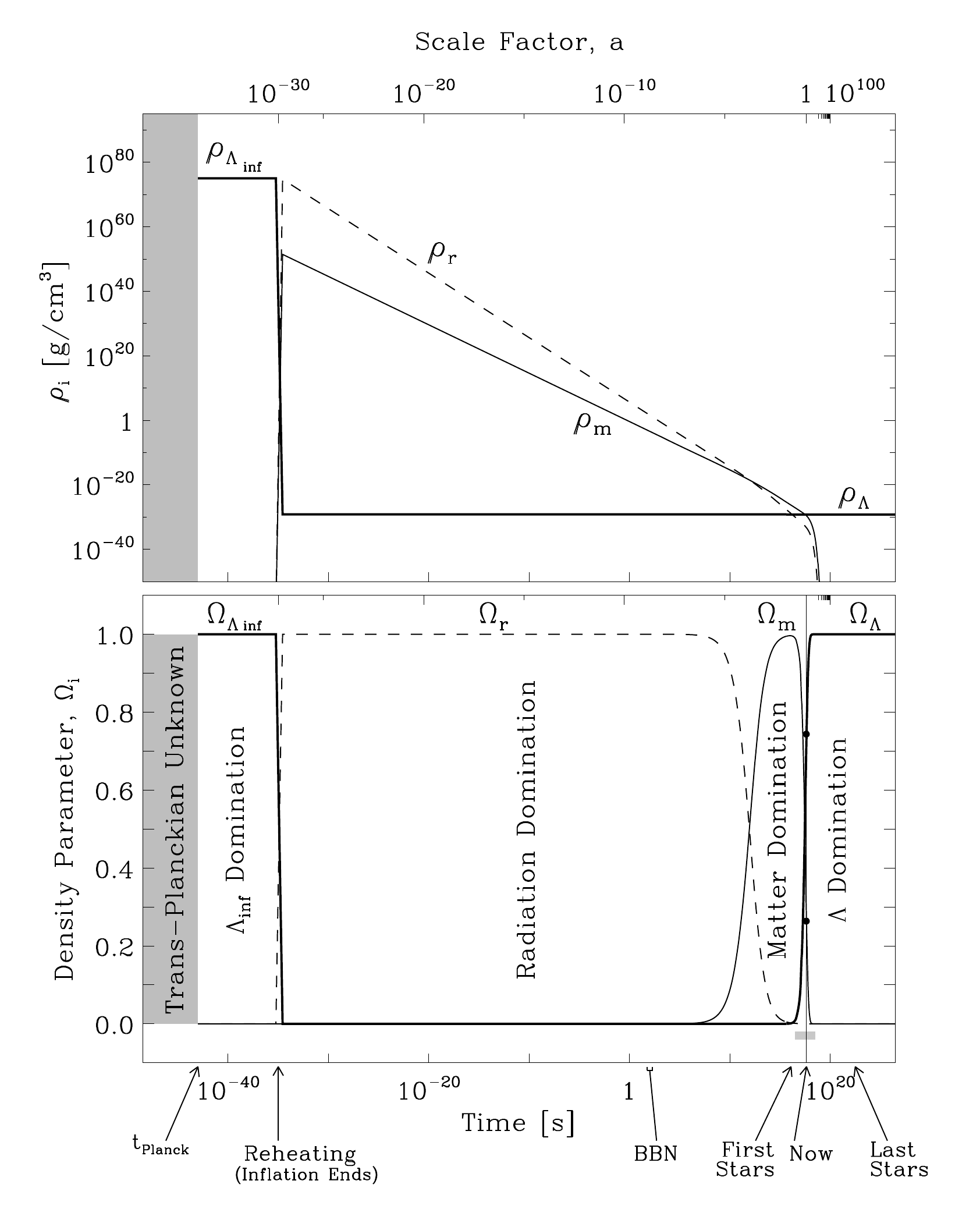}
\caption[The time dependence of the densities of the major components of the Universe.]
{The time dependence of the densities of the major components of the Universe.
Given the observed Hubble constant, $H_{o}$ and energy densities in the Universe today,  $\orado$, $\omo$, $\olo$
(radiation, matter and cosmological constant),
we use the Friedmann equation to plot the temporal evolution of the components of the
Universe in g/cm$^{3}$ (top panel), or  normalized to the time-dependent critical 
density $\rhoc = \frac{3 H(t)^{2}}{8\pi G}$ (bottom panel).
We assume an epoch of inflation at $\sim 10^{-35}$ seconds after the big bang and
 a false vacuum energy density $\rho_{\Lambda_{\rm inf}}$ between the Planck 
scale and $t_{GUT}$.
See Table \ref{table:C2importantimes} and Appendix A for details.
}
\label{fig:C2omegas}
\end{figure}

Here, we construct a similar argument to address the cosmic coincidence:
Why just now do we find ourselves in the relatively brief interval during which 
$\om \sim \ol$.
The temporal constraints on observers that we present are more empirical and 
specific than those used in Dicke's analysis, but the reasoning is similar.
Our conclusion is also similar:  a temporal observational selection effect can explain 
the apparent cosmic coincidence.  That is,  given the evolution of $\ol$ and $\om$ in our Universe, most observers 
in our Universe who have emerged on terrestrial planets will find $\ol \sim \om$. 
Rather than  being an unusual coincidence, it is what one should expect.

There are two distinct problems associated with the cosmological constant
\citep{Weinberg2000,Garriga2001,Steinhardt2003}. One is the coincidence problem that we address here.
The other is the smallness problem and has to do with the observed
energy density of the vacuum, $\rhol$. Why is $\rhol$ so small compared to the $\sim 10^{120}$ times larger value
predicted by particle physics? Anthropic solutions to this problem invoke a multiverse and
argue that galaxies would not form and there would be no life in a Universe, if $\rhol$ were larger than $\sim 100 $ times its 
observed value \citep{Weinberg1987,Martel1998,Garriga2001,Pogosian2007}. 
Such explanations for the smallness of $\rhol$ do not explain the temporal coincidence 
between the time of our observation and the time of the near-equality of $\om$ and $\ol$.
Here we address this temporal coincidence in our Universe, not the smallness problem in a multiverse.

\subsection{Evolution of the Energy Densities}
\label{sec:C2subjectivity}

\begin{figure}[!p]
\includegraphics[width=0.8\linewidth]{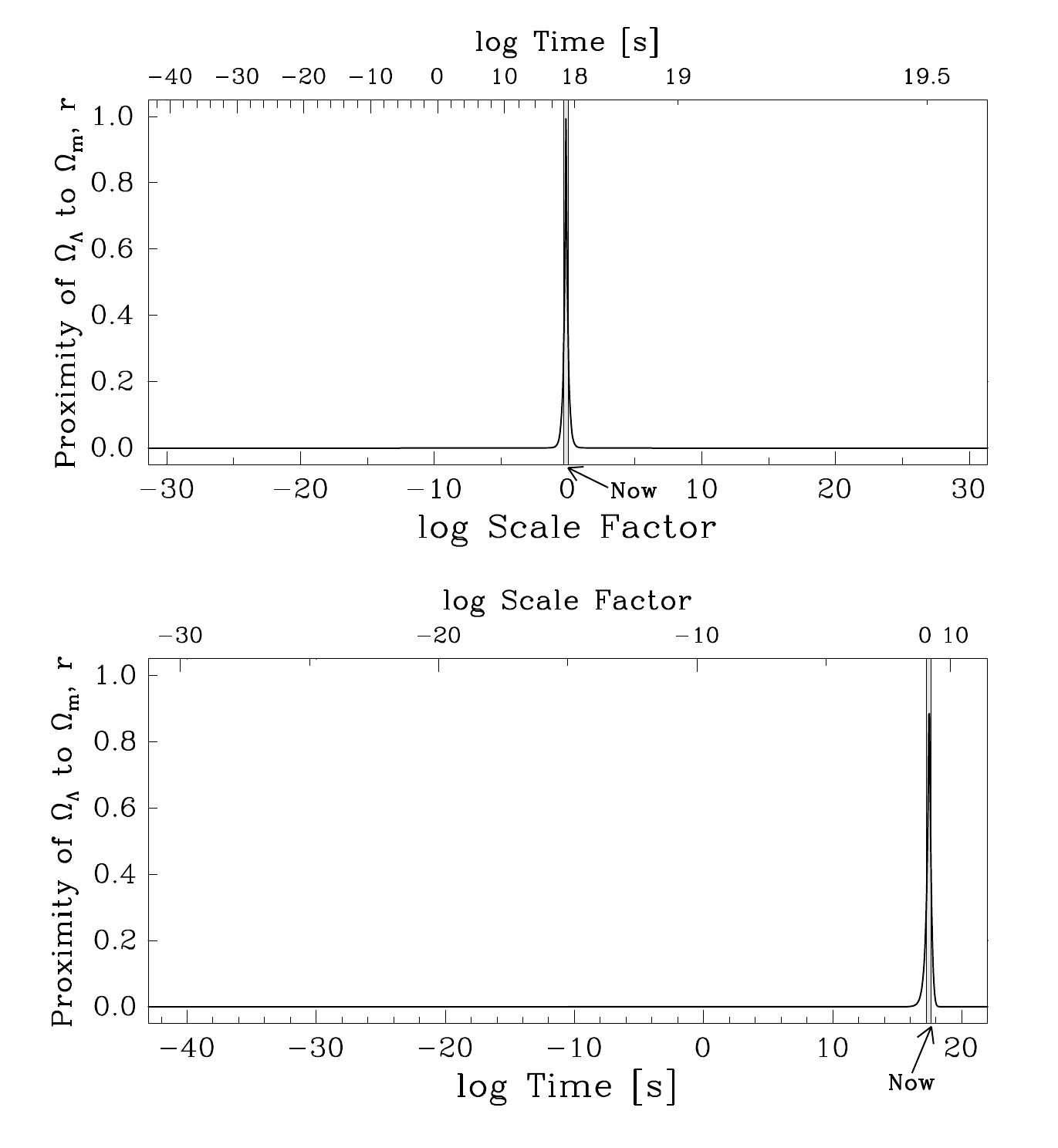}
\caption[The time dependence of the vacuum-matter proximity factor in Log space.]
{Plot of the  proximity factor $r$ (see Eq. \ref{eq:C2rdef}).
When the matter and vacuum energy densities of the Universe are the same,  $\om = \ol$, we have $r=1$. 
We currently observe $\omo \sim \olo$ and thus, $r \sim 1$.  
Our existence now when $r \sim 1$ appears to be an unlikely cosmic coincidence 
when the x axis is logarithmic in the 
scale factor (top panel) or logarithmic in time (bottom panel).
In the top panel, following \citet{Carroll2001a}, we have chosen a range of 
scale factors  with ``Now'' midway between the scale factor at the Planck time 
and the scale factor at the inverse Planck time $[\ap < a <  \ap^{-1}]$. 
The brief epoch shown in grey between the thin vertical lines is the epoch during which $r > r_{o}$
(where $r_{o} \approx 0.4$ is the currently observed value).
In the bottom panel the range shown on the x axis is $[\tp < t < 10^{22}]$ seconds.
The Planck time and Planck scale provide reasonably objective lower time limits.
The upper limits are somewhat arbitrary but contribute to the impression that $r \approx 0.4 \sim 1$ 
is an unlikely coincidence.
}
\label{fig:C2logproblem}
\end{figure}

Given the currently observed values for $H_{o}$ and the energy densities 
$\orado$, $\omo$ and $\olo$ in the Universe 
\citep{Spergel2006,Seljak2006}, the Friedmann equation tells us the evolution of the scale factor $a$, and the evolution
 of these energy densities.  These are plotted in Fig. \ref{fig:C2omegas}.
The history of the Universe can be divided chronologically into four distinct periods 
each dominated by a different form of energy: initially the false vacuum energy of inflation dominates, 
then radiation, then matter, and finally vacuum energy. Currently the 
Universe is making the transition from matter domination to vacuum energy domination.
In an expanding Universe, with an initial condition $\om > \ol > 0$, there will be some epoch in which
$\om \sim \ol$, since $\rhom$ is decreasing as $ \propto 1/a^{3}$ while $\rhol$ is a constant (see top panel of
Fig. \ref{fig:C2omegas} and Appendix A).  
Figure \ref{fig:C2omegas} also shows that the transition from matter domination 
to vacuum energy domination is occurring now. 
When we view this transition in the context of the time evolution of the Universe (Fig. \ref{fig:C2logproblem})
we are presented with the cosmic coincidence problem:
Why just now do we find ourselves at the relatively brief interval during which this transition happens?
\citet{Carroll2001a,Carroll2001b} and \citet{Dodelson2000} find this coincidence to be a remarkable result 
that is crucial to understand.
%
The cosmic coincidence problem is often regarded as an important unsolved problem
whose solution may help unravel the nature of dark energy (\citealt{Turner2001,Carroll2001b}).
The coincidence problem is one of the main motivations for the tracker
potentials of quintessence models 
\citep{Caldwell1998,Steinhardt1999,Zlatev1999,Wang2000,Dodelson2000,Armendariz-Picon2001,Guo2005}.
In these models the cosmological constant is replaced by a
more generic form of dark energy in which $\om$ and $\ol$ are in
near-equality for extended periods of time.  It is not clear that these models successfully
explain the coincidence without fine-tuning (see \citealt{Weinberg2000,Bludman2004}).

The interpretation of the observation $\omo \sim \olo$ as a remarkable coincidence in need 
of explanation depends on some assumptions that we quantify to determine  
how surprising this apparent coincidence is.
We begin this quantification by introducing a time-dependent proximity parameter, 
\be
r = \min\left[\frac{\ol}{\om},\frac{\om}{\ol}\right]
\label{eq:C2rdef}
\ee
which is equal to one when $\om = \ol$ and is close to zero when $\om >> \ol$ or $\om << \ol$.
The current value is $r_{o} \approx 0.4$.
In Figure \ref{fig:C2logproblem} we plot $r$ as a function of log(scale factor) in the upper panel and  
as a function of log(time) in the lower panel.  These logarithmic axes allow a large dynamic range that
makes our existence at a time when $r \sim 1$, appear to be an unlikely coincidence.
This appearance depends on the implicit assumption that we could make cosmological observations 
at any time with equal likelihood. More specifically, the implicit assumption is that 
the {\it a priori} probability distribution $P_{obs}$, of the times we could have made our observations, 
is uniform in log $t$, or log $a$, over the interval shown.  

Our ability to quantify the significance of the coincidence
depends on whether we assume that
$P_{obs}$ is uniform in time, log(time), scale factor or log(scale factor).
That is, our result depends on whether we assume:
$P_{obs}(t) = constant$, $P_{obs}(\log t) = constant$, $P_{obs}(a) = constant$ or 
$P_{obs}(\log a) = constant$. These are the most common possibilities, but there are others.
For a discussion of the relative merits of log and linear time scales and implicit uniform 
priors see Section \ref{sec:C2measure} and \citet{Jaynes1968}.

\begin{figure}[!p]
\includegraphics[width=0.8\linewidth]{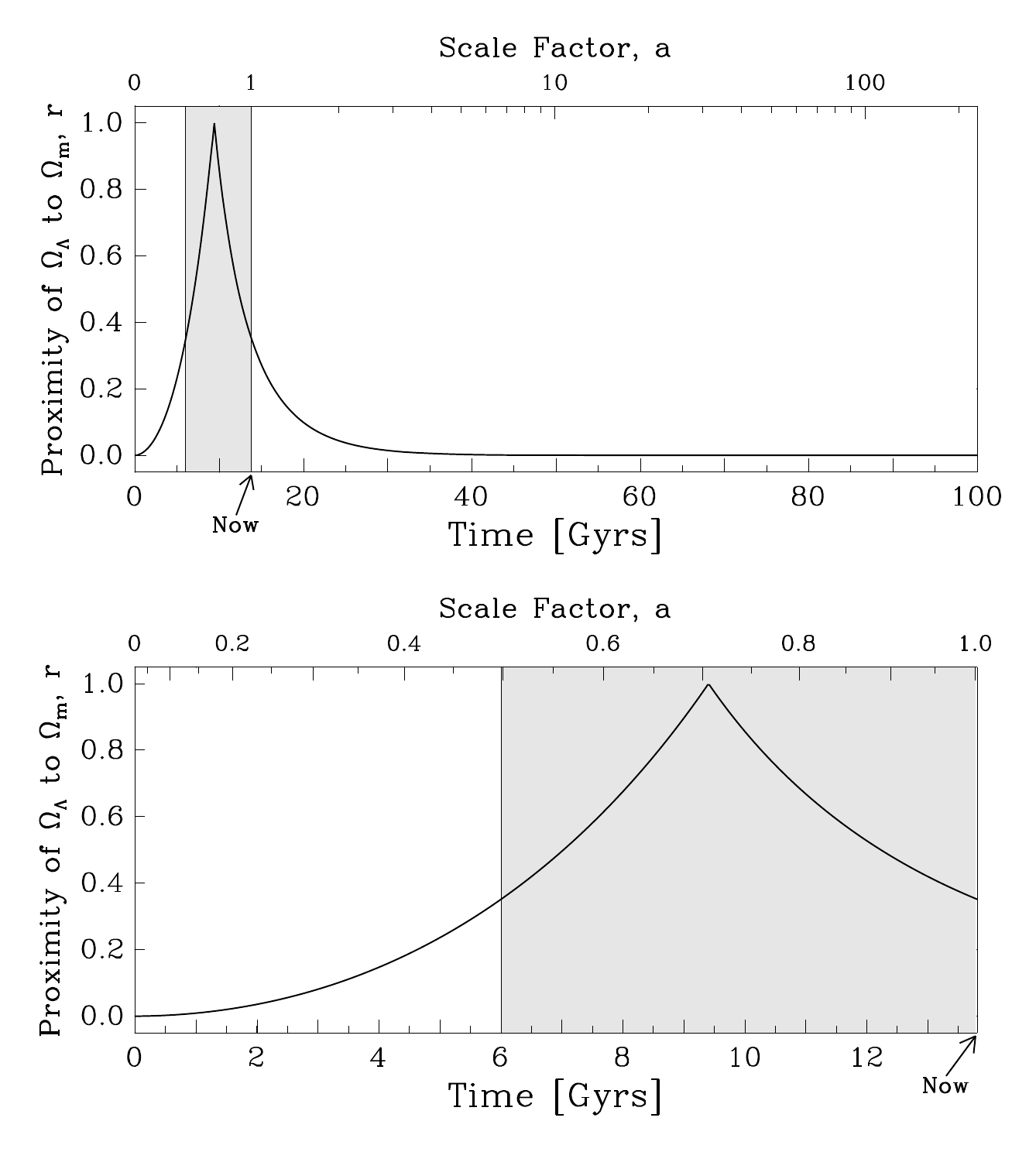}
\caption[The time dependence of the vacuum-matter proximity factor in linear space.]{
Plot of the  proximity factor $r$, as in the previous figure, but plotted here 
with a linear rather than a logarithmic time axis.  
The condition  $r > r_{o} \approx 0.4$ does not seem as unlikely as in the previous figure.
The range of time plotted also affects this appearance;
with the $[0,100]$ Gyr range of the top panel, the time interval highlighted in grey where $r > r_{o}$,
appears narrow and relatively unlikely.  In contrast, the $[0,13.8]$ Gyr range of the bottom panel
seems to remove the appearance of $r > r_{o}$ being an unlikely coincidence in need of 
explanation;  
for the first $\sim 6$ Gyrs we have
$r < r_{o}$ while in the subsequent $7.8$ Gyr we have $r > r_{o}$. 
How can $r > r_{o}$ be an unlikely coincidence when it has been true for most of the
history of the Universe?
}
\label{fig:C2linearnoproblem}
\end{figure}

In Fig. \ref{fig:C2linearnoproblem} we plot $r(t)$ on an axis linear in time where the 
implicit assumption is that the {\it a priori} probability distribution of our existence is uniform 
in $t$ over the intervals $[0,100]$ Gyr (top panel)
and $[0,13.8]$ Gyr (bottom panel). The bottom panel shows that the observation $r > 0.4$ could have been made 
anytime during the past 7.8 Gyr.  
Thus, our current observation that $r_{o} \approx 0.4$, does not appear to be a remarkable coincidence.
Whether this most recent 7.8 Gyr period is seen 
as ``brief'' (in which case there is an unlikely coincidence in need of explanation) 
or ``long'' (in which case there is no coincidence to explain) depends on whether we view 
the issue in log time (Fig. \ref{fig:C2logproblem}) or linear time (Fig. \ref{fig:C2linearnoproblem}).


A large dynamic range is necessary to present
the fundamental changes that occurred in the very early Universe, e.g., the transitions at the Planck time, 
inflation, baryogenesis, nucleosynthesis, 
recombination and the formation of the first stars.  Thus a logarithmic time axis is often preferred by 
early Universe cosmologists 
because it seems obvious, from the point of view of fundamental physics, that 
the cosmological clock ticks logarithmically.  
This defensible view and the associated logarithmic axis gives the impression that there is a 
coincidence in need of an explanation.
The linear time axis gives a somewhat different impression.
Evidently, deciding whether a coincidence is of some significance or only an accident is not easy \citep{Peebles1999}.
We conclude that although the importance of the cosmic coincidence problem is subjective, 
it is important enough to merit the analysis we perform here.

The interpretation of the observation $\omo \sim \olo$ as a coincidence in need of explanation 
depends on the {\it a priori}  (not necessarily uniform) probability distribution of our existence. 
That is, it depends on when cosmological observers can exist.
We propose that the cosmic coincidence problem can be more constructively evaluated by replacing these 
uninformed uniform  priors with the more realistic assumption that observers capable of 
measuring cosmological parameters are dependent on the emergence of high density
regions of the Universe called terrestrial planets, which require non-trivial amounts 
of time to form -- and that once these planets are in place, the observers themselves 
require non-trivial amounts of time to evolve.

In this paper we use the age distribution of terrestrial planets estimated by \citet{Lineweaver2001}
to constrain when in the history of the Universe, observers on terrestrial planets 
can exist.
In Section \ref{sec:C2computeprobability},  
we briefly describe this age distribution (Fig. \ref{fig:C2cosmologists}) and
show how it limits the existence of such observers to an interval in which $\om \sim \ol$ (Fig. \ref{fig:C2rollercoasterzoom}).
Using this age distribution as a temporal selection function, we compute the probability of an observer on 
a terrestrial planet observing $r \ge r_{o}$ (Fig. \ref{fig:C2pofr_nonorm}).
In Section \ref{sec:C2robust} we discuss the robustness of our result and find
(Fig.\ \ref{fig:C2varyingtev}) that this result is relatively robust
if the time it takes an observer to evolve on a terrestrial planet is less than $\sim 10$ Gyr. 
In Section \ref{sec:C2discussion} we discuss and summarize our results, and compare it to previous work
to resolve the cosmic coincidence problem \citep{Garriga2000,Bludman2001}.

\section{How We Compute the Probability of Observing $\om \sim \ol$}
\label{sec:C2computeprobability}

\subsection{The Age Distribution of Terrestrial Planets and New Observers}

\begin{figure}[!p]
\includegraphics[width=\linewidth]{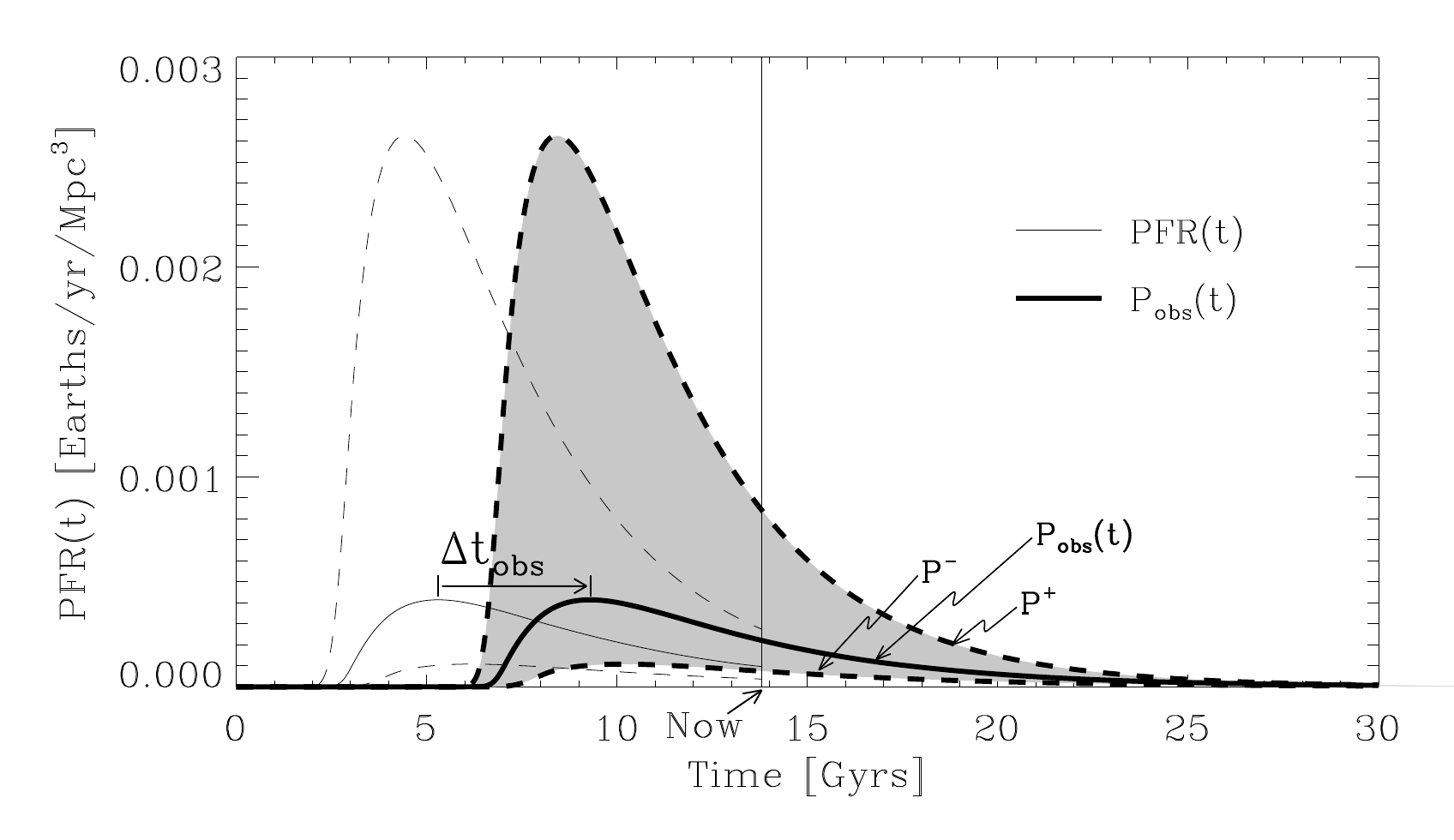}
\caption[Terrestrial planet and observer distributions.]
{The terrestrial planet formation rate $PFR(t)$, derived in \citet{Lineweaver2001} is an 
estimate of the age distribution of terrestrial planets in the Universe and is shown here as a 
thin solid line. Estimated uncertainty is given by the thin dashed lines. 
To allow time for the evolution of observers on terrestrial planets, we shift this distribution by $\Delta t_{obs}$ 
to obtain an estimate of the age distribution of observers:  
$P_{obs}(t) = PFR(t-\Delta t_{obs})$ (thick solid line). 
The grey band represents the error estimate on $P_{obs}(t)$ which is the shifted error 
estimates on $PFR(t)$.  
In the case shown here $\Delta t_{obs} = 4$ Gyr, which is how long it took life on Earth to 
emerge, evolve and be able to measure the composition of the Universe.  
To obtain the numerical values on the y axis, we have followed \citet{Lineweaver2001} and assumed 
that one out of one hundred stars is orbited by a terrestrial planet.
We have smoothly extrapolated the $PFR(t)$ of \citet{Lineweaver2001} into the future.
This time dependence and our subsequent analysis does not depend on whether the probability for
terrestrial planets to produce observers is high or low.
}
\label{fig:C2cosmologists}
\end{figure}

The mass histogram of detected extrasolar planets
peaks at low masses: $dN/dM \propto M^{-1.7}$, suggesting that low
mass planets are abundant \citep{Lineweaver2003b}.
Terrestrial planet formation may be a common feature of star formation
(\citealt{Wetherill1996a,Chyba1999,Ida2005}).
Whether terrestrial planets are common or rare,  they will have an age distribution proportional to the
star formation rate -- modified by the fact that in the first $\sim 2$ billion years of star formation, 
metallicities are so low that the material for terrestrial 
planet formation will not be readily available.
Using these considerations,  \citet{Lineweaver2001} estimated the age 
distribution of terrestrial planets --  how many Earths are produced by the 
Universe per year, per $Mpc^{3}$ (Figure \ref{fig:C2cosmologists}). 
If life 
emerges rapidly on terrestrial planets \citep{Lineweaver2002} 
then this age distribution is the age distribution of biogenesis in the Universe. 
However, we are not just interested in any life; we would like to know the distribution in time 
of when independent observers first emerge and are able to measure $\om$ and $\ol$, as we are able to do now.
If life originates and evolves preferentially 
on terrestrial planets, then the \citet{Lineweaver2001} estimate of the age distribution of 
terrestrial planets is an {\it a priori} input which can guide our expectations of when we 
(as members of a hypothetical group of terrestrial-planet-bound observers) 
could have been present in the Universe.
It takes time (if it happens at all) for life to emerge on a new terrestrial planet 
and evolve into cosmologists who can 
observe $\om$ and $\ol$.  
Therefore, to obtain the age distribution of new independent observers
able to measure the composition of the Universe for the first time,
we need to shift the age distribution 
of terrestrial planets by some characteristic time, $\Delta t_{obs}$ required for observers to evolve.
On Earth, it took $\Delta t_{obs} \sim 4 $ Gyr for this to happen.  
Whether this is characteristic of life elsewhere in the Universe is uncertain 
(\citealt{Carter1983,Lineweaver2003a}).
For our initial analysis we use $\Delta t_{obs} = 4$ Gyr as a nominal time to evolve observers.
In Section \ref{sec:C2deltat} we allow $\Delta t_{obs}$ to vary from 0-12 Gyr to see how sensitive
our result is to these variations.
Fig. \ref{fig:C2cosmologists} shows the age distribution of terrestrial planet 
formation in the Universe shifted by $\Delta t_{obs} = 4$ Gyr. This curve, labeled ``$P_{obs}$''  
is a crude prior for 
the temporal selection effect of when independent observers can first measure $r$. 
Thus, if the evolution of biological equipment capable of doing cosmology takes about $\Delta t_{obs} \sim 4 $
Gyr, the ``$P_{obs}$'' in Fig. \ref{fig:C2cosmologists}
shows the age distribution of the first 
cosmologists on terrestrial planets able to look at the Universe and determine the overall 
energy budget, just as we have recently been able to do.

\begin{figure}[!p]
\includegraphics[width=\linewidth]{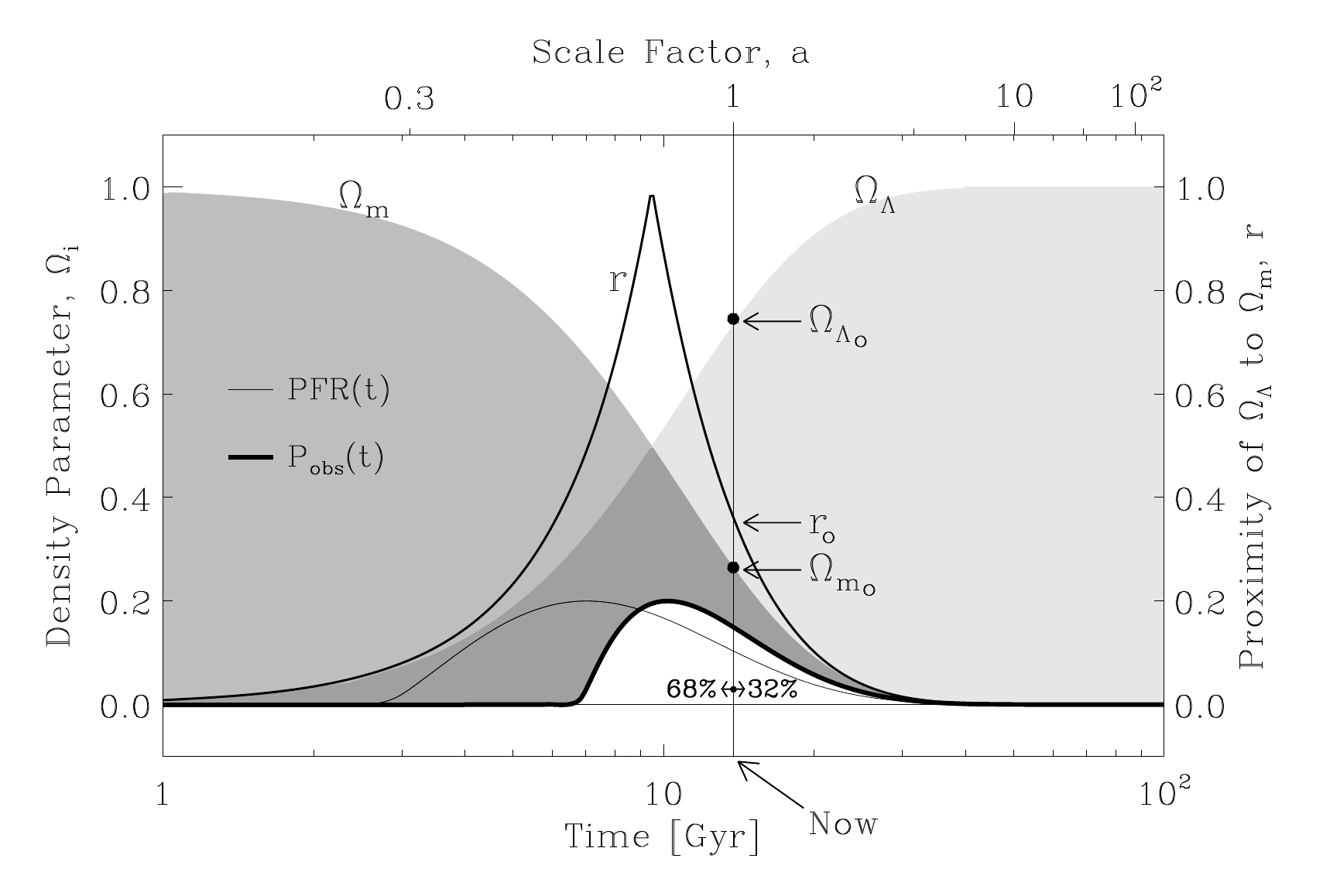}
\caption[Comparing the proximity parameter and the temporal distribution of observers.]
{Zoom-in of the portion of Fig. \ref{fig:C2omegas} between 1 and 100 billion years after
the big bang, containing the relatively narrow window of time in which $\om \sim \ol$. 
The 99 Gyr time interval displayed here is indicated in Fig. \ref{fig:C2omegas}
by the small grey rectangle above the ``Now'' label.
The proximity parameter $r(t)$ (Eq. \ref{eq:C2rdef}, Figs. \ref{fig:C2logproblem} \& \ref{fig:C2linearnoproblem})
is superimposed.
The thin solid line shows the age distribution of terrestrial planets in the Universe
while the thick solid line is the lateral displacement of this distribution by $\Delta t_{obs} = 4$ Gyr.
These distributions were presented in Fig. \ref{fig:C2cosmologists}, but here the time axis is logarithmic.
We interpret $P_{obs}$ as the frequency distribution of new observers 
able to measure $\om$ and $\ol$ for the first time.
Since $r(t)$ peaks at about the same time as $P_{obs}(t)$, large values of $r$ 
will be observed more often than small values. 
}
\label{fig:C2rollercoasterzoom}
\end{figure}

\subsection{The Probability of Observing $\om \sim \ol$.}

In Fig. \ref{fig:C2rollercoasterzoom} we zoom into the portion of Fig. 
\ref{fig:C2omegas} containing the relatively narrow window of time in which $\om \sim \ol$.
We plot $r(t)$ to show where $r \sim 1$ and we also plot the age distribution of planets and the 
age distribution of recently emerged cosmologists from Fig. \ref{fig:C2cosmologists}.
The white area under the thick $P_{obs}(t)$ curve provides an estimate of the time distribution 
of new observers in the Universe.
We interpret $P_{obs}(t)$ as the probability distribution of the times at which new, independent observers 
are able to measure $r$ for the first time. 

\citet{Lineweaver2001} estimated that the Earth is relatively young compared to other terrestrial planets
in the Universe.  It follows under the simple assumptions of our analysis that most 
terrestrial-planet-bound observers will emerge earlier than we have.
We compute the fraction $f$ of observers who have emerged earlier than we have,
\be
f = \frac {\int_{0}^{t_{o}} P_{obs}(t) \; dt}{\int_{0}^{\infty} P_{obs}(t) \; dt} \approx 68\%
\label{eq:C2older}
\ee
and find that $68\%$ emerge earlier while $32\%$ emerge later.  These numbers are indicated
in Fig. \ref{fig:C2rollercoasterzoom}.

\subsection{Converting $P_{obs}(t)$  to $P_{obs}(r)$}
\label{sec:C2convert}
We have an estimate of the distribution in time of observers, $P_{obs}(t)$, and we have
the proximity parameter $r(t)$.  We can then 
convert these to a probability $P_{obs}(r)$, of observed values of $r$. 
That is, we change variables and convert the $t-$dependent
probability to an $r-$dependent probability:  $P_{obs}(t)  \rightarrow P_{obs}(r)$.
We want the probability distribution of the $r$ values first observed by new observers in the Universe. 
Let the probability of observing $r$ in the interval $dr$ be $P_{obs}(r)dr$.  
This is equal to the probability of observing $t$ in the interval $dt$,
which is $P_{obs}(t)dt$ 

\begin{figure}[!p]
\includegraphics[width=0.8\linewidth]{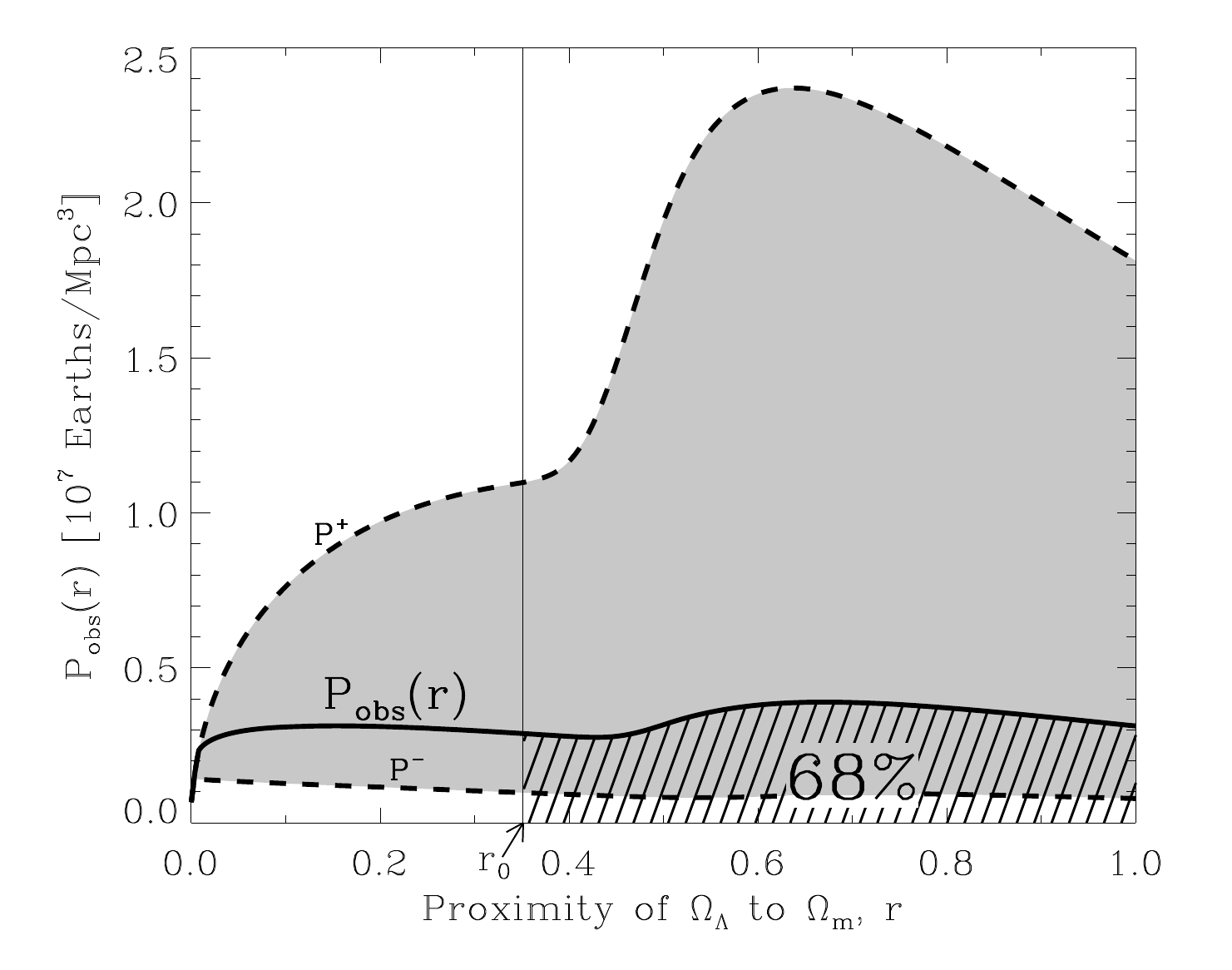}
\caption[Probability of new observers on terrestrial planets observing a given $r$.]
{Probability of new observers on terrestrial planets observing a given 
$r$ (Eq. \ref{eq:C2prob2}).  
Given our estimate of the age distribution of new cosmologists in the Universe $P_{obs}(t)$,
the probability of observing $\om$ and $\ol$ as close together as
they are, or closer, is the integral given in Eq. (\ref{eq:C2mainresult}),
shown here as the hashed area labeled $68\%$. 
The dashed lines labeled $P^{+}$ and $P^{-}$ are from replacing $P_{obs}(t)$ 
in Eq. \ref{eq:C2prob2} with the curves labeled $P^{+}$ and $P^{-}$ 
in Fig. \ref{fig:C2cosmologists}.
}
\label{fig:C2pofr_nonorm}
\end{figure}

Thus,
\be
P_{obs}(r)\;dr = P_{obs}(t)\;dt
\label{eq:C2prob}
\ee
or equivalently
\be
P_{obs}(r) = \frac{P_{obs}(t)}{dr/dt}
\label{eq:C2prob2}
\ee
where
$P_{obs}(t) = PFR(t-\Delta t_{obs})$
is the temporally shifted age distribution of terrestrial planets 
and $dr/dt$ is the slope of $r(t)$. Both are shown in Fig. \ref{fig:C2rollercoasterzoom}.
The distribution $P_{obs}(r)$ is shown in Fig. \ref{fig:C2pofr_nonorm} along with the upper and lower
confidence limits on $P_{obs}(r)$ obtained by inserting the upper and lower confidence limits
of $P_{obs}(t)$  (denoted ``$P^{+}$'' and ``$P^{-}$'' 
in Fig. \ref{fig:C2cosmologists}), into Eq. \ref{eq:C2prob2} in
place of $P_{obs}(t)$.

The probability of observing $r > r_{o}$ is,
\be
P(r > r_{o}) = \int_{r_{o}}^{1} P_{obs}(r) \;dr = \int_{t^{\prime}}^{t_{o}} P_{obs}(t) \; dt  \approx 68\%
\label{eq:C2mainresult}
\ee
where $t^{\prime}$ is the time in the past when $r$ was equal to its present value, i.e., 
$r(t^{\prime}) = r(t_{o}) = r_{o} \approx 0.4$.  We have $t^{\prime} = 6$ Gyr and $t_{o} = 13.8$ Gyr 
(see bottom panel of Fig. \ref{fig:C2linearnoproblem}).
This integral is shown graphically in Fig. \ref{fig:C2pofr_nonorm} as the hatched area
underneath the ``$P_{obs}(r)$'' curve, between $r=r_{o}$ and $r = 1$.  
We interpret this as follows: of all observers that have emerged on terrestrial planets, 
68\% will emerge when $r > r_{o}$ and thus will find $r > r_{o}$.
The $68\%$ from Eq. \ref{eq:C2older} is only the same as the $68\%$ from Eq. \ref{eq:C2mainresult}
because all observers who emerge earlier than we did, did so more recently than 7.8 billion years
ago and thus, observe $r > r_{o}$  (Fig. \ref{fig:C2rollercoasterzoom}).

We obtain estimates of the uncertainty on this $68\%$ estimate by computing analogous integrals 
underneath the curves labeled $P^{+}$ and $P^{-}$ in Fig. \ref{fig:C2pofr_nonorm}.
These yield $82\%$ and $59\%$ respectively.
Thus, under the assumptions made here, $68 ^{+14}_{-10}\%$ of the observers in the Universe will find
$\ol$ and $\om$ even closer to each other than we do.
This suggests that a temporal selection effect due to the constraints on the emergence of
observers on terrestrial planets provides a plausible solution to the cosmic coincidence problem.
If observers in our Universe evolve predominantly on Earth-like planets
(see the ``principle of mediocrity'' in \citet{Vilenkin1995a}), we should not be surprised to find
ourselves on an Earth-like planet and we should not be surprised to find $\olo \sim \omo$.

\section{How Robust is this $68\%$ Result?}
\label{sec:C2robust}

\subsection{Dependence on the timescale for the evolution of observers}
\label{sec:C2deltat}

A necessary delay, required for the biological evolution of
observing equipment -- e.g. brains, eyes, telescopes, makes the observation of 
recent biogenesis unobservable 
\citep{Lineweaver2002,Lineweaver2003a}. That is, no observer in the Universe can wake up
to observerhood and find that their planet is only a few hours old.
Thus, the timescale for the evolution of observers, $\Delta t_{obs} > 0$. 

\begin{figure}[!p]
\includegraphics[width=0.8\linewidth]{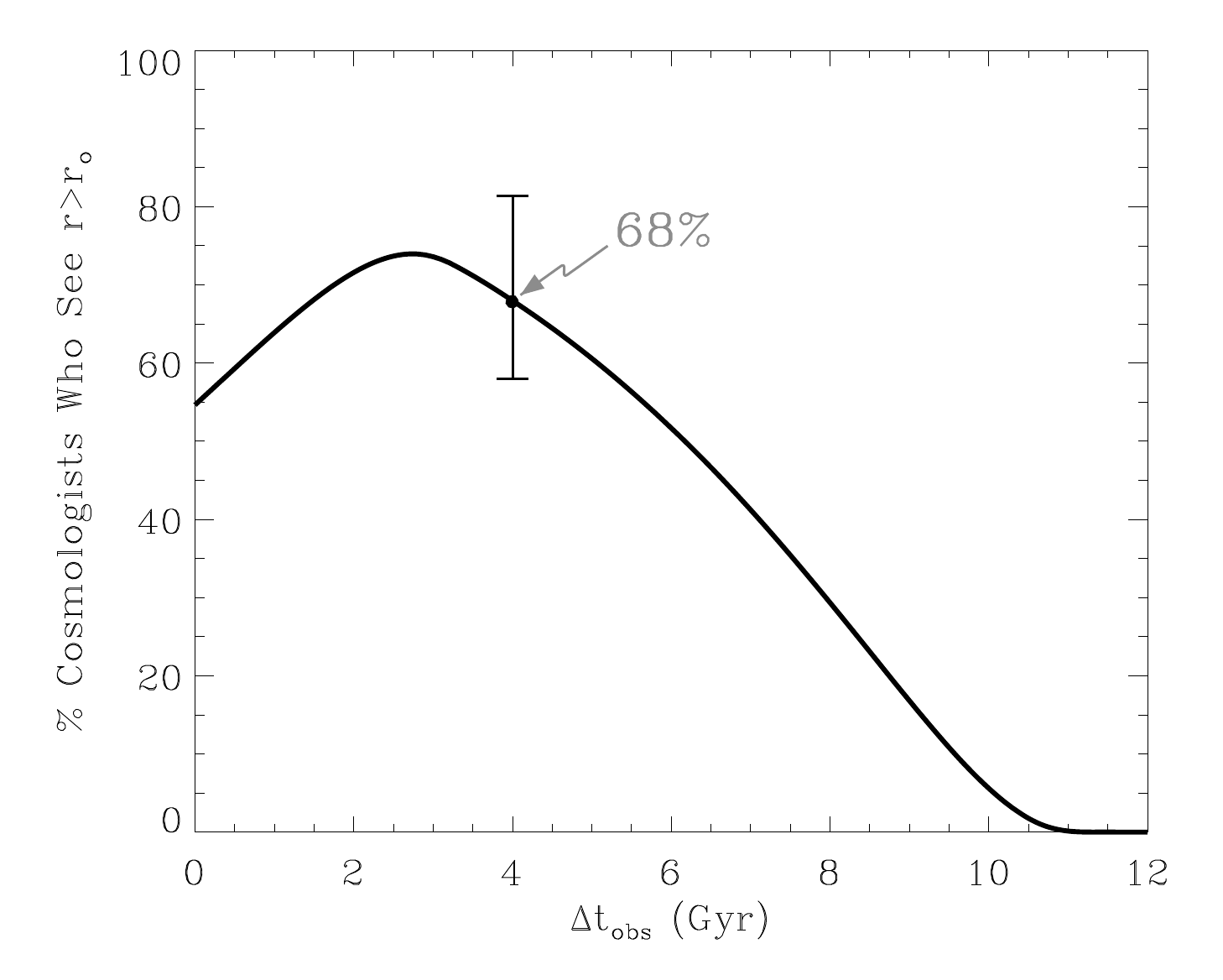}
\caption[Percentage of cosmologists who see $r>r_{o}$ as a function of $\Delta t_{obs}$.]
{Percentage of cosmologists who see $r > r_{o}$ as a function of the time  
$\Delta t_{obs}$, it takes observers
to evolve on a terrestrial planet. Since we have only vague notions about how long
it takes observers to evolve on a planet, we vary $\Delta t_{obs}$ between
0 and 12 billion years and show how the probability  $P(r>r_{o})$ 
 of observing $ r > r_{o}$ (Eq. \ref{eq:C2mainresult}) varies as a function of $\Delta t_{obs}$.
The $68 ^{+14}_{-10}\%$ point plotted is the result from 
Fig. \ref{fig:C2pofr_nonorm} where $\Delta t_{obs} = 4$ Gyr.
If $\Delta t_{obs} = 0$,  we use the thin solid line in Fig. \ref{fig:C2rollercoasterzoom}
as $P_{obs}(t)$ rather than the thick
solid line and we obtain $55\%$. 
}
\label{fig:C2varyingtev}
\end{figure}

Our $68 ^{+14}_{-10}\%$ result was calculated under the assumption that 
evolution from a new terrestrial planet to an observer takes $\Delta t_{obs} \sim 4$ Gyr.
To determine how robust our result is to variations in $\Delta t_{obs}$,
we perform the analysis of Sec. \ref{sec:C2computeprobability}
for  $0 < \Delta t_{obs} < 12$ Gyr. 
The results are shown in Fig. \ref{fig:C2varyingtev}.  Our $68 ^{+14}_{-10}\%$ result is
the data point plotted at $\Delta t_{obs} = 4$ Gyr.
If life takes $\sim 0$ Gyr to evolve to observerhood, once a terrestrial planet is in place, 
$P_{obs}(t) \approx PFR(t)$
and $55\%$ of new cosmologists would observe an $r$ value larger than the $r_{o} \approx 0.4$
that we actually observe today.
If observers typically take twice as long as we did to evolve ($\Delta t_{obs} \sim 8$ Gyr), there is still a 
large chance ($\sim 30\%$) of observing $r > r_{o}$.  
If $\Delta t_{obs} > 11$ Gyr, $P_{obs}(t)$ in Fig. \ref{fig:C2rollercoasterzoom}
peaks substantially after $r(t)$ peaks, and the percentage 
of cosmologists who see $r > r_{o}$, is close to zero (Eq. \ref{eq:C2mainresult}).
Thus, if the characteristic time it takes for life to emerge
and evolve into cosmologists is $\Delta t_{obs} \lsim 10$ Gyr,  our
analysis provides a plausible solution to the cosmic coincidence problem.  

The Sun is more massive than $ 94 \%$ of all stars.  Therefore $94\%$ of stars live longer
than the $t_{\odot}  \approx 10$ Gyr main sequence lifetime of the Sun.  
This is mildly anomalous and it is plausible
that the Sun's mass has been anthropically selected.
For example, perhaps stars as massive as the Sun are needed to provide the UV photons to 
jump start and energize the molecular evolution that leads to life.
If so, then $\sim 10$ Gyr is a rough upper limit to the amount of time a terrestrial planet with simple life has 
to produce observers.  Even if the characteristic time for life to evolve into observers is much longer
than $10 $ Gyr, as concluded by \cite{Carter1983}, this UV requirement that life-hosting stars have 
main sequence lifetimes $\lsim 10$ Gyr would lead to the extinction of most extraterrestrial life
before it can evolve into observers.  This would lead to observers waking to observerhood to find
the age of their planet to be a large fraction of the main sequence lifetime of their star;
the time they took to evolve would satisfy $\Delta t_{obs} \lsim 10 $ Gyr, and they would observe that
 $r \sim 1$ and that other observers are very rare.
Such is our situation.

If we assume that we are typical observers \citep{Vilenkin1995b,Vilenkin1995a,Vilenkin1996a,Vilenkin1996b} and that 
the coincidence problem must be resolved by an observer selection effect \citep{Bostrom2002}, 
then we can conclude that the typical time 
it takes observers to evolve on terrestrial planets is less than $10$ Gyr  ($\Delta t_{obs} < 10$ Gyr).

\subsection{Dependence on the age distribution of terrestrial planets}

The $P_{obs}(t)$ used here (Fig. \ref{fig:C2rollercoasterzoom})
is based on the star formation rate  (SFR) computed in \citet{Lineweaver2001}.
There is general agreement that the SFR has been declining since redshifts $z \sim 2$.
Current debate centers around whether that decline has only been since $z \sim 2$ or whether the SFR has been declining
from a much higher redshift (\citealt{Lanzetta2002,Hopkins2006,Nagamine2006,Thompson2006}).
Since \citet{Lineweaver2001} assumed a relatively high value for the SFR at redshifts above 2, this
led to a relatively high estimate of the metallicity of the Universe at $z \sim 2$, which corresponds
to a relatively short delay ($\sim 2$ Gyr) between the big bang and the first terrestrial planets.
For the purposes of this analysis, the early-SFR-dependent uncertainty in the $\sim 2 $ Gyr delay is degenerate with, but much
smaller than, the uncertainty of $\Delta t_{obs}$.
Thus the variations of $\Delta t_{obs}$ discussed above subsume the 
SFR-dependent uncertainty in $P_{obs}(t)$.

\begin{figure}[!p]
\includegraphics[width=\linewidth]{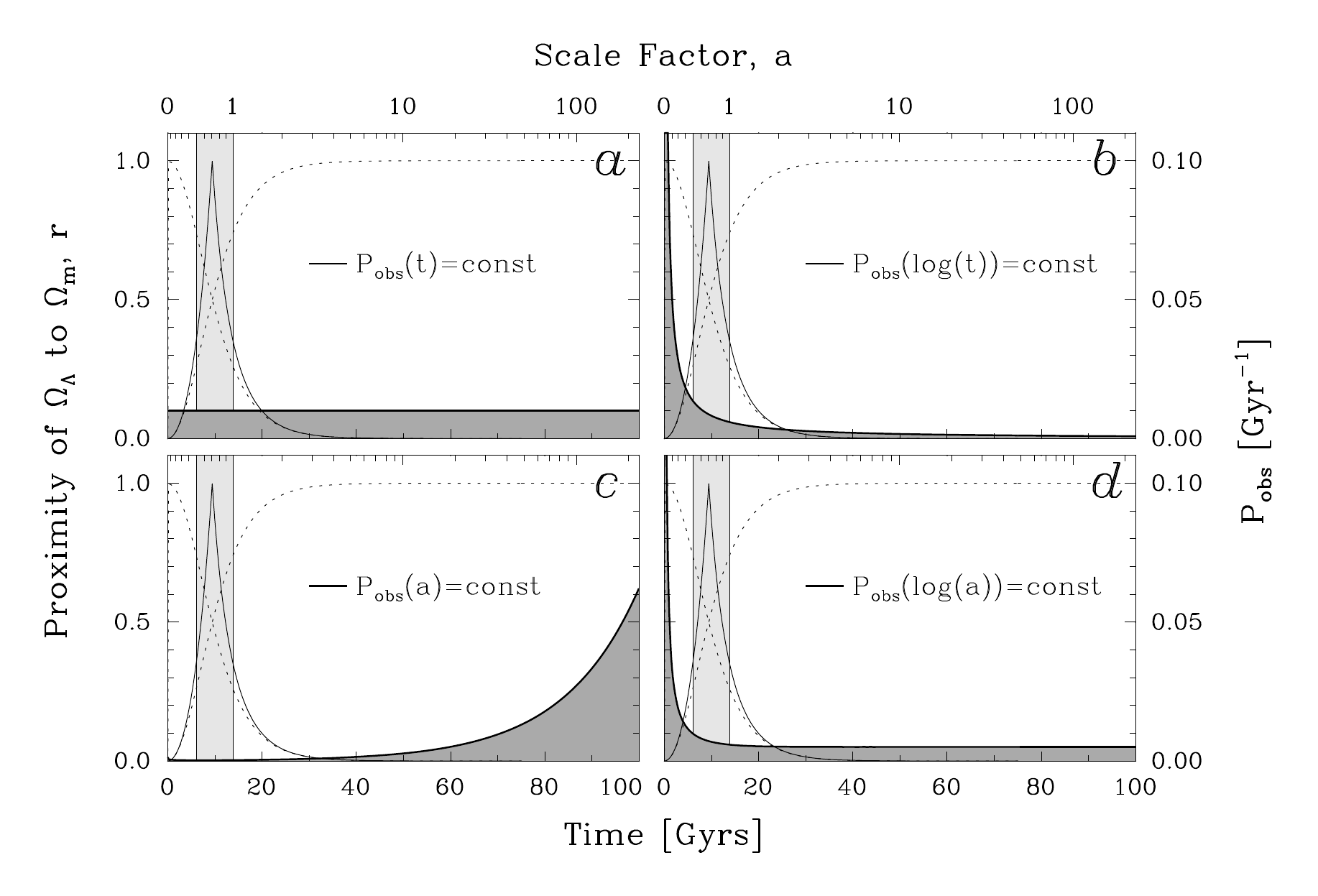} 
\caption[Dependence of the cosmic coincidence problem on measure.]{The expected
observed value of $r$ depends strongly on the assumed
distribution of observers over time $t$. This
figure demonstrates a variety of uniform observer distributions $P_{obs}$
which, if used, result in the cosmic coincidence problem that the
observed value of $r$ is unexpectedly high. 
The $P_{obs}$ that are functions
of $log(a)$ or $log(t)$ have been normalized to the interval
$t_{rec}$ to $100$ Gyr.
Panel {\it a}) is the same as the top panel of Fig. \ref{fig:C2linearnoproblem}.
The probabilities that an observer would fall within the vertical light grey band ($r > r_{o}$)
 in Panels {\it a,b,c} and {\it d} are $8\%,7\%,0.2\%$ and $6\%$
respectively, and are given in the first row of Table \ref{table:C2minmaxProbs}.
} 
\label{fig:C2naivedistrib}
\end{figure}

\subsection{Dependence on Measure}
\label{sec:C2measure}

In Figs. \ref{fig:C2logproblem}  \& \ref{fig:C2linearnoproblem}
we illustrated how the importance of the cosmic coincidence depends on the range over 
which one assumes that the observation of $r$ could have occurred.
This involved choosing the range $\Delta x$ shown on the x axis in 
Figs. \ref{fig:C2logproblem}  \& \ref{fig:C2linearnoproblem}.
We also showed how the apparent significance of the coincidence depended on how one expressed that range, i.e., 
logarithmic in Fig. \ref{fig:C2logproblem} and linear in Fig. \ref{fig:C2linearnoproblem}.  
The coincidence seems most compelling when $\Delta x$ is the largest and
the problem is presented on a logarithmic $x$ axis.  This dependence is  a specific example of a ``measure'' 
problem (\citealt{Aguirre2005,Aguirre2006}).

The measure problem is illustrated in Fig. \ref{fig:C2naivedistrib},
where we plot four different uniform distributions of observers on a linear time axis.
In Panel {\it a}) $P_{obs}(t)= $ constant. That is, we assume that observers could find themselves
anywhere between $t_{rec} = 380,000$ yr  and $100$ Gyr after the big bang, with uniform probability (dark grey).
In {\it b}), we make the different assumption that observers are distributed uniformly in log(t) over the same
range in time. This means for example that
the probability of finding yourself between 0.1 and 1 Gyr is the same as between 1 and 10 Gyr.
We plot this as a function of linear time and find that the distribution of observers  (dark grey) is highest 
towards earlier times.

To quantify and explore these dependencies further, in Table \ref{table:C2minmaxProbs}, 
we take the duration when $r > r_{o}$ (call this interval $\Delta x_{r}$)
and divide it by various larger ranges $\Delta x$ (a range of time or scale factor).
Thus, when the probability  $P ( r > r_{o})  = \frac{\Delta x_{r}}{\Delta x}$
is $<< 1$, there is a low probability that one would find oneself in the interval $\Delta x_{r}$ and
the cosmic coincidence is compelling.  However,  when $P( r > r_{o}) \sim 1$ 
the coincidence is not significant.

In the four panels {\it a,b,c} and {\it d} of Fig.  \ref{fig:C2naivedistrib} the probability of us
observing $r \ge r_{o}$ (finding ourselves in the light grey area) is respectively 
$8\%, 7\%, 0.2\%$ and $6\%$.   These values are given in the first row of  
Table \ref{table:C2minmaxProbs} along with analogous values when 
11 other ranges for $\Delta x$ are considered.
Probabilities corresponding to the four panels of Figs. \ref{fig:C2logproblem}  \& \ref{fig:C2linearnoproblem}
are shown in bold in Table \ref{table:C2minmaxProbs}.
Our conclusion is that this simple ratio method of measuring the significance of a coincidence yields results 
that can vary by many orders of magnitude depending on the range ($\Delta x$) and 
measure (e.g. linear or logarithmic) chosen.
The use of the non-uniform $P_{obs}(t)$ shown in Fig. \ref{fig:C2cosmologists}
is not subject to these ambiguities in the choice of range and measure.

\section{Discussion \&  Summary}
\label{sec:C2discussion}

Anthropic arguments to resolve 
the coincidence problem include \citet{Garriga2000} and \citet{Bludman2001}. 
Both use a semi-analytical formalism (\citealt{Gunn1972,Press1974,Martel1998}) to compute 
the number density of objects that collapse into large galaxies.
This is then used as a measure of the number density of intelligent observers. 
Our work complements these semi-analytic models by using observations of the star formation rate
to constrain the possible times of observation. Our work also extends this previous work by including the 
effect of $\Delta t_{obs}$, the time it takes observers to 
evolve on terrestrial planets. This inclusion puts an important limit on the validity of anthropic solutions
to the coincidence problem. 

\cite{Garriga2000} is probably the work most similar to ours.  They take $\rhol$ 
as a random variable in a multiverse model with a prior probability distribution.  
For a wide range of $\rhol$ (prescribed by a prior based on 
inflation theory) they find approximate equality between the time of galaxy formation $t_{G}$,
the time when $\Lambda$ starts to dominate the energy density of the Universe $t_{\Lambda}$ and now $t_{o}$.
That is, they find that, within one order of magnitude, $t_{G} \sim t_{\Lambda}  \sim t_{o}$. 
Their analysis is more generic but approximate in that it 
addresses the coincidence for a variety of values of $\rhol$  to an order of magnitude precision.
Our analysis is more specific and empirical in that we condition on our Universe and use the \cite{Lineweaver2001}
star-formation-rate-based estimate of the age distribution of terrestrial planets to reach our main result ($68 \%$).

To compare our result to that of \cite{Garriga2000}, we limit their analysis to the $\rhol$ observed in 
our Universe ($\rhol = 6.7 \times 10^{-30} g/cm^{3}$)
and differentiate their cumulative number of galaxies which have assembled up to a given time (their Eq. 9). We
find a broad time-dependent distribution for galaxy formation which is the analog of our more empirical and narrower
(by a factor of 2 or 3) $P_{obs}(t)$.

We have made the most specific anthropic explanation of the cosmic coincidence using the age distribution of
terrestrial planets in our Universe and found this explanation fairly robust to the
largely uncertain time it takes observers to evolve.
Our main result is an understanding of the cosmic coincidence as a
temporal selection effect if observers emerge preferentially
on terrestrial planets in a characteristic time $\Delta t_{obs} < 10 $ Gyr.
Under these plausible conditions,
we, and any observers in the Universe who have evolved on terrestrial planets, 
should not be surprised to find $\omo \sim \olo$.

{\bf Acknowledgements}
We would like to thank Paul Francis and Charles Jenkins for helpful discussions. CE acknowledges a UNSW 
School of Physics post graduate fellowship.\\

\section*{Appendix A: Evolution of Densities}

Recent cosmological observations have led to the new standard $\Lambda$CDM model in which
the density parameters of radiation, matter and vacuum energy are currently observed to be 
$\orado \approx 4.9 \pm 0.5 \times 10^{-5}$, $\omo \approx 0.26 \pm  0.03$ 
and $\olo \approx 0.74 \pm 0.03$ respectively and
Hubble's constant is $H_{o} = 71 \pm 3 \; km s^{-1} Mpc^{-1}$
\citep{Spergel2006,Seljak2006}.

The energy densities in relativistic particles 
(``radiation'' i.e., photons, neutrinos, hot dark matter), 
non-relativistic particles (``matter'' i.e., baryons,cold dark matter) 
and in vacuum energy scale differently \citep{Peacock1999},

\be
\rho_{i} \propto a^{-3(w_{i}+1)}.
\ee 

Where the different equations of state are, $\rho_{i} = w_{i}\; p$ where $w_{radiation} = 1/3$, $w_{matter} = 0$ 
and $w_{\Lambda} = -1$ \citep{Linder1997}.
That is, as the Universe expands, these different forms of energy density dilute at different rates.

\bea
\rhor & \propto & a^{-4} \\
\rhom & \propto & a^{-3} \\
\rhol & \propto & a^{0} 
\eea

Given the currently observed values for $\orad$, $\om$ and $\ol$, the Friedmann 
equation for a standard flat cosmology tells us the evolution of the scale factor 
of the Universe, and the history of the energy densities:
\bea \label{eq:C2friedmann}
\left( \frac{\dot{a}}{a} \right)^{2} &=& \frac{8 \pi G}{3} (\rhor + \rhom + \rhol)\\
                                     &=& \frac{8 \pi G}{3}(\rhoro  a^{-4}+ \rhomo a^{-3} + \rhol  a^{0})\\ 
                                     &=& ( \orado a^{-4}+ \omo a^{-3} + \olo  a^{0}) 
\eea
where we have $\rho_{crit} = \frac{3 H(t)^{2}}{8\pi G}$ and $\Omega_{i} = \frac{\rho_{i}}{\rho_{crit}}$.
The upper panel of Fig. \ref{fig:C2omegas} illustrates these different
dependencies on scale factor and time
in terms of densities while the lower panel shows the corresponding normalized density
parameters. 
A false vacuum energy $\rho_{\Lambda_{inf}}$ is assumed between the Planck scale and the GUT scale. 
In constructing this density plot and setting a value for $\Omega_{\Lambda_{inf}}$ we have used the 
constraint that at the GUT scale, all the energy densities add up to 
$\rho_{\Lambda_{inf}}$ which remains constant at earlier times.

\section*{Appendix B: Tables}

\begin{table}[!hp]
\begin{center}
\scriptsize
\begin{tabular}{l l|l l}
\hline
Event						&   Symbol              &      \multicolumn{2}{c}{Time after Big Bang}  \\ 
							&                       &     seconds                 &   Gyr                \\
\hline
\hline
Planck time, beginning of time                                &  $\tp$                &  $5.4 \times 10^{-44}$   &$1.7 \times 10^{-60}$    \\ 
end of inflation, reheating, origin of matter, thermalization &$t_{reheat}$           &  $[10^{-43},10^{-33}]$   &$[10^{-60}, 10^{-50}]$   \\ 
energy scale of Grand Unification Theories (GUT)              &$t_{GUT}$              &  $10^{-33}$              &$10^{-50}$               \\
matter-anti-matter annihilation, baryogenesis                 & $t_{baryogenesis}$    &  $[10^{-33}, 10^{-12}]$  &$[10^{-50},10^{-29}]$    \\    
electromagnetic and weak nuclear forces diverge               & $t_{electroweak}$     &  $10^{-12}$              &$10^{-29}$               \\
light atomic nuclei produced                                  &  $t_{BBN}$            &  $[100, 300]$            &$[3, 9] \times 10^{-15}$ \\
 radiation-matter equality$^1$                                &  $t_{r-m}$            &  $8.9 \times 10^{11}$    &$2.8 \times 10^{-5}$     \\
 recombination$^1$ (first chemistry)                          & $t_{rec}$            &  $1.2 \times 10^{13}$    &$0.38 \times 10^{-3}$    \\
first thermal disequilibrium                                  & $t_{1st therm-dis}$  &  $1.2 \times 10^{13}$    &$0.38 \times 10^{-3}$    \\ 
 first stars, Pop III, reionization$^1$                       & $t_{1st stars}$      &  $1 \times 10^{16}$      &$0.4$                    \\
 first terrestrial planets$^2$                                & $t_{1st Earths}$     &  $8 \times 10^{16}$      &$2.5$                    \\ 
last time $r$ had same value as today                         & $t_{rnow}$           &  $1.9 \times 10^{17}$    &$6.1$                    \\ 
formation of the Sun, Earth$^3$                               & $t_{Sun}$,$t_{Earth}$ &  $2.9 \times 10^{17}$    &$9.1$                    \\
 matter-$\Lambda$ equality$^1$                                & $t_{m-\Lambda}$     &  $3.0 \times 10^{17}$    &$9.4$                    \\
now                                                           & $t_{o}$               &  $4.4 \times 10^{17}$    &$13.8$                   \\
 last stars die$^4$                                           & $t_{last stars}$      &  $10^{22}$               &$10^{6}$                 \\
protons decay$^4$                                             & $t_{proton decay}$    &  $10^{45}$               &$10^{29}$                \\ 
 super massive black holes consume matter$^4$                 & $t_{black holes}$    &  $10^{107}$              &$10^{91}$                \\
 maximum entropy (no gradients to drive life)$^4$             &$t_{heat death}$       &  $10^{207}$              &$10^{191}$               \\
\hline
\end{tabular}
\end{center}
\caption[Important times in the history of the Universe.]{Important Times in the History of the Universe. References: \\
(1) \citealt{Spergel2006}, http://map.gsfc.nasa.gov/ \\
(2) \citealt{Lineweaver2001} \\
(3) \citealt{Allegre1995} \\
(4) \citealt{Adams1997}
}
\label{table:C2importantimes}
\end{table}

\begin{table}[!p]
\begin{center}
\small
\begin{tabular}{l l|l l l l}
\hline
\multicolumn{2}{c}{Range  $\Delta x \;^a$} \vline       & \multicolumn{4}{c}{  $P ( r > r_{o})$ [\%]     } \\
$x_{min}$		& $x_{max}$		& $t$			& $\log(t)$			& $a$ 		& $\log(a)$ \\
\hline \hline
%
$t_{rec}$          & $100$ Gyr $^b$   &${\bf 8}\;^c$        & $7$                    & $0.2$                & $6$ \\
$\tp$              & $t_{last stars}$ &$8 \times 10^{-4}$   & ${\bf 0.6}\; ^d$       & $10^{-10^{4}}$        & $10^{-3}$ \\
$\tp$              & $t_{o}$          & ${\bf 60}\;^c$      & $0.6 $                 & $50$                  & $1$ \\ 
$\tp$              & ${\tp^{-1}}$     & $30$                & $0.3$                  & $30$                  & ${\bf 0.5}\;^d$ \\
$\tp$              & $t_{heat death}$ & $10^{-188}$         & $0.1$                  & $10^{-10^{189}}$      & $10^{-188}$ \\ 
$t_{rec}$          & $t_{protondecay}$& $10^{-26}$         & $1$                     & $10^{-10^{27}}$       & $10^{-26}$ \\ 
$t_{rec}$          &$t_{black holes}$ & $10^{-88}$         & $0.4$                   & $10^{-10^{89}}$       & $10^{-88}$ \\  
$t_{rec}$          & $t_{heat death}$ & $10^{-188}$        &$0.2$                    & $10^{-10^{189}}$      & $10^{-188}$ \\  
$t_{1st stars}$    & $t_{last stars}$ &$8 \times 10^{-4}$  &$6$                      & $10^{-10^{4}}$        & $10^{-3}$ \\
$t_{1st stars}$    & $t_{protondecay}$& $10^{-26}$         & $1$                     & $10^{-10^{27}}$       & $10^{-26}$ \\ 
$t_{1st stars}$    & $t_{black holes}$& $10^{-88}$         & $0.4$                   & $10^{-10^{89}}$       & $10^{-88}$ \\  
$t_{1st stars}$    & $t_{heat death}$ & $10^{-188}$        & $0.2$                   & $10^{-10^{189}}$      & $10^{-188}$ \\
\hline
\end{tabular} 
\end{center}
\caption[The probability $P(r>r_{o})$ assuming even distributions in time, Log(time), scalefactor and Log(scalefactor).]{
The probability $P(r > r_{o})$ of observing $r > r_{o}$ assuming a uniform distribution of observers $P_{obs}$
in linear time, log(time), scale factor and log(scale factor) within the range $\Delta x$ listed. \\
$^a$ See Table 1 for the times corresponding to columns  1 and 2.\\
$^b$ The four values in the top row correspond to Fig. \ref{fig:C2naivedistrib}.\\
$^c$ The two values shown in bold in the $t$ column correspond to the two panels of Fig. \ref{fig:C2linearnoproblem}.\\
$^d$ These values correspond  to the two panels of Fig.\ref{fig:C2logproblem}.
}
\label{table:C2minmaxProbs}
\end{table}

%
\pagestyle{fancy}
\chapter{Dark Energy Dynamics Required to Solve the Cosmic Coincidence}
\label{chap3}

\begin{center}
\emph{
Tonight I have a date on Mars. \\
Tonight I'm gonna get real far. \\
I'll be leaving Earth behind me.
}
\end{center}
\begin{flushright}
- Encounter, ``Date on Mars'' \hspace*{2cm}
\end{flushright}
\vspace{1cm}

\section{Introduction}

In 1998, using supernovae Ia as standard candles, \citet{Riess1998} and 
\citet{Perlmutter1999} revealed a recent and continuing epoch of cosmic acceleration - strong 
evidence that Einstein's cosmological constant $\Lambda$, or something else with comparable negative 
pressure $p_{de} \sim -\rho_{de}$, currently dominates the energy density of the universe 
\citep{Lineweaver1998}. $\Lambda$ is usually interpreted as the energy of zero-point quantum 
fluctuations in the vacuum \citep{ZelDovich1967,Durrer2007} with a constant equation of state 
$w \equiv p_{de}/\rho_{de} = -1$.
This necessary additional energy component, construed as $\Lambda$ or otherwise, has become 
generically known as ``dark energy'' (DE).

A plethora of observations have been used to constrain the free parameters of the
new standard cosmological model, \lcdm, in which $\Lambda$ \emph{does} play the role of the dark energy.
Hinshaw et al.\ \cite{Hinshaw2006} find that the universe is expanding at a rate of $H_0 = 71 \pm 4\ km/s/Mpc$; 
that it is spatially flat and therefore critically dense 
($\Omega_{tot 0} = \frac{\rho_{tot 0}}{\rho_{crit0}} = \frac{8 \pi G}{3 H_0^2} \rho_{tot 0} = 1.01 \pm 0.01$); 
and that the total density is comprised of contributions from 
vacuum energy ($\Omega_{\Lambda 0} = 0.74 \pm 0.02$), 
cold dark matter (CDM; $\Omega_{CDM 0} = 0.22 \pm 0.02$),
baryonic matter ($\Omega_{b 0} = 0.044 \pm 0.003$) and
radiation ($\Omega_{r 0} = 4.5 \pm 0.2 \times 10^{-5}$).
Henceforth we will assume that the universe is flat ($\Omega_{tot 0} = 1$) as predicted 
by inflation and supported by observations.

Two problems have been influential in moulding ideas about dark energy,
specifically in driving interest in alternatives to \lcdm.
The first of these problems is concerned with the smallness of the dark energy
density \citep{ZelDovich1967,Weinberg1989,Cohn1998}.  
Despite representing more than $70\%$ of the total energy of the universe, the 
current dark energy density is $\sim 120$ orders of magnitude smaller than 
energy scales at the end of inflation (or $\sim 80$ orders of magnitude smaller than 
energy scales at the end of inflation if this occurred at the GUT rather than Planck 
scale) \citep{Weinberg1989}. Dark energy 
candidates are thus challenged to explain why the observed DE density is so small. 
The standard idea, that the dark energy is the energy of zero-point quantum 
fluctuations in the true vacuum, seems to offer no solution to this problem.

The second cosmological constant problem \cite{Weinberg2000b,Carroll2001b,Steinhardt2003} 
is concerned with the near coincidence between the current cosmological matter density 
($\rho_{m0} \approx 0.26 \times \rho_{crit0}$) and the dark energy density 
($\rho_{de0} \approx 0.74 \times \rho_{crit0}$). In the standard \lcdm\ model, the 
cosmological window during which these components have comparable density is 
short (just $1.5$ e-folds of the cosmological scalefactor $a$)
since matter density dilutes as $\rho_{m} \propto a^{-3}$ while vacuum density 
$\rho_{de}$ is constant \citep{Lineweaver2007}.
Thus, even if one explains why the DE density is much less than the Planck density (the smallness 
problem) one must explain why we happen to live during the time when $\rho_{de} \sim \rho_m$.

The likelihood of this coincidence depends on the range of times during which we suppose we might 
have lived.
In works addressing the smallness problem, \citet{Weinberg1987,Weinberg1989,Weinberg2000} 
considered a multiverse consisting of a large number of big bangs, each with a different value 
of $\rho_{de}$. There he asked, suppose that we could have arisen in any one of these universes; 
What value of $\rho_{de}$ should we expect our universe to have? 
While Weinberg supposed we could have arisen in another universe, we are simply supposing that 
we could have arisen in another time. We ask, what time $t_{obs}$, and corresponding densities 
$\rho_{de}(t_{obs})$ and $\rho_{m}(t_{obs})$ should we expect to observe?
Weinberg's key realization was that not every universe was equally probable: those with smaller 
$\rho_{de}$ contain more Milky-Way-like galaxies and are therefore more hospitable 
\citep{Weinberg1987,Weinberg1989}. Subsequently, he, and other authors used the relative 
number of Milky-Way-like galaxies to estimate the distribution of observers as a function of 
$\rho_{de}$, and determined that our value of $\rho_{de}$ was indeed likely 
\citep{Efstathiou1995,Martel1998,Pogosian2007}. 
Our value of $\rho_{de}$ could have been found to 
be unlikely and this would have ruled out the type of multiverse being considered. 
Here we apply the same reasoning to the cosmic coincidence problem. Our observerhood 
could not have happened at any time with equal probability \citep{Lineweaver2007}. By 
estimating the temporal distribution of observers we can determine whether the observation of 
$\rho_{de} \sim \rho_{m}$ was likely. If we find $\rho_{de} \sim \rho_{m}$ to be unlikely while 
considering a particular DE model, that will enable us to rule out that DE model.

In a previous paper \citep{Lineweaver2007}, we tested \lcdm\ in this way and found that 
$\rho_{de} \sim \rho_{m}$ is expected. In the present paper we apply this test to dynamic dark 
energy models to see what dynamics is required to solve the coincidence problem when the temporal
distribution of observers is being considered.


The smallness of the dark energy density has been anthropically explained in multiverse models
with the argument that in universes with much larger DE components, DE driven acceleration starts 
earlier and precludes the formation of galaxies and large scale structure. Such universes are 
probably devoid of observers \citep{Weinberg1987,Martel1998,Pogosian2007}.
A solution to the coincidence problem in this scenario was outlined by \citet{Garriga1999} who showed 
that if $\rho_{de}$ is low enough to allow galaxies to form, then observers in those galaxies will 
observe $r \sim 1$.

\begin{figure}[!p]
       \begin{center}
               \includegraphics[width=0.7\linewidth]{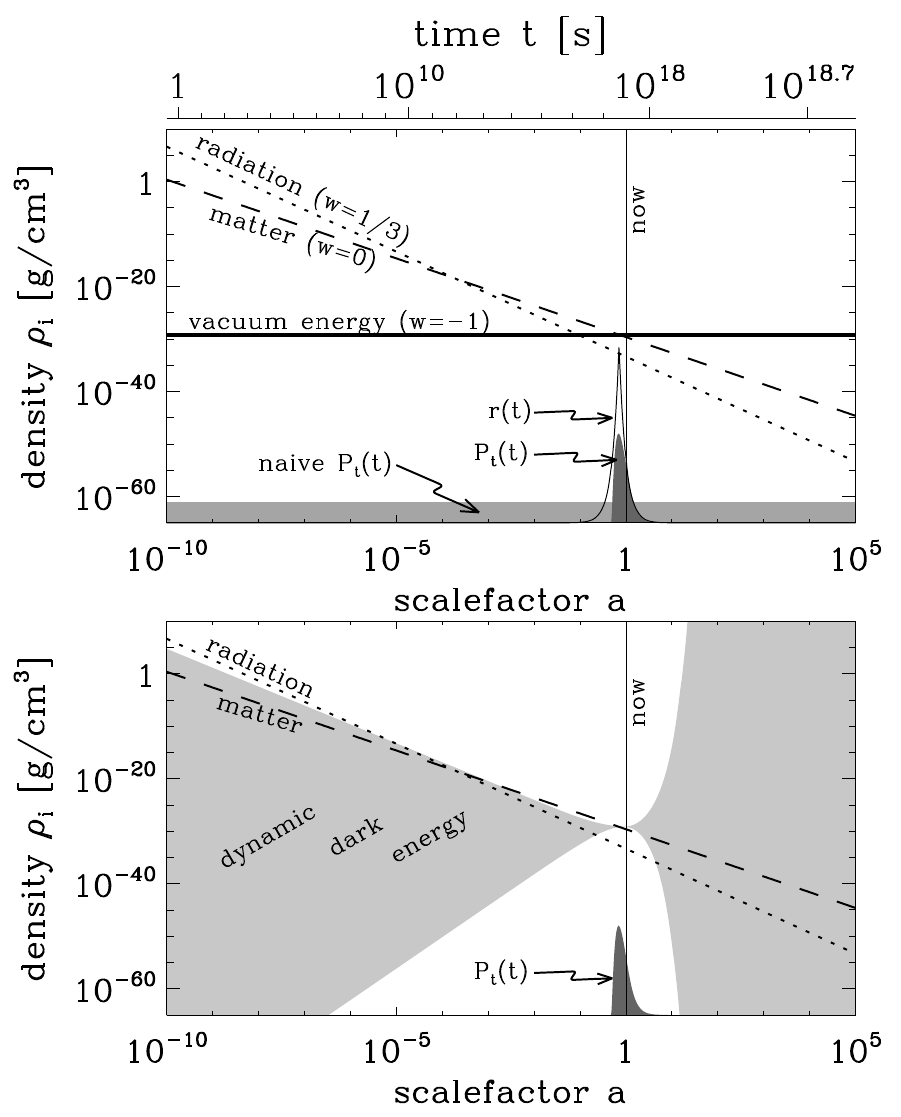}
               \caption[The coincidence problem and dynamic dark energy.]
               {{\bf(Top)} The history of the energy density of the 
               universe according to standard \lcdm. The dotted line shows the
               energy density in radiation (photons, neutrinos and other relativistic
               modes). The radiation density dilutes as $a^{-4}$ as the universe
               expands. The dashed line shows the density in ordinary non-relativistic
               matter, which dilutes as $a^{-3}$. The thick solid line shows the energy 
               of the vacuum (the cosmological constant) which has remained constant since 
               the end of inflation. The thin solid peaked curve shows the proximity $r$ of 
               the matter density to the vacuum energy density (see Eq. \ref{eq:C3definer}).
               The proximity $r$ is only $\sim 1$ for a brief period in the $log(a)$ 
               history of the cosmos. Whether or not there is a coincidence problem
               depends on the distribution $P_t(t)$ for $t_{obs}$. If one naively
               assumes that we could have observed any epoch with equal probability (the 
               light grey shade) then we should not expect to observe $r$ as large as
               we do. If, however, $P_t(t)$ is based on an estimate of the temporal 
               distribution of observers (the dark grey shade) then $r_0 \approx 0.35$ is not 
               surprising, and the coincidence problem is solved under \lcdm\ \citep{Lineweaver2007}. 
               {\bf(Bottom)} The dark energy density history is modified in DDE models.
               Observational constraints on the dark energy density history are 
               represented by the light grey shade (details in Section \ref{C3constraints}).}
               \label{fig:C3illustrated_coincidence}
       \end{center}
\end{figure}

To quantify the time-dependent proximity of $\rho_m$ and $\rho_{de}$, we define a 
proximity parameter,
	\begin{equation} \label{eq:C3definer}
		r \equiv \min \left[ \frac{\rho_{de}}{\rho_m}, \frac{\rho_m}{\rho_{de}} \right],
	\end{equation}
which ranges from $r \approx 0$, when many orders of magnitude separate the two densities, 
to $r = 1$, when the two densities are equal. 
The presently observed value of this parameter is $r_0 = \frac{\rho_{m 0}}{\rho_{de 0}} \approx 0.35$. 
In terms of $r$, the coincidence problem is as follows. If we naively presume that the time of our 
observation $t_{obs}$ has been drawn from a distribution of times $P_t(t)$ spanning many decades of 
cosmic scalefactor, we find that the expected proximity parameter is $r \approx 0 \ll 0.35$. 
In the top panel of Fig.\ \ref{fig:C3illustrated_coincidence} we use a naive distribution for 
$t_{obs}$ that is constant in $\log(a)$ to illustrate how observing $r$ as large as $r_0 \approx 0.35$ 
seems unexpected.

In \citet{Lineweaver2007} we showed how the apparent severity 
of the coincidence problem strongly depends upon the distribution $P_t(t)$ from which $t_{obs}$ is 
hypothesized to have been drawn. Naive priors for $t_{obs}$, such as the one illustrated 
in the top panel of Fig.\ \ref{fig:C3illustrated_coincidence}, lead to naive conclusions.
Following the reasoning of \citet{Weinberg1987,Weinberg1989,Weinberg2000} we interpret $P_t(t)$ as 
the temporal distribution of observers. 
The temporal and 
spatial distribution of observers has been estimated using large ($10^{11} M_{\sol}$) galaxies 
\citep{Weinberg1987,Efstathiou1995,Martel1998,Garriga1999} and terrestrial planets 
\citep{Lineweaver2007} as tracers. The top panel of Fig.\ \ref{fig:C3illustrated_coincidence} shows 
the temporal distribution of observers $P_t(t)$ from \citet{Lineweaver2007}.

A possible extension of the concordance cosmological model that may explain the observed 
smallness of $\rho_{de}$ is the generalization of dark energy candidates to include dynamic 
dark energy (DDE) models such 
as quintessence, phantom dark energy, k-essence and Chaplygin gas. In these models the dark 
energy is treated as a new matter field which is approximately homogenous, and evolves as the 
universe expands. DDE evolution offers a mechanism for the decay of $\rho_{de}(t)$ from the expected
Planck scales ($10^{93}$ g/cm$^3$) in the early universe ($10^{-44}$ s) to the small value we 
observe today ($10^{-30}$ g/cm$^3$). The light 
grey shade in the bottom panel of Fig.\ \ref{fig:C3illustrated_coincidence} represents contemporary 
observational constraints on the DDE density history.
Many DDE models are designed to solve the coincidence problem 
by having $\rho_{de}(t) \sim \rho_{m}(t)$ for a large fraction of the history/future of the universe
\citep{Amendola2000,Dodelson2000,Sahni2000,Chimento2000,Zimdahl2001,Sahni2002,Chimento2003,Ahmed2004,
Franca2004,Mbonye2004,delCampo2005,Guo2005,Olivares2005,Pavon2005,Scherrer2005,Zhang2005,delCampo2006,
Franca2006,Feng2006,Nojiri2006,Amendola2006,Amendola2007,Olivares2007,Sassi2007}.
With $\rho_{de} \sim \rho_{m}$ for extended or repeated periods the hope is to ensure that $r \sim 1$ 
is expected.

Our main goal in this paper is to take into account the temporal distribution of observers to 
determine when, and for how long, a DDE model must have $\rho_{de} \sim \rho_{m}$ in order to solve 
the coincidence problem? Specifically, we extend the work of \citet{Lineweaver2007} to find out for
which cosmologies (in addition to \lcdm) the coincidence problem is solved when the temporal
distribution of observers is considered.
In doing this we answer the question, Does a dark energy model fitting contemporary 
constraints on the density $\rho_{de}$ and the equation of state parameters, necessarily solve 
the cosmic coincidence?
Both positive and negative answers have interesting consequences. An answer in the affirmative 
will simplify considerations that go into DDE modeling: any DDE model in agreement with current cosmological 
constraints has $\rho_{de} \sim \rho_m$ for a significant fraction of observers.
An answer in the negative would yield a new 
opportunity to constrain the DE equation of state parameters \emph{more strongly} than contemporary 
cosmological surveys. 

A different coincidence problem arises when the time of observation is conditioned on 
and the parameters of a model are allowed to slide. The tuning of parameters and 
the necessity to include ad-hoc physics are large problems for many current dark energy models.
This paper does not address such issues, and the interested reader is referred to \citet{Hebecker2001},
\citet{Bludman2004} and \citet{Linder2006}. In the coincidence problem addressed here we let the time 
of observation vary to see if $r(t_{obs}) \ge 0.35$ is unlikely according to the model.

In Section \ref{C3solving} we present several examples of DDE models
used to solve the coincidence problem. An overview of observational constraints on 
DDE is given in Section \ref{C3constraints}. 
In Section \ref{C3observers} we estimate the temporal distribution of observers. 
Our main analysis is presented in Section \ref{C3results}. Our main result - that the coincidence
problem is solved for all DDE models fitting observational constraints - is
illustrated in Fig.\ \ref{fig:C3severities}. Finally, in Section \ref{C3conclusion}, we end with a 
discussion of our results, their implications and potential caveats. 


\section{Dynamic Dark Energy Models in the Face of the Cosmic Coincidence} \label{C3solving}

\begin{figure*}[!p]
       \begin{center}
               \includegraphics[width=0.7\linewidth]{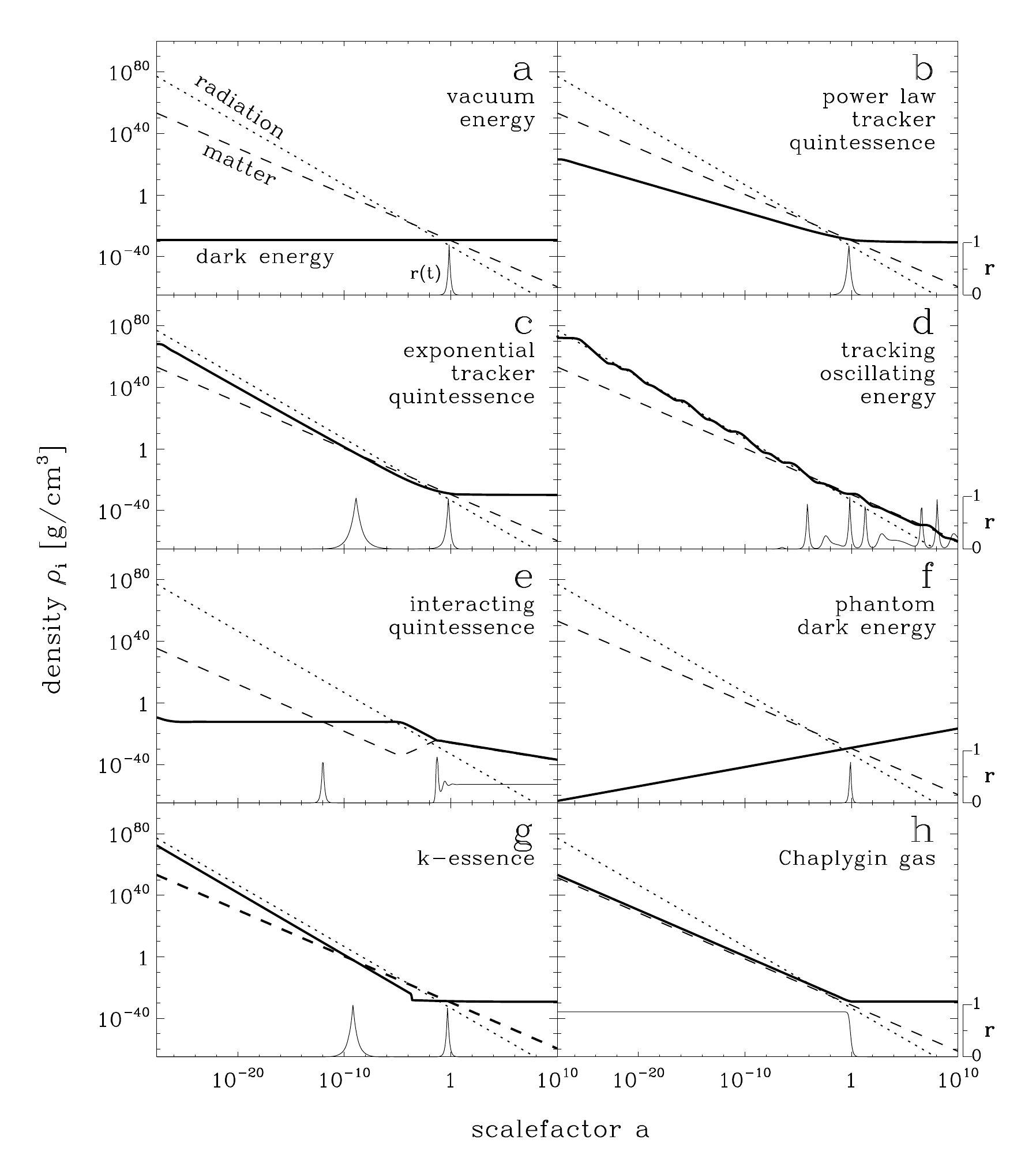}
               \caption[Families of dark energy models.]
               			{The energy density history of the universe according to \lcdm\ (panel a), and seven
				DDE models selected from the literature (see text for references). In each panel the
				radiation and matter densities are the dotted and dashed lines respectively. The DE
				density is given by the thick black line. The
				proximity parameter $r$ is given by the thin black line at the base of each panel.
				Of the DDE models shown here, tracker quintessence and k-essence (panels b, c and g) have $r \sim 1$
				for a small fraction of the life of the universe (whether the abscissa is $t$, $\log(t)$, $a$,
				$\log(a)$, or any other of a large number of measures).
				On the other hand, tracking oscillating energy, interacting quintessence, phantom DE 
				and Chaplygin gas (panels d, e, f and h) exhibit $r \sim 1$
				for a large fraction of the life of the universe. 
				For the phantom DE example (panel f) this is true 
				in $t$, but not 
				in $a$ or $\log(a)$.
				In phantom models the future universe grows super-exponentially to $a=\infty$ (a ``big-rip'') 
				shortly after matter-DE equality. Thus the universe spends a large fraction of \emph{time} 
				with $r \sim 1$, however this is is not seen in $\log(a)$-space. 
				For each of the models in this figure, numerical values for free parameters were chosen
				to crudely fit observational constraints and are given in Appendix \ref{C3paramvals}.}
               \label{fig:C3quintessence_energy}
       \end{center}
\end{figure*}

Though it is beyond the scope of this article to provide a complete review of DDE (see 
\citet{Copeland2006,Szydlowski2006}), here we give a few representative examples in 
order to set the context and motivation of our work.
Fig.\ \ref{fig:C3quintessence_energy} illustrates density histories typical of tracker quintessence, 
tracking oscillating energy, interacting quintessence, phantom dark energy, k-essence, and Chaplygin 
gas. They are discussed in turn below.

\subsection{Quintessence}

In quintessence models the dark energy is interpreted as a homogenous scalar field with 
Lagrangian density ${\mathcal L}(\phi,X) = \frac{1}{2} \dot{\phi}^2 - V(\phi)$
\citep{Ozer1987,Ratra1988,Ferreira1998,Caldwell1998,Steinhardt1999,Zlatev1999,Dalal2001}.
The evolution of the quintessence field and of the cosmos depends on the postulated
potential $V(\phi)$ of the field and on any postulated interactions. In general, quintessence 
has a time-varying equation of state 
$w = \frac{p_{de}}{\rho_{de}} = \frac{\dot{\phi}^2 / 2 - V(\phi)}{\dot{\phi}^2 / 2 + V(\phi)}$.
Since the kinetic term $\dot{\phi}^2 / 2$ cannot be negative, the equation of state is restricted 
to values $w \ge -1$. Moreover, if the potential $V(\phi)$ is non-negative then $w$ is also 
restricted to values $w \le +1$.

If the quintessence field only interacts gravitationally then energy density evolves as 
$\frac{\delta \rho_{de}}{\rho_{de}} = -3(w+1) \frac{\delta a}{a}$ and the restrictions 
$-1 \le w \le +1$ mean $\rho_{de}$ decays (but never faster than $a^{-6}$) or remains constant 
(but never increases). 

\subsubsection{Tracker Quintessence}

Particular choices for $V(\phi)$ lead to interesting attractor solutions which can be exploited 
to make $\rho_{de}$ scale (``track'') sub-dominantly with $\rho_r + \rho_m$.

The DE can be forced to transit to a $\Lambda$-like ($w \approx -1$) state at any time by 
fine-tuning $V(\phi)$. In the $\Lambda$-like state $\rho_{de}$ overtakes $\rho_m$ and dominates 
the recent and future energy density of the universe. We illustrate tracker quintessence in Fig.\ 
\ref{fig:C3quintessence_energy} using a power law potential $V(\phi) = M \phi^{-\alpha}$ (panel b)
\citep{Ratra1988,Caldwell1998,Zlatev1999} and an exponential potential $V(\phi) = M \exp(1/\phi)$ 
(panel c) \citep{Dodelson2000}.

The tracker paths are attractor solutions of the equations governing the evolution of the field. 
If the tracker quintessence field is initially endowed with a density off the tracker path (e.g.\ 
an equipartition of the energy available at reheating) its density quickly approaches and joins 
the tracker solution.

\subsubsection{Oscillating Dark Energy}
\citet{Dodelson2000} explored a quintessence potential with oscillatory perturbations 
$V(\phi) = M \exp(- \lambda \phi) \left[1 + A \sin(\nu \phi)\right]$. They refer to
models of this type as tracking oscillating energy. Without the
perturbations (setting $A=0$) this potential causes exact tracker behaviour: the 
quintessence energy decays as $\rho_r + \rho_m$ and never dominates. With the perturbations
the quintessence energy density oscillates about $\rho_r + \rho_m$ as it decays (Fig.\ 
\ref{fig:C3quintessence_energy}d). The quintessence energy 
dominates on multiple occasions and its equation of state varies continuously between 
positive and negative values. One of the main motivations for tracking oscillating energy 
is to solve the coincidence problem by ensuring that $\rho_{de} \sim \rho_{m}$ or $\rho_{de} \sim \rho_{r}$
at many times in the past or future.

It has yet to be seen how such a potential might
arise from particle physics. Phenomenologically similar cosmologies have been discussed 
in \citet{Ahmed2004,Yang2005,Feng2006}.

\subsubsection{Interacting Quintessence}
Non-gravitational interactions between the quintessence field and matter fields might allow energy 
to transfer between these components. Such interactions are not forbidden by any known symmetry
\cite{Amendola2000b}. The primary motivation for the exploration of interacting dark energy models 
is to solve the coincidence problem. In these models the present matter/dark energy density 
proximity $r$ may be constant \citep{Amendola2000,Zimdahl2001,Amendola2003,Franca2004,Guo2005,
Olivares2005,Pavon2005,Zhang2005,Franca2006,Amendola2006,Amendola2007,Olivares2007} or slowly 
varying \citep{delCampo2005,delCampo2006}. 
 
We plot a density history of the interacting quintessence model of \citet{Amendola2000} in Fig.\ 
\ref{fig:C3quintessence_energy}e. This model is characterized by a DE potential $V(\phi)=A \exp[B \phi]$ 
and DE-matter interaction term $Q = -C \rho_m \dot{\phi}$, specifying the rate at which energy 
is transferred to the matter fields. The free parameters were tuned such that radiation domination 
ends at $a=10^{-5}$ and that $r_{t \rightarrow \infty} = 0.35$.

\subsection{Phantom Dark Energy}
The analyses of \citet{Riess2004} and \citet{Wood-Vasey2007} have mildly ($\sim 1 \sigma$) 
favored a dark energy equation of state $w_{de} < -1$. 
These values are unattainable by standard quintessence models
but can occur in phantom dark energy models \citep{Caldwell2002}, in which kinetic energies
are negative. The energy density in the phantom field \emph{increases} with 
scalefactor, typically leading to a future ``big rip'' singularity where the scalefactor 
becomes infinite in finite time. Fig.\ \ref{fig:C3quintessence_energy}f shows the density 
history of a simple phantom model with a constant equation of state $w=-1.25$. The big rip
($a=\infty$ at $t=57.5$ Gyrs) is not seen in $\log(a)$-space.

\citet{Caldwell2003} and \citet{Scherrer2005} have explored how phantom models may solve 
the coincidence problem: since the big rip is triggered by the onset of DE domination, 
such cosmologies spend a significant fraction of their total time with $r$ large. 
For the phantom model with $w=-1.25$ (Fig.\ \ref{fig:C3quintessence_energy}f) \citet{Scherrer2005}
finds $r > 0.1$ for $12\%$ of the total lifetime of the universe. Whether this solves the 
coincidence or not depends upon the prior probability distribution $P_t(t)$ for the time of 
observation. \citet{Caldwell2003} and \citet{Scherrer2005} implicitly assume that the temporal 
distribution of observers is constant in time (i.e.\ $P_t(t)=\textrm{constant}$). For this prior 
the coincidence problem \emph{is} solved because the chance of observing $r \ge 0.1$ is large 
($12\%$). Note that for the ``naive $P_t(t)$'' prior shown in Fig.\ \ref{fig:C3illustrated_coincidence}, 
the solution of \citet{Caldwell2003} and \citet{Scherrer2005} fails because $r > 0.1$ is brief in 
$\log(a)$-space. It fails in this way for many other choices of $P_t(t)$ including, for example, 
distributions constant in $a$ or $\log(t)$.

\subsection{K-Essence}
In k-essence the DE is modeled as a scalar field with non-canonical kinetic energy 
\citep{Chiba2000,Armendariz-Picon2000,Armendariz-Picon2001,Malquarti2003}. Non-canonical
kinetic terms can arise in the effective action of fields in string and supergravity 
theories. Fig.\ \ref{fig:C3quintessence_energy}g shows a density history typical of 
k-essence models. This particular model is from \citet{Armendariz-Picon2001} and \citet{Steinhardt2003}. 
During radiation domination the k-essence field tracks radiation sub-dominantly (with 
$w_{de}=w_r=1/3$) as do some of the other models in Fig.\ \ref{fig:C3quintessence_energy}. 
However, 
no stable tracker solution exists for $w_{de}=w_m (=0)$. Thus after radiation-matter equality, 
the field is unable to continue tracking the dominant component, and is driven to another 
attractor solution (which is generically $\Lambda$-like with $w_{de} \approx -1$). 
The onset of DE domination was recent in k-essence models because matter-radiation equality 
prompts the transition to a $\Lambda$-like state. K-essence thereby 
avoids fine-tuning in any particular numerical parameters, but the Lagrangian 
has been constructed ad-hoc.

\subsection{Chaplygin Gas}
A special fluid known as Chaplygin gas motivated by braneworld cosmology may be able to 
play the role of dark matter \emph{and} the dark energy \citep{Bento2002,Kamenshchik2001}. 
Generalized Chaplygin gas has the equation of state $p_{de}=-A \rho_{de}^{-\alpha}$ 
which behaves like pressureless dark matter at early times ($w_{de} \approx 0$ when $\rho_{de}$ 
is large), and like vacuum energy at late times ($w_{de} \approx -1$ when $\rho_{de}$ is small). 
In Fig.\ \ref{fig:C3quintessence_energy}h we show an example with 
$\alpha=1$.

\subsection{Summary of DDE Models}
Two broad classes of DDE models emerge from our comparison:
\begin{enumerate}
\item In \lcdm, tracker quintessence and k-essence models, 
the dark energy density is vastly different from the matter density
for most of the lifetime of the universe (panels a, b, c, g of Fig.\ \ref{fig:C3quintessence_energy}). 
The coincidence problem can only be solved if the probability distribution $P_t(t)$ for the time of 
observation is narrow, and overlaps significantly with an $r \sim 1$ peak. If $P_t(t)$ 
is wide, e.g.\ constant over the life of the universe in $t$ or $\log(t)$, then
observing $r \sim 1$ would be unlikely in these models and the coincidence problem \emph{is not} 
resolved.
\item Tracking oscillating energy, interacting quintessence, phantom models 
and Chaplygin gas models (panels d, e, f, h of Fig.\ \ref{fig:C3quintessence_energy}) employ 
mechanisms to ensure that $r \sim 1$ for large fractions of the life of the universe. In these 
models the coincidence problem may be solved for a wider range of $P_t(t)$ including, depending
on the DE model, distributions that are constant over the whole life of the universe 
in $t$, $\log(t)$, $a$ or $\log(a)$.
\end{enumerate}
The importance of an estimate of the distribution $P_t(t)$ is highlighted: such an 
estimate will either rule out models of the first category because they do not solve the coincidence
problem, or demotivate models of the second because their mechanisms are unnecessary to solve the
coincidence problem. This analysis does not address the problems associated with fine-tuning, 
initial conditions or ad hoc mechanisms of many DDE models \citep{Hebecker2001,Bludman2004,Linder2006}.
 
We leave this line of enquiry temporarily to discuss contemporary \emph{observational} constraints on the dark 
energy density history, because we wish to test what DE dynamics are required to solve the
coincidence, beyond those which models must exhibit to satisfy standard cosmological observations.


\section{Current Observational Constraints on Dynamic Dark Energy} \label{C3constraints}

\begin{figure*}[!p]
       \begin{center}
               \includegraphics[width=\linewidth]{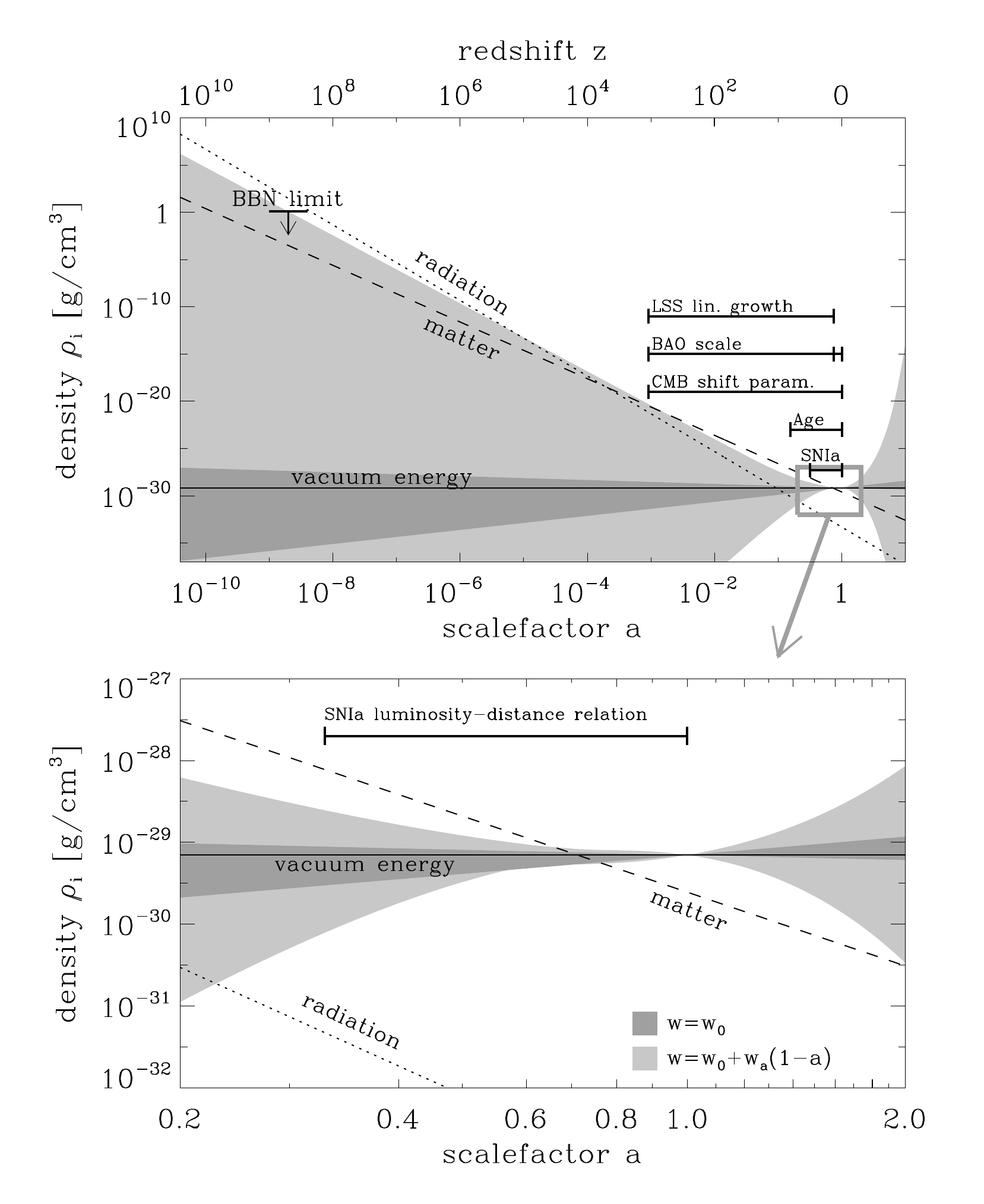}
               \caption[Observational constraints on dark energy dynamics.]
               	{The energy densities of radiation $\rho_r$, matter $\rho_m$ and 
               	the cosmological constant $\rho_{\Lambda}$ are shown as a function of
	           scalefactor, by the dotted, dashed, and solid lines respectively. 
	           Cosmological probes of dark energy include SNIa, CMB, BAO, the LSS linear
	           growth factor and constraints from BBN (see text). Each of these probes is sensitive 
	           to the effects of dark energy over different redshift intervals, as indicated.
	           The light grey band envelopes $w_0$-$w_a$-parameterized DDE models allowed at 
	           $<2 \sigma$ by \citet{Davis2007} (the contour in $w_0-w_a$ space is shown 
	           explicitly in Fig.\ \ref{fig:C3severities}). The dark grey band envelopes 
	           $w_0$-parameterized DDE models ($w_a = 0$ assumed) allowed at $<2 \sigma$ by 
	           \citet{Wood-Vasey2007}. The constraint is $w = -1.09 \pm 0.16$ at $2 \sigma$.
	           }
               \label{fig:C3SN1aConstraints}
       \end{center}
\end{figure*}

\subsection{Supernovae Ia}

Observationally, possible dark energy dynamics is explored almost solely using measurements of 
the cosmic expansion history.
Recent cosmic expansion is directly probed by using type Ia supernova 
(SNIa) as standard candles \citep{Riess1998,Perlmutter1999}. Each observed SNIa provides an
independent measurement of the luminosity distance $d_l$ to the redshift of the supernova $z_{SN}$.
The luminosity distance to $z_{SN}$ is given by
\be \label{eq:C3lumdist}
d_l(z_{SN}) = (1+z_{SN}) \frac{c}{H_0} \int_{z=0}^{z_{SN}} \frac{d z}{E(z)}
\ee 
where
\bea
E(z) & = & \frac{H(z)}{H_0} \label{eq:C3e} \\
     & = & \left[ \Omega_{r 0}(1+z)^4 + \Omega_{m 0}(1+z)^3 + \Omega_{de 0} 
     \frac{\rho_{de}(z)}{\rho_{de 0}}\right]^{\frac{1}{2}} \nonumber
\eea 
and thus depends on $H_0$, $\Omega_{m 0}$, and the evolution of the dark energy $\rho_{de}(z) / \rho_{de 0}$. 
The radiation term, irrelevant at low redshifts, can be dropped from Equation \ref{eq:C3e}. 
$\Omega_{de 0}$ is a dependent parameter due to flatness ($\Omega_{de 0} = 1-\Omega_{m 0}$).
Contemporary datasets include $\sim 200$ supernovae at redshifts $z_{SN} \le 2.16$ ($a \ge 0.316$) 
\citep{Astier2006,Riess2007,Wood-Vasey2007} and provide an effective continuum of constraints on the expansion 
history over that range \citep{Wang2005,Wang2006}. 
The redshift range probed by SNIa is indicated in both panels of Fig.\ \ref{fig:C3SN1aConstraints}. 

\subsection{Cosmic Microwave Background}

The first peak in the cosmic microwave background (CMB) temperature power spectrum corresponds 
to density fluctuations 
on the scale of the sound horizon at the time of recombination. Subsequent peaks correspond to 
higher-frequency harmonics. The locations of these peaks in $l$-space depend on the comoving 
scale of the sound horizon at recombination, and the angular distance to recombination.
This is summarized by the so-called CMB shift parameter $R$ \citep{Efstathiou1999,Elgaroy2007} 
which is related to the cosmology by
\be \label{eq:C3cmbshift}
 	R = \sqrt{\Omega_{m0}} \int_{z=0}^{z_{rec}} \frac{d z}{E(z)}
\ee
where $z_{rec} \approx 1089$ \citep{Spergel2006} is the redshift of recombination. 
The 3-year WMAP data gives a shift parameter $R=1.71 \pm 0.03$ \citep{Davis2007,Spergel2006}.
Since the dependence of Equation \ref{eq:C3cmbshift} on $H_0$ and $\Omega_{m 0}$ differs from that 
of Equation \ref{eq:C3lumdist}, measurements of the CMB shift parameter can be used to break 
degeneracies between $H_0$, $\Omega_{m 0}$ and DE evolution in the analysis of SNIa.
In the top panel of Fig.\ \ref{fig:C3SN1aConstraints} we represent the CMB observations using a 
bar from $z=0$ to $z_{rec}$.

\subsection{Baryonic Acoustic Oscillations and Large Scale Structure}

As they imprinted acoustic peaks in the CMB, the baryonic oscillations at recombination were
expected to leave signature wiggles - baryonic acoustic oscillations (BAO) - in the power spectrum 
of galaxies \citep{EisensteinHu1998}. These were detected with significant confidence in the 
SDSS luminous red galaxy power spectrum \citep{Eisenstein2005}. The expected BAO scale depends on the scale of 
the sound horizon at recombination, and on transverse and radial scales at the mean redshift $z_{BAO}$, of 
galaxies in the survey.
\citet{Eisenstein2005} measured the quantity 
\be
A(z_{BAO}) = \frac{\sqrt{\Omega_{m 0}}}{E(z_{BAO})^{\frac{1}{3}}} \left[ \frac{1}{z_{BAO}} \int_{z=0}^{z_{BAO}} \frac{dz}{E(z)} \right]^{\frac{2}{3}}
\ee
to have a value $A(z_{BAO}=0.35) = 0.469 \pm 0.017$,
thus constraining the matter density and the dark energy evolution parameters in a configuration
which is complomentary to the CMB shift parameter and the SNIa luminosity distance relation. 
Ongoing BAO projects have been designed specifically to produce stronger constraints on the 
dark energy equation of state parameter $w$. For example, WIGGLEZ \citep{Glazebrook2007} will use a 
sample of high-redshift galaxies to measure the BAO scale at $z_{BAO} \approx 0.75$. As well as 
reducing the effects of non-linear clustering, this redshift is at a larger angular distance, making 
the observed scale more sensitive to $w$.
Constraints from the BAO scale depend on the evolution of the universe from $z_{rec}$ to 
$z_{BAO}$ to set the physical scale of the oscillations. They also depend on the evolution of the
universe from $z_{BAO}$ to $z=0$, since the observed angular extent of the oscillations depends on 
this evolution. The bar representing BAO scale observations in the top panel of Fig.\ 
\ref{fig:C3SN1aConstraints} indicates both these regimes.

The amplitude of the BAOs - the amplitude of the large scale structure (LSS) power spectrum - is 
determined by the amplitude of the power spectrum at recombination, and how much those fluctuations 
have grown (the transfer function) between $z_{rec}$ and $z_{BAO}$. By comparing the recombination 
power spectrum (from CMB) with the galaxy power spectrum, the LSS linear growth factor can be measured 
and used to constrain the expansion history of the universe (independently of the BAO scale) over 
this redshift range. In practice, biases hinder precise normalization of the galaxy power spectrum, 
weakening this technique. The range over which this technique probes the DE is indicated in Fig.\ 
\ref{fig:C3SN1aConstraints}.

\subsection{Ages}

Cosmological parameters from SN1a, CMB, LSS, BAO and other probes allow us to calculate the current 
age of the universe to be $13.8 \pm 0.1$ \citep{Hinshaw2006} assuming \lcdm. Uncertainties on the age 
calculated in this way grow dramatically if we drop the assumption that the DE is vacuum energy ($w=-1$). 

An independent lower limit on the current age of the universe is found by estimating the ages of the 
oldest known globular clusters \citep{Hansen2004}. 
These observations rule out models which predict the universe to be younger than 
$12.7 \pm 0.7 \; \textrm{Gyrs}$ ($2 \sigma$ confidence):
\bea
t_0 & = & H_0^{-1} \int_{z=0}^{\infty} \frac{dz}{(1+z) E(z)} \\
 	& \gsim & 12.7 \pm 0.7 \; \textrm{Gyrs}. \nonumber
\eea
Other objects can also be used to set this age limit \citet{Lineweaver1999}, but generally less 
successfully due to uncertainties in dating techniques.

Assuming \lcdm, an age of $12.7$ Gyrs corresponds to a redshift of $z \approx 5.5$. Contemporary age 
measurements are sensitive to the dark energy content from $z \approx 5.5$ to $z=0$. In the top panel
of Fig.\ \ref{fig:C3SN1aConstraints} we show this redshift interval. The evolution and energy content 
of the universe before $12.7$ Gyrs ago is not probed by these age constraints.

\subsection{Nucleosynthesis}

In addition to the constraints on the expansion history (SN1a, CMB, BAO and $t_0$) we 
know that $\rho_{de}/\rho_{tot} < 0.045$ (at $2 \sigma$ confidence) during Big Bang Nucleosynthesis (BBN)
\citep{Bean2001b}. Larger dark energy densities imply a higher expansion rate at that epoch 
($z \sim 6 \times 10^{8}$) which would result in a lower neutron to proton ratio, conflicting with the 
measured helium abundance, $Y_{\textrm{He}}$.

\subsection{Dark Energy Parameterization}

Because of the variety of proposed dark energy models, it has become usual to summarize
observations by constraining a parameterized time-varying equation of state. Dark 
energy models are then confronted with observations in this parameter space.
The unique zeroth order parameterization of $w$ is $w=w_0$ (a constant), with $w=-1$ 
characterizing the cosmological constant model. 
The observational data can be used to constrain the first derivative of $w$. 
This additional dimension in the DE parameter space may be useful in distinguishing
models which have the same $w_0$. From an observational standpoint, the 
obvious choice of 1st order parameterization is $w(z)=w_0+\frac{dw}{dz}z$ \citep{diPietro2003}. 
This is rarely used today since currently considered DDE models are poorly portrayed by 
this functional form. The most popular parameterization is $w(a)=w_0+w_a(1-a)$ 
\cite{Albrecht2006,Linder2006b}, which does not diverge at high redshift. 

\citet{Linder2005} have argued that the extension of this approach to second order, e.g.\ 
$w(a)=w_0+w_a(1-a)+w_{aa}(1-a)^2$, is not motivated by current DDE models. Moreover, they 
have shown that next generation observations are unlikely to be able to distinguish the
quadratic from a linear expansion of $w$. \citet{Riess2007} have illustrated this recently 
using new SN1a. 

An alternative technique for exploring the history of dark energy is to constrain $w(z)$ or 
$\rho_{de}(z)$ in independent redshift bins. This technique makes fewer assumptions about 
the specific shape of $w(z)$. In the absence of any strongly motivated parameterization of 
$w(z)$ this bin-wise method serves as a good reminder of how little we actually know from
observation.
Using luminosity distance measurements from SNIa, DE evolution has been constrained in this way
in $\Delta z \sim 0.5$ bins out to redshift $z_{SN} \sim 2$ \citep{Wang2004b,Huterer2005,Riess2007}.
In the future, BAO measurements at various redshifts may contribute to these constraints, 
however $z_{BAO}$ will probably never be larger than $z_{SN}$.
Moreover, because the recombination redshift 
$z_{rec} \approx 1089$ is fixed, only the cumulative effect (from $z=z_{rec}$ to $z=0$) of the 
DE can be measured with the CMB and LSS linear growth factor. With only this single data point 
above $z_{SN}$, the bin-wise technique effectively degenerates to a parameterized analysis at 
$z > z_{SN}$.

\subsection{Summary of Current DDE Constraints}

If one assumes the popular $w_0 - w_a$ parameterization until last 
scattering, then all cosmological probes can be combined to constrain $w_0$ and 
$w_a$. In a recent analysis of SN1a, CMB and BAO observations, \citet{Davis2007} found 
$w_0 = -1.0 \pm 0.4$ and $w_a = -0.4 \pm 1.8$ at $2 \sigma$ confidence (the
contour is shown in Fig.\ \ref{fig:C3severities}). Using the same 
observations, \citet{Wood-Vasey2007} assumed $w_a = 0$ and found $w = w_0 = -1.09 \pm 0.16$
($2 \sigma$).

The evolution of $\rho_{de}$ is related to $w$ by covariant energy conservation 
\citep{Carroll2004book}
\be
\frac{\delta \rho_{de}}{\rho_{de}} = - 3 \left( w(a)+1 \right) \frac{\delta a}{a}.
\ee
The dark energy density corresponding to the $w_0 - w_a$ parameterization of $w$ is
thus given by 
\be
\rho_{de}(z) = \rho_{de 0}\ e^{3 w_a (a - 1)}\ a^{-3 (1 + w_0 + w_a)}.
\ee

The cosmic energy density history is illustrated in Fig.\ \ref{fig:C3SN1aConstraints}. Radiation and
matter densities steadily decline as the dotted and dashed lines. With the DE equation of 
state parameterized as $w(a)=w_0+w_a(1-a)$, its density history is constrained to the light-grey 
area \citep{Davis2007}. If the evolution of $w$ is negligible, i.e.\ we condition on $w_a \approx 0$,
then $w(a) \approx w_0$ and the DE density history 
lies within the dark-grey band \citep{Wood-Vasey2007}. If the dark energy is pure vacuum energy 
(or Einstein's cosmological constant) then $w=-1$ and its density history is given by the 
horizontal solid black line. 


\section{The Temporal Distribution of Observers} \label{C3observers}

The energy densities $\rho_r$, $\rho_m$ and $\rho_{de}$, and the proximity parameter $r$ 
we imagine we might 
have observed, depend on the distribution $P_t(t)$ from which
we imagine 
our time of observation $t_{obs}$ 
has been drawn. 
What we can expect to 
observe must be restricted by the conditions necessary for our presence as observers \citep{Carter1974}.
Thus, for example, it is meaningless to suppose we might have lived during inflation, or during radiation
domination, or before the first atoms \citep{Dicke1961}. 

We can, however, suppose that we are randomly selected cosmology-discovering observers, and we can expect 
our observations of $\rho_m$ and $\rho_{de}$ to be typical of observations made by such observers.
This is Vilenkin's principle of mediocrity \cite{Vilenkin1995a}. Accordingly, the distribution 
$P_t(t)$ for the time of observation $t_{obs}$ is proportional to the temporal distribution of 
cosmology-discovering observers (referred to henceforth as simply ``observers''). Thus to solve the coincidence 
problem one must show that the proximity parameter we measure, $r_0$, is typical of those measured by 
other observers.

The most abundant elements in the cosmos are hydrogen, helium, oxygen and carbon \citep{Pagel1997}. 
In the past decade $>200$ extra solar planets have been observed via doppler, transit or microlensing
methods. 
Extrapolation of current patterns in planet mass and orbital period are consistent with the idea that 
planetary systems like our own are common in the universe \citep{Lineweaver2003b}. 
All this does not necessarily imply that observers are 
common, but it does support the idea 
that terrestrial-planet-bound carbon-based 
observers, even if rare, may be the \emph{most common} observers. In the following estimation of $P_t(t)$ 
we consider only observers bound to terrestrial planets.

\subsection{First the Planets...} 

\begin{figure}[!p]
       \begin{center}
               \includegraphics[width=0.7\linewidth]{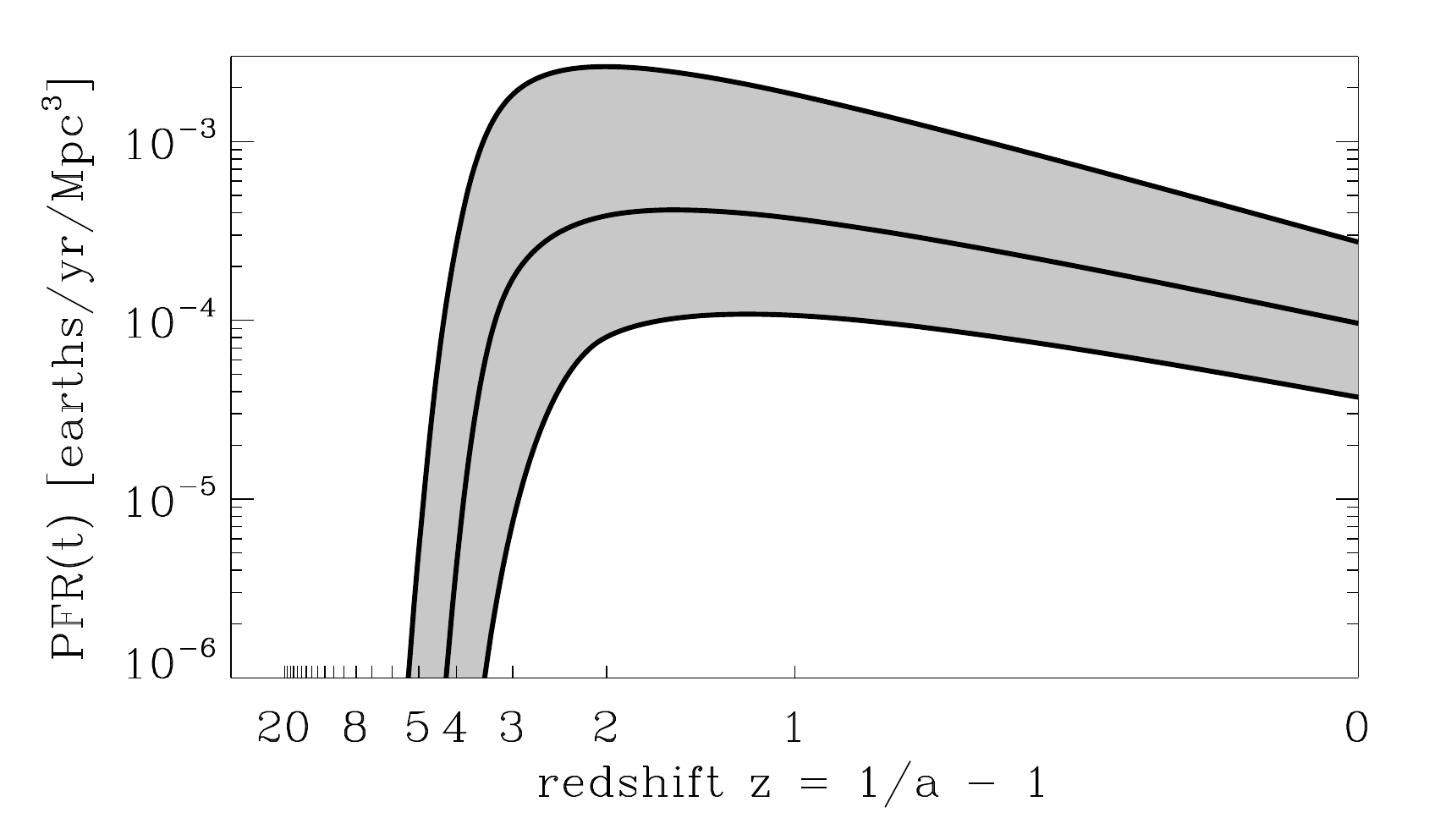}
               \caption[First the planets...]{The terrestrial planet formation rate as estimated by \citet{Lineweaver2001}.
				It is based on a compilation of SFR measurements and has been corrected for 
				the low metallicity of the early universe, which prevents the terrestrial planet 
				formation rate from rising as quickly as the stellar formation rate at $z \gsim 4$.}
               \label{fig:C3tpfhz}
       \end{center}
\end{figure}

\citet{Lineweaver2001} estimated the terrestrial planet formation rate (PFR) by making 
a compilation of measurements of the cosmic star formation rate (SFR) and suppressing a 
fraction of the early stars $f(t)$ to correct for the fact that the metallicity was too low for 
those early stars to host terrestrial planetary systems,
	\begin{equation}
		PFR(t) = const \times SFR(t) \times f(t).
	\end{equation}
In Fig.\ \ref{fig:C3tpfhz} we plot the PFR reported by \citet{Lineweaver2001} as a function of 
redshift, $z=\frac{1}{a}-1$. As illustrated in the figure, there is large uncertainty in the normalization of the 
formation history. Our analysis will not depend on the normalization of this function so this 
uncertainty \emph{will not} propagate into our analysis. There are also uncertainties 
in the location of the turnover at high redshift, and in the slope of the formation history 
at low redshift - both of these \emph{will} affect our results.

\begin{figure}[!p]
       \begin{center}
               \includegraphics[width=0.7\linewidth]{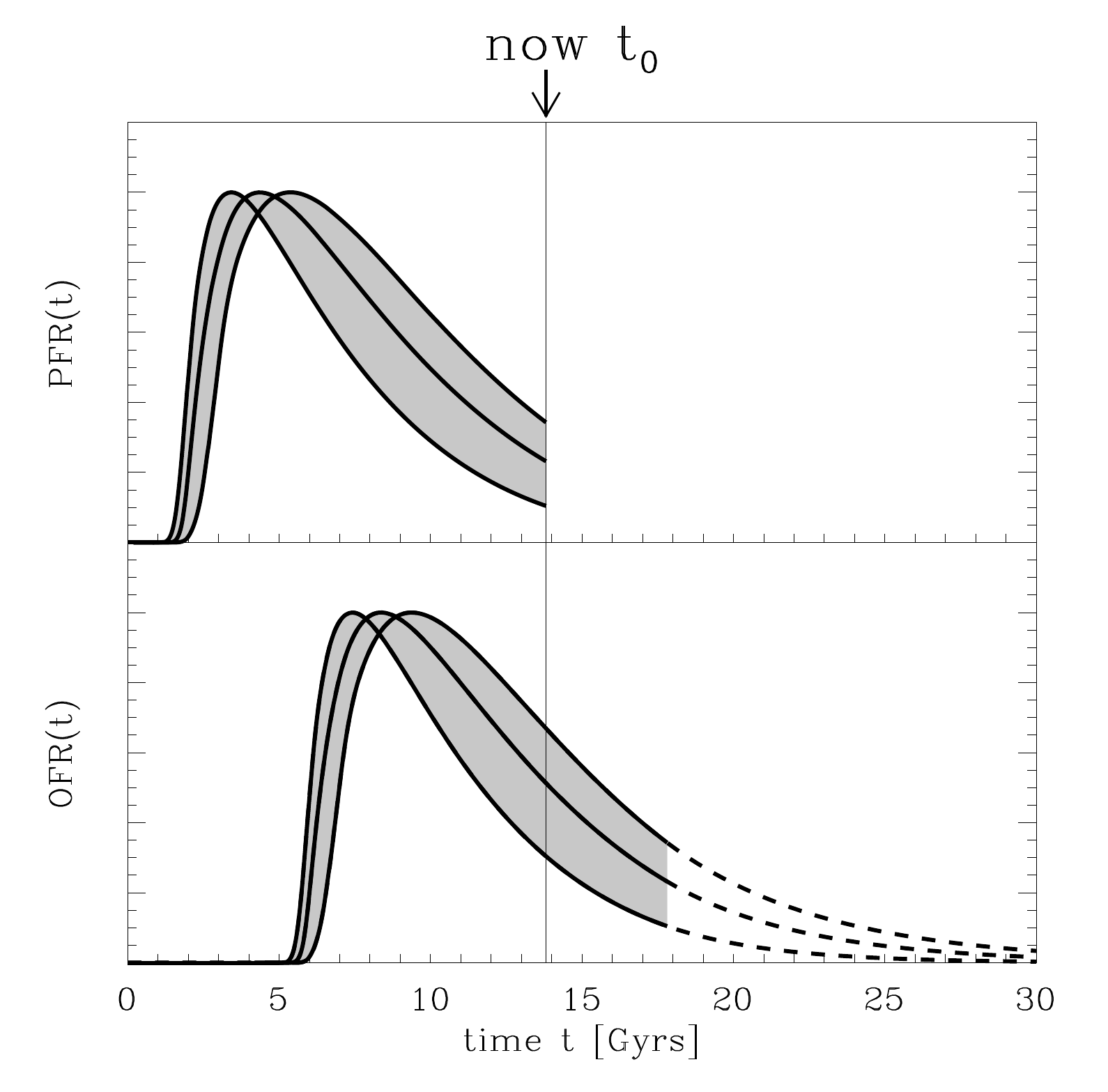}
               \caption[... then first observers.]{The terrestrial planet formation from Fig.\ \ref{fig:C3tpfhz} is shown here 
				as a function of time. The transformation from redshift to time is cosmology dependent. 
				To create this figure we have used best-fit values for the DDE parameters, $w_0=-1.0$ 
				and $w_a=-0.4$ \citep{Davis2007}. The y-axis is linear (c.f.\ the logarithmic axis in Fig.\ \ref{fig:C3tpfhz}) 
				and the family of curves have been re-normalized to highlight the sources of uncertainty 
				important for this analysis: uncertainty in the width of the function, and in the 
				location of its peak.
				The observer formation rate (OFR) is calculated by shifting the planet formation 
				rate by some amount $\Delta t_{obs}$ ($=4\ \textrm{Gyrs}$) to allow the planet to 
				cool, and the possible emergence of observers. These distributions are closed by 
				extrapolating exponentially in $t$.}
               \label{fig:C3tpfht}
       \end{center}
\end{figure}

The conversion from redshift to time depends on the 
particular cosmology, through the Friedmann equation,
	\begin{eqnarray}
		\left( \frac{da}{dt} \right)^2 & = & H(a)^2 a^2 \\
			& = & H_0^2 \bigg[ \Omega_{r 0}a^{-2} + \Omega_{m 0}a^{-1} + {} \biggr. \nonumber \\
			& & \biggl. \Omega_{de 0}\ \exp [3 w_a (a-1)]\ a^{-3 w_0 -3 w_a -1} \biggr]. \nonumber
	\end{eqnarray}
In Fig.\ \ref{fig:C3tpfht} we plot the PFR from Fig.\ \ref{fig:C3tpfhz} as a function of time assuming 
the best fit parameterized DDE cosmology.

\subsection{... then First Observers} 

After a star has formed, some non-trivial amount of time $\Delta t_{obs}$ will pass before observers,
if they arise at all, arise on an orbiting rocky planet. This time allows planets to form and cool and, 
possibly, biogenesis and the emergence observers. $\Delta t_{obs}$ is constrained to be shorter than the 
life of the host star. 
If we consider that our $\Delta t_{obs}$ has been drawn from a probability distribution $P_{\Delta t_{obs}}(t)$.
The observer formation rate (OFR) would then be given by the convolution
	\begin{equation} \label{eq:C3OFR}
		OFR(t) = const \times \int_{0}^{\infty} PFR(\tau) P_{\Delta t_{obs}}(t - \tau) d\tau.
	\end{equation}
	
In practice we know very little about $P_{\Delta t_{obs}}(t)$. It must be very nearly zero below
about $\Delta t_{obs} \sim 0.5\ \textrm{Gyrs}$ - this is the amount of time it takes for terrestrial 
planets to cool and the bombardment rate to slow down. Also, it must be near zero above the lifetime 
of a small ($0.1 M_{\odot}$) star (above $\sim 500\ \textrm{Gyrs}$). 
If we assume 
that our $\Delta t_{obs}$ is typical, then $P_{\Delta t_{obs}}(t)$ has
significant weight around 
$\Delta t_{obs}=4\ \textrm{Gyrs}$ - the amount of time it has taken for us to evolve here on Earth. 

A fiducial choice, where \emph{all} observers emerge $4\ \textrm{Gyrs}$ after the formation 
of their host planet, is $P_{\Delta t_{obs}}(t)=\delta(t-4\ \textrm{Gyrs})$. This choice results in 
an OFR whose shape is the same as the PFR, but is shifted $4\ \textrm{Gyrs}$ into the future,
	\begin{equation}
			OFR(t) = const \times PFR(t - 4\ \textrm{Gyrs})
	\end{equation}
(see the lower panel of Fig.\ \ref{fig:C3tpfht}).	
Even for non-standard $w_0$ and $w_a$ values, this fiducial OFR aligns closely with the $r(t)$ peak 
and the effect of a wider $P_{\Delta t_{obs}}$ is generally to increase the severity of the coincidence 
problem by spreading observers outside the $r(t)$ peak. Hence using our fiducial $P_{\Delta t_{obs}}$ 
(which is the narrowest possibility) will lead to conclusions which are conservative in that they 
underestimate the severity of the cosmic coincidence. If another choice for $P_{\Delta t_{obs}}$ could 
be justified, the cosmic coincidence would be more severe than estimated here. We will discuss this 
choice in Section \ref{C3conclusion}.

The OFR is then extrapolated into the future using a decaying exponential with respect to $t$ (the 
dashed segment in the lower panel of Fig.\ \ref{fig:C3tpfht}). The observed SFH is consistent with 
a decaying exponential. 
We have tested other choices (linear \& 
polynomial decay) and our results do not depend strongly on the shape of the extrapolating function 
used. 

The temporal distribution of observers $P_t(t)$ is proportional to the observer formation rate,
	\begin{equation}
		P_t(t) = const \times OFR(t).
	\end{equation}

This observer distribution is similar to the one used by \citet{Garriga1999} to treat the coincidence 
problem in a multiverse scenario. By comparison, our $OFR(t)$ distribution starts later because we have 
considered the time required for the build up of metallicity, and because we have included an evolution 
stage of $4\ \textrm{Gyrs}$. Our distribution also decays more quickly than theirs does.
Some of our cosmologies suffer big-rip singularities in the future. In these cases we
truncate $P_t(t)$ at the big-rip.


\section{Analysis and Results: Does fitting contemporary constraints necessarily solve the cosmic coincidence?} \label{C3results}

\begin{figure}[!p]
       \begin{center}
               \includegraphics[width=\linewidth]{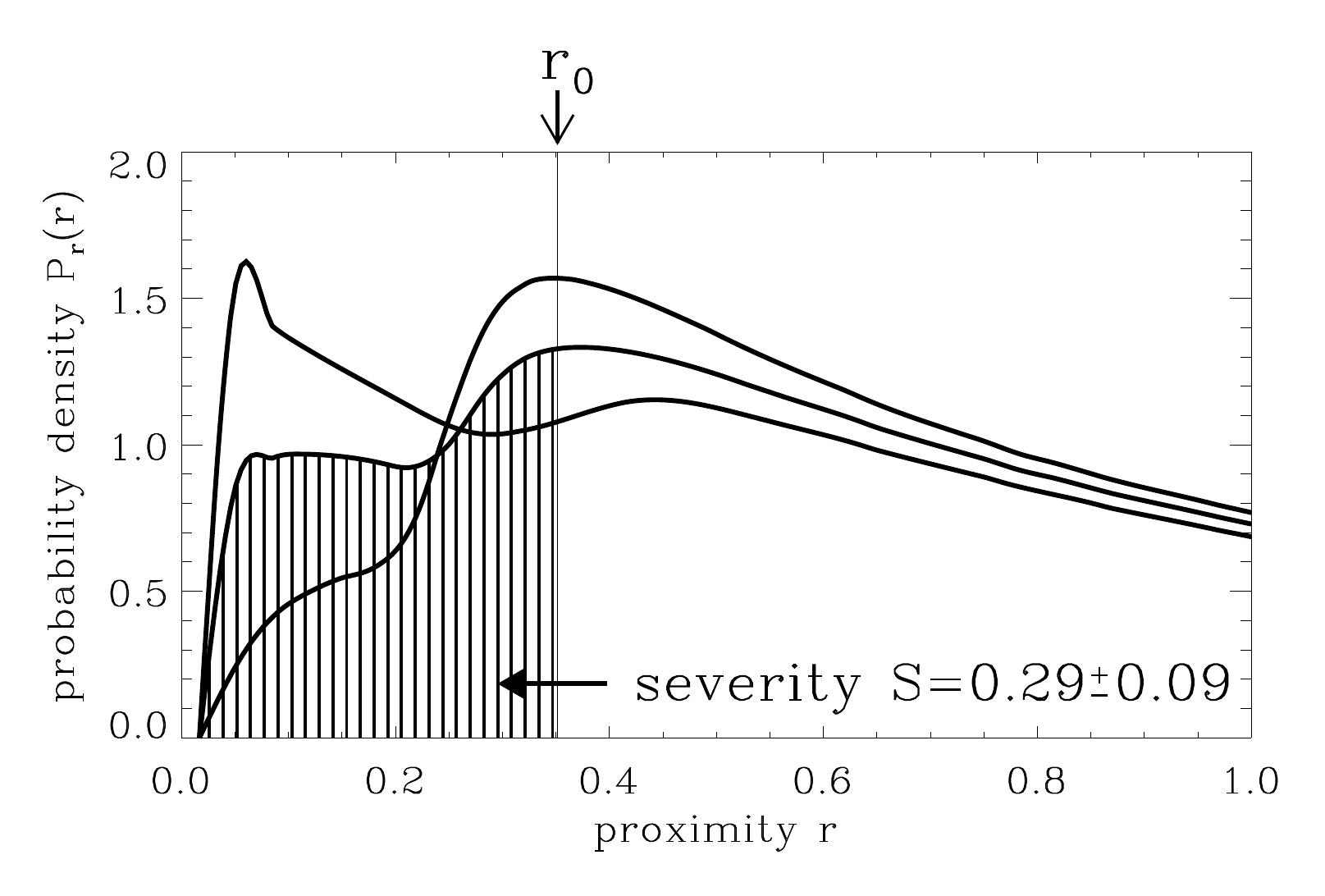}
               \caption[Probability distribution in $r$ for example dynamic dark energy model]
               {The predicted distribution of observations of $r$ is plotted for 
               the parameterized DDE model which best-fits cosmological observations: $w_0=-1.0$ 
               	and $w_a=-0.4$. The proximity parameter we observe 
				$r_0 =\frac{\rho_{m 0}}{\rho_{de 0}} \approx 0.35$ is typical in this cosmology since only
				$29 \%$ of observers (vertical striped area) observe $r<0.35$. The upper and lower 
				limits on this value resulting from 
				uncertainties in the SFR are $38 \%$ and $20 \%$ respectively. Thus the severity 
				of the cosmic coincidence in this model is $S=0.29\pm0.09$. This model does not suffer 
				a coincidence problem.}
               \label{fig:C3pofr}
       \end{center}
\end{figure}

For a given model the proximity parameter observed by a 
typical observer is described by a probability distribution $P_r(r)$ calculated as
	\begin{equation}
		P_r(r) = \sum \frac{dt}{dr} P_t(t(r)).
	\end{equation}
The summation is
over contributions from all solutions of $t(r)$ (typically,
any given value of $r$ occurs at multiple times during the lifetime of the Universe).
In Fig.\ \ref{fig:C3pofr} we plot $P_r(r)$ for the $w_0=-1.0$, $w_a=-0.4$ cosmology. In this case, 
observers are distributed over a wide range of $r$ values, with $71\%$
seeing $r>r_0$, and $29\%$ seeing $r<r_0$.

We define the severity $S$ of the cosmic coincidence problem as the 
probability that a randomly selected observer measures a proximity parameter 
$r$ lower than we do:
	\begin{equation}
		S = P(r < r_0) = 1 - P(r > r_0) = \int_{r=0}^{r_0} P_r(r) dr.
	\end{equation}
For the $w_0=-1.0$, $w_a=-0.4$ cosmology of Figs.\ $\ref{fig:C3tpfht}$ and $\ref{fig:C3pofr}$, the 
severity is $S=0.29 \pm 0.09$. This model does not suffer a coincidence problem since $29\%$
of observers would see $r$ lower than we do.
If the severity of the cosmic coincidence would be near $0.95$ ($0.997$) in a particular model, then 
that model would suffer a $2 \sigma$ ($3 \sigma$) coincidence problem and the value of $r$ we observe 
really would be unexpectedly high. 

We calculated the severities $S$ for cosmologies spanning a large region 
of the $w_0-w_a$ plane and show our results in Fig.\ \ref{fig:C3severities} using contours of
equal $S$. The severity of the coincidence problem is low (e.g.\ $S \lsim 0.7$) for most combinations
of $w_0$ and $w_a$ shown. There is a coincidence problem, where the severity is high ($S \gsim 0.8$), 
in two regions of this parameter space. These are indicated in Fig.\ \ref{fig:C3severities}. 

Some features in Fig.\ \ref{fig:C3severities} are worth noting: 
\begin{itemize}
\item Dominating the left of the plot, the severity of the coincidence increases
towards the bottom left-hand corner. This is because as $w_0$ and $w_a$ become more 
negative, the $r$ peak becomes narrower, and is observed by fewer observers. 
\item There is a strong vertical dipole of coincidence severity centered at $(w_0=0, w_a=0)$. 
For $(w_0 \approx 0, w_a > 0)$ there is a large coincidence problem because in such 
models we would be currently witnessing the very closest approach between DE and 
matter, with $\rho_{de} \gg \rhom$ for all earlier and later times (see Fig.\ \ref{fig:C3severityexamples}c). 
For $(w_0 \approx 0, w_a < 0)$ there is an anti-coincidence problem because in those models 
we would be currently witnessing the DDE's furthest excursion from the matter
density, with $\rho_{de}$ and $\rhom$ in closer proximity for all relevant earlier 
and later times, i.e., all times when $P_t(t)$ is non-negligible.
\item There is a discontinuity in the contours running along $w_a = 0$ for 
phantom models ($w_0 < -1$). The distribution $P_t(t)$ is truncated by 
big-rip singularities in strongly phantom models (provided they remain phantom; $w_a > 0$). 
This truncation of late-time observers means that early observers who witness large 
values of $r$ represent a greater fraction of the total population. 
\end{itemize}
 
To illustrate these features, Fig.\ \ref{fig:C3severityexamples} shows the density histories 
and observer distributions for four specific examples selected from the $w_0-w_a$ plane of 
Fig.\ \ref{fig:C3severities}.

We find that \emph{all} observationally allowed combinations of $w_0$ and $w_a$ result in low 
severities ($S < 0.4$), i.e., there are large ($> 60\%$) probabilities of observing the 
matter and vacuum density to be at least as close to each other as we observe them to be.


\section{Discussion} \label{C3conclusion}

\begin{figure*}[!p]
       \begin{center}
               \includegraphics[width=\linewidth]{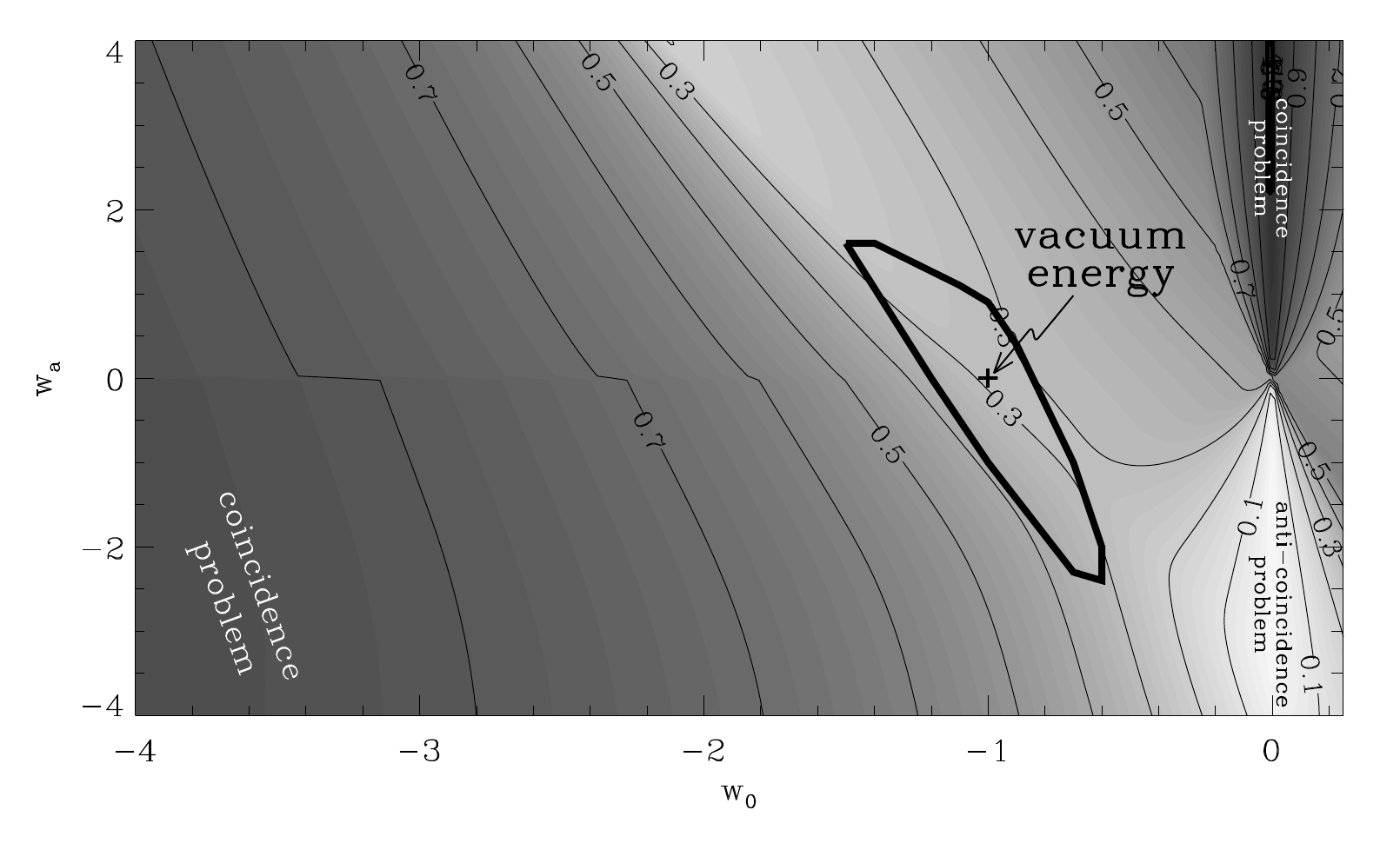}
               \caption[The severity of the cosmic coincidence as a function of phase space.]
               {Here we plot contours of equal severity $S$ in $w_0-w_a$ parameter 
               space. $S$ is the fraction of observers who see $r < r_0$. If $S$ is large, 
               a large percentage of observers should see $r$ lower than we do - those models suffer 
               coincidence problems.
               The thick black contour represents the observational constraints on $w_0$ 
               and $w_a$ from \citet{Davis2007} ($2 \sigma$ confidence and marginalized over other uncertainties). 
               In \citet{Lineweaver2007} we showed that the severity of the coincidence problem is 
               low for \lcdm\ (indicated by the ``\textbf{+}''). 
               Values of $w_0$ and $w_a$ that result in a mild coincidence problem (e.g.\ $S \gsim 0.7$) 
               are already strongly excluded by observations. This leads to our main 
               result: none of the models in the observationally allowed regime 
               suffer a cosmic coincidence problem when our estimate of the temporal 
               distribution of observers $P_{obs}(t)$ is used as a selection function.
               }
               \label{fig:C3severities}
       \end{center}
\end{figure*}

\begin{figure*}[!p]
       \begin{center}
               \includegraphics[width=\linewidth]{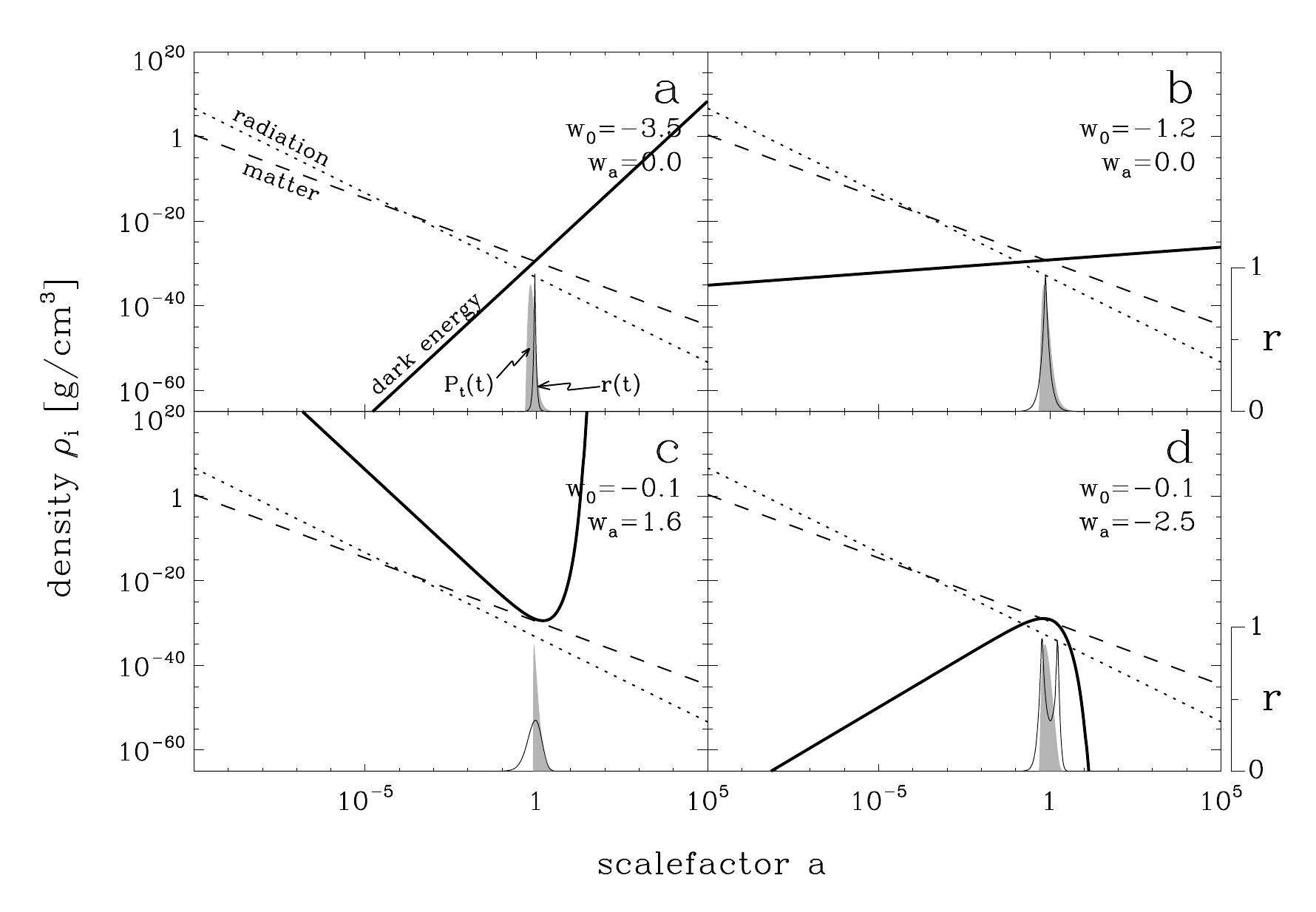}
               \caption[Four points off the previous plot.]
               {History of the energy densities in radiation (dotted line), matter (dashed line)
               and dark energy (thick black line) for four parameterized DE models from Fig.\ 
               \ref{fig:C3severities}. The proximity parameter $r$ (thin black line) and the temporal
               distribution of observers $P_t(t)$ (grey shade) are also given.
               {\bf Panel a} shows a phantom model with a constant equation of state
               $w=-3.5$. In this model the phantom density increases quickly and the $r(t)$ peak is 
               narrow. As a result, a large fraction of observers live while the matter and dark 
               energy densities are vastly different ($r \approx 0$) and there is a mild coincidence
               problem ($S \approx 0.8$). This might be used to rule-out the model shown in 
               Panel a, except that it is already strongly excluded by direct cosmological observations (refer 
               to Fig.\ \ref{fig:C3severities}). 
               {\bf Panel b} shows a phantom model which lies within the 
               observationally allowed $2\sigma$ region. There is no coincidence problem in 
               this model ($S \approx 0.4$). 
               {\bf Panel c} shows a model in which there is a 
               coincidence problem ($S \approx 0.95$). This models lies within the cluster of
               contours in the upper right-hand corner of Fig.\ \ref{fig:C3severities}. In this model
               the dark energy dominates the past \emph{and} future energy budget. Again however, the 
               coincidence problem can tell us nothing new, as this model is already strongly excluded 
               by observations.
               {\bf Panel d} shows a model in which there is an anti-coincidence problem. This models lies 
               within the cluster of contours in the lower right-hand corner of Fig.\ \ref{fig:C3severities}. 
               In this model the dark energy and matter densities are more similar ($r$ is greater) in the 
               recent past and near future (although $r \rightarrow 0$ further into the past or future). 
               According to the observer distribution $P_t(t)$ most observers live near the current epoch, 
               during $r>0.35$, with just $7 \%$ living during $r<0.35$ ($S=0.07$) in this particular model. 
               One might argue that this model can be ruled out because our value of $r$ is anomalously 
               small. However, this model too is already strongly excluded by observations.
	           }
               \label{fig:C3severityexamples}
       \end{center}
\end{figure*}

It was not clear what DDE dynamics were required to solve the 
coincidence problem. Our analysis might 
have resulted in new constraints on the values of $w_0$ and $w_a$, by simply demanding that 
we do not live during a special time in which $r \sim 1$. 
There are regions of $w_0-w_a$ parameter space that can be ruled out in this manner (see
Fig.\ \ref{fig:C3severities}) however those points are already strongly excluded by observational
constraints on $w_0$ and $w_a$.
Therefore, the cosmic coincidence problem can not be used as a tool to further 
constrain DDE since the problem is solved for all DDE models satisfying observational constraints 
on $w_0$ and $w_a$.

The main result of our analysis is that any DDE model in agreement with current cosmological 
constraints has $\rho_{de} \sim \rho_m$ for a significant fraction of observers.

Interacting quintessence models in which the proximity parameter asymptotes to a constant at 
late times 
\citep{Amendola2000,Zimdahl2001,Amendola2003,Franca2004,Guo2005,Olivares2005,Pavon2005,
Zhang2005,Franca2006,Amendola2006,Amendola2007,Olivares2007}
have been proposed as a solution to the coincidence problem. More recently, 
\citet{delCampo2005,delCampo2006} have argued for a broader class of interacting quintessence 
models that ``soften'' the coincidence problem by predicting a very slowly varying 
(though not constant) proximity parameter. 
Our analysis finds that $r$ need not asymptote to a constant, nor evolve particularly slowly, 
partially undermining the motivations for these interacting quintessence models.

\citet{Caldwell2003} and \citet{Scherrer2005} have proposed that the coincidence problem may 
be solved by phantom models in which there is a future big-rip singularity because such 
cosmologies spend a significant fraction of their lifetimes in $r \sim 1$
states. In our work $P_t(t)$ is terminated by big-rip singularities in ripping models. In 
non-ripping models, however, the distribution is effectively terminated by the declining star 
formation rate. Therefore the big-rip gives phantom models only a marginal advantage over other 
models. This marginal advantage manifests as the discontinuity along $w_a=0$ on the left side of 
Fig.\ \ref{fig:C3severities}.

We could improve our analysis, in the sense of getting tighter coincidence constraints 
(larger severities), if we used a less conservative $P_{\Delta t_{obs}}$. We used the
most conservative choice - a delta function - because the present understanding of the
time it takes to evolve into observers is too poorly developed to motivate any other form of 
$P_{\Delta t_{obs}}$.
Another possible improvement is the DE equation of state parameterization. We used the
current standard, $w=w_0+w_a(1-a)$, which may not parameterize some models well for
very small or very large values of $a$.

We conclude that DDE models need not be fitted with exact tracking or oscillatory 
behaviors specifically to solve the coincidence by generating long or repeated periods of 
$\rho_{de} \sim \rho_{m}$. 
Also, particular interactions guaranteeing $\rho_{de} \sim \rho_m$ for 
long periods are not well motivated.
Moreover phantom models have no significant 
advantage over other DDE models with respect to the coincidence problem discussed here.


\section*{ACKNOWLEDGMENTS}
CE acknowledges a UNSW School of Physics postgraduate fellowship. CE thanks the ANU's RSAA 
for its kind hospitality, where this research was carried out.


\section*{Appendix A: Numerical Values for Parameters of Models Illustrated in Fig.\ \ref{fig:C3quintessence_energy}} \label{C3paramvals}

\begin{table}[!hp]
   \begin{center}
   \small
   \begin{tabular}{l l l}
      \hline
      Model                            & Parameter                & Value \\
      \hline
      \hline
      power law tracker quintessence   & $\alpha$                 & $2$ \\
                                       & $M$                      & $1.4 \times 10^{-124}$ \\
      exponential tracker quintessence & $M$                      & $1.3 \times 10^{-124}$ \\
      tracking oscillating energy      & $M$                      & $1.8 \times 10^{-126}$ \\
                                       & $\lambda$                & $4$ \\
                                       & $A$                      & $0.99$ \\
                                       & $\nu$                    & $2.7$ \\
      interacting quintessence         & $A$                      & $1.4 \times 10^{-119}$ \\
                                       & $B$                      & $9.7$ \\
                                       & $C$                      & $16$ \\
      Chaplygin gas                    & $\alpha$                 & $1$ \\
                                       & $A$                      & $2.8 \times 10^{-246}$ \\                                      
      \hline
   \end{tabular}
   \end{center}
   \caption[Free parameters of the DDE models illustrated in Fig.\ \ref{fig:C3quintessence_energy}.]
   {Free parameters of the DDE models
   illustrated in Fig.\ \ref{fig:C3quintessence_energy}. These values were chosen such 
   that observational constraints are crudely satisfied. These are by no means the 
   only combinations fitting observations. These values are intended for the purposes of
   illustration in Fig.\ \ref{fig:C3quintessence_energy}. Units are Planck units.}
   \label{tab:C3paramvals}
\end{table}






%

%
\pagestyle{fancy}
\chapter{Comparing the Sun to Other Stars: Searching for Life Tracers Amongst the Solar Properties}
\label{chap4}

\begin{center}
\emph{
Here comes the Sun, \\
here comes the Sun, \\
and I say it's alright.}
\end{center}
\begin{flushright}
- The Beatles, ``Here Comes the Sun'' \hspace*{2cm}
\end{flushright}
\vspace{1cm}



\section{Introduction}

In the past decade 
the first several hundred extra-solar planets have been detected using various 
techniques. These techniques are all biased towards massive planets in 
small orbits (so-called hot Jupiters). The next generation of 
projects (including NASA's Terrestrial Planet Finder and ESA's Darwin Project) 
is eagerly anticipated, with the discovery of Earth-like planets orbiting within
habitable zones expected. In the mean time, the search for 
solar twins (stars with properties most like those of the Sun) in stellar surveys, continues. 
The reason behind our fascination with Earth-like planets and Sun-like stars is 
that without a thorough understanding of the requirements of life, a reasonable 
strategy in the search for extra terrestrial life is to look in environments that we 
know are capable of hosting life, i.e.\ those planets and stars which are most like 
our own.

If the emergence of life and observers on a planet depends on special
properties of the planet's host star, then we would expect our Sun to exhibit
those properties and stand out when compared to a sample of randomly 
selected stars. In this way, the comparison of the Sun to other stars may be a way of 
identifying stellar properties important to the origin of life and the evolution of 
observers.

What we stand to gain by looking for anomalous properties in the Sun are 
statements about the dependence of life on various stellar properties (the 
reliability of which can be quantified). Such information could be used to
improve searches for life in the universe by focussing them on the most 
important properties.

An early example of this type of work is \citet{Gonzalez1999a,Gonzalez1999b}
who proposed, based on his findings that the Sun was more massive than 
$91\%$ of all stars, that life may exist preferentially around high-mass stars.
%

Taken together however, previous work of this type is inconsistent in its
conclusions. While \citet{Gonzalez1999a,Gonzalez1999b,Gonzalez2001}  
suggested the Sun to be anomalous, \citet{Gustafsson1998,Allende2006} 
found it to be typical.
These discrepancies result from inconsistent use of language, stellar sample
selection and inconsistency in the choice of stellar/solar properties compared. 

In order to clarify these issues we have undertaken a joint $11$-parameter 
$\chi^2$ analysis that compares the Sun to representative samples of stars in $11$ 
independent  properties plausibly related to life and habitability. The analysis 
quantifies the degree of (a)typicality of the Sun and draws conclusions about 
the legitimacy of postulated links between particular properties and habitability.

\section{Selection of Solar Properties and Stellar Samples}

Since the purpose of our analysis is to identify significantly anomalous 
solar properties (or a lack thereof), it is important that our selection criteria is not 
dependent on any prior knowledge we may have about the (a)typicality of the Sun 
with respect to its properties. Intentionally selecting one (or a few) parameters
in which the Sun is known to be anomalous would pre-load the result.

Suppose, for example, that the Sun has previously been compared to 
representative samples of stars in $20$ uncorrelated parameters, none of which 
play any significant role in determining habitability. It should be expected that the 
Sun is a $2\sigma$ ($95$-percentile) outlier in one in these parameters (call it
$X$) just by pure chance. If the present analysis were conducted using just $X$, 
naive that $X$ had been selected, the results would erroneously suggest that 
$X$ was related to habitability. 

On the other hand, the indiscriminant inclusion of properties unlikely to have 
any connection to habitability could dilute a legitimate signal. Suppose that $Y$ 
is a stellar property, and that only stars in the upper $2.5\%$ with respect to $Y$ 
are capable of hosting life (the upper $2.5\%$ are amongst the $95$-percentile 
outliers). The $2\sigma$ signal can be diluted away by including 
$\sim 1/0.05 = 20$ other parameters (and can be reduced to a $90\%$ signal by 
including just $2$ other parameters).

No single stellar survey contains unbiased measurements in as many parameters
as we are interested in, and for this reason we have used different stellar samples
for each parameter. The benefit of being able to choose the best available sample 
for each parameter comes at a cost, which is that we must eliminate any correlated
parameters from our analysis. 

With the above considerations in mind, we have included $11$ maximally 
uncorrelated stellar properties all of which are plausibly related to habitability and
for which a sufficiently large unbiased sample of stellar values exists.
The included properties come from a full set of $23$ candidates (refer to 
\citet{Robles2008a} for the full list and correlation analysis). Below we give a brief
description of the $11$ included properties, along with a brief description of their 
relevance to habitability and a description of the stellar samples we have used for 
each. 
\begin{enumerate}
	\item Mass: The mass of a star is arguably the most important property of a star. 
		It determines luminosity, temperature and main sequence longevity, in turn 
		influencing conditions and stability in the circumstellar habitable zone. Low 
		mass stars are intrinsically dim, so large stellar samples are biased towards 
		high-mass stars. We have used the nearest $125$ stars from the RECONS 
		compilation \citep{Henry2006}, which is complete to $7.1\ pc$.
	\item Age: If the evolution of observers takes (on average) much longer than 
		typical lifetime of a star \citep{Carter1983} then observers may be expected to
		arise preferentially around older stars. We construct an age distribution for stars
		in the galaxy using galactic star formation history from \citet{RochaPinto2000}.
		Their star formation history is based on chromospheric ages of $552$ dwarf
		stars at up to $200\ pc$, and has been corrected for scale-height, stellar 
		evolution and volume incompleteness.
		We also consider the cosmic age distribution, using the cosmic star formation 
		history from \citet{Hopkins2006}.
	\item Metallicity [Fe/H]: A star's iron content is a good proxy for its abundance 
		in other elements heavier than helium. With regard to this analysis, metallically 
		is correlated with abundance of the ingredients for terrestrial planets 
		(O, Fe, Si and Mg) and life (C, O, N and S). We use the sample of $453$ FGK 
		stars of \citet{Grether2006,Grether2007} selected from the Hipparcos catalogue.
		This sample is complete to $25\ pc$ for stars within the spectral range 
		$F7$-$K3$ and absolute magnitude of $M_{V} \ge 8.5$ \citep{Reid2002} and 
		contains metallicities derived from a range of spectroscopic and photometric 
		surveys.
	\item Carbon-to-oxygen ratio [C/O]: The relative abundance of carbon to 
		oxygen impacts the abundance of oxygen (and the balance of REDOX 
		chemistry) in the circumstellar habitable zone. If [C/O] is higher than
		$1$, most oxygen forms carbon monoxide which is subsequently cleared 
		by stellar winds leaving a chemically reducing habitable zone 
		\citep{Kuchner2005}. For our stellar distribution in [C/O] we use $256$ stars 
		from \citet{Gustafsson1999,Reddy2003,Bensby2006}.
	\item Magnesium-to-silicon ratio [Mg/Si]: After [Fe/H] and [C/O], the 
		magnesium to silicon ratio is the next most important elemental abundance
		ratio, also impacting terrestrial planet chemistry.
		Our stellar distribution in [Mg/Si] consists of $231$ stars from 
		\citet{Reddy2003,Bensby2005}.
	\item Rotational velocity $v \sin i$: The rotational velocity of a star is related
		to the angular momentum of the protoplanetary disk. A low rotational 
		velocity may be correlated with the presence of planets \citep{Soderblom1983}.
		We use the subset of $276$ stars in the $0.9$-$1.1\msol$ mass range from
		the sample of \citet{Valenti2005}. By cutting near the solar value in mass
		we minimize the effects of a known correlation between stellar mass and 
		$v \sin i$ which becomes significant at higher masses.
	\item Eccentricity of the star's galactic orbit $e$: The galactic orbit of a star
		determines the stellar environments that the star passes through. Stars with
		highly eccentric orbits pass closer to the galactic center where the risk of a 
		nearby supernova, and the flux of potentially harmful radiation is higher.
		We use the distribution of stellar eccentricities from $1987$ stars witin 
		$40\ pc$ as computed by \citet{Nordstrom2004}.
	\item Maximum height away from the galactic plane $Z_{\textrm{max}}$: 
		As the Sun oscillates through the thin disk of the galaxy objects in the Oort
		cloud may be disrupted by tidal gravitational forces. It is plausible that the 
		frequency of such disruptions (for which $Z_{\textrm{max}}$ is a proxy) 
		influences the frequency of impacts on planets in the habitable zone.
		For $Z_{\textrm{max}}$ we use the same stellar sample as we did for eccentricity.
	\item Mean galactocentric radius $R_{\textrm{Gal}}$: The mean galactocentric 
		radius is the most important of a star's galactic orbital properties. It determines
		(to a greater degree than eccentricity for typical eccentricities) the minimum 
		approach to the galactic center, the risk of nearby supernova, and the flux 
		of potentially harmful radiation from the galactic center.
		Our distribution of $R_{\textrm{Gal}}$ is based on the model of 
		\cite{Bahcall1980} assuming a scale length of $3.0\pm0.4\ kpc$ 
		\citep{Gould1996}.
	\item The stellar mass of the star's host galaxy $M_{\textrm{gal}}$: This 
		influences the overall (galactic) metallically, including the metallically of 
		the star's neighbourhood.
		We construct a distribution of $M_{\textrm{gal}}$ based on the $K$-band
		luminosity function of \cite{Loveday2000} and assume a constant 
		stellar-mass-to-light ratio of $0.5$ \citep{Bell2001}. It is important to note 
		that we are interested in the distribution of stars in $M_{\textrm{gal}}$, not 
		the distribution of galaxies in $M_{\textrm{gal}}$. 
	\item The stellar mass of the star's host group of galaxies $M_{\textrm{group}}$: 
		This is correlated with the density of the 
		galactic environment. This influences the stability of the host galaxy.
		We use the $B$-band luminosity distribution of galactic groups from the 
		2dFGRS Percolation-Inferred Galaxy Groups catalogue \citep{Eke2004} 
		and assume a constant stellar-mass-to-light ratio of 1.5 \citep{Bell2001}.
		As for host galaxy mass, we are interested in the distribution of stars in 
		$M_{\textrm{group}}$, not the distribution of groups in $M_{\textrm{group}}$.
\end{enumerate}

\section{Analysis and Results}

The solar values $x_{\odot ,i}$ of the selected properties $i=1,11$ are taken (or
derived from related properties) from the literature. The solar values are shown in
Table \ref{tab:C4solarvalues}.

\begin{table}[!hbp]
   \begin{center}
   \begin{tabular}{l l l l l}
      \hline
      $i$		& Parameter $[units]$			& Solar Value $x_{\odot ,i}$		& Median Value $\mu_{1/2,i}$	& $\sigma_{68,i}$ \\
      \hline
      \hline
      $1$		& Mass $[\msol]$				& $1^a$						& $0.33$					& $0.37$ \\
      $2$		& Age $[Gyr]$					& ${4.9^{3.1}_{2.7}}^b$			& $5.4$					& $3.25$ \\
      $3$		& $[\textrm{Fe}/\textrm{H}]$		& $0^a$						& $-0.08$					& $0.20$ \\
      $4$		& $[\textrm{C}/\textrm{O}]$			& $0^a$						& $0.07$					& $0.09$ \\
      $5$		& $[\textrm{Mg}/\textrm{Si}]$		& $0^a$						& $0.01$					& $0.04$ \\
      $6$		& $v \sin i$ $[km\ s^{-1}]$			& $1.28^c$					& $2.51$					& $1.27$ \\
      $7$		& $e$						& $0.036\pm0.002^d$			& $0.10$					& $0.05$ \\
      $8$		& $Z_{\textrm{max}}$ $[kpc]$		& $0.104\pm0.006^e$			& $0.14$					& $0.10$ \\
      $9$		& $R_{\textrm{Gal}}$ $[kpc]$		& $7.62\pm0.32^f$				& $4.9$					& $5.03$ \\
      $10$		& $M_{\textrm{Gal}}$ $[\msol]$		& ${10^{10.55\pm0.16}}^g$		& $10^{10.2}$				& $0.47^{\alpha}$ \\
      $11$		& $M_{\textrm{group}}$ $[\msol]$	& ${10^{10.91\pm0.07}}^h$		& $10^{11.1}$				& $0.47^{\alpha}$ \\
      \hline
   \end{tabular}
   \end{center}
   \caption[Comparing the solar values of $11$ parameters to corresponding stellar distributions in those parameters.]
   	{The solar values of the $11$ properties included in this analysis. \\
   	a. By definition. \\
	b. Chromospheric age of the Sun from \citet{Wright2004}. \\
	c. Rotational velocity at the surface is $v = 1.63\ km\ s^{-1}$ \citep{Valenti2005}. We 
	calculate a $v \sin i$ which may be compared to our stellar $v \sin i$ distribution by 
	apply a factor of $\frac{\pi}{4}$ (the average over of $\sin i$ over a $3$-sphere; this 
	simulates randomizing the inclination at which the Sun is viewed). \\
	d. The eccentricity of the Sun's galactic orbit is calculated from solar motion reported 
	in \citet{Dehnen1998}. \\
	e. Found by integrating the solar orbit in the galactic potential (initial motion as 
	reported in \citet{Dehnen1998}). \\
	f. \citet{Eisenhauer2005} \\
	g. Derived in the same way as our distribution: by applying a $0.5$ stellar-mass-to-light
	ratio to the $K$-band luminosity. The $K$-band luminosity is inferred from the $V$-band
	measured by \citet{Courteau1999} using the mean color of an Sbc spiral galaxy from
	the 2MASS Large Galaxy Atlas \citep{Jarrett2003} and the color conversion prescription 
	of \citet{Driver1994}.  \\
	h. Derived in the same way as our distribution: by applying a $1.5$ stellar-mass-to-light
	ratio to the $B$-band luminosity of 2PIGG and Local Group galaxies from \cite{Courteau1999}. \\
	$\alpha$. These distributions span several orders of magnitude and are analyzed 
	in $\log_{10}$-space. The $\sigma_{68,i}$ values for $M_{\textrm{Gal}}$ and 
	$M_{\textrm{group}}$ represent widths measured in $\log_{10}$-space.
	}
   \label{tab:C4solarvalues}
\end{table}

We find the median $\mu_{1/2,i}$ and the $16^{\textrm{th}}$ and $84^{\textrm{th}}$ 
percentiles for each of the distributions. $68\%$ of a distribution is between the 
$16^{\textrm{th}}$ and $84^{\textrm{th}}$ percentiles, and in the case of a Gaussian 
distribution this interval corresponds to the $\pm1\sigma$ interval. 

Our distributions are generally asymmetric. Since we are calculating the deviation of 
the Sun from the mean stellar sample we are most interested in the width of the 
distributions in the direction of the solar value. We define $\sigma_{68,i}$ as the
difference between the median and the $16^{\textrm{th}}$ or $84^{\textrm{th}}$ percentile, 
depending on whether the solar value is below or above the median.

\begin{figure*}[!p]
  \begin{center}
  	\includegraphics[width=0.7\linewidth]{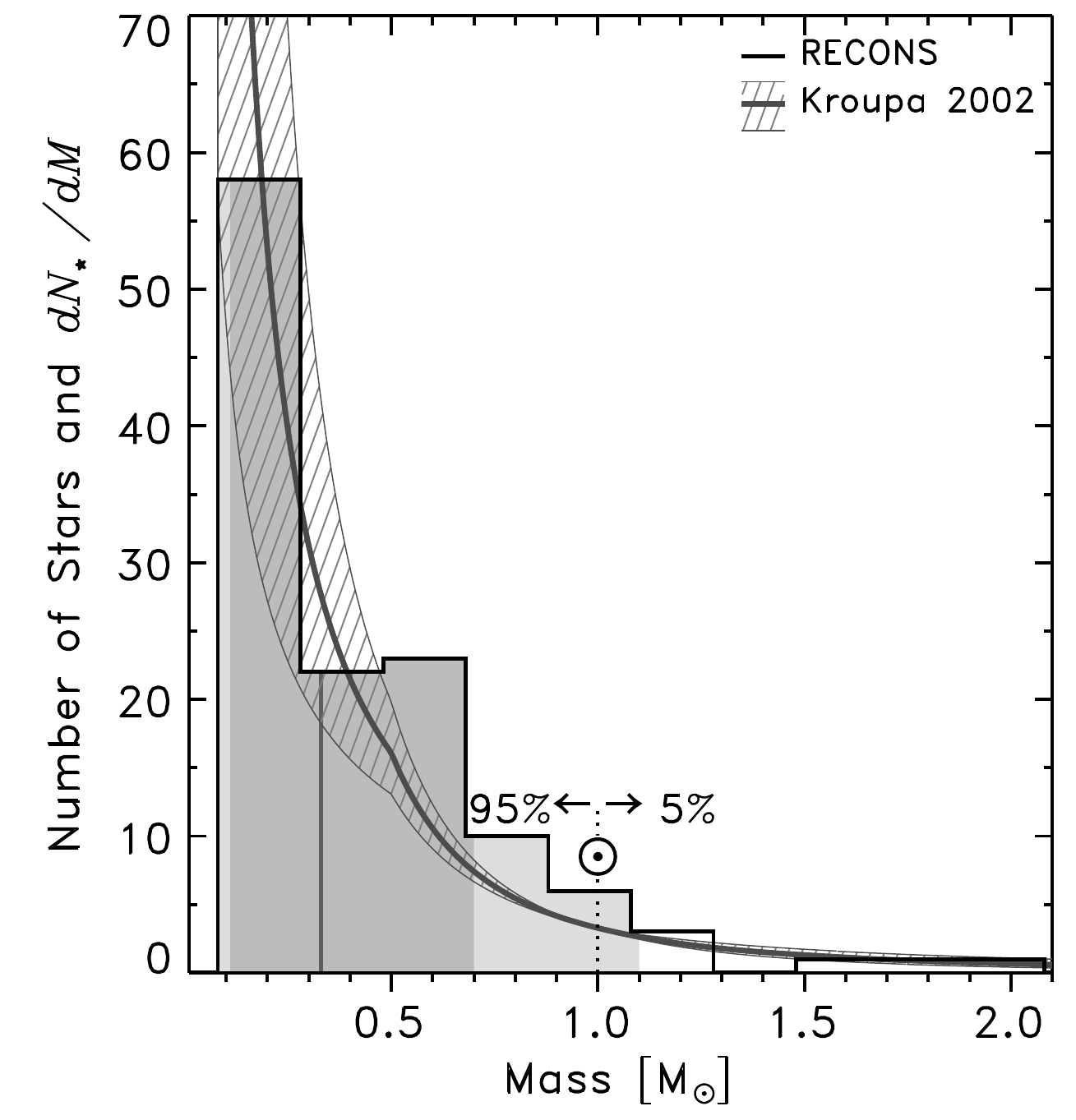}
    	\caption[Comparing the Sun to other stars in mass.]
	{
	The histogram shows the masses of the volume complete sample of $125$ stars 
	from RECONS \citep{Henry2006}. The median of the distribution is indicated by the 
	vertical grey line, and the $68\%$ and $95\%$ intervals are represented by the dark 
	grey and light grey shades respectively.
	The Sun, indicated by the ``$\odot$'', is found to be more massive than $95 \pm 2 \%$ 
	of stars in the Universe when the RECONS sample is used to represent the 
	cosmic distribution of stellar mass.
	We have also over-plotted the initial mass function \citep{Kroupa2002}, which we 
	have normalized to $125$-stars for the purposes of this plot. The initial mass function
	may be expected to trace the stellar mass function, since stars with masses lower than 
	$\sim 1 M_{\odot}$ have main sequence lifetimes longer than the age of the 
	Universe. We find good agreement between the histogram and the IMF: the Sun
	is more massive than $94 \pm 2 \%$ of all stars when the IMF is used instead of 
	the RECONS sample.
    	}
    	\label{fig:C4josef1}
  \end{center}
\end{figure*}

\begin{figure*}[!p]
  \begin{center}
  	\includegraphics[width=0.7\linewidth]{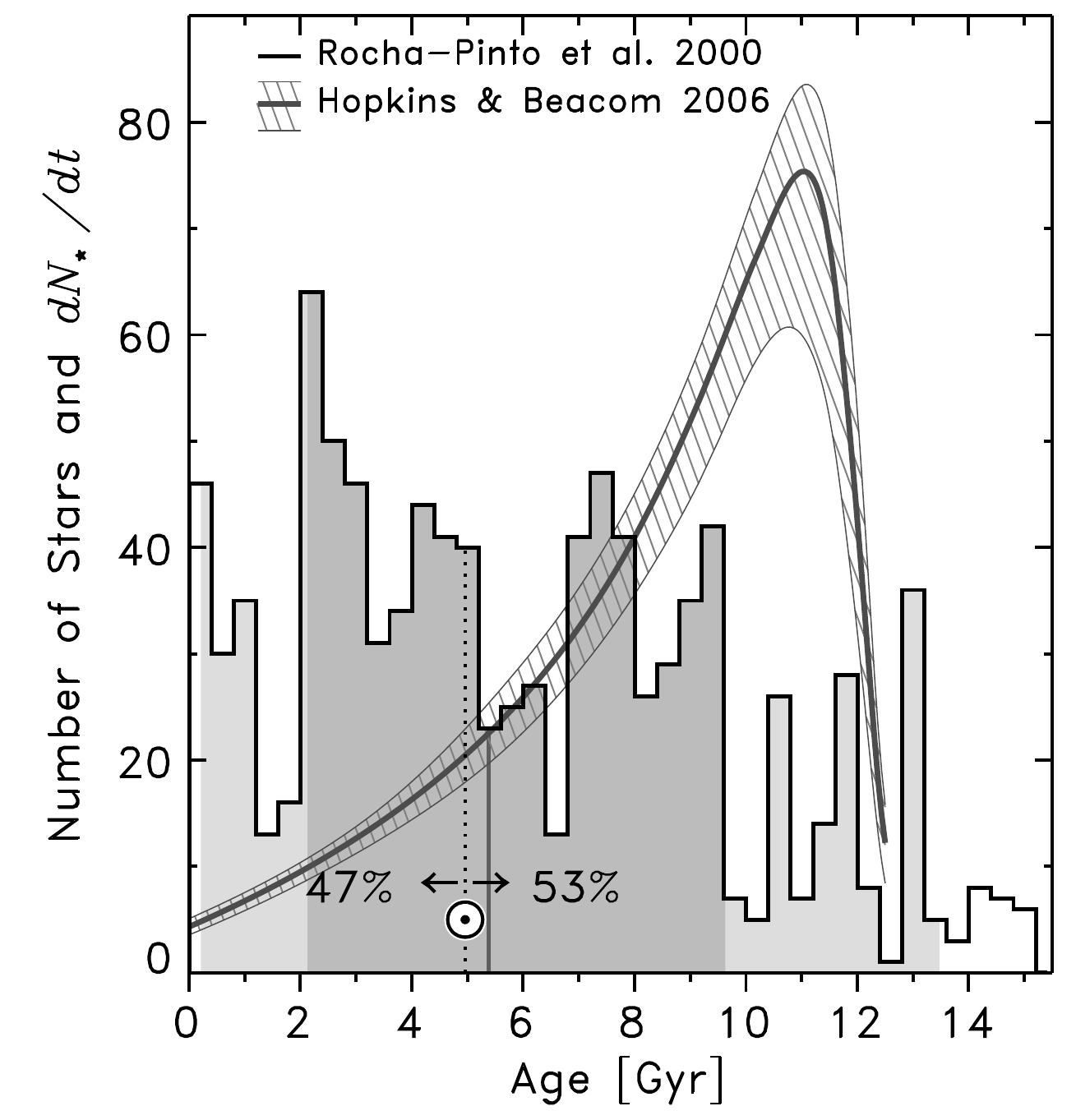}
    	\caption[Comparing the Sun to other stars in age.]
	{
	The distribution of ages of stars in our galaxy (the histogram) is inferred from the 
	galactic star formation history of \citet{RochaPinto2000}. The Sun is found to be 
	younger than $53 \pm 2 \%$ of stars in the disk of the galaxy.
	The distribution of cosmic stellar ages from the cosmic star formation history of 
	\cite{Hopkins2006} (over-plotted) peaks at around $10\ Gyrs$, representing the 
	burst of star formation associated with giant ellipticals $1$-$4\ Gyrs$ after the 
	big bang. The Sun is younger than $86 \pm 5\%$ of stars in the Universe.
    	}
    	\label{fig:C4josef2}
  \end{center}
\end{figure*}

\begin{figure*}[!p]
  \begin{center}
  	\includegraphics[width=0.7\linewidth]{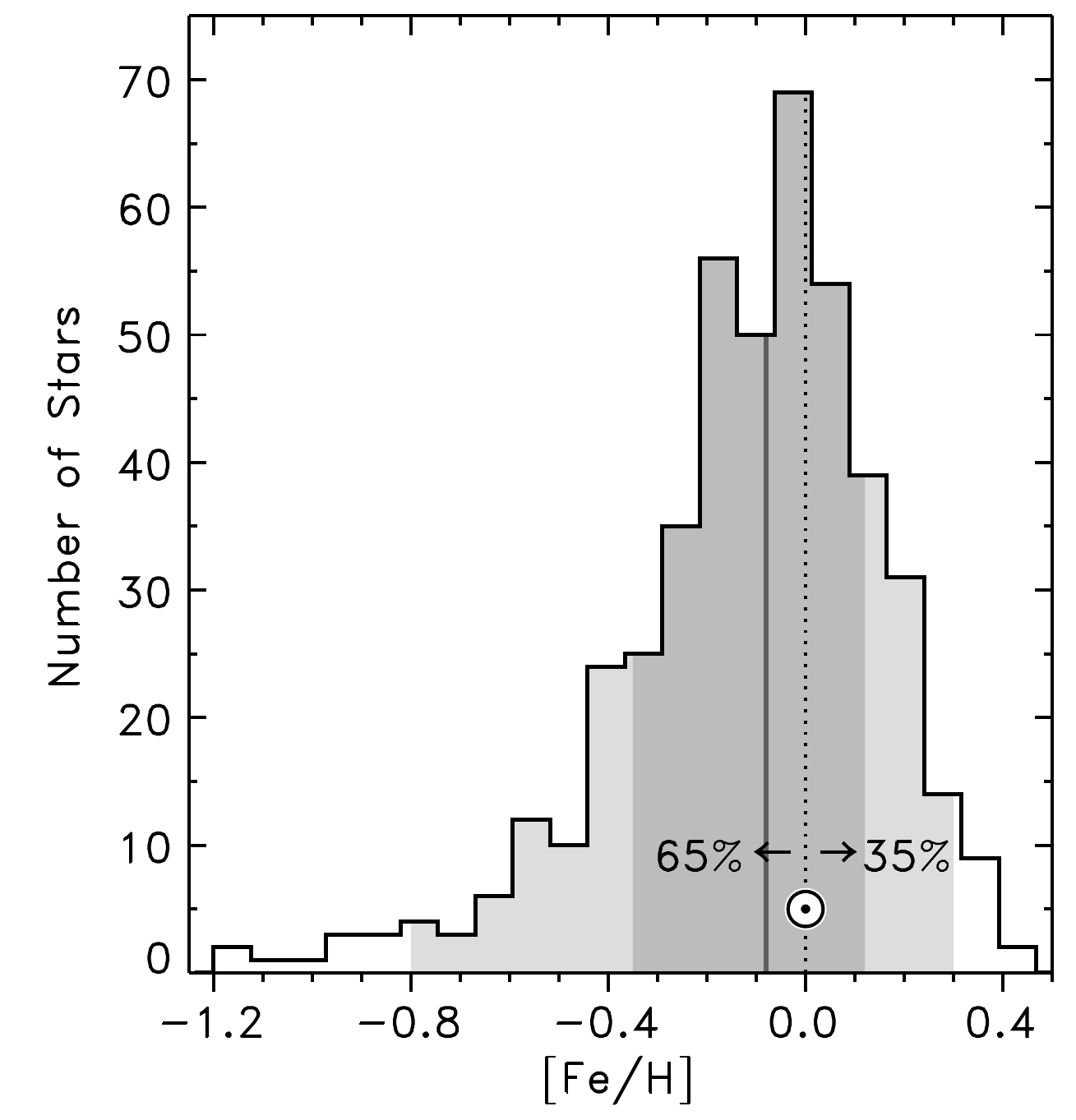}
    	\caption[Comparing the Sun to other stars in metallicity.]
	{
	The distribution of stellar metallicities [Fe/H], represented by $453$ FGK 
	Hipparcos stars selected by \citet{Grether2007}. The Sun's metallicity is higher 
	than $65 \pm 2 \%$ of other stars.
    	}
    	\label{fig:C4josef3}
  \end{center}
\end{figure*}

The medians $\mu_{1/2,i}$ and widths $\sigma_{68,i}$ of the stellar distributions are
shown in the last two columns of Table \ref{tab:C4solarvalues}. Figures \ref{fig:C4josef1}, 
\ref{fig:C4josef2} and \ref{fig:C4josef3} show the stellar distributions in first three properties: 
mass, age and metallicity. These figures have been included as examples. The 
equivalent figures for parameters $4$-$11$ are omitted from this summary, but can be
found in the reviewed publication of this work \citep{Robles2008a}.

We compute the combined deviation of the solar properties from those of an 
average star using the $\chi^2$ statistic. 
\begin{eqnarray}
	\chi^2_{\odot} = \sum^{N=11}_{i=1} \frac{(x_{\odot ,i} - \mu_{1/2,i})^2}{\sigma^2_{68,i}} \label{eq:C4chi2odot}
\end{eqnarray}
While uncertainties in the solar values are propagated directly, we employ a bootstrap 
\citep{Efron1979} to account for the uncertainties in $\mu_{1/2,i}$ and $\sigma_{68,i}$ 
due to small number statistics. This involves randomly repeatedly resampling the distributions 
(from the original samples, but allowing the same star to be drawn multiple times), and 
calculating the corresponding $\mu_{1/2,i}$, $\sigma_{68,i}$ and $\chi^2_{\odot}$ 
each time. 

We find the combined $11$-parameter solar chi-square to be
\begin{eqnarray}
	\chi^2_{\odot}  = 8.39\pm0.96.
\end{eqnarray}

The probability of selecting, at random, a star which is more normal than the Sun 
(a star with a lower $\chi^2$) can be calculated using the standard $\chi^2$ distribution 
for $11$ degrees of freedom.
\begin{eqnarray}
	P(< \chi^2_{\odot} | N=11 ) = 0.32 \pm 0.09
\end{eqnarray}
 
The probability calculated above relies on the assumption that our $11$ properties are 
normally distributed. Since several of our distributions are poorly approximated by 
Gaussians (for example age, see Figure \ref{fig:C4josef2}), we have performed a Monte 
Carlo simulation \citep{Metropolis1949} to calculate a more accurate value of $P$. 
In the Monte Carlo simulation a large number of stars are randomly selected from 
our distributions and their $\chi^2$ are calculated (according to equation 
\ref{eq:C4chi2odot}). In this way we find that the probability of randomly selecting
a star more normal than the Sun is
\begin{eqnarray}
	P_{MC}(< \chi^2_{\odot} | N=11 ) = 0.29 \pm 0.11.
\end{eqnarray}

\section{Discussion}

\begin{figure*}[!p]
  \begin{center}
    \begin{tabular}{rr}
      \includegraphics[scale=0.5]{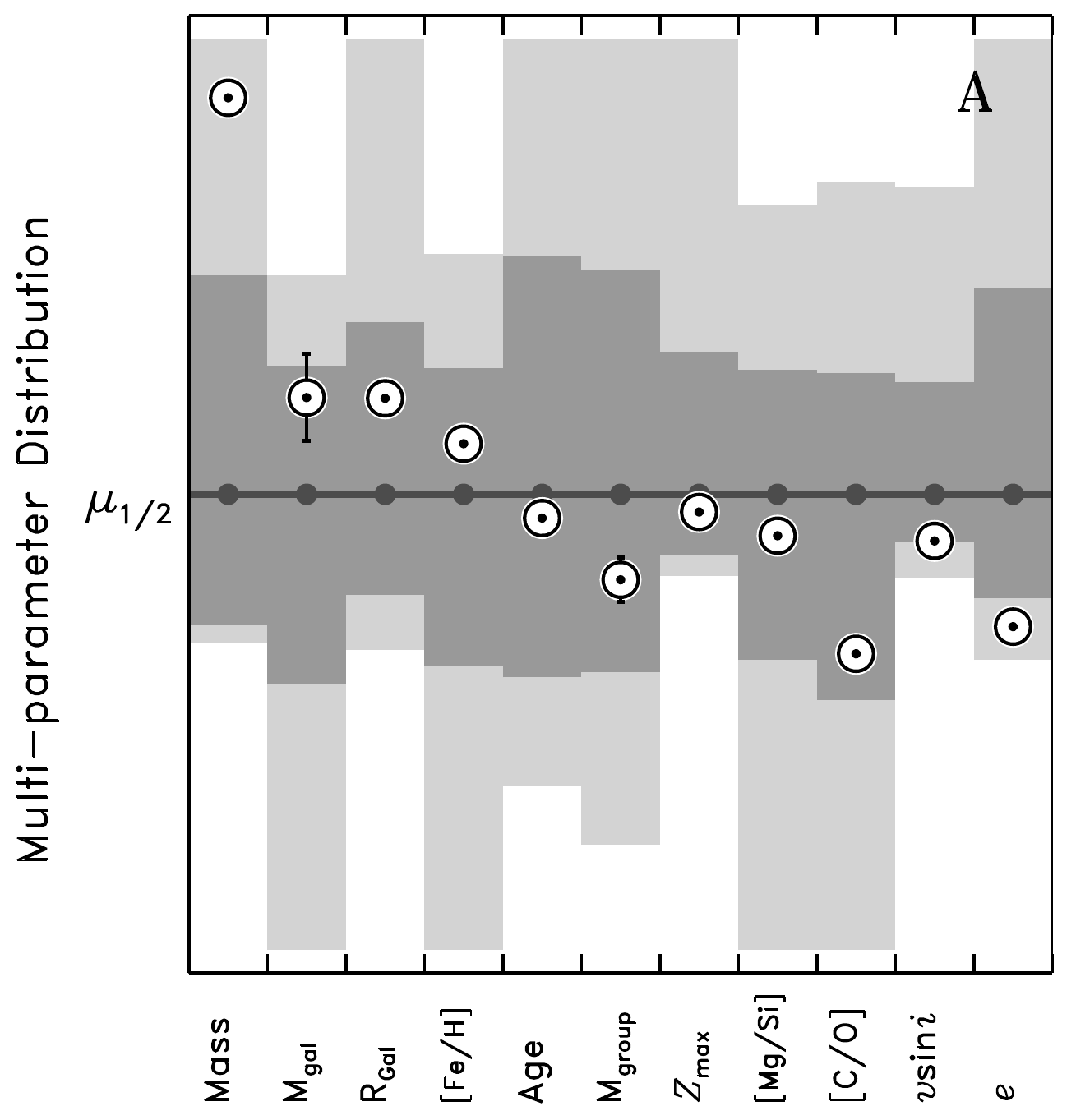} & \includegraphics[scale=0.5]{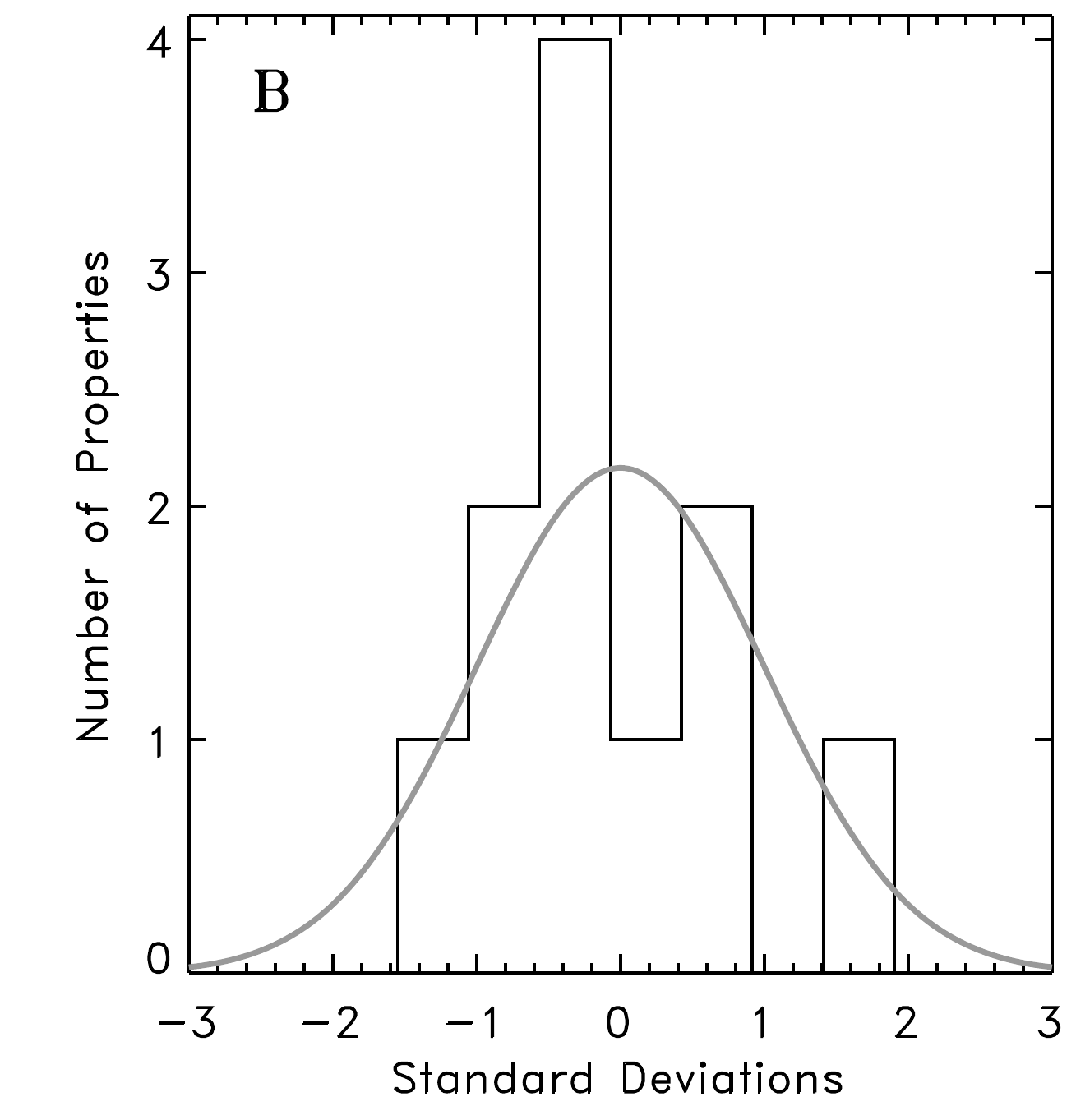} \\
      \includegraphics[scale=0.5]{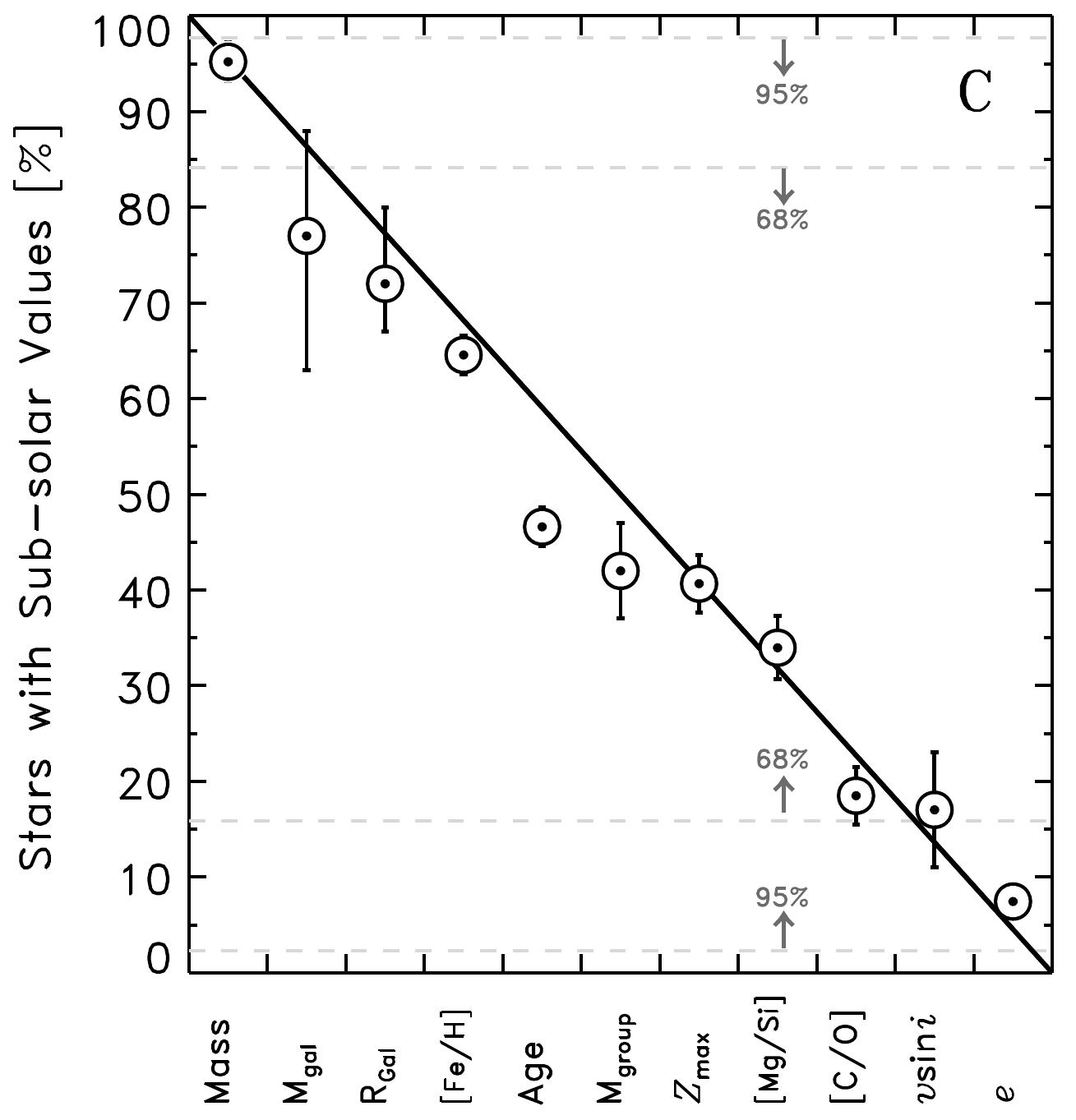} & \includegraphics[scale=0.5]{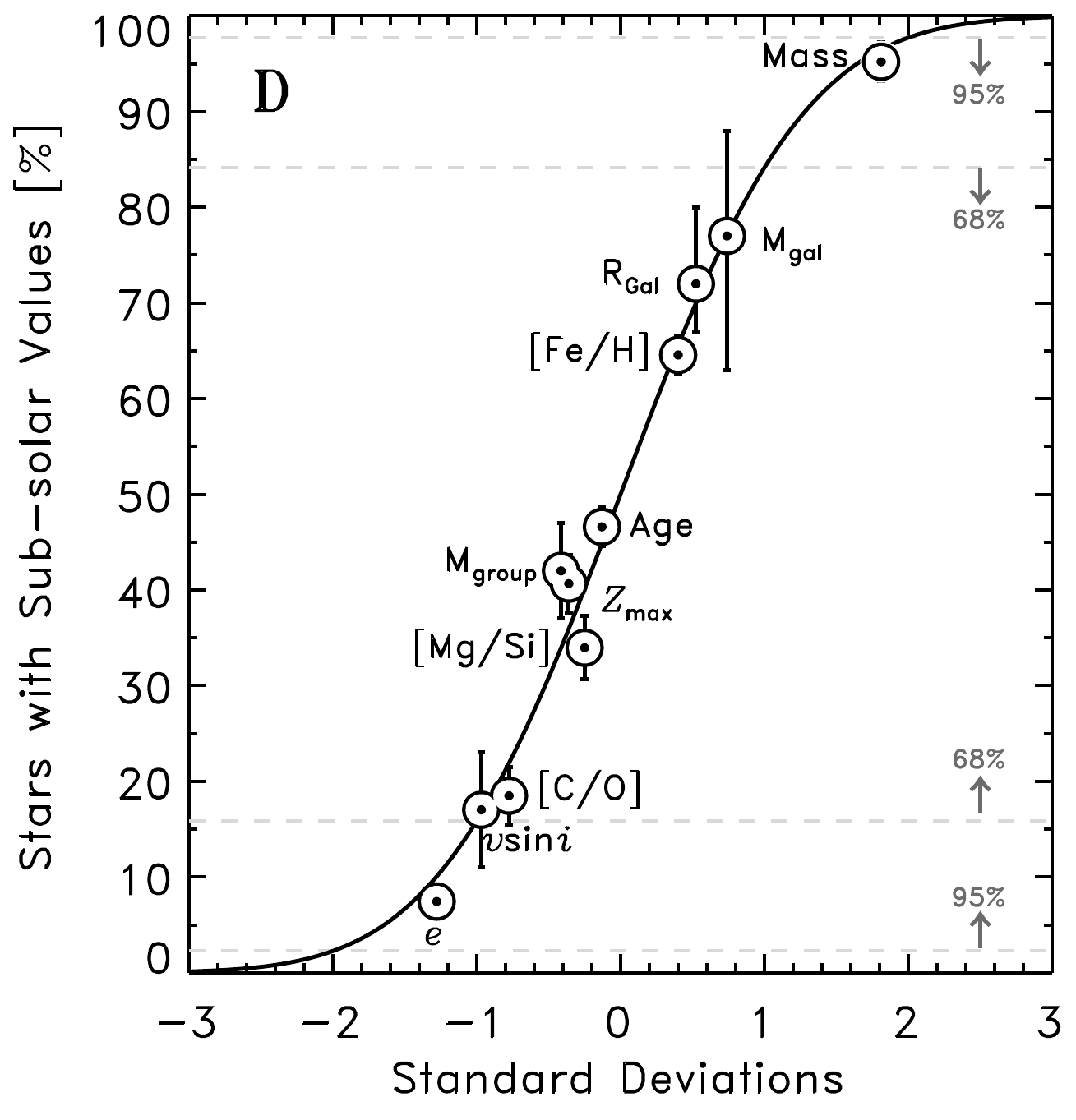} \\
    \end{tabular}
    \caption[The results of our $11$ parameter analysis comparing the Sun to other stars.]{
    Various representations of our main results. \\
    A. Comparing the solar values to the $68\%$ (dark grey) and $95\%$ (light grey) 
    intervals for each parameter. The horizontal line represents the median of
    each of the distributions, and the vertical axis is normalized to the longer of the two 
    $95\%$ intervals for each of the distributions. \\
    B. A histogram of the distribution of parameters in standard deviation compared to
    a gaussian. \\
    C. The percentage of stars $n\%$ with sub-solar values for each of the parameters 
    in decreasing order of $n\%$. The line represents the expected arrangement of
    points ``$\odot$'' for randomized parameter values. Significant deviations from the 
    line may indicate unexpected anomalies, none of which are seen here. \\
    D. Each of the solar properties is plotted in $n\%$-standard deviation space. If the
    distributions were Gaussian the points would fall along the solid line.    
    }
    \label{fig:C4josef13}
  \end{center}
\end{figure*}

Of the $11$ stellar properties included in this study, the Sun is most anomalous in
its mass (more massive than $95 \pm 2\%$ stars in our sample), in the eccentricity of its 
galactic orbit (which is lower than $93 \pm 1\%$ of stars in our sample) 
and in its rotational velocity (which is slower than $83 \pm 7\%$ of stars in our sample). 
Panel A of figure \ref{fig:C4josef13} illustrates how the solar values compare to the 
$68\%$ and $95\%$ intervals for each of the parameters. The levels of solar 
anomaly for each of the properties are shown in Panel D as a percentile, and in 
terms of the number of standard deviations from the median.

When mass, eccentricity and rotational velocity are combined with our other 
parameters, which are also plausibly related to habitability, the Sun looks entirely 
mediocre. The probability of selecting a star, at random, which is more typical than 
the Sun with respect to these parameters is only $23\pm11\%$. 

This result undermines suggestions that an anthropic explanation is required
for the Sun's unusually large mass. The alternative explanation, which is defended 
by our results is that in measuring the properties of the Sun we have come across a 
few mild outliers - as few and as mild as we should expect for a random star, given 
the number of properties measured. A convenient visualization of this is given in 
Panel C of figure \ref{fig:C4josef13}. The $11$ properties are arranged in decreasing 
order of $n\%$, where $n\%$ is the percentage of stars with sub-solar values in a 
given property. When arranged in this way, we expect the parameters to be near 
the line given by
\begin{eqnarray}
	n_{j,\textrm{expected}}\% = \left[ 1 - \frac{(j-1/2)}{N} \right] \times 100\%
\end{eqnarray}
where $N=11$ is the number of parameters.
Any anomalies that cannot be attributed to noise would appear up as points 
significantly far from the line.

We have repeated the analysis without mass and eccentricity and find the Sun to
be unexpectedly average: only $7 \pm 4 \%$ of stars are more average than the 
Sun with respect to the remaining $9$ parameters. This supports the proposition
that the anomalies observed in mass and eccentricity are expected by pure 
chance. By including just mass and eccentricity, we find that the Sun is more 
anomalous than $94 \pm 4 \%$ of stars. This would be some evidence ($2\sigma$) 
for an anthropic explanation for one or both of these properties, if they had not
been selected to ensure this result.

Our parameter selection criteria selected a larger number of properties plausibly
related to life. Without strong evidence suggesting that any are more important 
than any other, they have been treated equally. A potential extension of this work
would be the inclusion of factors to weight the plausibility (based on planetary 
formation models and biology) that a property is related to life.

\pagestyle{fancy}
\chapter{A Larger Estimate of the Entropy of the Universe}
\label{chap5}

\begin{center}
\emph{
We are all \\
made of stars. \\
We were created in the birth of stars.
}
\end{center}
\begin{flushright}
- Encounter, ``Starborn'' \hspace*{2cm}
\end{flushright}
\vspace{1cm}



%

%
\newif\ifTRACKCHANGES
\TRACKCHANGESfalse
\ifTRACKCHANGES
   \newcommand{ \removethis}[ 1]{ \textcolor{red}{ #1 }}
   \newcommand{ \addthis}[ 1]{ \textcolor{blue}{ #1 }}
\else
   \newcommand{ \removethis}[ 1]{}
   \newcommand{ \addthis}[ 1]{#1}
\fi

%

\newboolean{colver}
\setboolean{colver}{true}

\section{Introduction} \label{C5introduction}

The entropy budget of the universe is important because its increase is associated
with all irreversible processes, on all scales, across all 
facets of nature: gravitational clustering, accretion disks, supernovae, stellar fusion, 
terrestrial weather, and chemical, geological and biological processes 
\citep{Frautschi1982,LineweaverEgan2008}.

Recently, \citet{Frampton2008} and \citet{Frampton2008b} 
reported the entropy budget of the observable universe. Their budgets (listed aside 
others in Table \ref{tab:C5currententropy}) estimate the total entropy of the observable 
universe to be $S_{\mathrm{obs}} \sim 10^{102} k - 10^{103} k$, dominated by the entropy of 
supermassive black holes (SMBHs) at the centers of galaxies.  
That the increase of entropy has not yet been capped by some limiting value, 
such as the holographic bound \citep{tHooft1993,Susskind1995} at 
$S_{\mathrm{max}} \sim 10^{123} k$ \citep{Frampton2008}, is the reason dissipative 
processes are ongoing and that life can exist. 

In this paper, we improve the entropy budget by using recent observational data
and quantifying uncertainties.
The paper is organized as follows. In what remains of the Introduction, we describe 
two different schemes for quantifying the increasing entropy of the universe, and we 
comment on caveats involving the identification of gravitational entropy.
Our main work is presented in Sections \ref{C5matter} and  \ref{C5scheme2}, where we 
calculate new entropy budgets within each of the two accounting schemes. 
We finish in Section \ref{C5discussion} with a discussion touching on the time evolution 
of the budgets we have calculated, and ideas for future work.

Throughout this paper we assume flatness ($\Omega_k = 0$) as predicted by 
inflation \citep{Guth1981,Linde1982} and supported by observations 
\citep{Spergel2007}. Adopted values for other cosmological parameters are 
$h = 0.705 \pm 0.013$, 
$\omega_b = \Omega_b h^2 = 0.0224 \pm 0.0007$,
$\omega_m = \Omega_m h^2 = 0.136 \pm 0.003$ \citep{Seljak2006}, 
and
$T_{\mathrm{CMB}} = 2.725 \pm 0.002\ K$ (\citealt{Mather1999}; quoted uncertainties are $1\sigma$).


\subsection{Two Schemes for Quantifying the Increasing Entropy of the Universe} \label{C5frwgsl}

Modulo statistical fluctuations, the generalized second law of thermodynamics 
holds that the entropy of the universe (including Bekenstein-Hawking entropy in 
the case of any region hidden behind an event horizon), must not decrease 
with time \citep{Bekenstein1974,Gibbons1977}. Within the FRW framework, the 
generalized second law can be applied in at least two obvious ways: 
\begin{enumerate}
	\item The total entropy in a sufficiently large comoving volume of the universe 
		does not decrease with cosmic time,
		\begin{equation}
			d S_{\mathrm{comoving\ volume}} \ge 0. \label{eq:C5one}
		\end{equation}
	\item The total entropy of matter contained within the cosmic event horizon 
		(CEH) plus the entropy of the CEH itself, does not decrease with cosmic 
		time,
		\begin{equation}
			d S_{\mathrm{CEH\ interior}} + d S_{\mathrm{CEH}} \ge 0. \label{eq:C5two}
		\end{equation}
\end{enumerate}

In the first of these schemes, the system is bounded by a closed comoving surface.
The system is effectively isolated because large-scale homogeneity and 
isotropy imply no net flows of entropy into or out of the comoving volume. The 
time-slicing in this scheme is along surfaces of constant cosmic time. 
Event horizons of black holes are used to quantify the entropy of black holes, 
however the CEH is neglected since the assumption of 
large-scale homogeneity makes it possible for us to keep track of the entropy of matter 
beyond it. A reasonable choice for the comoving volume in this scheme is the comoving 
sphere that presently corresponds to the observable universe, i.e., the 
\ifthenelse{\boolean{colver}} { 
gray area in Figure \ref{fig:C5FRW}. }{
dark gray area in Figure \ref{fig:C5FRW}. }
Correspondingly, in Section \ref{C5matter} we calculate the present entropy budget of 
the observable universe and we do not include the CEH.

The second scheme is similar to the first in that we time-slice along surfaces of 
constant cosmic time. However, here the system 
\ifthenelse{\boolean{colver}} { 
(yellow shade in Figure \ref{fig:C5FRW}) }{
(light gray shade in Figure \ref{fig:C5FRW}) }
is bounded by the time-dependent CEH instead of a comoving 
boundary. Migration of matter across the CEH is not negligible, and the CEH 
entropy \citep{Gibbons1977} must be included in the budget to account for 
this (e.g.\ \citealt{Davis2003}). The present entropy of the CEH 
and its interior is calculated in Section \ref{C5scheme2}.

%
%
\ifthenelse{\boolean{colver}} {
	\begin{figure*}[!p]
       		\begin{center}
               		\includegraphics[width=0.9\linewidth]{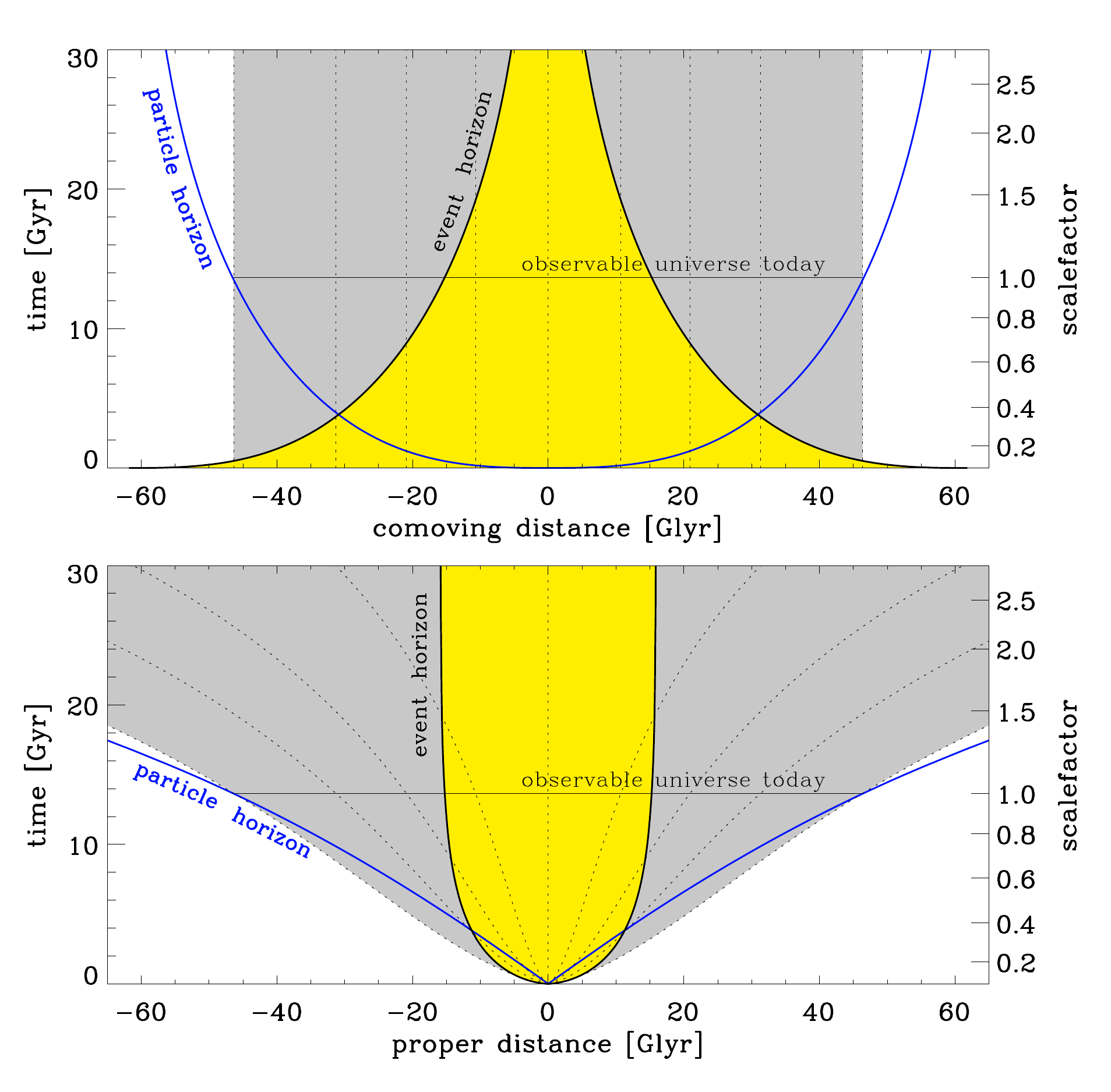}
               		\caption[The particle horizon and the cosmic event horizon as functions of time.]
			{These two panels show the particle horizon (see Equation \ref{eq:C5robs} 
               		and Figure \ref{fig:C5dehh0}) and the cosmic event horizon (see Equation \ref{eq:C5deh}) 
               		as a function of time. The difference between the two panels is the
               		spatial coordinate system used: the $x$-axis in the bottom panel is proper
               		distance $D$ and in the top panel it is comoving distance $\chi \equiv \frac{D}{a}$,
               		where $a$ is the cosmic scalefactor.
               		The origin is chosen so that our galaxy is the central vertical dotted line.
               		The other dotted lines represent distant galaxies, which are approximately 
               		comoving and recede as the universe expands. The region inside the 
			particle horizon is the observable universe. The comoving volume that 
			corresponds to the observable universe today, about $13.7\ Gyr$ 
               		after the big bang, is filled gray. In scheme 1, the entropy within this 
               		comoving volume increases (or remains constant) with time. 
               		Alternatively, in scheme 2 the entropy within the event horizon (the region 
               		filled yellow), plus the entropy of the horizon itself, increases (or remains 
               		constant) with time.}
               		\label{fig:C5FRW}
       		\end{center}
	\end{figure*}
}{
	\begin{figure*}[!p]
       		\begin{center}
               		\includegraphics[width=0.9\linewidth]{15_chapter5_v2_apj/frw_top_two_panels_bw.pdf}
               		\caption[The particle horizon and the cosmic event horizon as functions of time.]
			{These two panels show the particle horizon (see Equation \ref{eq:C5robs} 
               		and Figure \ref{fig:C5dehh0}) and the cosmic event horizon (see Equation \ref{eq:C5deh}) 
               		as a function of time. The difference between the two panels is the
               		spatial coordinate system used: the $x$-axis in the bottom panel is proper
               		distance $D$ and in the top panel it is comoving distance $\chi \equiv \frac{D}{a}$,
               		where $a$ is the cosmic scalefactor.
               		The origin is chosen so that our galaxy is the central vertical dotted line.
               		The other dotted lines represent distant galaxies, which are approximately 
               		comoving and recede as the universe expands. The region inside the 
			particle horizon is the observable universe. The comoving volume that 
			corresponds to the observable universe today, about $13.7\ Gyr$ 
               		after the big bang, is filled dark gray. In scheme 1, the entropy within this 
			comoving volume increases (or remains constant) with time. 
               		Alternatively, in scheme 2 the entropy within the event horizon (the light gray 
			region), plus the entropy of the horizon itself, increases (or remains 
               		constant) with time. \textit{[See the electronic edition of the journal for a 
			 version of this figure.]}}
               		\label{fig:C5FRW}
       		\end{center}
	\end{figure*}
}

\subsection{Entropy and Gravity}

It is widely appreciated that non-gravitating systems of particles evolve 
toward homogenous temperature and density distributions. The corresponding 
increase in the volume of momentum-space and position-space occupied by the
constituent particles represents an increase in entropy.
On the other hand, strongly 
gravitating systems become increasingly lumpy. With ``lumpyness'' naively akin to 
``orderliness'', it is not as easy to see that the total entropy increases. In these systems
the entropy is shared among numerous components, all of which must be considered.



For example, approximately collisionless long-range gravitational interactions between
stars result in dynamical relaxation of galaxies (whereby bulk motions are dissipated and 
entropy is transferred to stars in the outer regions of the galaxy; \citealt{LyndenBell1967}) 
and stellar evaporation from galaxies (whereby stars are ejected altogether, carrying with 
them energy, angular momentum and entropy, and allowing what remains behind to 
contract; e.g.\ \citealt{Binney2008}). 
In more highly dissipative systems, i.e., accretion disks, non-gravitational 
interactions (viscosity and/or magnetorotational instability; \citealt{Balbus2002}) transfer 
angular momentum and dissipate energy and entropy.

In addition to these considerations, entropy also increases when gravitons are produced. 
A good example is the in-spiral of close binaries, such as the Hulse-Taylor 
binary pulsar system \citep{HulseTaylor1975,Weisberg2005}. Gravitational waves
emitted from the system extract orbital energy (and therefore entropy) allowing the system 
to contract. 

The entropy of a general gravitational field is still not known. 
\citet{Penrose1987,Penrose1979,Penrose2004}
has proposed that it is related to the Weyl curvature tensor $W_{\mu \nu \kappa \lambda}$. 
In conformally flat spacetimes (such as an ideal FRW universe), the Weyl curvature vanishes 
and gravitational entropy is postulated to vanish (to limits imposed by quantum uncertainty). In 
clumpy spacetimes the Weyl curvature takes large values and the gravitational entropy is high. 
While Ricci curvature $R_{\mu \nu}$ vanishes in the absence of matter, Weyl curvature may 
still be non-zero (e.g.\ gravitational waves traveling though empty space) and the corresponding 
gravitational entropy may be non-zero.

If these ideas are correct then the low gravitational entropy of the early universe comes from 
small primordial gravitational perturbations. Gravitational entropy then increases with the growing 
amplitude of linear density fluctuations parameterized through the matter power spectrum $P(k)$. 
The present gravitational entropy, however, is expected to be dominated by the nonlinear 
overdensities (with large Weyl tensors) which have formed since matter-radiation equality. 

In extreme cases, gravitational clumping leads to the formation of black holes. The
entropy of black holes is well known \citep{Bekenstein1973,Hawking1976,Strominger1996}. 
The entropy of a Schwarzschild black hole is given by 
\begin{eqnarray}
	S_ {\mathrm{BH}} = \frac{k c^3}{G \hbar} \frac{A}{4}  = \frac{4 \pi k G}{c \hbar} M^2 \label{eq:C5areaent}
\end{eqnarray}
where $A = \frac{16 \pi G^2 M^2}{c^4}$ is the event-horizon area and $M$ is the black hole
mass. 



Because gravitational entropy is difficult to quantify, we only include it in the two extremes:
the thermal distribution of gravitons and black holes.



\section{The Present Entropy of the Observable Universe} \label{C5matter} \label{C5scheme1}

The present entropy budget of the observable universe was estimated most recently by 
\citet{Frampton2008} and \citet{Frampton2008b}. Those papers and earlier work 
\citep{Kolb1981,Frautschi1982,Penrose2004,Bousso2007} identified the largest contributors 
to the entropy of the observable universe as black holes, followed distantly by the cosmic 
microwave background (CMB) and the neutrino background. The last column of Table 
\ref{tab:C5currententropy} contains previous estimates of the entropy in black holes, the CMB and 
neutrinos, as well as several less significant components.

Sections \ref{C5bar} -- \ref{C5smbh} below describe the data and assumptions 
used to calculate our entropy densities (given in Column 2 of Table \ref{tab:C5currententropy}). 
Our entropy budget for the observable universe (Column 3 of Table \ref{tab:C5currententropy}) 
is then found by multiplying the entropy density by the volume of the observable 
universe $V_{\mathrm{obs}}$,
\begin{eqnarray}
	S_{i} & = & s_{i} V_{\mathrm{obs}}
\end{eqnarray}
where $s_i$ is the entropy density of component $i$. The volume of the observable 
universe is (see Appendix)
\begin{eqnarray}
	V_{\mathrm{obs}} & = & 43.2 \pm 1.2 \xt{4}\ \mathrm{Glyr}^3 \nonumber \\
		& = & 3.65 \pm 0.10 \xt{80}\ m^3.
\end{eqnarray}




\begin{sidewaystable*}[!tp]
\small
\begin{tabular}{l l l r}
	Component 			
		& Entropy Density $s$ $[k\ m^{-3}]$
		& Entropy $S$ $[k]$ 	
		& Entropy $S$ $[k]$ (previous work)						
		\\
	\hline               
	\hline
	SMBHs
		& $8.4^{+8.2}_{-4.7} \xt{23}$
		& $3.1^{+3.0}_{-1.7} \xt{104}$
		& $10^{101}[1]$, $10^{102}[2]$, $10^{103}[3]$
		\\
	Stellar BHs $(2.5 - 15\ M_{\odot})$
		& $1.6\xt{17^{+0.6}_{-1.2}}$
		& $5.9\xt{97^{+0.6}_{-1.2}}$		
		& $10^{97}[2]$, $10^{98}[4]$
		\\
	Photons
		& $1.478\pm0.003\xt{9}$
		& $5.40\pm0.15\xt{89}$				
		& $10^{88}[1,2,4]$, $10^{89}[5]$	
		\\
	Relic Neutrinos
		& $1.411\pm0.014\xt{9}$
		& $5.16\pm0.15\xt{89}$
		& $10^{88}[2]$,$10^{89}[5]$		
		\\	
	\addthis{WIMP} Dark Matter
		& $5\xt{7 \pm 1}$
		& $2\xt{88 \pm 1}$	
		& $-$												
		\\
	Relic Gravitons
		& $1.7\xt{7^{+0.2}_{-2.5}}$
		& $6.2\xt{87^{+0.2}_{-2.5}}$				
		& $10^{86}[2,3]$												
		\\
	ISM and IGM
		& $20 \pm 15$ 
		& $7.1 \pm 5.6 \xt{81}$	
		& $-$												
		\\
	Stars
		& $0.26 \pm 0.12$  
		& $9.5 \pm 4.5 \xt{80}$	
		& $10^{79}[2]$
		\\
	\hline
	{\bf Total}
		& \boldmath$8.4^{+8.2}_{-4.7}\xt{23}$
		& \boldmath$3.1^{+3.0}_{-1.7}\xt{104}$ 	
		& \boldmath$10^{101}[1]$, $10^{102}[2]$, $10^{103}[3]$
		\\
	\hline
	\addthis{Tentative Components:}
		&
		&
		&
		\\
	\addthis{Massive Halo BHs $(10^5\ M_{\odot})$}
		& \addthis{$10^{25}$}
		& \addthis{$10^{106}$}	
		& \addthis{$10^{106}[6]$}
		\\
	\addthis{Stellar BHs $(42 - 140\ M_{\odot})$}
		& \addthis{$8.5\xt{18^{+0.8}_{-1.6}}$}
		& \addthis{$3.1\xt{99^{+0.8}_{-1.6}}$}
		& \addthis{$-$}
\end{tabular}
\caption[Entropy budget of the current observable Universe.]
	{Our budget is consistent with previous estimates from the literature with the exception that
	SMBHs, which dominate the budget, contain at least an order of magnitude more 
	entropy as previously estimated, due to the contributions of black holes $100$ times larger
	than those considered in previous budgets.
	Uncertainty in the volume of the observable universe (see Appendix) has been included in the quoted 
	uncertainties. \removethis{Stellar black holes in the mass range $42-140\ M_{\odot}$ (marked with an 
	$*$) are included tentatively since their existence is speculative.} \addthis{Massive halo black holes at 
	$10^5\ M_{\odot}$ and stellar black holes in the range $42-140\ M_{\odot}$ are included tentatively 
	since their existence is speculative. They are not counted in the budget totals.}
	Previous work: 
	[1] \citet{Penrose2004},
	[2] \citet{Frampton2008},
	[3] \citet{Frampton2008b},
	[4] \citet{Frautschi1982},
	[5] \citet{Kolb1981},
	\addthis{[6] \citet{Frampton2009b}.}
	}
\label{tab:C5currententropy}
\end{sidewaystable*}

\subsection{Baryons} \label{C5bar}

%

For a non-relativistic, non-degenerate gas the specific entropy (entropy per baryon) is given 
by the Sakur-Tetrode equation (e.g.\ \citealt{Basu1990})
\begin{eqnarray}
	(s/n_b) = \frac{k}{n_b} \sum_i n_i \ln \left[ Z_i(T) (2 \pi m_i k T)^{\frac{3}{2}} e^{\frac{5}{2}} n_i^{-1} h^{-3} \right], \label{eq:C5sakurtetrode}
\end{eqnarray}
where $i$ indexes particle types in the gas, $n_i$ is the $i$th particle type's number density, 
and $Z_i(T)$ is its internal partition function. \citet{Basu1990} found specific entropies between $11\ k$ 
and $21\ k$ per baryon for main-sequence stars of approximately solar mass. For components of the 
interstellar medium (ISM) and intergalactic medium (IGM) they found specific entropies between 
$20\ k$ ($H_{2}$ in the ISM) and $143\ k$ (ionized hydrogen in the IGM) per baryon. 

The cosmic entropy density in stars $s_{*}$ can be estimated by multiplying the specific entropy 
of stellar material by the cosmic number density of baryons in stars $n_{b*}$:
\begin{eqnarray}
	s_{*} = (s/n_b)_{*} n_{b*} = (s/n_b)_{*} \frac{\rho_{*}}{m_{p}} = (s/n_b)_{*} \left[ \frac{3 H^2}{8 \pi G} \frac{\Omega_{*}}{m_p} \right]. 
\end{eqnarray}
Using the stellar cosmic density parameter $\Omega_{*} = 0.0027 \pm 0.0005$ \citep{Fukugita2004}, 
and the range of specific entropies for main-sequence stars around the solar mass 
(which dominate stellar mass), we find 
\begin{eqnarray}
	%
	%
	s_{*} & = & 0.26\pm0.12\ k\ m^{-3}, \\
	S_{*} & = & 9.5 \pm 4.5 \xt{80}\ k. \label{eq:C5sstars}
\end{eqnarray}
Similarly, the combined energy density for the ISM and IGM is $\Omega_{\mathrm{gas}} = 0.040 \pm 0.003$ 
\citep{Fukugita2004}, and by using the range of specific entropies for ISM and IGM components, we 
find
\begin{eqnarray}
	%
	%
	s_{\mathrm{gas}} & = & 20\pm{15}\ k\ m^{-3}, \\
	S_{\mathrm{gas}} & = & 7.1\pm 5.6 \xt{81}\ k. \label{eq:C5sismigm}
\end{eqnarray}
The uncertainties in Equations (\ref{eq:C5sstars}) and (\ref{eq:C5sismigm}) are dominated by uncertainties in the
mass weighting of the specific entropies, but also include uncertainties in 
$\Omega_{*}$, $\Omega_{\mathrm{gas}}$ and the volume of the observable universe. 

\subsection{Photons} \label{C5bpg}

The CMB photons are the most significant non-black hole 
contributors to the entropy of the observable universe. The distribution of CMB photons is 
thermal \citep{Mather1994} with a present temperature of 
$T_{\gamma} = 2.725 \pm 0.002\ K$ \citep{Mather1999}.

The entropy of the CMB is calculated using the equation for a black body (e.g.\ \citet{Kolb1990}),
\begin{eqnarray}
	s_{\gamma} 	& = & \frac{2 \pi^2}{45} \frac{k^4}{c^3 \hbar^3} g_{\gamma} T_{\gamma}^{3} \label{eq:C5radent} \\
				& = & 1.478 \pm 0.003 \xt{9}\ k\ m^{-3}, \nonumber \\
	S_{\gamma}	& = & 2.03 \pm 0.15 \xt{89}\ k, \label{eq:C5sgamma}
\end{eqnarray}
where $g_{\gamma} = 2$ is the number of photon spin states. The uncertainty in Equation (\ref{eq:C5sgamma}) 
is dominated by uncertainty in the size of the observable universe.

The non-CMB photon contribution to the entropy budget (including starlight and heat emitted
by the ISM) is somewhat less, at around $10^{86} k$ \citep{Frautschi1982,Bousso2007,Frampton2008}. 

\subsection{Relic Neutrinos} \label{C5nusec}


The neutrino entropy cannot be calculated directly since the temperature of cosmic neutrinos has
not been measured. Standard treaties of the radiation era (e.g.\ \citealt{Kolb1990,Peacock1999}) 
describe how the present temperature (and entropy) of massless relic neutrinos can be calculated 
from the well known CMB photon temperature. Since this background physics is required 
for Sections \ref{C5relicgrav} and \ref{C5darkmatter}, we summarize it briefly here.

A simplifying feature of the radiation era (at least at known energies $\lsim 10^{12} eV$) is that the 
radiation fluid evolves adiabatically: the entropy density decreases as the cube of the increasing 
scalefactor $s_{\mathrm{rad}} \propto a^{-3}$. The evolution is adiabatic because reaction rates in the fluid 
are faster than the expansion rate $H$ of the universe. It is convenient to write the entropy density as
\begin{eqnarray}
	s_{\mathrm{rad}} & = & \frac{2 \pi^2}{45} \frac{k^4}{c^3 \hbar^3} g_{*S} T_{\gamma}^{3} \propto a^{-3} \label{eq:C5srad}
\end{eqnarray}
where $g_{*S}$ is the number of relativistic degrees of freedom in the fluid (with $m < kT/c^2$) 
given approximately by
\begin{eqnarray}
	g_{*S}(T) & \approx & \sum_{
	{\scriptsize \begin{array}{c}
	bosons,\ i
	\end{array}}
	} g_{i} \left( \frac{T_{i}}{T_{\gamma}} \right)^{3} + \sum_{
	{\scriptsize \begin{array}{c}
	fermions,\ j
	\end{array}}
	} \frac{7}{8} g_{j} \left( \frac{T_{j}}{T_{\gamma}} \right)^{3}.
\end{eqnarray}
For photons alone, $g_{*S} = g_{\gamma} = 2$, and thus Equation (\ref{eq:C5srad}) becomes Equation (\ref{eq:C5radent}). 
For photons coupled to an electron-positron component, such as existed before electron-positron 
annihilation,
$g_{*S} = g_{\gamma} + \frac{7}{8} g_{e\pm} = 2 + \frac{7}{8} 4 = \frac{11}{2}$. 

As the universe expands, massive particles annihilate, heating the
remaining fluid. The effect on the photon temperature is quantified by inverting Equation (\ref{eq:C5srad}), 
\begin{eqnarray}
	T_{\gamma} 
				& \propto & a^{-1} g_{*S}^{-1/3}.  \label{eq:C5gheat}
\end{eqnarray}
The photon temperature decreases less quickly than $a^{-1}$ because $g_{*S}$ decreases 
with time. 
Before electron-positron $e^{\pm}$ annihilation the temperature of the photons was the same 
as that of the almost completely decoupled neutrinos. After $e^{\pm}$ annihilation, heats only 
the photons, the two temperatures differ by a factor $C$,
\begin{eqnarray}
	T_{\nu} = C\ T_{\gamma}. \label{eq:C5nucool}
\end{eqnarray}
A reasonable approximation $C \approx (4/11)^{1/3}$ is derived by assuming that only
photons were heated during $e^{\pm}$ annihilation, where $4/11$ is the ratio of 
$g_{*S}$ for photons to $g_{*S}$ for photons, electrons, and positrons. 

Corrections are necessary at the $10^{-3}$ level because neutrinos had not completely
decoupled at $e^{\pm}$ annihilation \citep{Gnedin1998}. 
The neutrino entropy density is computed assuming a thermal distribution with 
$T_{\nu} = (4/11)^{1/3} T_{\gamma}$, and we assign a $1\%$ uncertainty. 


\begin{eqnarray}
	s_{\nu} 	& = & \frac{2 \pi^2}{45} \frac{k^4}{c^3 \hbar^3} g_{\nu} \left( \frac{7}{8} \right) T_{\nu}^{3} \nonumber \\
			& = & 1.411 \pm 0.014 \xt{9}\ k\ m^{-3}, \label{eq:C5nu0ent}
\end{eqnarray}	
where $g_{\nu} = 6$ ($3$ flavors, $2$ spin states each).  The total neutrino entropy in the 
observable universe is then
\begin{eqnarray}
	S_{\nu}	& = & 5.16 \pm 0.14 \xt{89}\ k \label{eq:C5nuentobs}
\end{eqnarray}
with an uncertainty dominated by uncertainty in the volume of the observable universe.

Neutrino oscillation experiments have demonstrated that neutrinos are massive by measuring 
differences between the three neutrino mass eigenstates \citep{Cleveland1998,Adamson2008,Abe2008}. 
At least two of the mass eigenstates are heavier than $\sim 0.009\ eV$. 
Since this is heavier than their current relativistic energy ($\frac{k}{2}\ C\ T_{\gamma} = 0.0001\ eV$; 
computed under the assumption that they are massless) at least two of the three masses are presently 
non-relativistic. 

Expansion causes non-relativistic species to cool as $a^{-2}$ instead of $a^{-1}$, which would
result in a lower temperature for the neutrino background than suggested by Equation (\ref{eq:C5nucool}). 
The entropy density (calculated in Equation \ref{eq:C5nu0ent}) and entropy (calculated in Equation \ref{eq:C5nuentobs}) are 
unaffected by the transition to non-relativistic cooling since the cosmic expansion of relativistic and 
non-relativistic gases are both adiabatic processes (the comoving entropy is conserved, so in either 
case $s \propto a^{-3}$).


We neglect a possible increase in neutrino entropy due to their infall into gravitational potentials 
during structure formation. If large, this will need to be considered in future work.

\subsection{Relic Gravitons} \label{C5relicgrav}


A thermal background of gravitons is expected to exist, which decoupled from 
the photon bath around the Planck time, and has been cooling as $T_{\mathrm{grav}} \propto a^{-1}$ 
since then. The photons cooled less quickly because they were heated by the annihilation 
of heavy particle species (Equation \ref{eq:C5gheat}). Thus we can relate the current graviton 
temperature to the current photon temperature
\begin{eqnarray}
	T_{\mathrm{grav}} & = & \left(\frac{g_{*S}(t_0)}{g_{*S}(t_{\mathrm{planck}})}\right)^{1/3} T_{\gamma}, \label{eq:C5gravtemp1} 
\end{eqnarray}
where $g_{*S}(t_{\mathrm{planck}})$ is the number of relativistic degrees of freedom at the Planck
time and $g_{*S}(t_0) = 3.91$ today (this is appropriate even in the case of massive neutrinos
because they decoupled from the photon bath while they were still relativistic). Given the 
temperature of background gravitons, their entropy can be calculated as 
\begin{eqnarray}
	s_{\mathrm{grav}} 	& = & \frac{2 \pi^2}{45} \frac{k^4}{c^3 \hbar^3} g_{\mathrm{grav}} T_{\mathrm{grav}}^{3} \label{eq:C5gravent}
\end{eqnarray}
where $g_{\mathrm{grav}} = 2$. 

Figure \ref{fig:C5gstar} shows $g_{*S}$ as a function of temperature. The function is well 
known for temperatures below about $10^{12} eV$, but is not known at higher 
temperatures. Previous estimates of the background graviton entropy have assumed 
$g_{*S}(t_{\mathrm{planck}}) \sim g_{*S}(10^{12} eV) = 106.75$ \citep{Frampton2008,Frampton2008b}, 
but this should be taken as a lower bound on $g_{*S}(t_{\mathrm{planck}})$ yielding an upper 
bound on $T_{\mathrm{grav}}$ and $s_{\mathrm{grav}}$.

To get a better idea of the range of possible graviton temperatures and entropies, we 
have adopted three values for $g_{*S}(t_{\mathrm{planck}})$. As a minimum likely value we use
$g_{*S}=200$ 
\ifthenelse{\boolean{colver}} { 
(Figure \ref{fig:C5gstar}, thick blue line), }{
(Figure \ref{fig:C5gstar}, solid gray line), }
which includes the minimal set of additional particles suggested by 
supersymmetry. As our middle value we use $g_{*S}=350$, corresponding to the 
linear extrapolation of $g_{*S}$ in $\log(T)$ to the Planck scale 
\ifthenelse{\boolean{colver}} { 
(Figure \ref{fig:C5gstar}, gray line). }{
(Figure \ref{fig:C5gstar}, dotted gray line). }
And as a maximum likely value we use $g_{*S}=10^{5}$, corresponding to an exponential 
extrapolation 
\ifthenelse{\boolean{colver}} { 
(Figure \ref{fig:C5gstar}, thin blue line). }{
(Figure \ref{fig:C5gstar}, dashed gray line). }
\ifthenelse{\boolean{colver}} {
	\begin{figure*}[!tp]
       		\begin{center}
               		\includegraphics[width=\linewidth]{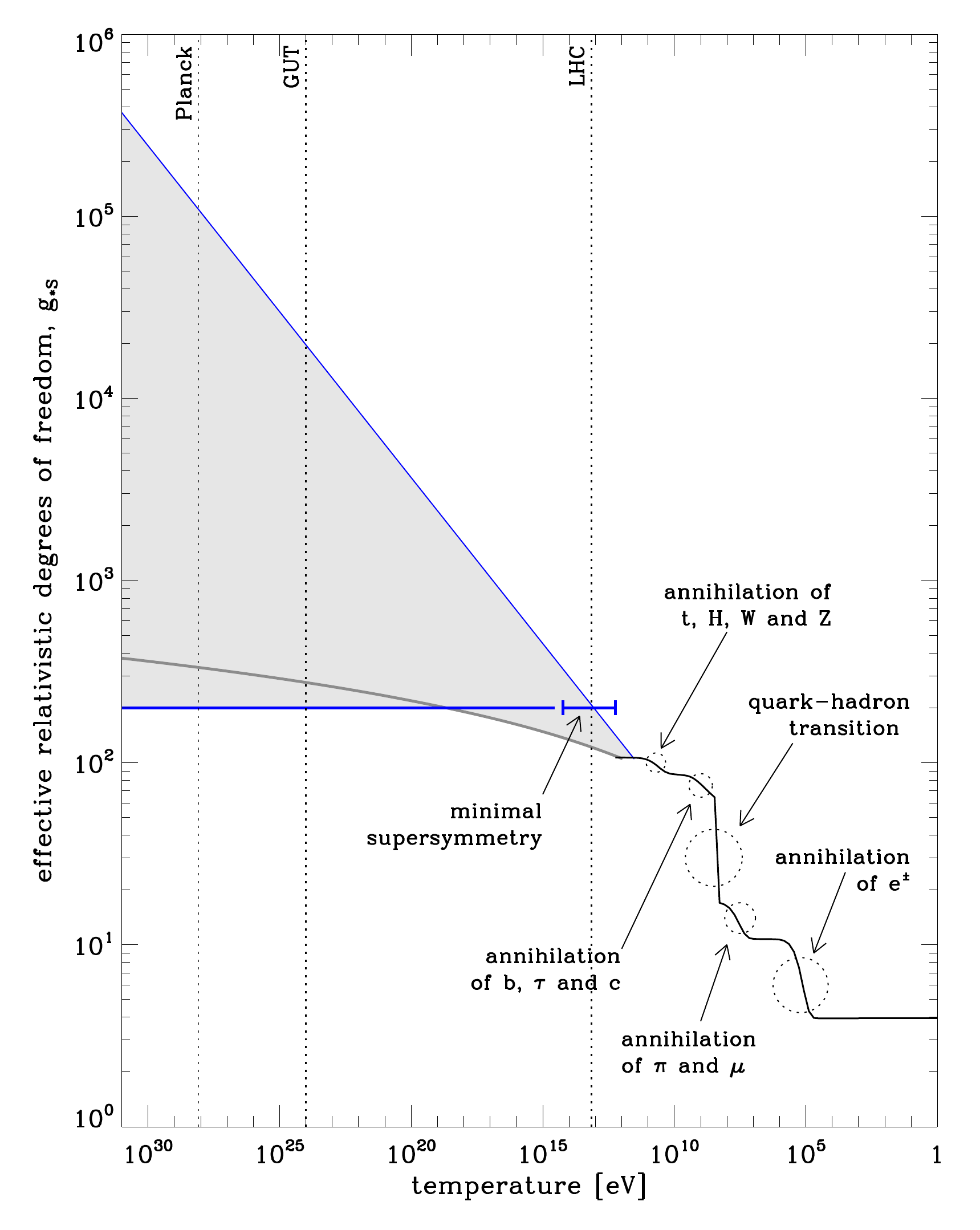}
               		\caption[Extrapolating the number of relativistic degrees of freedom.]
			{Number of relativistic degrees of freedom $g_{*S}$ as a function
               		of temperature, computed using the prescription given by \citet{Coleman2003}. 
               		All the particles of the standard model are relativistic at $T \gsim 10^{12}\ eV$ 
               		and $g_{*S}(10^{12}\ eV)=106.75$. The value of $g_{*S}$ is not known above $T \sim 10^{12}$. 
               		To estimate plausible ranges of values, we extrapolate $g_{*S}$ linearly (gray
			line) and exponentially (thin blue line) in $\log(T)$. The minimum contribution 
               		to $g_{*S}$ from supersymmetric partners is shown (blue bar) and taken to 
			indicate a minimum likely value of $g_{*S}$ at higher temperatures (thick blue line).
              		}
               		\label{fig:C5gstar}
       		\end{center}
	\end{figure*}
}{
	\begin{figure*}[!tp]
       		\begin{center}
               		\includegraphics[width=\linewidth]{15_chapter5_v2_apj/g_star_predictor_pretty_log_bw.pdf}
               		\caption[Extrapolating the number of relativistic degrees of freedom.]
			{Number of relativistic degrees of freedom $g_{*S}$ as a function
               		of temperature, computed using the prescription given by \citet{Coleman2003}. 
               		All the particles of the standard model are relativistic at $T \gsim 10^{12}\ eV$ 
               		and $g_{*S}(10^{12}\ eV)=106.75$. The value of $g_{*S}$ is not known above $T \sim 10^{12}$. 
               		To estimate plausible ranges of values, we extrapolate $g_{*S}$ linearly (dotted 
               		line) and exponentially (dashed line) in $\log(T)$. The minimum contribution 
               		to $g_{*S}$ from supersymmetric partners is shown and taken to indicate a 
               		minimum likely value of $g_{*S}$ at higher temperatures. \textit{[See the electronic 
			edition of the journal for a color version of this figure.]}}
               		\label{fig:C5gstar}
       		\end{center}
	\end{figure*}
}

%
%
%
The corresponding graviton temperatures today are (Equation \ref{eq:C5gravtemp1})
\begin{eqnarray}
		T_{\mathrm{grav}} & = & 0.61^{+0.12}_{-0.52} \textrm{ K}.
\end{eqnarray}
Inserting this into Equation (\ref{eq:C5gravent}) we find the entropy in the relic graviton 
background to be 
\begin{eqnarray}
	s_{\mathrm{grav}}	& = & 1.7 \xt{7^{+0.2}_{-2.5}}\ k\ m^{-3},  \\
	S_{\mathrm{grav}}	& = & 6.2 \xt{87^{+0.2}_{-2.5}}\ k.
\end{eqnarray}


It is interesting to note the possibility of applying Equation (\ref{eq:C5gravtemp1}) in 
reverse, i.e., calculating the number of relativistic degrees of freedom at the Planck time 
using future measurements of the graviton background temperature.

\subsection{Dark Matter} \label{C5darkmatter}


The most compelling interpretation of dark matter is as a weakly interacting superpartner
(or weakly interacting massive particle, WIMP). 
According to this idea, dark matter particles decoupled from the radiation background at some 
energy above the particle mass. 

If this interpretation is correct, the fraction of relativistic background entropy in dark matter 
at the time dark matter decoupled $t_{\mathrm{dm\ dec}}$ is determined by the fraction of relativistic 
degrees of freedom that were associated with dark matter at that time (see Equation \ref{eq:C5srad}). 
\begin{eqnarray}
	s_{\mathrm{dm}} = \frac{g_{*S\ \mathrm{dm}}(t_{\mathrm{dm\ dec}})}{g_{*S\ \mathrm{non-dm}}(t_{\mathrm{dm\ dec}})} s_{\mathrm{non-dm\ rad}} \label{eq:C5sdm}
\end{eqnarray}
This can be evaluated at dark matter decoupling, or any time thereafter, since both $s_{\mathrm{dm}}$
and $s_{\mathrm{non-dm\ rad}}$ are adiabatic ($\propto a^{-3}$).



We are unaware of any constraint on the number of superpartners that may collectively 
constitute dark matter. The requirements that they are only weakly interacting, and that they 
decouple at a temperature above their mass, are probably only satisfied by a few (even one) 
species. Based on these arguments, we assume $g_{*S\ dm}(t_{\mathrm{dm\ dec}}) \lsim 20$ and 
$g_{*S}(t_{\mathrm{dm\ dec}}) \gsim 106.75$ which yields the upper limit 
\begin{eqnarray}
	\frac{g_{*S\ \mathrm{dm}}(t_{\mathrm{dm\ dec}})}{g_{*S}(t_{\mathrm{dm\ dec}})} \lsim \frac{1}{5}. \label{eq:C5sdmhi}
\end{eqnarray}

On the other hand there may be many more degrees of freedom than suggested by minimal 
supersymmetry. By extrapolating $g_{*S}$ exponentially beyond supersymmetric scales 
(to $10^{15}\ eV$), we find $g_{*S}(t_{\mathrm{dm\ dec}}) \lsim 800$. In the simplest case, dark matter 
is a single scalar particle so $g_{*S\ \mathrm{dm}}(t_{\mathrm{dm\ dec}}) \gsim 1$ and we take as 
a lower limit
\begin{eqnarray}
	\frac{g_{*S\ \mathrm{dm}}(t_{\mathrm{dm\ dec}})}{g_{*S\ \mathrm{non-dm}}(t_{\mathrm{dm\ dec}})} \gsim \frac{1}{800}. \label{eq:C5sdmlo}
\end{eqnarray}



Inserting this into Equation (\ref{eq:C5sdm}) at the present day gives 
\begin{eqnarray}
	s_{\mathrm{dm}} & = & 5\xt{7\pm1}\ k\ m^{-3},
\end{eqnarray}
where we have used the estimated limits given in Equations (\ref{eq:C5sdmhi}) and (\ref{eq:C5sdmlo}) 
and taken $s_{\mathrm{non-dm\ rad}}$ to be the combined entropy of neutrinos and radiation today
(Equations \ref{eq:C5radent} and \ref{eq:C5nu0ent}). The corresponding estimate for the total dark matter 
entropy in the observable universe is
\begin{eqnarray}
	S_{\mathrm{dm}} & = & 2\xt{88\pm1}\ k.
\end{eqnarray}

As with our calculated neutrino entropy, our estimates here carry the caveat that we have not 
considered changes in the dark matter entropy associated with gravitational structure formation.

\subsection{Stellar Black Holes}

In the top panel of Figure \ref{fig:C5sbhentropy} we show the stellar initial mass function (IMF) 
parameterized by
\begin{eqnarray}
	\frac{d n_{\mathrm{initial}}}{d \log(M)} \propto \left(\frac{M}{M_{\odot}}\right)^{\alpha + 1},
\end{eqnarray}
with $\alpha = -1.35$ at $M < 0.5 M_{\odot}$ and $\alpha = -2.35^{+0.65}_{-0.35}$ at $M \ge 0.5 M_{\odot}$
\citep{Elmegreen2007}. We also show the present distribution of main-sequence stars, 
which is proportional to the initial distribution for $M \lsim 1 M_{\odot}$,
but which is reduced by a factor of $(M/M_{\odot})^{-2.5}$ for heavier stars \citep{Fukugita2004}. 
\begin{equation}
	\frac{d n_{\mathrm{present}}}{d \log(M)} = \left\{ 
	\begin{array}{ll} 
		\frac{d n_{\mathrm{initial}}}{d \log(M)}, 								& \mbox{for $M < 1 M_{\odot}$} \\
		\frac{d n_{\mathrm{initial}}}{d \log(M)} \left( \frac{M}{M_{\odot}} \right)^{-2.5}, 	& \mbox{for $M \ge 1 M_{\odot}$}
	\end{array}
	\right. .
\end{equation}
The initial and present distributions are normalized using the present cosmic density of 
stars, $\Omega_* = 0.0027 \pm 0.0005$ \citep{Fukugita2004}.
\ifthenelse{\boolean{colver}} {
	\begin{figure*}[!p]
       		\begin{center}
               		\includegraphics[width=\linewidth]{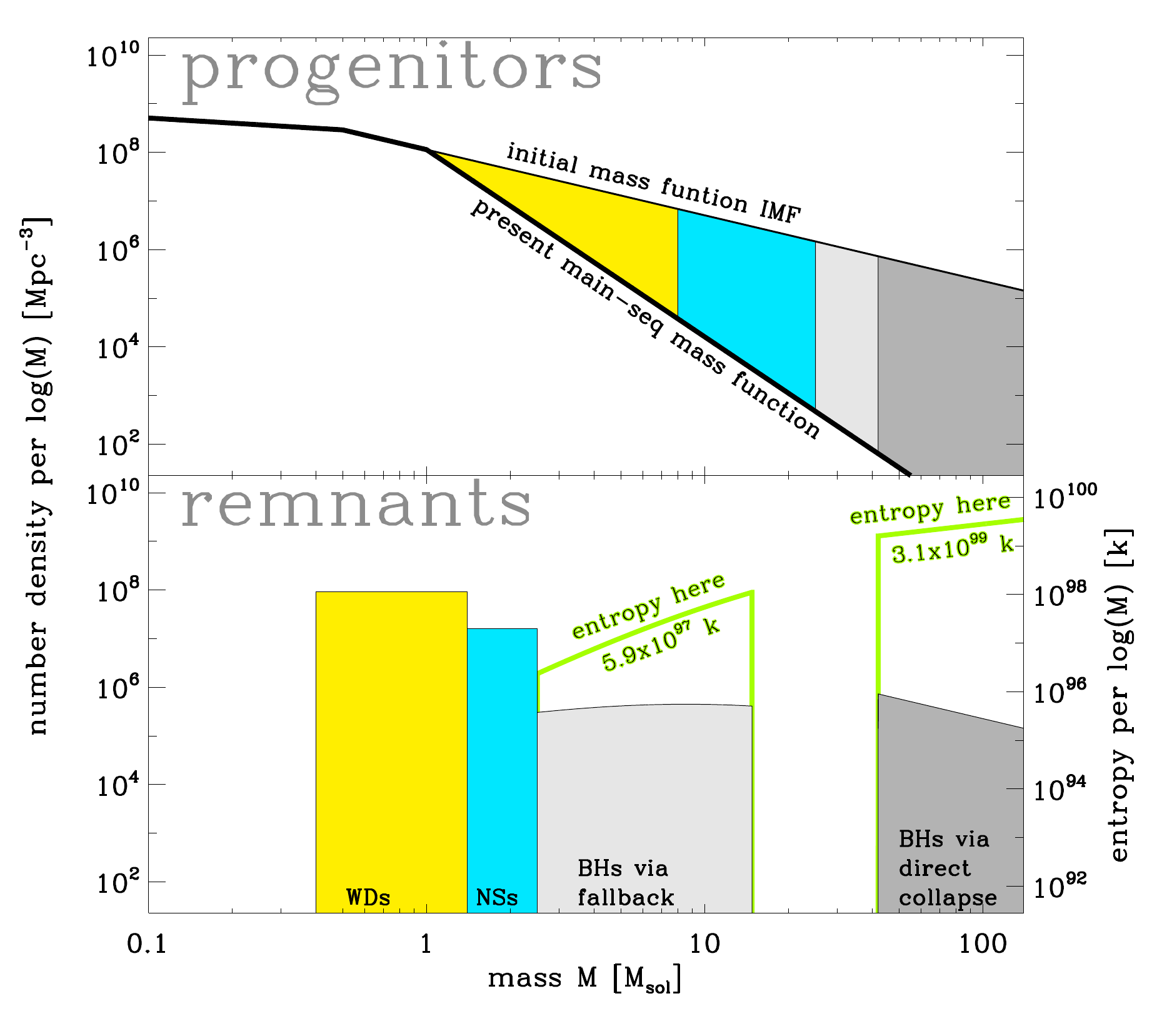}
               		\caption[The distribution of stellar black holes.]
			{Progenitors in the IMF (top panel) evolve into the distribution of remnants in 
               		the bottom panel. The shape of the present main-sequence mass function differs from 
               		that of the IMF (top panel) by the stars that have died leaving
               		white dwarfs (yellow), neutron stars (blue), and black holes (light and dark gray). 
               		The present distribution of remnants is shown in the bottom panel. Black 
               		holes in the range $2.5 M_{\odot} \lsim M \lsim 15 M_{\odot}$ (light gray) have been 
               		observationally confirmed. They form from progenitors in the range 
               		$25 M_{\odot} \lsim M \sim 42 M_{\odot}$ via core collapse supernova and fallback, and we
			calculate their entropy to be $5.9\xt{97^{+0.6}_{-1.2}} k$. Progenitors above 
			about $42\ M_{\odot}$ may evolve directly to black holes without significant 
			loss of mass (dark gray) and may carry much more entropy, but this population has 
			not been observed. The green curve, whose axis is on the right, shows 
			the mass distribution of stellar black hole entropies in the observable universe.}
               		\label{fig:C5sbhentropy}
       		\end{center}
	\end{figure*}
}{
	\begin{figure*}[!p]
       		\begin{center}
               		\includegraphics[width=\linewidth]{15_chapter5_v2_apj/full_imf_2panel_bw.pdf}
               		\caption[The distribution of stellar black holes.]
			{Progenitors in the IMF (top panel) evolve into the distribution of remnants in 
               		the bottom panel. The shape of the present main-sequence mass function differs from 
               		that of the IMF (top panel) by the stars that have died leaving
               		white dwarfs (dotted), neutron stars (striped), and black holes (light and dark gray). 
               		The present distribution of remnants is shown in the bottom panel. Black 
               		holes in the range $2.5 M_{\odot} \lsim M \lsim 15 M_{\odot}$ (light gray) have been 
               		observationally confirmed. They form from progenitors in the range 
               		$25 M_{\odot} \lsim M \sim 42 M_{\odot}$ via core collapse supernova and fallback, and we 
               		calculate their entropy to be $5.9\xt{97^{+0.6}_{-1.2}} k$. Progenitors above 
               		about $42\ M_{\odot}$ may evolve directly to black holes without significant 
               		loss of mass (dark gray) and may carry much more entropy, but this population has
               		not been observed. The gray curve, whose axis is on the right, shows 
			the mass distribution of stellar black hole entropies in the observable universe. 
			\textit{[See the electronic edition of the journal for a color version of this figure.]}}
               		\label{fig:C5sbhentropy}
       		\end{center}
	\end{figure*}
}

\ifthenelse{\boolean{colver}} { 
The yellow fill }{
The dotted fill }
in the top panel represents stars of mass $1 M_{\odot} \lsim M \lsim 8 M_{\odot}$, 
which died leaving white dwarf remnants of mass $M \lsim 1.4 M_{\odot}$ 
\ifthenelse{\boolean{colver}} { 
(yellow fill, bottom panel). }{
(dotted fill, bottom panel). }
\ifthenelse{\boolean{colver}} { 
The blue fill }{
The striped fill }
represents stars of mass $8 M_{\odot} \lsim M \lsim 25 M_{\odot}$, 
which died and left neutron star remnants of mass $1.4 M_{\odot} \lsim M \lsim 2.5 M_{\odot}$. 
The light gray area represents stars of mass $25 M_{\odot} \lsim M \lsim 42 M_{\odot}$
which became black holes of mass $2.5 M_{\odot} \lsim M \lsim 15 M_{\odot}$ via supernovae 
(here we use the simplistic final-initial mass function of \citet{Fryer2001}). 
Stars larger than $\sim 42 M_{\odot}$ collapse directly to black holes, without supernovae, 
and therefore retain most of their mass (dark gray regions; \citealt{Fryer2001,Heger2005}).

Integrating Equation (\ref{eq:C5areaent}) over stellar black holes in the range $M \le 15 M_{\odot}$ 
(the light gray fill in the bottom panel of Figure \ref{fig:C5sbhentropy}) we find
\begin{eqnarray}
	s_{\mathrm{SBH}\ (M<15M_{\odot})} & = & 1.6\xt{17^{+0.6}_{-1.2}}\ k\ m^{-3}, \label{eq:C5sbhentropy} \\
	S_{\mathrm{SBH}\ (M<15M_{\odot})} & = & 5.9\xt{97^{+0.6}_{-1.2}}\ k,
\end{eqnarray}
which is comparable to previous estimates of the stellar black hole entropy (see 
Table \ref{tab:C5currententropy}). 
Our uncertainty is dominated by uncertainty in the slope of the IMF, but also includes 
uncertainty in the normalization of the mass functions and uncertainty in the volume of 
the observable universe.

If the IMF extends beyond $M \gsim 42 M_{\odot}$ 
as in Figure \ref{fig:C5sbhentropy}, then these higher mass black holes (the dark gray
fill in the bottom panel of Figure \ref{fig:C5sbhentropy}) may contain more
entropy than black holes of mass $M < 15\ M_{\odot}$ (Equation \ref{eq:C5sbhentropy}). For example, if the 
Salpeter IMF is reliable to $M=140\ M_{\odot}$ (the Eddington limit and the edge of Figure
\ref{fig:C5sbhentropy}), then black holes in the mass range $42$ - $140\ M_{\odot}$ would 
contribute about $3.1\xt{99^{+0.8}_{1.6}}\ k$ to the entropy of the observable universe. 
Significantly less is known about this potential population, 
and should be considered a tentative contribution in Table \ref{tab:C5currententropy}.

\subsection{Supermassive Black Holes} \label{C5smbh}

Previous estimates of the SMBH entropy \citep{Penrose2004,Frampton2008,Frampton2008b} 
have assumed a typical SMBH mass and a number density and yield 
$S_{\mathrm{SMBH}} = 10^{101}-10^{103} k$. Below we use the SMBH mass function as measured 
recently by \citet{Graham2007}. 
Assuming a three-parameter Schechter function
\begin{eqnarray}
	\frac{d n}{d \log(M)} = \phi_* \left( \frac{M}{M_*} \right)^{\alpha + 1} \exp{\left[ 1 - \left(\frac{M}{M_*}\right) \right]}
\end{eqnarray} 
(number density per logarithmic mass interval) they find 
$\phi_* = 0.0016 \pm 0.0004\ \mathrm{Mpc}^{-3}$, 
$M_* = 2.9 \pm 0.7 \xt{8}\ M_{\odot}$, and 
$\alpha = -0.30 \pm 0.04$. 
The data and best-fit model are shown in black in Figure \ref{fig:C5bhentropy}. 

\ifthenelse{\boolean{colver}} {
	\begin{figure*}[!p]
       		\begin{center}
               		\includegraphics[width=\linewidth]{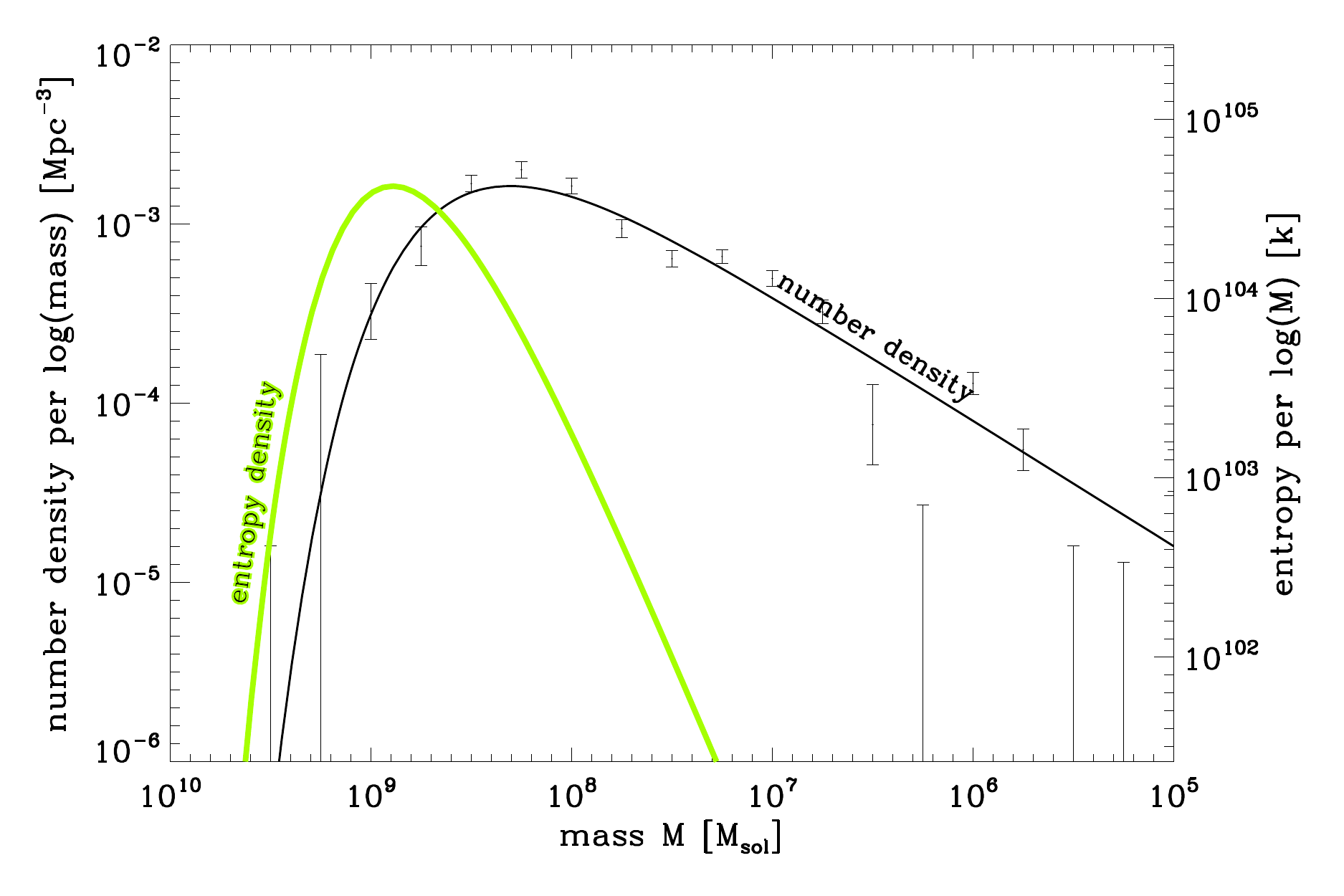}
               		\caption[The distribution of supermassive black holes.]
			{The black curve,
               		whose axis is on the left, is the SMBH mass function from \citet{Graham2007}, i.e.,
               		the number of supermassive black holes per $\mathrm{Mpc}^{3}$ per logarithmic mass interval. 
               		The green curve, whose axis is on the right, shows the mass distribution 
			of SMBH entropies in the observable universe.}
               		\label{fig:C5bhentropy}
       		\end{center}
	\end{figure*}
}{
	\begin{figure*}[!p]
       		\begin{center}
               		\includegraphics[width=\linewidth]{15_chapter5_v2_apj/SMBH_mass_function_bw.pdf}
               		\caption[The distribution of supermassive black holes.]
			{The black curve,
               		whose axis is on the left, is the SMBH mass function from \citet{Graham2007}, i.e.,
               		the number of supermassive black holes per $\mathrm{Mpc}^{3}$ per logarithmic mass interval. 
               		The gray curve, whose axis is on the right, shows the mass distribution 
			of SMBH entropies in the observable universe. \textit{[See the electronic 
			edition of the journal for a version of this figure.]}}
               		\label{fig:C5bhentropy}
       		\end{center}
	\end{figure*}
}

We calculate the SMBH entropy density by integrating Equation (\ref{eq:C5areaent}) over
the SMBH mass function,
\begin{eqnarray}
	s 	
		& = & \frac{4 \pi k G}{c \hbar}\ \int M^2 \left(\frac{d n}{d \log(M)}\right)\ d\log(M).
\end{eqnarray}
The integrand is plotted using 
\ifthenelse{\boolean{colver}} { 
a green line }{
a gray line }
in Figure \ref{fig:C5bhentropy} showing that
the contributions to SMBH entropy are primarily due to black holes around 
$\sim 10^9 M_{\odot}$. 
The SMBH entropy is found to be 
\begin{eqnarray}
	s_{\mathrm{SMBH}} & = & 8.4^{+8.2}_{-4.7} \xt{23}\ k\ m^{-3}, \\
	S_{\mathrm{SMBH}} & = & 3.1^{+3.0}_{-1.7} \xt{104}\ k.
\end{eqnarray}
The uncertainty here includes uncertainties in the SMBH mass function and uncertainties
in the volume of the observable universe.
This is at least an order of magnitude larger than previous estimates (see Table \ref{tab:C5currententropy}). 
The reason for the difference is that the \citep{Graham2007} SMBH mass function contains 
larger black holes than assumed in previous estimates.

\ifthenelse{\boolean{colver}} {
	\begin{figure*}[!p]
       		\begin{center}
               		\includegraphics[width=\linewidth]{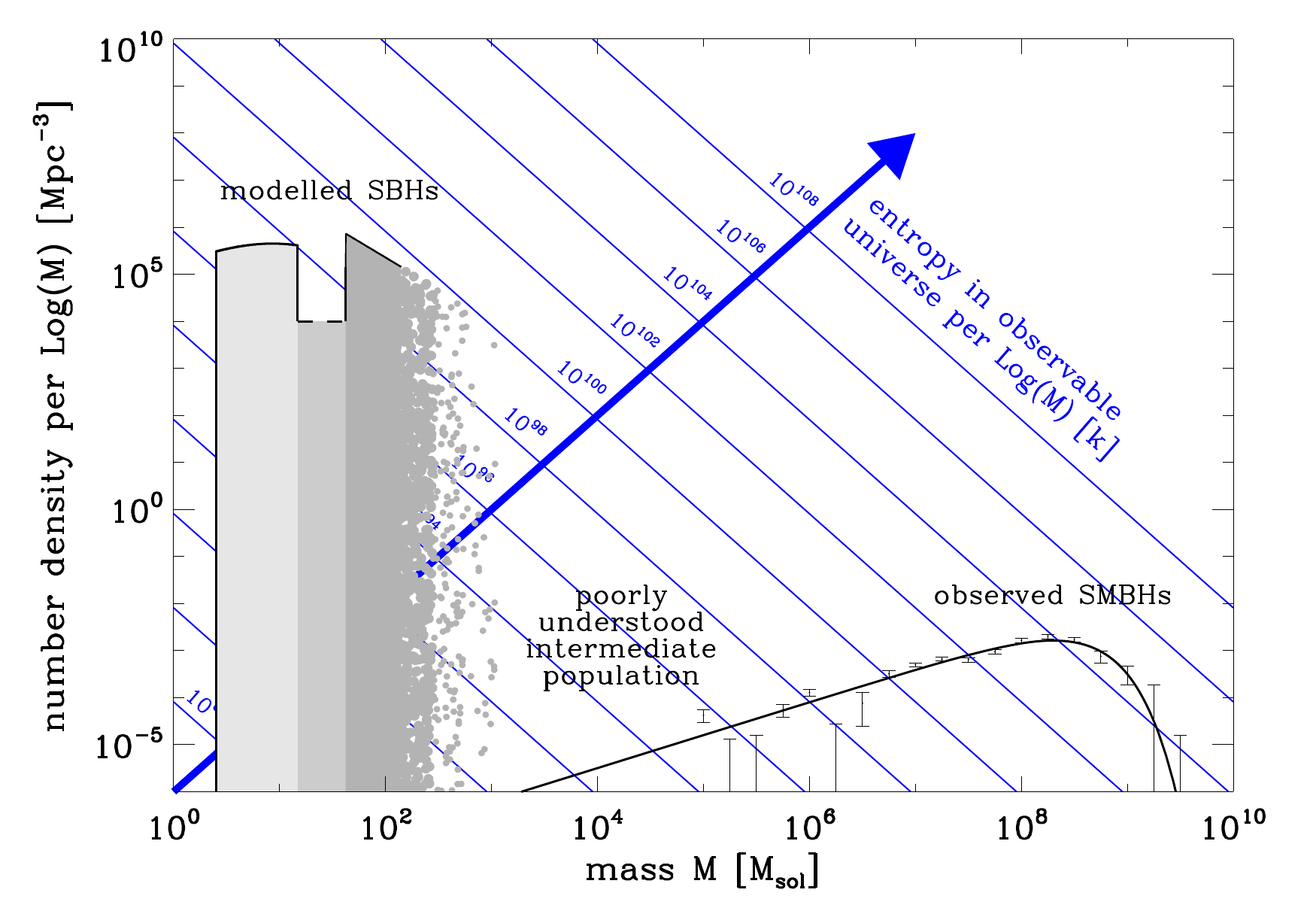}
               		\caption[Stellar black holes, intermediate black holes and supermassive black holes.]
			{Whether or not the total black hole entropy is dominated by SMBHs
               		depends on the yet-unquantified number of intermediate mass black holes.}
               		\label{fig:C5combinedbhentropy}
       		\end{center}
	\end{figure*}
}{
	\begin{figure*}[!p]
       		\begin{center}
               		\includegraphics[width=\linewidth]{15_chapter5_v2_apj/combination_fried_rice_bw.pdf}
               		\caption[Stellar black holes, intermediate black holes and supermassive black holes.]
			{Whether or not the total black hole entropy is dominated by SMBHs
               		depends on the yet-unquantified number of intermediate mass black holes. 
			\textit{[See the electronic edition of the journal for a color version of this figure.]}}
               		\label{fig:C5combinedbhentropy}
       		\end{center}
	\end{figure*}
}

\citet{Frampton2009,Frampton2009b} has suggested that intermediate mass black holes 
in galactic halos may contain more entropy than SMBHs in galactic cores. For example, according 
to the massive astrophysical compact halo object (MACHO) explanation of dark matter, intermediate 
mass black holes in the mass range $10^2$ - $10^5 \ M_{\odot}$ may constitute 
dark matter. Assuming $10^5 \ M_{\odot}$ black holes, these objects would contribute up to 
$10^{106}\ k$ to the entropy of the observable universe \citep{Frampton2009b}.
Whether or not this is so depends on the number density and mass distribution of this population. 
Figure \ref{fig:C5combinedbhentropy} combines Figures \ref{fig:C5sbhentropy} and \ref{fig:C5bhentropy} and 
shows what intermediate black hole number densities would be required.

\section{The Entropy of the CEH and its Interior} \label{C5scheme2} \label{C5ceh}

In this section we calculate the entropy budget for scheme 2 (refer  to discussion in Section \ref{C5frwgsl}). 
Scheme 2 differs from scheme 1 in two ways: first, along with the components previously considered 
(and listed in Table \ref{tab:C5currententropy}), here we consider the CEH as an 
additional entropy component; and second, the volume of interest is that within the event horizon 
not the particle horizon (or observable universe).

The proper distance to the CEH is generally time-dependent, increasing when the 
universe is dominated by an energy component with an equation of state $w>-1$ (radiation 
and matter) and remaining constant when the universe is dark energy dominated (assuming 
a cosmological constant, $w=-1$). Since our universe is presently entering dark energy domination, 
the growth of the event horizon has 
slowed, and it is almost as large now as it will ever become (bottom panel of Figure \ref{fig:C5FRW}). 
In the Appendix, we calculate the present radius and volume of the CEH
\begin{eqnarray}
	R_{\mathrm{CEH}} = 15.7 \pm 0.4\ \mathrm{Glyr},
\end{eqnarray}
\begin{eqnarray}
	V_{\mathrm{CEH}} & = & 1.62 \pm 0.12 \xt{4}\ \mathrm{Glyr}^3 \nonumber \\
		& = & 1.37 \pm 0.10 \xt{79}\ m^3.
\end{eqnarray}
We also calculate the present entropy of the CEH (following \citealt{Gibbons1977}),
\begin{eqnarray}
	S_{\mathrm{CEH}} & = & \frac{k c^3}{G \hbar} \frac{A}{4} \nonumber \\
		& = & \frac{k c^3}{G \hbar} \pi R_{\mathrm{CEH}}^2 \\
		& = & 2.6 \pm 0.3 \xt{122}\ k. \nonumber
\end{eqnarray}

Entropies of the various components within the CEH are calculated using the 
entropy densities $s_{i}$ from Section \ref{C5scheme1}: 
\begin{eqnarray}
	S_{i} & = & s_{i} V_{\mathrm{CEH}}
\end{eqnarray}
Table \ref{tab:C5scheme2budget} shows that the cosmic event horizon contributes almost $20$
orders of magnitude more entropy than the next largest contributor, supermassive black holes.

\begin{table*}[!p]
\begin{center}
\begin{tabular}{l  l}
	Component 			
		& Entropy $S$ $[k]$ 						
		\\
	\hline
	\hline
	Cosmic Event Horizon
		& $2.6\pm0.3\xt{122}$
		\\
	SMBHs
		& $1.2^{+1.1}_{-0.7} \xt{103}$
		\\
	Stellar BHs $(2.5-15\ M_{\odot})$
		& $2.2\xt{96^{+0.6}_{-1.2}}$		
		\\
	Photons
		& $2.03\pm0.15\xt{88}$				
		\\
	Relic Neutrinos
		& $1.93\pm0.15\xt{88}$
		\\
	\addthis{WIMP} Dark Matter
		& $6\xt{86\pm1}$	
		\\
	Relic Gravitons
		& $2.3\xt{86^{+0.2}_{-3.1}}$				
		\\
	ISM and IGM
		& $2.7 \pm 2.1 \xt{80}$	
		\\
	Stars
		& $3.5 \pm 1.7 \xt{78}$
		\\
	\hline
	{\bf Total}
		& \boldmath$2.6\pm0.3\xt{122}$
		\\
	\hline
	\addthis{Tentative Components:}
		&
		\\
	\addthis{Massive Halo BHs $(10^5\ M_{\odot})$}
		& \addthis{$10^{104}$}
		\\
	\addthis{Stellar BHs $(42-140\ M_{\odot})$}
		& \addthis{$1.2\xt{98^{+0.8}_{-1.6}}$}
\end{tabular}
\end{center}
\caption[The current entropy of the cosmic event horizon, and the entropy of its contents.]
	{This budget is dominated by the 
	cosmic event horizon entropy. 
	While the CEH entropy should be considered as an additional component in
	scheme 2, it also corresponds to the holographic bound \citep{tHooft1993} on the 
	possible entropy of the other components and may represent a significant overestimate.
	\removethis{Stellar black holes in the mass range $42-140\ M_{\odot}$ (marked with an $*$) are 
	included tentatively since their existence is speculative.} \addthis{Massive halo black holes at 
	$10^5\ M_{\odot}$ and stellar black holes in the range $42-140\ M_{\odot}$ are included 
	tentatively since their existence is speculative.}
	}
\label{tab:C5scheme2budget}
\end{table*}

\section{Discussion} \label{C5discussion}

The second law of thermodynamics holds that the entropy of an isolated system
increases or remains constant, but does not decrease.
This has been applied to the large-scale universe in at least two ways (Equation
\ref{eq:C5one} and \ref{eq:C5two}). The first scheme requires the entropy in a comoving 
volume of the universe to not decrease. The second scheme requires the entropy 
of matter contained within the event horizon, plus the entropy of the event horizon, 
to not decrease.

We have calculated improved estimates of the current entropy budget under scheme 1
(normalized to the current observable universe) and scheme 2. These are given in Tables 
\ref{tab:C5currententropy} and \ref{tab:C5scheme2budget}, respectively.




The entropy of dark matter has not been calculated previously. 
We find that dark matter contributes $10^{88\pm1}\ k$ to the entropy of the 
observable universe. 
We note that the neutrino and dark matter estimates do not include an increase 
due to their infall into gravitational potentials during structure formation. It is not
clear to us a priori whether this non-inclusion is significant, but it may be since 
both components are presently non-relativistic. 
This should be investigated in future work.

Previous estimates of the relic graviton entropy have assumed that only the
known particles participate in the relativistic fluid of the early universe at $t \gsim t_{\mathrm{planck}}$. 
In terms of the number of relativistic degrees of freedom, this means 
$g_{*S} \rightarrow 106.75$ at high temperatures. However, additional particles are 
expected to exist, and thus $g_{*S}$ is expected to become larger as 
$t \rightarrow t_{\mathrm{planck}}$. In the present work, we have calculated the relic graviton 
entropy corresponding to three high-energy extrapolations of $g_{*S}$ (constant, linear 
growth and exponential growth) and reported the corresponding graviton temperatures 
and entropies. 


In this paper, we have computed the entropy budget of the observable universe 
today $S_{\mathrm{obs}}(t=t_0)$. Figure \ref{fig:C5eoftimes1} illustrates the evolution of the entropy 
budget under scheme 1, i.e., the entropy in a comoving volume (normalized to the current 
observable universe). For simplicity, we have included only the most important 
components.
\ifthenelse{\boolean{colver}} {
	\begin{figure*}[!p]
       		\begin{center}
               		\includegraphics[width=\linewidth]{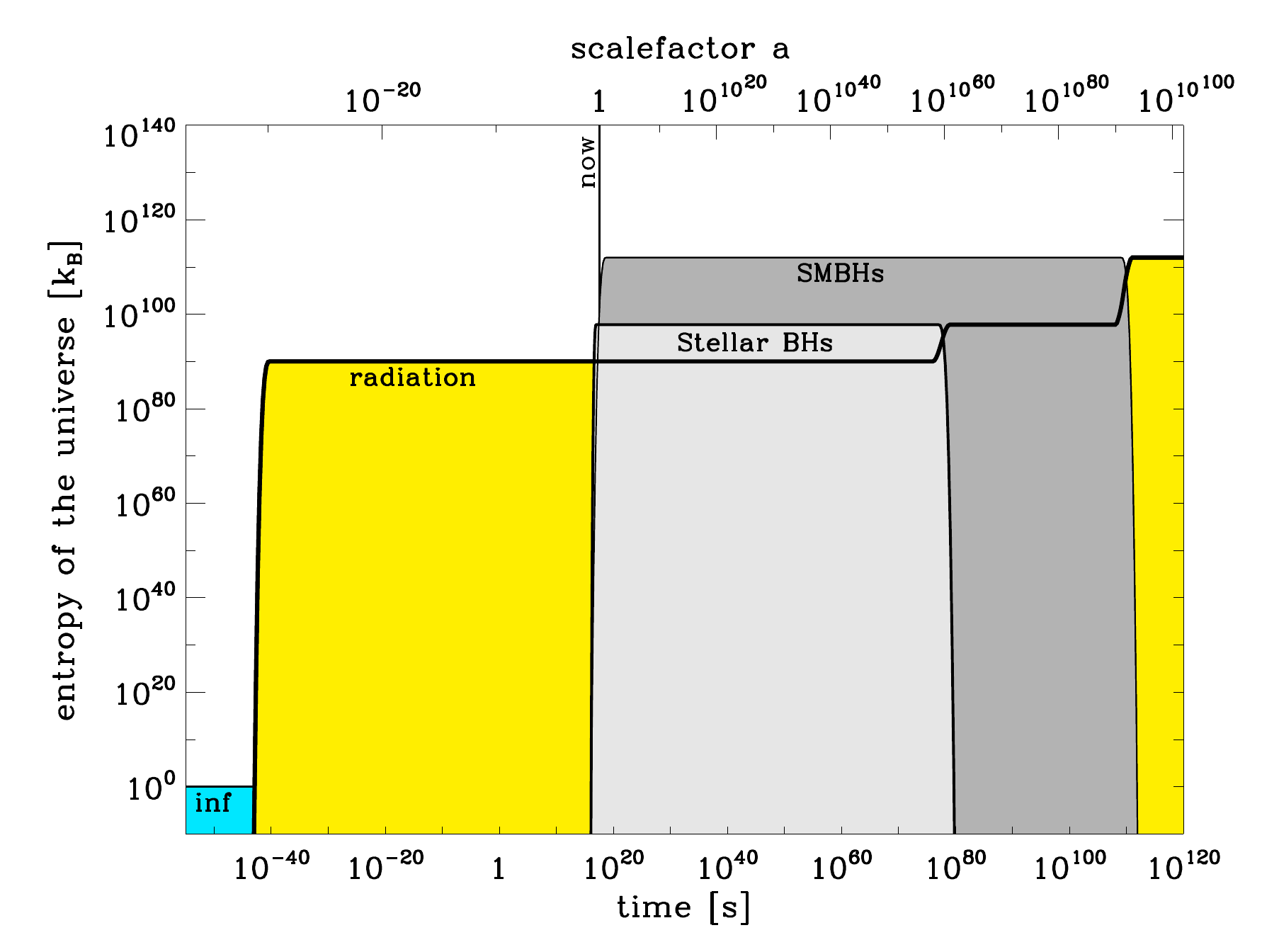}
               		\caption[The entropy history of a comoving volume normalized to the observable universe today.]
			{The entropy in a comoving volume (normalized to the 
               		present observable universe). This figure illustrates the time-dependence
               		of the scheme 1 entropy budget. N.B.\ $10^{10^{100}} = 1$ googolplex.}
               		\label{fig:C5eoftimes1}
       		\end{center}
	\end{figure*}
}{
	\begin{figure*}[!p]
       		\begin{center}
               		\includegraphics[width=\linewidth]{15_chapter5_v2_apj/eoftimes1_bw.pdf}
               		\caption[The entropy history of a comoving volume normalized to the observable universe today.]
			{The entropy in a comoving volume (normalized to the 
               		present observable universe). This figure illustrates the time-dependence
               		of the scheme 1 entropy budget. N.B.\ $10^{10^{100}} = 1$ googolplex. 
			\textit{[See the electronic edition of the journal for a color version of this figure.]}}
               		\label{fig:C5eoftimes1}
       		\end{center}
	\end{figure*}
}

At the far-left of the figure, we show a brief period of inflation. During this period
all of the energy is in the inflaton \citep{Guth1981,Linde1982}, which 
has very few degrees of freedom and low entropy 
\ifthenelse{\boolean{colver}} { 
(blue fill; \citealt{Linde2009,Steinhardt2009}).}{
(crosshatched fill; \citealt{Linde2009,Steinhardt2009}).}
Inflation ends 
with a period of reheating somewhere between the Planck scale ($10^{-45}s$) 
and the GUT scale ($10^{-35}s$), during which the inflaton's energy is transferred
into a relativistic fluid 
\ifthenelse{\boolean{colver}} { 
(yellow fill). }{
(dotted fill). }
During reheating, the entropy increases by
many orders of magnitude. After reheating, the constitution of the relativistic 
fluid continues to change, but the changes occur reversibly and do not increase 
the entropy.

After a few hundred million years ($\sim 10^{16}s$), the first stars form from collapsing 
clouds of neutral hydrogen and helium. Shortly thereafter the first black holes form. 
The entropy in stellar black holes (light gray) and SMBHs (dark
gray) increases rapidly during galactic evolution. The budget given in Table 
\ref{tab:C5currententropy} is a snapshot of the entropies at the present time ($4.3\xt{17}s$). 
Over the next $10^{26}s$, the growth of structures 
larger than about $10^{14}\ M_{\odot}$ will be halted by the acceleration of the 
universe. Galaxies within superclusters will merge and objects in the outer limits 
of these objects will be ejected. The final masses of SMBHs will 
be $\sim 10^{10} M_{\odot}$ \citep{Adams1997} with the entropy dominated by
those with $M \sim 10^{12} M_{\odot}$.

Stellar black holes will evaporate away into Hawking radiation in about $10^{80}s$
and SMBHs will follow in $10^{110}s$. The decrease in
black hole entropy is accompanied by a compensating increase in radiation entropy.
The thick black line in Figure \ref{fig:C5eoftimes1} represents the radiation entropy growing 
as black holes evaporate.
The asymptotic future of the entropy budget, under scheme 1, will be radiation 
dominated.

Figure \ref{fig:C5eoftimes2} illustrates the evolution of the entropy budget under 
scheme 2, i.e., the entropy within the CEH, plus the 
entropy of the CEH. 
\ifthenelse{\boolean{colver}} {
	\begin{figure*}[!p]
       		\begin{center}
               		\includegraphics[width=\linewidth]{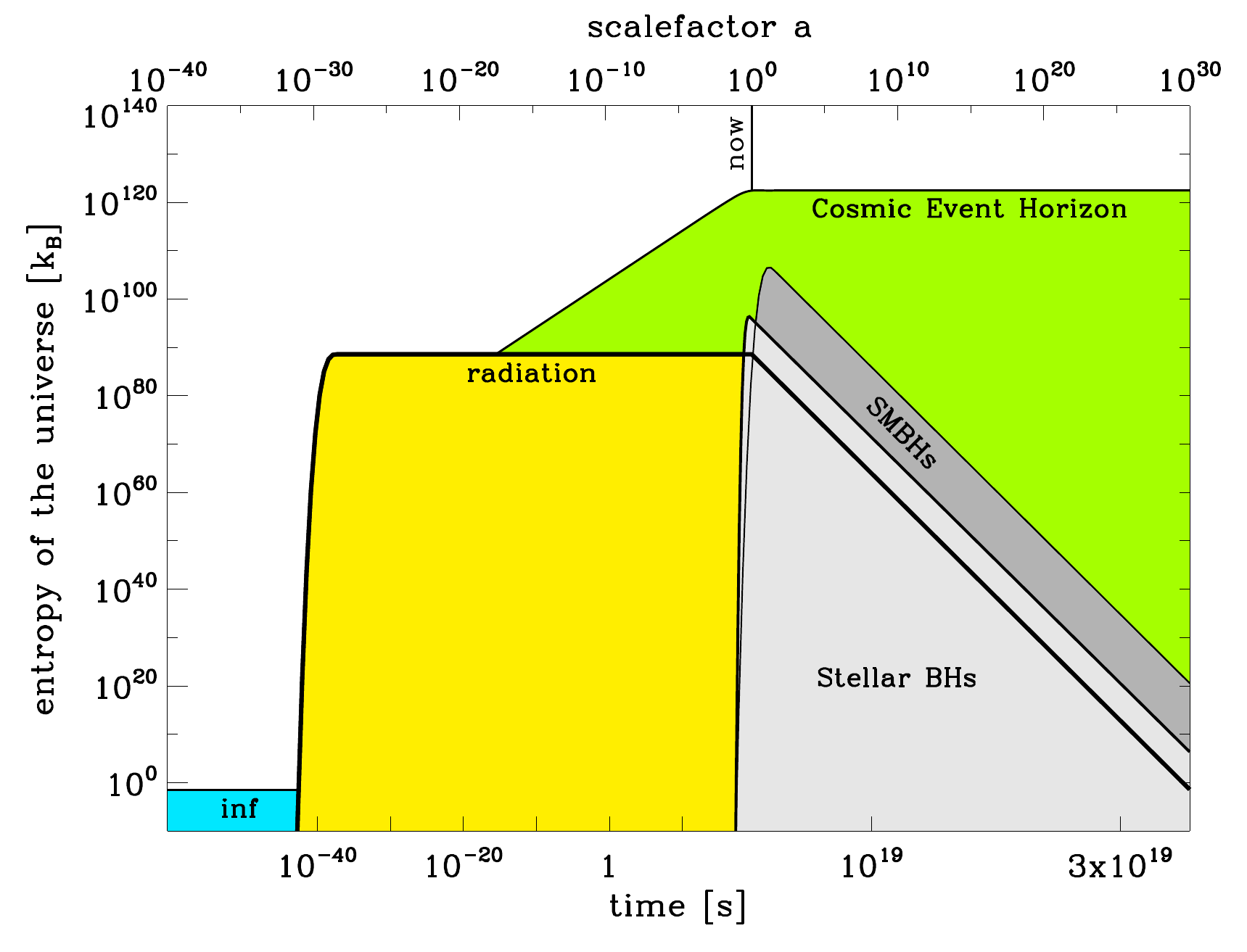}
               		\caption[The entropy history of the cosmic event horizon and its interior.]
			{Entropy of matter within the CEH, and the
               		entropy of the cosmic event horizon. This figure illustrates the time
               		dependence of the scheme 2 entropy budget. Note: the horizontal 
               		axis is shorter than in Figure \ref{fig:C5eoftimes1}.}
               		\label{fig:C5eoftimes2}
       		\end{center}
	\end{figure*}
}{
	\begin{figure*}[!p]
       		\begin{center}
               		\includegraphics[width=\linewidth]{15_chapter5_v2_apj/eoftimes2_bw.pdf}
               		\caption[The entropy history of the cosmic event horizon and its interior.]
			{Entropy of matter within the CEH, and the
               		entropy of the CEH. This figure illustrates the time
               		dependence of the scheme 2 entropy budget. Note: the horizontal 
               		axis is shorter than in Figure \ref{fig:C5eoftimes1}. \textit{[See the electronic 
			edition of the journal for a color version of this figure.]}}
               		\label{fig:C5eoftimes2}
       		\end{center}
	\end{figure*}
}

Whereas in scheme 1 we integrate over a constant comoving volume, here
the relevant volume is the event horizon. The event horizon is discussed in some 
detail in the Appendix. During radiation domination, the comoving 
radius of the CEH is approximately constant (the proper 
distance grows as $R_{\mathrm{CEH}} \propto a$) and in the dark energy dominated 
future, it is a constant proper distance ($R_{\mathrm{CEH}} = \mathrm{constant}$). 
The few logarithmic decades around the present time cannot be described
well by either of these.

Since the event horizon has been approximately comoving in the past, the 
left half of Figure \ref{fig:C5eoftimes2} is almost the same as in Figure \ref{fig:C5eoftimes1}
except that we have included the event horizon entropy 
\ifthenelse{\boolean{colver}} { 
(green fill). }{
(striped fill). } 
The event horizon entropy dominates this budget from about $10^{-16}s$.

After dark energy domination sets in, the CEH becomes a
constant proper distance. The expansion of the universe causes comoving 
objects to recede beyond the CEH. On average, the number 
of galaxies, black holes, photons etc.\ within our CEH decreases 
as $a^{-3}$. The stellar and SMBH entropy contained
within the CEH decreases accordingly (decreasing gray filled regions).

The decreasing black hole entropy (as well as other components not shown) 
is compensated by the asymptotically growing CEH entropy 
(demonstrated explicitly for a range of scenarios in \citealt{Davis2003}), and thus 
the second law of thermodynamics is satisfied. 
See \citet[in preparation]{EganLineweaver2009b} for further discussion of the time-dependence 
of the entropy of the universe.

\section*{Acknowledgments}

We are grateful for many useful discussions with Tamara Davis, 
Ken Freeman, \addthis{Pat Scott,} Geoff Bicknell, Mike Turner, Andrei Linde, and Paul 
Steinhardt. C.A.E. thanks Anna Fransson for financial support and the Research School of 
Astronomy and Astrophysics, Australian National University, for its hospitality during the 
preparation of this paper.

\section*{Appendix: The observable universe and the cosmic event horizon} \label{C5phceh}

Here we calculate the radius and volume of the observable universe (for use in Section \ref{C5scheme1});
and we calculate the radius, volume, and entropy of the CEH (for use in Section \ref{C5scheme2}).
We use numerical methods to track the propagation of errors from the cosmological parameters. 


The radius of the observable universe (or particle horizon) is
\begin{eqnarray}
	R_{\mathrm{obs}} = a(t) \int_{t'=0}^{t} \frac{c}{a(t')} dt'. \label{eq:C5robs}
\end{eqnarray}
Here $a(t)$ is the time-dependent scalefactor of the universe given by the Friedmann equation for a 
flat cosmology
\begin{eqnarray}
	\frac{da}{dt} = \sqrt{\frac{\Omega_{r}}{a^2} + \frac{\Omega_{m}}{a} + \frac{\Omega_{\Lambda}}{a^{-2}}}.
\end{eqnarray}
Hubble's constant and the 
matter density parameter are taken from \citet{Seljak2006}: 
$h = H / 100\ km\ s^{-1}\ Mpc^{-1} = 0.705 \pm 0.013$, $\omega_m = \Omega_m h^2 = 0.136 \pm 0.003$.
The radiation density is calculated from the observed CMB temperature, 
$T_{\mathrm{CMB}} = 2.725 \pm 0.002\ K$ \citep{Mather1999}, using
$\Omega_{r} = \frac{8 \pi G}{3 H^2} \frac{\pi^2 k^4 T^4}{15 c^5 \hbar^3}$.
The vacuum energy density parameter is determined by flatness, 
$\Omega_{\Lambda} = 1-\Omega_{r}-\Omega_{m}$.

A distribution of $R_{\mathrm{obs}}$ values is built up by repeatedly evaluating Equation (\ref{eq:C5robs})
at the present time (defined by $a(t_0)=1$) using 
cosmological parameters randomly selected from the allowed region of $h - \omega_{m} - T_{\mathrm{CMB}}$ 
parameter space (assuming uncorrelated Gaussian errors in these parameters). We find
\begin{eqnarray}
	R_{\mathrm{obs}} = 46.9 \pm 0.4\ \mathrm{Glyr}
\end{eqnarray}
with an approximately Gaussian distribution. The quoted confidence interval here, and elsewhere
in this Appendix, is $1\sigma$.
The volume of the observable universe $V_{\mathrm{obs}}$ is calculated using the normal formula for 
the volume of a sphere. 
\begin{eqnarray}
	V_{\mathrm{obs}} & = & 43.2 \pm 1.2 \xt{4}\ \mathrm{Glyr}^3 \nonumber \\ 
		& = & 3.65 \pm 0.10 \xt{80}\ m^3 
\end{eqnarray}
See Figure \ref{fig:C5vphveh}.
Uncertainty in $R_{\mathrm{obs}}$ and $V_{\mathrm{obs}}$ is predominantly due to uncertainty in $\omega_m$ 
however $h$ also makes a non-negligible contribution.

\ifthenelse{\boolean{colver}} {
	\begin{figure*}[!p]
       		\begin{center}
               		\includegraphics[width=\linewidth]{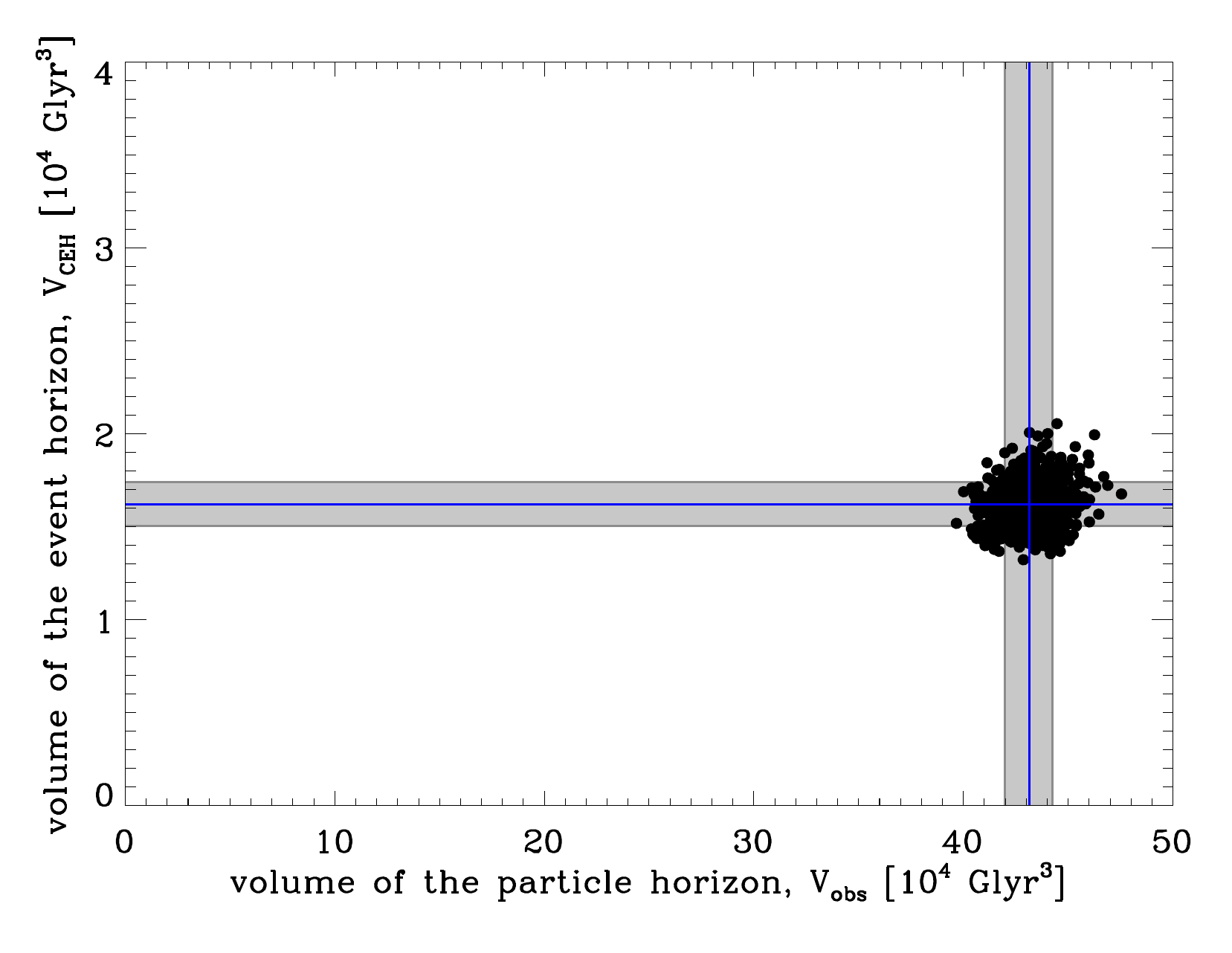}
               		\caption[The volume of the event horizon and the volume of the particle horizon.]
			{Eight hundred realizations of $V_{\mathrm{obs}}$ and $V_{\mathrm{CEH}}$ indicate the volume of the 
               		observable universe is $43.2 \pm 1.2 \xt{4}\ \mathrm{Glyr}^3$ (horizontal axis) and the volume of 
               		the cosmic event horizon is $V_{\mathrm{CEH}}=1.62 \pm 0.12 \xt{4}\ \mathrm{Glyr}^3$ (vertical axis). We 
               		note that there is only a weak correlation between uncertainties in the two volumes.}
               		\label{fig:C5vphveh}
       		\end{center}
	\end{figure*}
}{
	\begin{figure*}[!p]
       		\begin{center}
               		\includegraphics[width=\linewidth]{15_chapter5_v2_apj/vph_and_veh_bw.pdf}
               		\caption[The volume of the event horizon and the volume of the particle horizon.]
			{Eight hundred realizations of $V_{\mathrm{obs}}$ and $V_{\mathrm{CEH}}$ indicate the volume of the 
               		observable universe is $43.2 \pm 1.2 \xt{4}\ \mathrm{Glyr}^3$ (horizontal axis) and the volume of 
               		the cosmic event horizon is $V_{\mathrm{CEH}}=1.62 \pm 0.12 \xt{4}\ \mathrm{Glyr}^3$ (vertical axis). We 
               		note that there is only a weak correlation between uncertainties in the two volumes. 
			\textit{[See the electronic edition of the journal for a color version of this figure.]}}
               		\label{fig:C5vphveh}
       		\end{center}
	\end{figure*}
}


The radius of the CEH at time $t$ is given by integrating along a photon's 
world line from the time $t$ to the infinite future.
\begin{eqnarray} \label{eq:C5deh}
	R_{\mathrm{CEH}} = a(t_{now}) \int_{t=t_{now}}^{\infty} \frac{c}{a(t)} dt
\end{eqnarray}
This integral is finite because the future of the universe is dark energy dominated.
Using the same methods as for the observable universe, we find the present radius 
and volume of the CEH to be
\begin{eqnarray}
	R_{\mathrm{CEH}} = 15.7 \pm 0.4\ \mathrm{Glyr},
\end{eqnarray}
and
\begin{eqnarray}
	V_{\mathrm{CEH}} & = & 1.62 \pm 0.12 \xt{4}\ \mathrm{Glyr}^3, \nonumber \\ 
		& = & 1.37 \pm 0.10 \xt{79}\ m^3. 
\end{eqnarray}

\ifthenelse{\boolean{colver}} {
	\begin{figure*}[!p]
       		\begin{center}
               		\includegraphics[width=\linewidth]{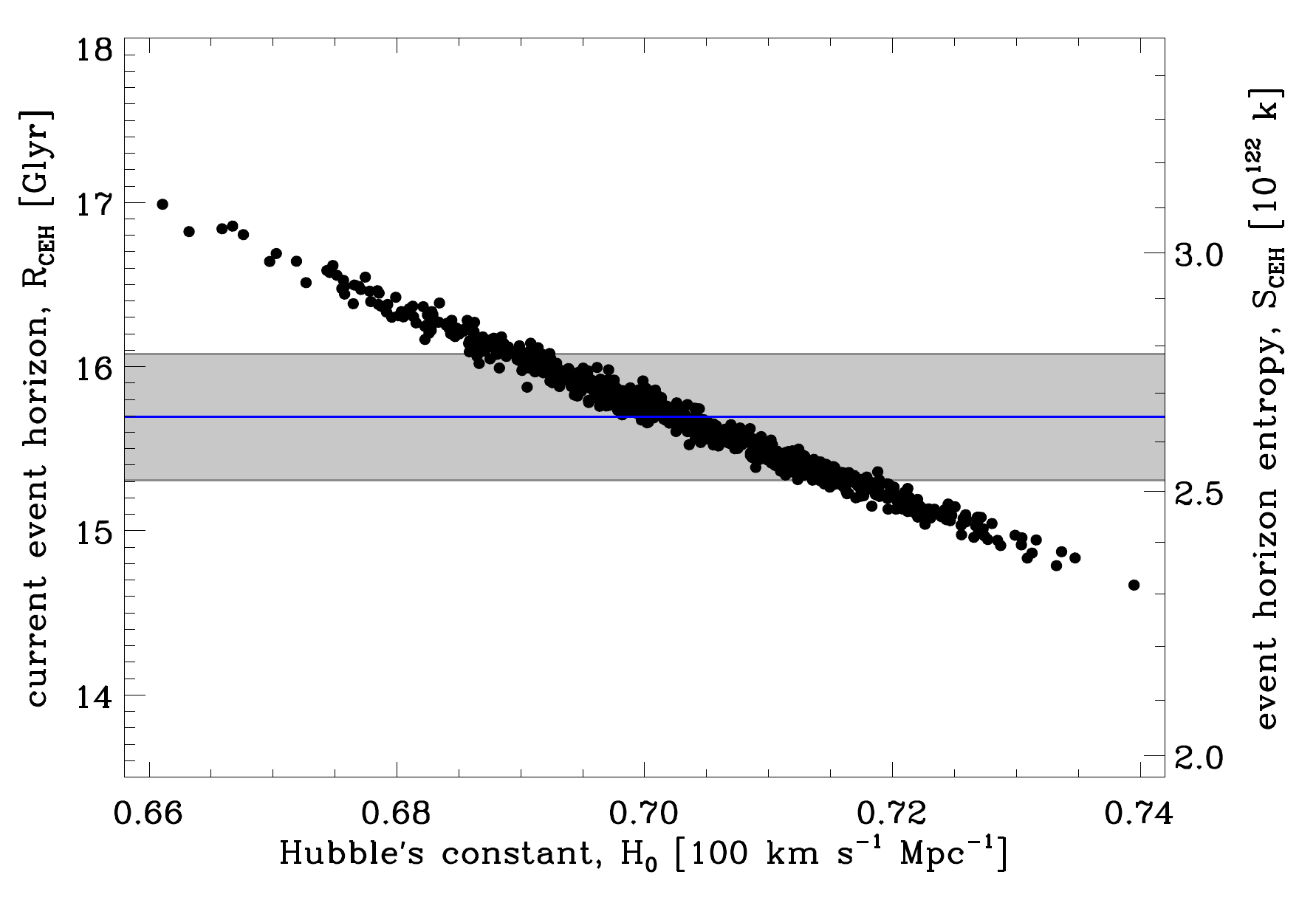}
               		\caption[The cosmic event horizon and its entropy showing uncertainty dependence on $h_0$.]
			{We find $S_{\mathrm{CEH}} = 2.6 \pm 0.3 \xt{122}\ k$, in agreement with previous 
               		estimates $S_{\mathrm{CEH}} \sim 10^{122}\ k$ \citep{Bousso2007}. Uncertainties in $S_{\mathrm{CEH}}$ 
               		come from uncertainties in $R_{\mathrm{CEH}}$, which are almost exclusively due to uncertainties 
			in $h$.}
               		\label{fig:C5dehh0}
       		\end{center}
	\end{figure*}
}{
	\begin{figure*}[!p]
       		\begin{center}
               		\includegraphics[width=\linewidth]{15_chapter5_v2_apj/deh_h0_bw.pdf}
               		\caption[The cosmic event horizon and its entropy showing uncertainty dependence on $h_0$.]
			{We find $S_{\mathrm{CEH}} = 2.6 \pm 0.3 \xt{122}\ k$, in agreement with previous 
               		estimates $S_{\mathrm{CEH}} \sim 10^{122}\ k$ \citep{Bousso2007}. Uncertainties in $S_{\mathrm{CEH}}$ 
               		come from uncertainties in $R_{\mathrm{CEH}}$, which are almost exclusively due to uncertainties 
			in $h$. \textit{[See the electronic edition of the journal for a color version of this figure.]}}
               		\label{fig:C5dehh0}
       		\end{center}
	\end{figure*}
}

\ifthenelse{\boolean{colver}} {
	\begin{figure*}[!p]
       		\begin{center}
               		\includegraphics[width=\linewidth]{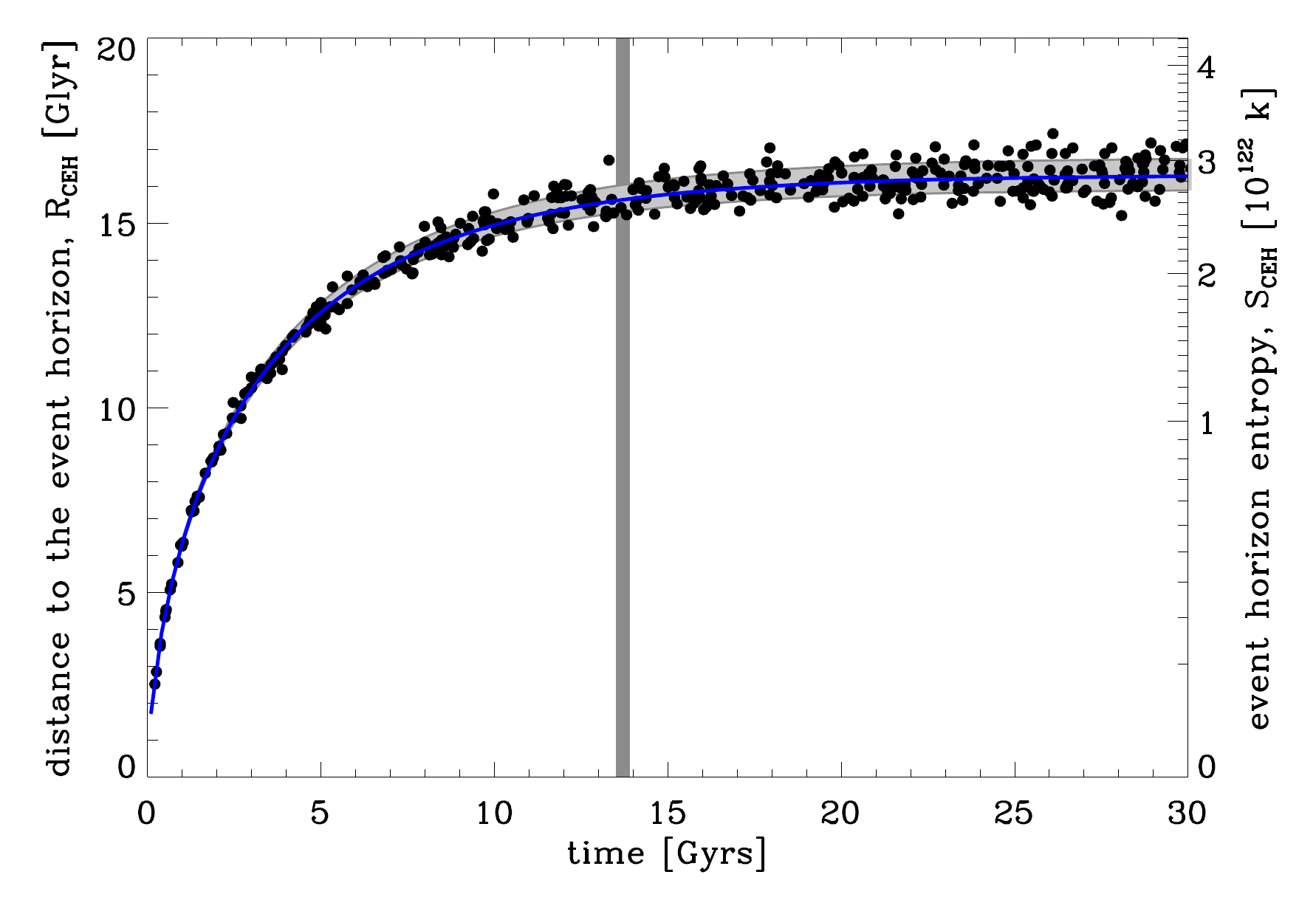}
               		\caption[The proper distance to the cosmic event horizon as a function of time.]
			{Proper distance to the event horizon is shown as a function
               		of time. The vertical gray line represents the present age of the universe (and its width, the 
               		uncertainty in the present age).
               		During dark energy domination, the proper radius, proper volume, and entropy 
               		of the CEH will monotonically increase, asymptoting to a constant.
			} \label{fig:C5deht}
       		\end{center}
	\end{figure*}
}{
	\begin{figure*}[!p]
       		\begin{center}
               		\includegraphics[width=\linewidth]{15_chapter5_v2_apj/deht_bw.pdf}
               		\caption[The proper distance to the cosmic event horizon as a function of time.]
			{Proper distance to the event horizon is shown as a function
               		of time. The vertical gray line represents the present age of the universe (and its width, the 
               		uncertainty in the present age).
               		During dark energy domination, the proper radius, proper volume, and entropy 
               		of the CEH will monotonically increase, asymptoting to a constant. 
			\textit{[See the electronic edition of the journal for a color version of this figure.]}}
               		\label{fig:C5deht}
       		\end{center}
	\end{figure*}
}

The entropy of the CEH is calculated using the Bekenstein-Hawking horizon 
entropy equation as suggested by \citet{Gibbons1977}. 
\begin{eqnarray}
	S_{\mathrm{CEH}} & = & \frac{k c^3}{G \hbar} \frac{A}{4} = \frac{k c^3}{G \hbar} \pi R_{\mathrm{CEH}}^2 \nonumber \\
		& = & 2.6 \pm 0.3 \xt{122}\ k
\end{eqnarray}
Uncertainty in the CEH radius, volume, and entropy are dominated by 
uncertainties in Hubble's constant (Figure \ref{fig:C5dehh0}).

The CEH monotonically increases, asymptoting to a constant radius and 
entropy slightly larger than its current value (see Figure \ref{fig:C5deht}). We calculate the
asymptotic radius, volume, and entropy to be
\begin{eqnarray}
	R_{\mathrm{CEH}}(t \rightarrow \infty) & = & 16.4 \pm 0.4\ \mathrm{Glyr} \nonumber \\   			
			& = & 1.55 \pm 0.04 \xt{26}\ m   			
\end{eqnarray}
\begin{eqnarray}
	V_{\mathrm{CEH}}(t \rightarrow \infty) & = & 1.84 \pm 0.15 \xt{4}\ \mathrm{Glyr}^3 \nonumber \\ 	
			& = & 1.56 \pm 0.13 \xt{79}\ m^3  		
\end{eqnarray}
\begin{eqnarray}
	S_{\mathrm{CEH}}(t \rightarrow \infty) & = & 2.88 \pm 0.16 \xt{122}\ k. 		
\end{eqnarray}

%


%
\pagestyle{fancy}
\chapter{How High Could the Entropy be and Will the Universe End in a Heat Death?}
\label{chap6}

\begin{center}
\emph{
If we should stay silent, \\
if fear should win our hearts, \\
our light will have long diminished, \\
before it reaches the farthest star.
}
\end{center}
\begin{flushright}
- VNV Nation, ``The Farthest Star'' \hspace*{2cm}
\end{flushright}
\vspace{1cm}

\section{Introduction} \label{C6smaxintro}

The increase of entropy (and use of free energy) drives all dissipative 
physical processes in the universe including gravitational clustering, accretion disks 
and supernovae, stellar fusion, terrestrial weather, chemical reactions, geological processes 
and terrestrial-planet-bound biology \citep{Frautschi1982,LineweaverEgan2008}.
%

\begin{figure*}[!p]
   	\begin{center}
		\includegraphics[width=0.7\linewidth]{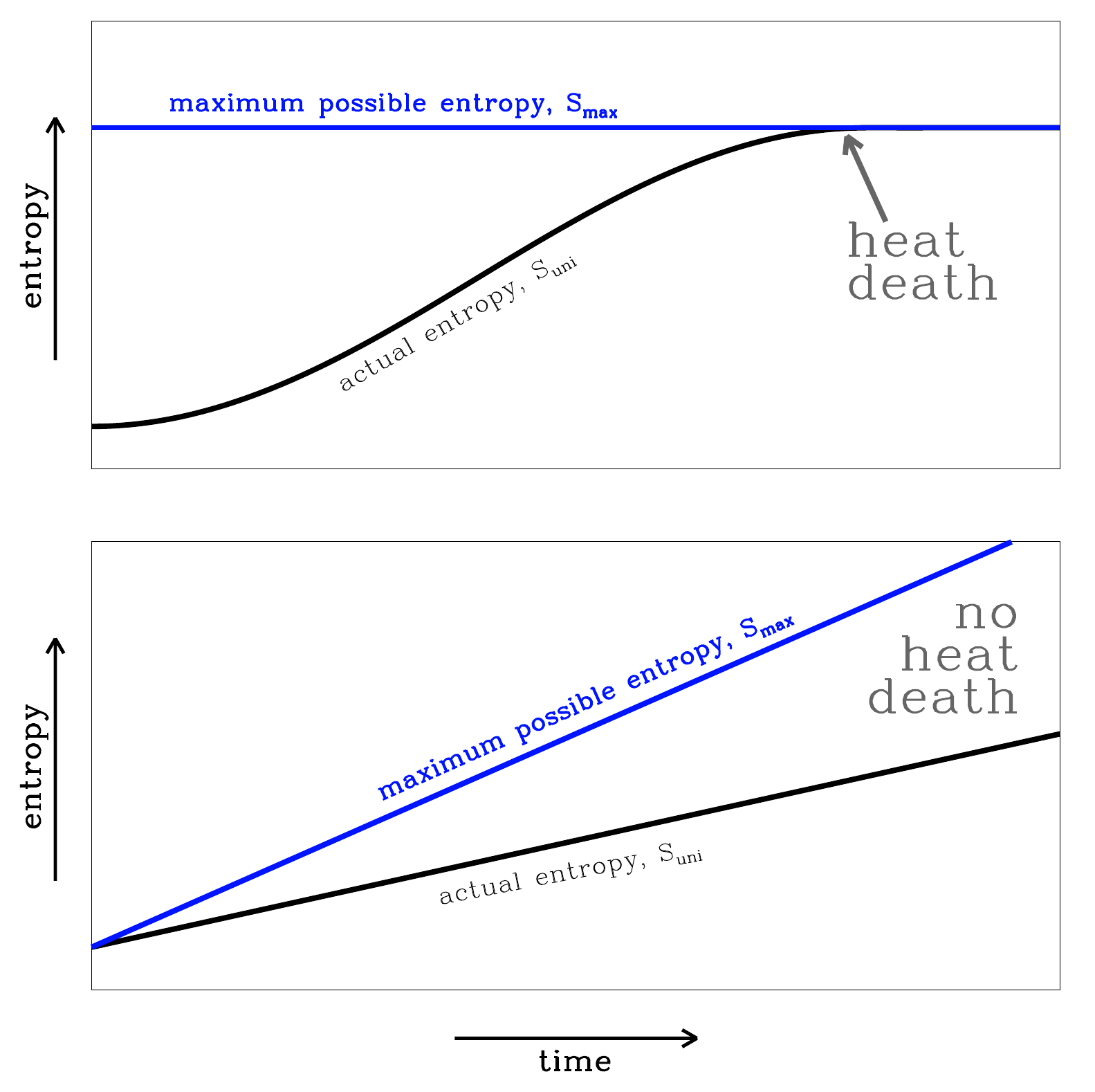}
   		\caption[Does the Universe end in a heat death?]
		{Whether the Universe eventually achieves maximum entropy 
		depends on the time dependence of the maximum entropy $S_{max}$ 
		and the actual entropy of the Universe, $S_{uni}$. There is some ambiguity 
		about how to best define $S_{max}$.}
		\label{fig:C6naive_entropy_gap}
	\end{center}
\end{figure*}

The long-term sustainability of dissipative processes (including life) depends on the 
availability of free energy in the future. If, for any reason, the entropy of the universe 
$S_{uni}$ achieves a value that cannot be further increased, then the universe 
enters a heat death (see figure \ref{fig:C6naive_entropy_gap}). This idea motivates the 
exploration of potential entropy growth and entropy limits.

\citet{Frampton2008,Frampton2008b} recently estimated that the present entropy of the 
observable Universe is $10^{102}\ k - 10^{103}\ k$ and that the maximum entropy the 
universe could have was $S_{max} \sim 10^{123}\ k$.  
Their maximum entropy was calculated by applying the holographic bound 
\citep{tHooft1993,Susskind1995} to the present volume of the observable Unvierse. 
That the increase of entropy has not yet been capped by some limiting value is the 
reason that dissipative processes are ongoing and that life can exist.

\begin{figure*}[!p]
       \begin{center}
               \includegraphics[scale=1.2]{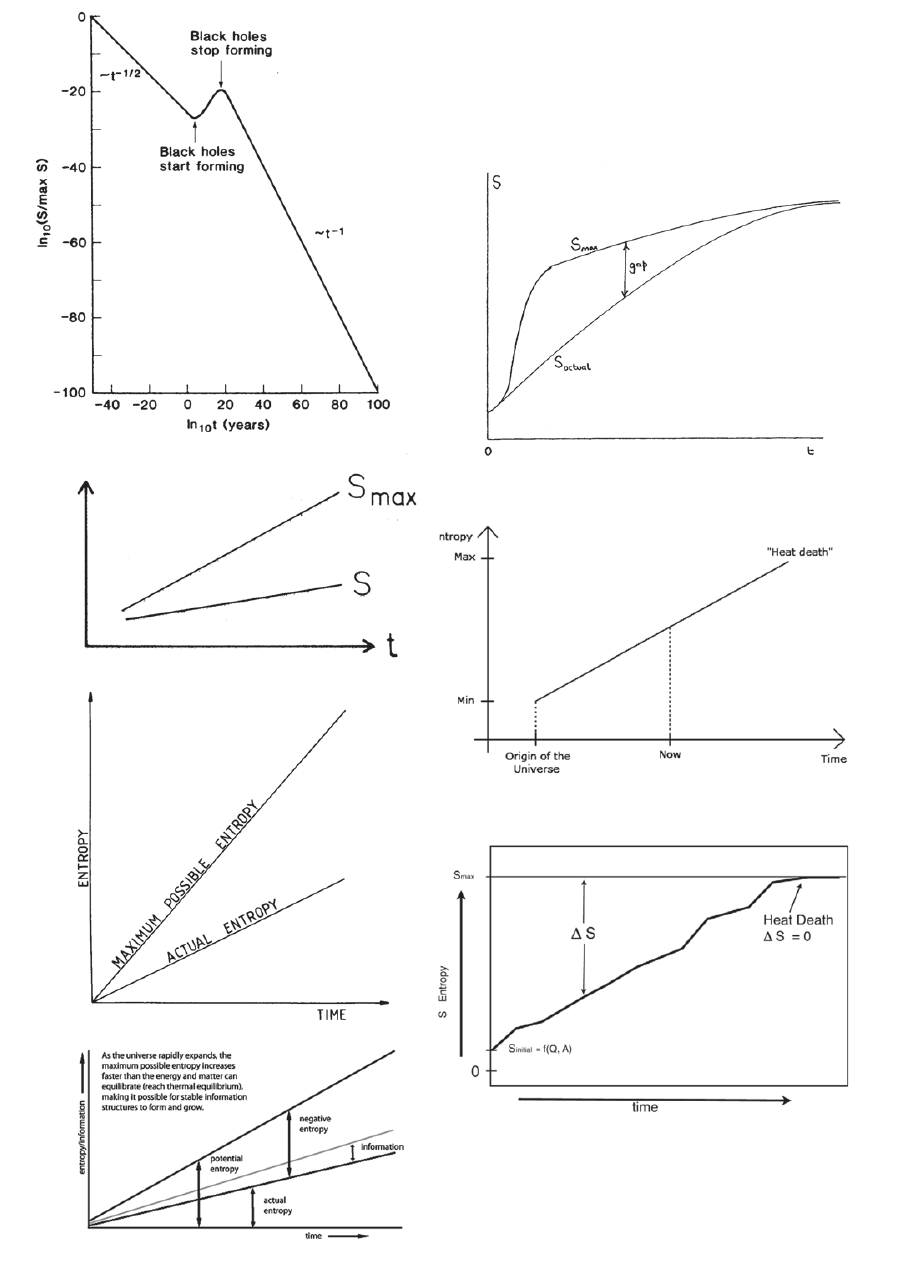}
               \caption[Disagreement in the literature on the issue of the eventual heat death.]
               {The figures in the left column are, starting at the top, from 
               \citet{Frautschi1982}, \citet{Frautschi1988}, \citet{Barrow1994} and \citet{Layzer2009}
               and show $S_{max}$ growing indefinitely (and faster than $S_{uni}$). In these 
               figures the universe does not end in a heat death and free energy is always 
               available to drive dissipative processes (including life).
               The figures in the right column are, starting at the top, from 
               \citet{Davies1994}, \citet{Thomas2009} (depicting the description given by
               \citet{Penrose2004}) and \citet{LineweaverEgan2008}. They show $S_{max}$ as
               a constant, or asymptoting to a constant, which is eventually reached by the 
               actual entropy of the universe $S_{max}$. The future depicted in these figures
               is very different to those in the left column: here the universe runs out of free
               energy and all dissipative processes cease. The goal of this work is to 
               understand the differences between these two points of view and help
               lead to a resolution of the fate of life in the Universe.} \label{fig:C6smaxlit}
       \end{center}
\end{figure*}

However there remains some ambiguity about how to best define the maximum 
entropy $S_{max}$ and whether or not the entropy of the Universe $S_{uni}$ will reach 
$S_{max}$. 
Figure (\ref{fig:C6smaxlit}) shows several illustrations depicting the relationship between 
$S_{uni}$ and $S_{max}$ that have appeared in the literature, books and popular
science over the past three decades.

\citet{Adams1997} provide an excellent overview of the processes which, according to
our current understanding, will dominate the future evolution of the universe, but in their 
brief discussion of the thermodynamic fate of a $\Lambda$CDM universe (now the standard
model of our Universe) they ultimately leave the question of whether our Universe will reach equilibrium, and
end in a heat death, open.

More recently (e.g.\ \citealt{Bousso2007,Mersini2008}), the maximum entropy of a de Sitter future 
has been discussed in the context of anthropic explanations for the low density of the dark 
energy (the so-called cosmological constant problem), but several issues remain to be clarified. 

Below are a number of considerations we have identified to help understand disagreement
in the literature. We resolve some, and some deserve further discussion. 
\begin{enumerate}
	\item {\bf Theoretically motivated $S_{max}$:} 
		If $S_{max}$ is defined using a theoretically motivated entropy bound such as the 
		Bekenstein bound \citep{Bekenstein1981} or the holographic bound 
		\citep{tHooft1993,Susskind1995} (both of which apply to weakly gravitating systems) 
		or using the more recent covariant entropy bound \citet{Bousso1999,Bousso2002} 
		(which may apply to strongly gravitating systems), then the result depends on which of 
		these is used. 
	\item {\bf Should we condition on the available energy?} In a spherical system 
		of radius $R$ which contains mass $M << \frac{R c^2}{2 G}$, the entropy is maximized 
		by converting that energy into massless radiation rather than a black hole (see 
		\citet{Page1981}). The resulting entropy is $S = \frac{M}{T} << R^{3/2}$ (in Planck units
		and dropping constants of order 1)
		and is much less than the holographic bound $S << R^2$ (same conventions). 
		This example illustrates that even while theoretically motivated bounds may hold,
		they are not always the best choice for $S_{max}$. If our system is constrained 
		by the amount of energy that is available, then a lower $S_{max}$ may apply. 
	\item {\bf Should we condition on the equation of state?} The energy in a comoving volume 
		$\rho_{\chi}$ is not conserved in an expanding universe (see e.g.\ \citep{Carroll2004book}).  
		Generally $\rho_{\chi} = \rho a^3 \propto a^{-3w}$ where $w$ is the equation of state, 
		with $w=1/3$ for radiation, $w=0$ for matter and $w=-1$ for dark energy. The amount 
		of energy available at a future time does depend on what form (what $w$) the energy
		is stored in. The question of the future of entropy production of the universe therefore 
		depends on whether or not we suppose that energy can be transferred between 
		different equations of state. 
		\citet{Harrison1995} pointed out that energy could be mined from the universe by tethering 
		distant galaxies. A network of tethered galaxies has a negative pressure and an 
		equation of state $w<0$ and is among the scenarios we may be 
		interested in considering in an analysis of possible future entropy production. 
	\item {\bf Normalization volume:} Whether the entropy of the universe is increasing 
		or not can depend on the definition of ``the Universe''. For example, while the entropy 
		in a comoving volume is constant during adiabatic expansion, the entropy in the 
		observable universe may grow due to the growth of the particle horizon (which bounds 
		the observable universe). 
	\item {\bf How efficiently can energy be collected?} If dissipative processes require the 
		collection of matter, then the transport costs (energy and entropy) need to be included 
		in the calculations.
	\item {\bf Other possible constraints:}
		When asking what the entropy of the universe could be, one is suggesting a 
		universe which is different to the real universe, but has not specified how it is different. 
		Some of the above points, such as ``Should we condition on the available energy?'' 
		and ``Should we condition on the equation of state?'' are examples of aspects which 
		are not clearly defined, but there may also be other, more subtle, issues. For example, 
		\citet{Frautschi1982} evaluates $S_{max}$ during the radiation era by supposing the 
		creation of a large black hole from radiation, but given that mechanisms for this did not 
		exist, the available free energy so calculated have little to do with reality, and may not 
		be of interest.
\end{enumerate}

In the present work we assume an FRW expanding universe 
with the concordance $\Lambda$CDM parameter values, $h_0=0.71$, $\Omega_{m}=0.27$ and 
$\Omega_{DE}=0.73$ \citep{Seljak2006}.

When we say ``the entropy of the universe, $S_{uni}$'' we mean the entropy in the sphere of 
comoving radius $46\ Glyr$ that is now the observable universe. Since the particle 
horizon grows in comoving coordinates, our $46\ Glyr$ comoving sphere was 
larger than the observable universe in the past and will not include the whole 
observable universe in the future.

Our aim is to investigate the issues we have listed in this introduction, and others 
that may arise. 

In Section \ref{C6maxentropy} we evaluate various entropy bounds $S_{max}$ 
that have been proposed in the literature. By applying these bounds on our
chosen volume, using a consistent cosmology, we gain some insight into their 
differences and similarities.
In Section \ref{C6smaxdiscussion} we explore a natural definition of $S_{max}$ as 
the entropy at which the universe has zero free energy $F$ given the available 
energy $U$ and exhaust temperatures $T_{exh}$. The conclusions of these 
preliminary investigations are also given in Section \ref{C6smaxdiscussion}.

\section{Different Versions of $S_{max}$} \label{C6maxentropy}

\subsection{The Holographic Bound}

The most discussed entropy bound is the holographic bound \citep{tHooft1993,Susskind1995}: 
the entropy within a sphere of radius $R$ will not exceed that of a black hole of radius $R$,
\begin{eqnarray}
	S_{sphere\ R} \le S_{Hol-Bound} = \frac{k c^3}{G \hbar} \pi R^2.
\end{eqnarray}

For a comoving volume of physical radius $R = \chi a$ the holographic bound grows as the
square of the cosmic scalefactor,
\begin{eqnarray}
	S_{Hol-Bound} = \frac{k c^3}{G \hbar} \pi \chi^2 a^2,
\end{eqnarray}
and becomes exponentially large in the future.

\begin{figure*}[!p]
       \begin{center}
               \includegraphics[width=0.8\linewidth]{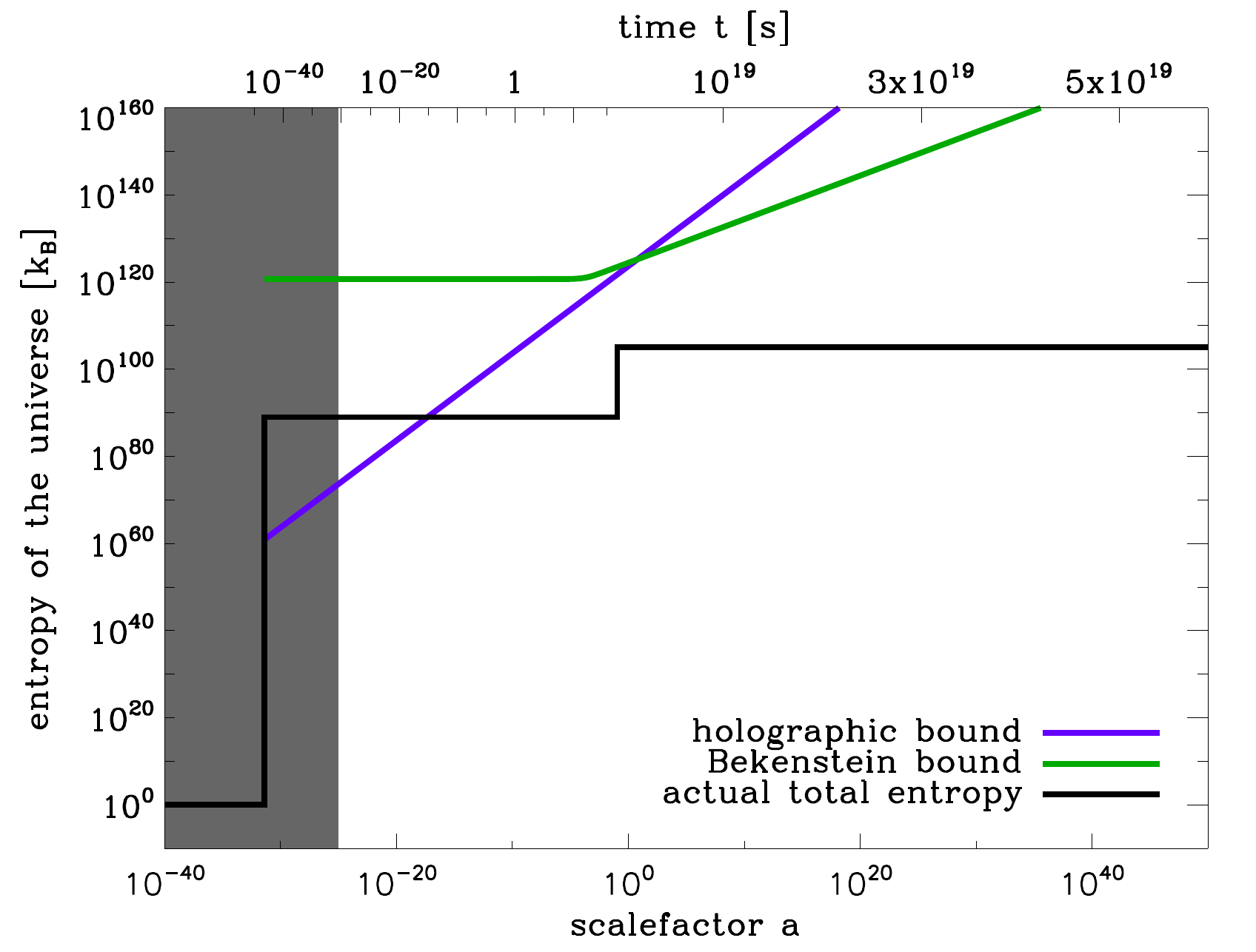}
               \caption[The holographic bound and the Bekenstein bound applied to a comoving 
               volume normalized to the current observable Universe.]
               {Two versions of $S_{max}$: the holographic bound and the Bekenstein bound.
               The actual entropy history of the universe from \citet{EganLineweaver2009} is plotted in black. The 
               holographic bound is violated by the actual entropy history of the universe during the 
               radiation epoch. We shade regions above the GUT temperature, where the evolution of
               the universe (and its entropy) becomes more speculative.
               }\label{fig:C6eofscale_firstplot}
       \end{center}
\end{figure*}

In figure \ref{fig:C6eofscale_firstplot} the holographic bound is applied to a sphere of comoving radius $46\ Glyr$
(the purple line). The actual entropy in the comoving sphere violates  the holographic bound 
during part the radiation era, as shown in the figure. This occurs because the entropy density $s$
of the Universe is approximately homogenous on large scales. The entropy in a volume $V$ is 
$S \propto V \propto length^3$ whereas the holographic bound grows as the surface area of
the volume $S_{Hol-Bound} \propto length^2$ and for a large enough volume $S > S_{Hol-Bound}$.
This violation has been used by \citet{Bousso2002} to motivate a covariant form of the 
holographic bound. At least in its original form, the holographic bound does not deliver a 
suitable maximum entropy $S_{max}$ for the universe.


\subsection{The Bekenstein Bound}

Historically preceding the holographic bound, the Bekenstein bound \citep{Bekenstein1981}
is the result of a gedankenexperiment in which a package of energy $E$, radius $R$ and 
entropy $S$ is deposited into a black hole. The bound,
\begin{eqnarray}
	S_{Bek-Bound} & = & 2 \pi \frac{k}{\hbar c} R E,
\end{eqnarray}
is required by the second law: the entropy $S$ lost into the black hole must not be larger than the 
increase in the horizon entropy of the black hole. 
Several papers have studied the related effect whereby entropy-containing-matter recedes
across the cosmic event horizon to extract similar bounds on the entropy density of matter
\citep{Davies1987,Bousso2001}. 

The plausibility of the Bekenstein bound was confirmed for numerous weakly gravitating 
systems \citep{Bekenstein2005}, but the bound is now known to fail for some gravitationally 
unstable systems (e.g.\ \citealt{Bousso2002}).

In the context of the flat-FRW universe, 
\begin{eqnarray}
	S_{Bek-Bound} & = & 2 \pi \frac{k}{\hbar c} (\chi a) \left(\rho \frac{4 \pi \chi^3 a^3}{3} \right) \nonumber \\
		& = & \frac{8 \pi^2}{3} \frac{k}{\hbar c} \chi^4 \rho a^4.
\end{eqnarray}
The Bekenstein bound is constant during the radiation era (when $\rho \propto a^{-4}$) and
increases during the matter and de Sitter eras ($\rho \propto a^{-3}$ and $\rho = const$ 
respectively). Compared to the Holographic bound, the Bekenstein bound is weaker when 
applied to regions larger than the Hubble sphere, and stronger when applied to regions 
smaller than the Hubble sphere. 
Notice that the Bekenstein bound is not violated in figure \ref{fig:C6eofscale_firstplot} (the green 
line).

\subsection{The Covariant Entropy Bound} \label{C6cebsmax}

A covariant formulation of the Holographic bound was advanced by \citet{Bousso1999}: 
the entropy $S$ on convergent light-sheets $L$ from a closed surface $B$ will 
not exceed $\frac{1}{4}$ the area of $B$,
\begin{eqnarray}\label{eq:C6covsbound}
	S[L(B)] \le \frac{k c^3}{G \hbar} \frac{A(B)}{4}.
\end{eqnarray} 
In flat spacetime convergent light-sheets from a closed surface $B$ cover the entire 
interior of $B$ and the covariant entropy bound (CEB) is the same as the original 
holographic bound: the entropy interior to $B$ cannot exceed $\frac{1}{4}$ the area 
of $B$. In general spacetimes the convergent light-sheets may not cover the interior 
of $B$. Specifically, the light sheets may be terminated by a singularity (such as the 
big bang) or they may stop converging and start to diverge (in which case they are 
truncated). In both of these cases the light sheets only cover part of the interior of $B$ 
and the CEB is weaker than the corresponding holographic bound.

\begin{figure*}[!p]
       \begin{center}
               \includegraphics[width=0.8\linewidth]{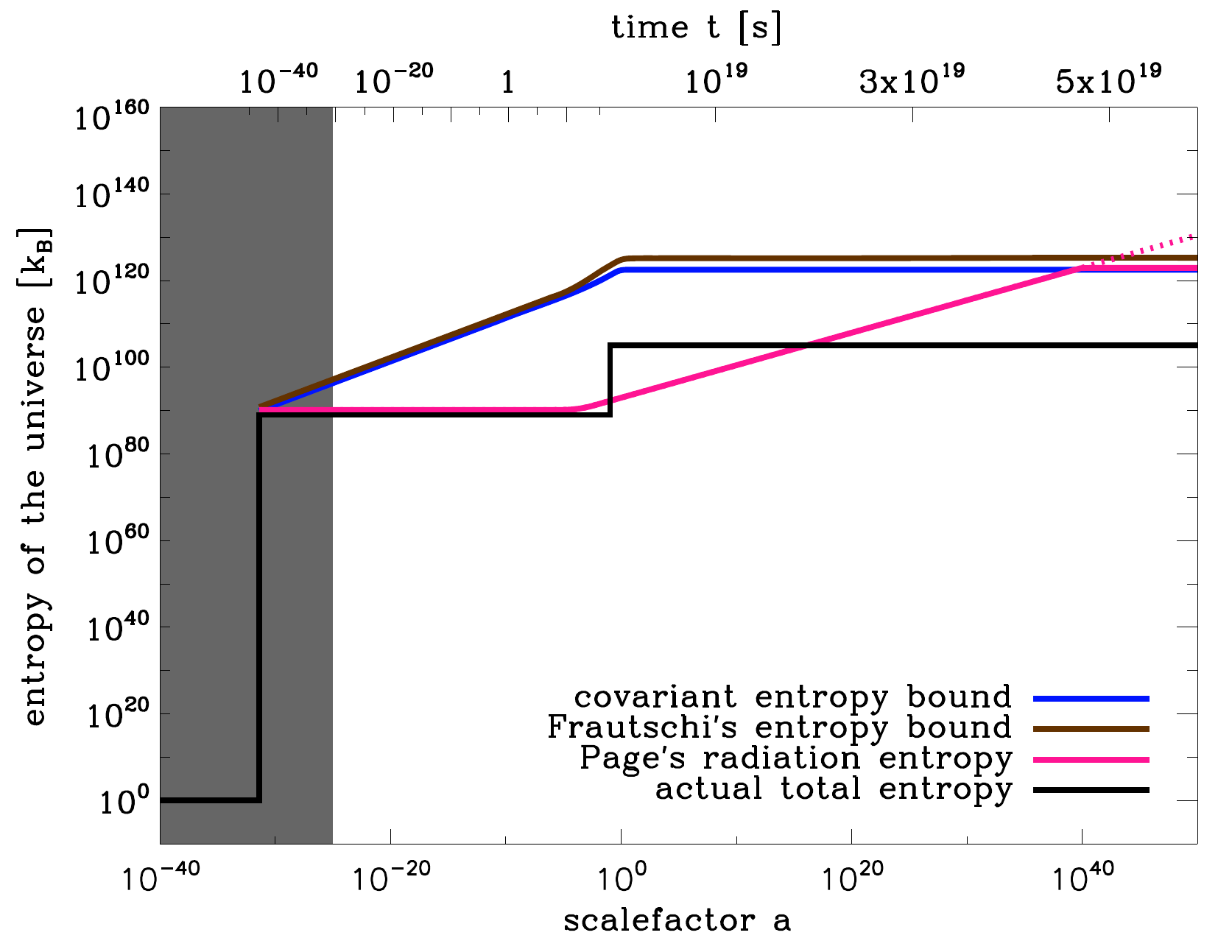}
               \caption[The covariant entropy bound, Frautshi's entropy bound and Page's
               radiation entropy applied to a comoving volume normalized to the current
               observable Universe.]
               {More versions of $S_{max}$: the covariant entropy bound, Frautschi's 
               maximum entropy and Page's evaporated matter entropy in orange with (solid) 
               and without (dotted) consideration for the de Sitter temperature. 
               The actual entropy of the universe is plotted in black. Page's evaporated matter 
               entropy is presently exceeded by the actual entropy history of the universe.}
               \label{fig:C6eofscale_secondplot}
       \end{center}
\end{figure*}

We calculate the covariant entropy bound in an expanding, flat, FRW universe by choosing 
the surface $B$ and light-sheet $L$ such that the entropy density bound on the light-sheets is 
strongest (following the prescription given in \citet{Bousso1999}). We find that the strongest 
bound on the comoving entropy density comes from the past-outgoing lightsheet of a closed 
spherical surface $B$ with a radius infinitessimaly larger than $c/H$. In this case the comoving 
entropy density is limited to
\begin{eqnarray}
	s_{\chi\ ceb} & \le & \frac{k c^2}{G \hbar} \frac{1}{4 \eta + 4 \eta^2 H + \frac{4 \eta^3 H^2}{3}} 
\end{eqnarray}
where $\eta \equiv \int_{0}^{t} \frac{dt}{a(t)}$ is the conformal time. The dark blue line in figure 
\ref{fig:C6eofscale_secondplot} shows this bound applied to a sphere of comoving radius 
$46\ Glyr$, i.e.\ 
\begin{eqnarray}
	S_{ceb} = s_{\chi\ ceb} \frac{4 \pi (46\ Glyr)^3}{3}. 
\end{eqnarray}
The bound is 
saturated by the entropy of radiation fields at the Planck time, increases during radiation 
and matter domination, and asymptotes to a constant during the de Sitter future,
\begin{eqnarray}
	s_{\chi \infty} & \le & \frac{k c^2}{G \hbar} \frac{3}{4 \eta_{\infty}^3 H_{\infty}^2} \\
	  & \le & 4.7 \xt{18}\ J\ K^{-1}\ m^{-3}. \nonumber
\end{eqnarray}

What Bousso has done is to introduce causal limitations to the regions on which 
the holographic bound can be applied, and intriguingly this seems to prevent the 
bound from being violated (at least in the cases studied in \citet{Bousso1999,Bousso2001} 
and in the concordance FRW universe here).

\subsection{Frautschi's Maximum Entropy} \label{C6frausmax}

\citet{Frautschi1982} identifies the maximum entropy inside a causal region
(particle horizon) as the entropy produced by the collection of all matter into a 
single black hole. 

Following \cite{Frautschi1982}, the mass available in any causal region for
the formation of the black hole is
\begin{eqnarray}
	M_{max\ BH} & = & \rho \frac{4 \pi}{3} \chi_{PH}^3 a^3 \label{eq:C6mmaxbh}
\end{eqnarray}
where $\chi_{PH}$ is the comoving radius of the particle horizon. The 
corresponding entropy is 
\begin{eqnarray}
	S_{max\ BH} 
				& = & \frac{4 \pi k  G}{c \hbar} \left[\rho \frac{4 \pi}{3} \chi_{PH}^3 a^3 \right]^2 \nonumber \\
				& = & \frac{64 \pi^3 k  G}{9 c \hbar} \rho^2 \chi_{PH}^6 a^6
\end{eqnarray}
We are interested in the total entropy in a comoving volume of radius $46\ Glyrs$.
In the early universe this comoving volume contains many adjacent particle 
horizons, each with potentially $S=S_{max\ BH}$. The corresponding maximum 
entropy for a comoving volume is thus
\begin{eqnarray}
	S_{Frautschi} & = & \frac{64 \pi^3 k  G}{9 c \hbar} \rho^2 \chi_{PH}^6 a^6 \frac{\chi^3}{\chi_{PH}^3} \nonumber \\
	 			& = & \frac{64 \pi^3 k  G}{9 c \hbar} \chi^3 \rho^2 \chi_{PH}^3 a^6
\end{eqnarray}


The brown line in figure \ref{fig:C6eofscale_secondplot} applies this bound to a sphere of 
comoving radius $46\ Glyr$. During radiation domination $\rho \propto a^{-4}$ and 
$\chi_{PH} \propto a$ so the limit on $s_{\chi}$ grows as $a$. During matter domination 
$\rho \propto a^{-3}$ and $\chi_{PH} \propto a^{1/2}$ so the limit on $s_{\chi}$ 
grows as $a^{3/2}$. During vacuum domination (but assuming black holes cannot
be made from dark energy), $\rho \propto a^{-3}$ and $\chi_{PH} \propto constant$ 
so the limit on $s_{\chi}$ is $\propto constant$.





The possibility of limited gravitational clustering was acknowledged by 
\citet{Frautschi1982}, but in the closed cosmology of the day his work on
$S_{max}$ led him and others to favor increasingly instability. The same 
idea, as presented here with updated cosmology, now predicts stability and 
a constant $s_{max\ BH}$. 

Note that the qualitative future of $s_{max\ BH}$ depends strongly on whether 
or not black holes can be made from dark energy. If we include the dark energy, 
then $\rho \propto constant$ and $\chi_{PH} \propto constant$ so the limit on 
$s_{\chi}$ grows as $a^6$ (not shown in figure \ref{fig:C6eofscale_secondplot}). 


Since black holes may radiate via the Hawking process they do not generally 
represent the maximum entropy state \cite{Page1981a,Frautschi1982}, i.e.\ it 
may be the case that Frautschi's bound is not only attainable (it is that by 
construction), but in sufficiently empty universes it will be surpassed by the 
evaporation of black holes. This is explored in the next section.

We note that Frautschi's idea may be unreliable in recent times. After matter
domination, the black holes suggested in equation \ref{eq:C6mmaxbh} have
radii larger than the Hubble sphere and are not Schwarzschild black holes. 
Further work on the interaction between large black holes and the FRW universe 
is needed to understand the entropy in such situations. 


\subsection{Page's Evaporated Matter}

Stellar black holes (with masses $\sim 4 M_{\odot}$) and SMBHs at the center
of galaxies (with masses $\sim 10^7 M_{\odot}$) emit Hawking radiation with
characteristic temperatures of $10^{-8}\ K$ and $10^{-14}\ K$ respectively. Both
these temperatures are far below that of the present CMB ($2.725\ K$) and 
consequently both classes of BHs currently absorb more radiation than they emit.
As the universe is starting to grow at an exponential rate, the CMB will be 
quickly redshifted below the temperatures of these black holes. 
$10^{-8}\ K$ should come when the universe is $330\ Gyrs$ old and 
$10^{-14}\ K$ when it is $550\ Gyrs$. 

By the time black holes stop growing by accretion, the background temperature
of the universe will be lower than the temperature of any black holes. They will 
begin to evaporate.
The formation of cluster-sized SMBHs up to $10^{12} M_{\odot}$ is expected 
(see e.g.\ \citep{Frampton2008b}). Subsequent evolution will depend on the 
Hawking process. Black holes this large will have temperatures of $10^{-19}\ K$, 
which will be hotter than the background after the universe is just $730\ Gyrs$ old.

It may be possible in principle to transmute matter in the universe into radiation
via black hole evaporation. In principle this could be done locally everywhere, in 
an arbitrarily short time, by using sufficiently small black holes. After evaporation, 
further entropy could be produced by re-thermalizing the Hawking radiation 
(e.g.\ by scattering off trace particles). The immediate transmutation and 
re-thermalization of all radiation, baryons and dark matter would result in a new 
blackbody background with temperature 
\begin{eqnarray}
	T_{\gamma} & = & \left[ \frac{15 \hbar^3 c^5}{\pi^2 k^4} 
		(\rho_{r} + \rho_{m}) \right]^{\frac{1}{4}}, \label{eq:C6pagetemp}
\end{eqnarray}
and entropy
\begin{eqnarray}
	\frac{S_{\gamma}}{V} & = & \frac{4 \pi^2 k^4}{45 \hbar^3 c^3} T_{\gamma}^3 \nonumber \\
		& = & \frac{4}{3} \left[ \frac{\pi^2 k^4 c^3}{15 \hbar^3} \right]^{\frac{1}{4}} (\rho_r + \rho_m)^{\frac{3}{4}}
\end{eqnarray}

This was identified by \citet{Page1981a} as an upper limit to the entropy of the
universe. Since $(\rho_r + \rho_m)$ decreases less quickly than $a^{-4}$ 
more entropy is produced per comoving volume if the transmutation and 
re-thermalization is done later rather than sooner. \citet{Page1981a} suggested 
that the entropy in a comoving region of the universe could be made arbitrarily 
large in this way.

The orange line in figure \ref{fig:C6eofscale_secondplot} shows the entropy produced 
by the immediate transmutation and re-thermalization of all radiation, baryons and dark 
matter in the universe.
The current maximum entropy of the universe calculated in this way is
\begin{eqnarray}
	S_{\gamma} & = & 9.2 \xt{92} k.
\end{eqnarray}
This is much lower than the actual current entropy ($3.1^{+3.0}_{-1.7} \xt{104} k$; \citealt{EganLineweaver2009}). 
This is because BHs and SMBHs are much colder than the CMB; in environments as hot as 
the present universe BHs and SMBHs do not spontaneously evaporate.

\subsection{Page's Evaporated Matter - de Sitter Limited}

A minor adjustment of Page's idea is required in the presence of a de Sitter cosmic
horizon. At $t \sim 1000 Gyrs$ the temperature in equation \ref{eq:C6pagetemp} falls 
below the de Sitter temperature $T_{deS}$. However, the de Sitter radiation would 
prevent any (re-)emission of radiation at temperatures lower than $T_{deS}$.

Thus the final minimum temperature of the evaporated black hole radiation is 
\begin{eqnarray}
	T_{\gamma} & = & Max \left[ \left[ \frac{15 \hbar^3 c^3}{\pi^2 k^4} 
		(\rho_{r} + \rho_{m}) \right]^{\frac{1}{4}}, T_{deS} \right],
\end{eqnarray}
and the entropy density is 
\begin{eqnarray}
	\frac{S_{\gamma}}{V} & = & Min \left[
			\frac{4}{3} \left[ \frac{\pi^2 k^4}{15 \hbar^3 c^3} \right]^{\frac{1}{4}} (\rho_r + \rho_m)^{\frac{3}{4}}, 
			\frac{4}{3} \frac{\rho_{r} + \rho_{m}}{T_{deS}} c^2
			\right].
\end{eqnarray}

Figure \ref{fig:C6eofscale_secondplot} shows this entropy bound applied to a sphere of 
comoving radius $46\ Glyr$ (the dotted orange line). Since the density $(\rho_r + \rho_m)$
tends to decay as $a^{-3}$ in late times, the maximum entropy in a comoving volume 
tends to a constant.

\section{Discussion} \label{C6smaxdiscussion}

The definition of ``the universe'' as a comoving volume is a practical answer to 
{\it  issue 4}. We have not checked whether using this definition reveals any 
more or less than if we had defined the universe to be the region within the 
particle horizon (i.e.\ the time-dependent observable universe), or the event 
horizon.

%


Our interest in the entropy history is primarily driven by the question of the long-term
sustainability of dissipative processes including life. These processes depend on the
availability of free energy (not energy). Dissipative processes deplete free energy by 
degrading an amount of high-grade (low entropy) energy into an equivalent amount of 
low-grade (high entropy) energy. As an example of this, dissipative weather action 
diffuses high-grade energy received from the Sun ($\sim 6000 K$ photons) into 
low-grade energy which is re-transmitted to space ($20$ times as many photons at 
$\sim 300 K$) \citep{LineweaverEgan2008}.

\subsection{Free Energy and an $S_{max}$ Defined by Zero Free Energy}

Consider a system with total energy $U$ and entropy $S_{sys}$, which is connected 
via a heat engine to an infinite exhaust bath at temperature $T_{exh}$. The system, 
which is used to fuel the engine has free energy
\begin{eqnarray}
	F_{sys}  & = & U_{sys} - T_{exh} S_{sys}, \label{eq:C6freeenergy}
\end{eqnarray}
which is generally less than its total energy. That is to say, the engine may extract all 
but $T_{exh} S_{sys}$ of the energy. The unextractable energy is necessarily expelled 
as exhaust and guarantees that the entropy of products is at least as large as the entropy of 
the consumed fuel.
\begin{eqnarray}
	S_{exh} = Q / T_{exh} \ge (T_{exh} S_{sys}) / T_{exh} = S_{sys}
\end{eqnarray} 
For real systems the exhaust bath may not be infinite and the temperature $T_{exh}$ 
may not be constant. In this case \ref{eq:C6freeenergy} should be replaced by an integral 
equation. 

%

The most natural definition of $S_{max}$ is one at which the system has zero free energy. 
\begin{eqnarray}
	F 			& \sim & U - T S \nonumber \\
	S_{max} 		& \equiv & S(F=0) \nonumber \\
				& \sim & \frac{U}{T} \label{eq:C6sut}
\end{eqnarray}
The universe is not in a heat death today because $S < S_{max}$ (and so $F \ne 0$).
Nevertheless, the amount of energy $U$ in a comoving volume, and the minimum 
available exhaust temperature $T$, are both finite and so there is some finite entropy 
$S_{max}$ which, if $S = S_{max}$ today, there would presently be no free energy. 

\begin{figure*}[!p]
   	\begin{center}
		\includegraphics[width=0.8\linewidth]{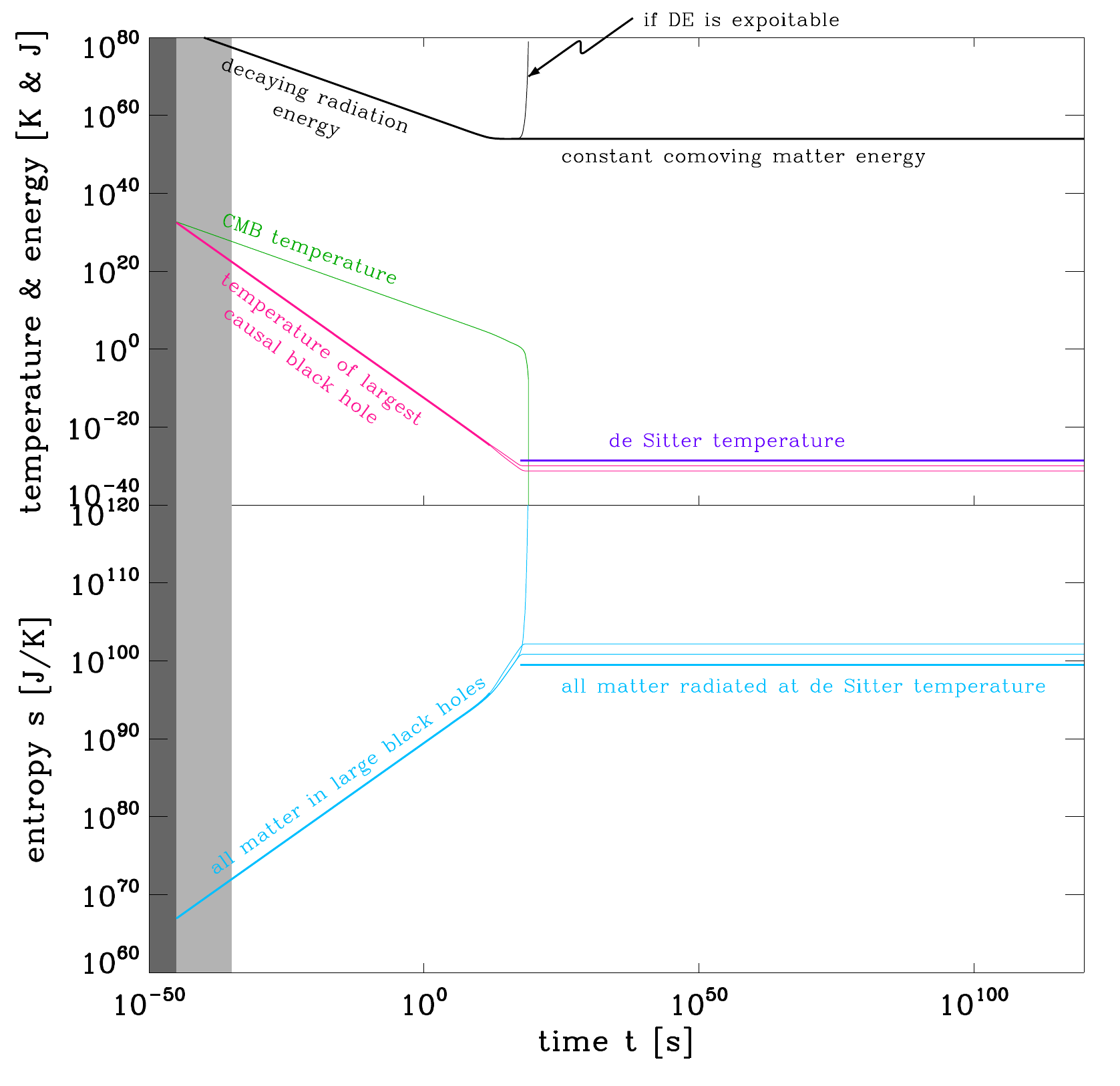}
   		\caption[A maximum entropy based on the available energy and minimum exhaust temperature.]
		{The most useful definition of $S_{max}$ is the entropy at which there is 
		no more free energy, $S_{max} \sim U/T_{exh}$. In this figure we show the 
		time evolution of $U$ and $T_{exh}$. The volume we are considering is the 
		comoving volume that currently corresponds to the observable universe. The 
		energy $U$ is taken to be the total radiation and matter in this volume (thick black). 
		If we include dark energy as a potential fuel $U$ then the energy in the comoving
		volume rapidly increases around the present time (thin black). Three candidates 
		for $T_{exh}$ are explored: the CMB background temperature (in green), largest 
		causal black holes (in thick pink) and the de Sitter background (thick purple). 
		We use thin pink lines for dubious black holes (these have Schwarzschild
		radii larger than the hubble sphere).
		The maximum entropy $S_{max}$ (shown in thick blue) is calculated using 
		$T_{max\ BH}$ at early times (which is lower than $T_{CMB}$) and $T_{deS}$ at 
		late times.}
		\label{fig:C6definingsmax}
	\end{center}
\end{figure*}

$S=S_{max}$ means that no free energy \emph{is} available, but it does not mean that 
no free energy can \emph{become} available. Both quantities on the right-hand-side of 
equation \ref{eq:C6sut} potentially change with time. Figure \ref{fig:C6definingsmax} shows the
evolution of $U$ (thick black; taken to be energy in radiation and matter in a comoving 
volume) and three potential exhaust temperatures,
\begin{eqnarray}
	T_{exh} = \left\{
		\begin{array}{l}
			T_{CMB}, \\
			T_{max\ BH}, \\
			T_{deS}.
		\end{array}
	\right.
\end{eqnarray}
The first potential exhaust that we have considered is the cosmic microwave background 
(CMB; shown in green), the second are large black holes (the largest causal black hole is 
shown in thick pink; refer to Section \ref{C6frausmax}), and the third potential exhaust is the 
the de Sitter background (thick purple).

Today the lowest available exhaust temperature is that of a large black hole (not that of the CMB). 
If we suppose that large causal black holes could have existed during the radiation era, then
those black holes would have been the lowest available exhausts then, and dissipative 
processes might have been driven by the CMB-black hole temperature gradient. In the near 
future the CMB temperature will drop below the de Sitter temperature, and that will become 
the lowest available exhaust.

The maximum entropy $S_{max}$ is calculated using equation \ref{eq:C6sut} and is shown in 
thick blue in the bottom panel of Figure \ref{fig:C6definingsmax}. Since $U$ and $T$ become 
constant in the future, and the entropy within the considered comoving volume $S_{uni}$ can 
only increase, the amount of free energy in the comoving volume decreases in the future 
(see equation \ref{eq:C6freeenergy}). 

The $S_{max}$ that we have discussed in this section is defined in terms of the available 
energy and exhaust temperatures (equation \ref{eq:C6sut}). It is quantitatively similar to the maximum 
entropy of \citet{Frautschi1982,Frautschi1988} by construction (see Section \ref{C6frausmax}).
It is also quantitatively similar to the covariant entropy bound of \citet{Bousso1999} (applied to 
the same volume; see Section \ref{C6cebsmax}), although the reason for this is not clear to us.

In this section we used the conserved matter in a comoving volume and the constant de Sitter 
temperature to show that maximum entropy in the comoving volume becomes approximately 
constant in the future. 
The bleak implication for dissipative processes is that there is a finite amount of free energy 
in any comoving volume. On the other hand, the fraction of $U$ that is not free (due to the 
cumulative increase in entropy to the present day) is only  
 \begin{eqnarray}
	\frac{S T_{exh}}{U} \lsim 10^{-10}.
 \end{eqnarray}

As mentioned in the introduction (issue 3) future entropy 
production may depend on whether or not the equation of state of matter can 
be changed. Using figure \ref{fig:C6definingsmax} it is easy now to see the effect this might have.
If energy was converted from matter into a form with a lower equation of state $w < w_{matter} = 0$ 
then the energy $U$ (the black line in Figure \ref{fig:C6definingsmax}) would increase into the 
future instead of becoming constant. The exhaust temperature (de Sitter temperature) would 
not change significantly, since it depends primarily on the dark energy density. As a 
consequence the maximum entropy, and the free energy would increase. On the other hand,
if matter were converted into a form with $w>0$ (such as radiation, $w_{rad}=1/3$) would cause 
$U$ to decrease into the future. Again, $T_{deS}$ remains unchanged and $S_{max}$ and
and $F$ would decrease with time.

\pagestyle{fancy}
\chapter{Conclusions}

\begin{center}
\emph{
As the music finishes again, \\
have you thought about, \\
how we almost never lift our eyes, \\
from the ground, \\
to the black space above the clouds.
}
\end{center}
\begin{flushright}
- Kent, ``View From a Castle in the Sky'' \hspace*{2cm}
\end{flushright}
\vspace{1cm}

\section*{The Cosmic Coincidence Problem}

Matter and dark energy are observed to presently contribute to the total cosmic energy density in the ratio
\begin{eqnarray}
	r_0 \equiv \min \left[ \frac{\rho_{m0}}{\rho_{de0}},\frac{\rho_{m0}}{\rho_{de0}} \right] \approx 0.4.
\end{eqnarray}
Since the matter and dark energy densities dilute at different rates during the expansion of the Universe,
we are faced with the cosmic coincidence problem: Why are the current matter and dark energy densities
the same order of magnitude today? In other words, why is $r$ (as defined above), so large?

We have used the temporal distribution of terrestrial planets in the Universe to estimate the temporal 
range during which terrestrial-planet-bound observers are likely to arise. Using this observer distribution
we have quantified the severity of the cosmic coincidence problem by computing the probability $P(r>0.4)$ 
of observing values of $r$ larger than $0.4$.

Assuming the standard $\Lambda$CDM density histories for $\rho_m$ and $\rho_{de}$ we find 
\begin{eqnarray}
	P(r>0.4) = 68^{+14}_{-10}\%.
\end{eqnarray}
Given the temporal distribution of terrestrial planets, terrestrial-planet-bound observers have a large 
probability of observing the matter and dark energy densities to be at least as close as we measure 
them to be.

The same method is applied under the assumption of dynamic dark energy, with an equation of state 
parameterized by $w_0$ and $w_a$. We find that some regions of $w_0$-$w_a$ parameter space can 
be discriminated against on grounds of the coincidence problem. I.e. for some regions of $w_0$-$w_a$ 
parameter space, the probability of observing values of $r>0.4$ is very low (Figure \ref{fig:C3severities}).
However those regions are already strongly excluded by observations. 

Our main result is an understanding of the coincidence problem as a temporal selection effect if observers
emerge preferentially on terrestrial planets which is found to hold under any model of dark energy fitting 
current observational constraints. The cosmic coincidence problem is therefore removed as a factor 
motivating dark energy models.

\section*{Searching for Life Tracers Amongst the Solar Properties}

We have compared the Sun to representative stellar samples in $11$ properties.
The properties were selected based a plausible relation to life, availability of a representative stellar
sample for comparison, and such that they were maximally uncorrelated. No properties were added to, 
or removed from the analysis based on previous information about whether or not the Sun was anomalous 
in those properties.

Those selected properties were 
\begin{enumerate}
	\item mass,
	\item age, 
	\item metallicity, 
	\item carbon-to-oxygen ratio, 
	\item magnesium-to-silicon ratio, 
	\item rotational velocity, 
	\item galactic orbital eccentricity, 
	\item maximum height above galactic plane, 
	\item mean galactocentric radius, 
	\item host galaxy mass and
	\item host group mass. 
\end{enumerate}

Our main results are:
\begin{itemize}
	\item Mass and galactic orbital eccentricity are the most anomalous properties of those included in
	our study. The Sun is more massive than $95\pm2\%$ of nearby stars and has a Galactic orbit
	which is more circular than $93\pm1$ of FGK stars within $40\ pc$.
	\item When the $11$ parameters are considered together, the probability of selecting a star, at
	random, which is more anomalous than the Sun, is just $29\pm11\%$. 
\end{itemize}
The observed ``anomalies'' in mass and galactic orbital eccentricity are consistent with statistical noise
(refer to Figure \ref{fig:C4josef13}). This contrasts with previous work suggesting anthropic explanations 
for the Sun's high mass.

To our knowledge, this is the most comprehensive comparison of the Sun to other stars.

\section*{The Entropy of the Present and Future Universe}

We present budgets of the entropy of the observable Universe (Table \ref{tab:C5currententropy})
and of the cosmic event horizon and its interior (Table \ref{tab:C5scheme2budget}). 
To our knowledge these are the most comprehensive and quantitative budgets of the present entropy 
of the Universe. The components included are 
\begin{enumerate}
	\item the cosmic event horizon (only applicable to the latter budget),
	\item supermassive black holes, 
	\item tentative stellar black holes in the range $42$-$140\ M_{\odot}$, 
	\item confirmed stellar black holes in the range $2.5$-$15\ M_{\odot}$, 
	\item photons, 
	\item relic neutrinos, 
	\item dark matter, 
	\item relic gravitons, 
	\item interstellar and intergalactic media and
	\item stars.
\end{enumerate}

The present entropy of the observable universe is found to be 
\begin{eqnarray}
	S_{obs} = 3.1^{+3.0}_{-1.7} \xt{104}\ k,
\end{eqnarray}
and is dominated by supermassive black holes. This is to be compared with previous estimates in the 
range $10^{101}\ k$ to $10^{103}\ k$. Our larger value arises from the inclusion of a new measurement 
of the supermassive black hole mass function.

The present entropy of the cosmic event horizon is calculated to be 
\begin{eqnarray}
	S_{CEH} = 2.6\pm0.3\xt{122}\ k,
\end{eqnarray}
dwarfing that of its interior, 
\begin{eqnarray}
	S_{CEH\ int}1.2^{+1.1}_{-0.7}\xt{103}\ k.
\end{eqnarray} 

Figure \ref{fig:C5combinedbhentropy} illustrates the possible role played by intermediate mass black holes. 
The time evolution of these two budgets is discussed (see Section \ref{C5discussion} and Figures 
\ref{fig:C5eoftimes1} and \ref{fig:C5eoftimes2}).

Entropy bounds from the literature are applied to a comoving volume normalized to the present 
observable Universe (Fgures \ref{fig:C6eofscale_firstplot} and \ref{fig:C6eofscale_secondplot}). 
While the Bekenstein bound and the holographic bound become arbitrarily large, Bousso's covariant 
entropy bound on this volume increases monotonically and asymptotes to a constant around 
$10^{123}\ k$. As does a simplistic bound based on the assumption that the comoving matter density is 
constant and that the future background temperature is the de Sitter temperature. According to both of
these bounds the free energy in the comoving volume is also finite and a heat death is therefore 
expected (either in a finite time or asymptotically).


%
%
\cleardoublepage
%

\appendix
\addcontentsline{toc}{chapter}{Appendices}
\chapter*{Appendices}
\chapter{Entropy and the Free Energy Prerequisites for Life in the Universe} 
\label{appendix1}

\begin{center}
	Charles H. Lineweaver \\
	Chas A. Egan
\end{center}

\section{The Irreversible History of Entropy}
\label{sec:A1intro}

\subsection{Pedagogical Pitfalls}

Although an undergraduate education in the physical sciences contains no explicit warnings against thinking about biology, 
most physics graduates come out believing that the most fundamental aspects of the 
universe are dead things in equilibrium obeying conservative forces.
Frictionless pendulums may be simple, but when studied for too long, students begin to believe that they really exist.
They don't.
Friction is not just an optional accessory inserted into simple equations to make life difficult.
Friction, dissipation and the unequal sign in the second law of thermodynamics is what makes life possible.
The first law of thermodynamics (energy conservation) precedes the second (entropy increase) in textbooks, 
but there is no evidence that this precedence reflects any natural order of things in the universe.

Physicists are taught that whatever biology is about, it can be reduced to chemistry; and that whatever the chemists are up to, it can be
reduced to physics.  However,  physics as we know it can also be viewed as a subset of biology since 
all physicists are the products of biological evolution.

Much has been made of our current inability to  unify general relativity and quantum mechanics to arrive at a theory of everything. 
Although the murky relationship between gravity and entropy may provide key insights into the theory of everything,
it has received much less attention.  
Although gravitational collapse plays the most important role in converting the initial low entropy of the universe 
into the dissipative structures we see all around us (including ourselves), 
gravity is almost universally ignored in thermodynamics textbooks.

\begin{quote}
``We do not yet know if the second law applies to gravitational interactions. 
Is the second law valid only from a given (or ``slowly'' varying) gravitational state? 
Can we include gravitation?'' (Prigogine 1980, p. 196)
\end{quote}

In this paper, we attempt to make sense out of the relationship between life, gravity and the second law of thermodynamics.
In Section \ref{sec:A1intro} we briefly review the history of attempts by physicists to understand life. In Section \ref{sec:A1SunEarth}
we describe how free energy and low entropy radiation from the Sun maintains the low entropy structures of Earth.
We review the entropy of photons in an expanding universe in Section \ref{sec:A1Sexpansion} and consider the relationship between gravity and entropy 
in Section \ref{sec:A1gravityentropy}. 
We conclude by discussing the heat death of the universe (Section \ref{sec:A1heatdeath}).
Our goal is to understand more clearly how gravitational collapse is the source of free energy for life 
in the universe.  Appendices contain mathematical details.

\subsection{Physicists and Life}

When iconoclastic physicists move out of equilibrium and think generally about the question
``What is life?'', the concepts of entropy and free energy play central roles.
In the first half of the 19th century, Carnot(1824), Clausius(1867) and others came to understand that although energy is conserved and
can not be destroyed, useful work -- or extractable free energy -- could be destroyed. Irreversible processes are destroying 
free energy all the time.
Ludwig Boltzmann (1886) was concerned about entropy and the distinction between energy and free energy:

\begin{quote}
``The general struggle for existence of animate beings is therefore not a struggle for raw materials --- these, for organisms, are air, water and soil
all abundantly available -- nor for energy which exists in plenty in any body in the form of heat (albeit unfortunately not transformable), but a struggle
for entropy, which becomes available through the transition of energy from the hot sun to the cold earth.''
\end{quote}

In Section \ref{sec:A1SunEarth} we describe and quantify this ``transition of energy from the hot sun to the cold earth''.
Although Boltzmann explicitly talks about ``animate beings'', the same thing could be said about any far from equilibrium
dissipative structure:  convection cells, hurricanes, eddies, vortices and accretion disks
around black holes (Glansdorff \& Pirgogine 1971, Nicolis \& Prigogine 1977, Prigogine 1980). 
Life is a subset of this general class of dissipative structure (Schneider \& Kay 1994, 1995, Lineweaver 2006, Schneider \& Sagan 2006).

In ``What is Life?'' (Schroedinger 1944) made it clear that Boltzmann's animate beings were not struggling
for entropy.  If they were struggling at all, it was to get rid of entropy, or to absorb negentropy:

\begin{quote}
``What an organism feeds upon is negative entropy. Or, to put it less paradoxically, the essential thing in 
metabolism is that the organism succeeds in freeing itself from all the entropy it cannot help producing while alive.''
\end{quote}

In the notes for a later edition (1956) Schroedinger apologizes to his physicist colleagues and admits that
instead of negative entropy, he should have been talking about free energy.
There is general agreement that life on Earth (and elsewhere) depends on the non-equilibrium of the universe 
and requires free energy to live. 

\begin{quote}
``[T]he one unequivocal thing we know about life is that it always dissipates energy and creates entropy in order to maintain its structure.''
(Andersen and Stein 1987).
\end{quote}

\begin{figure*}[!p]
  \begin{center}
    \includegraphics[scale=0.50]{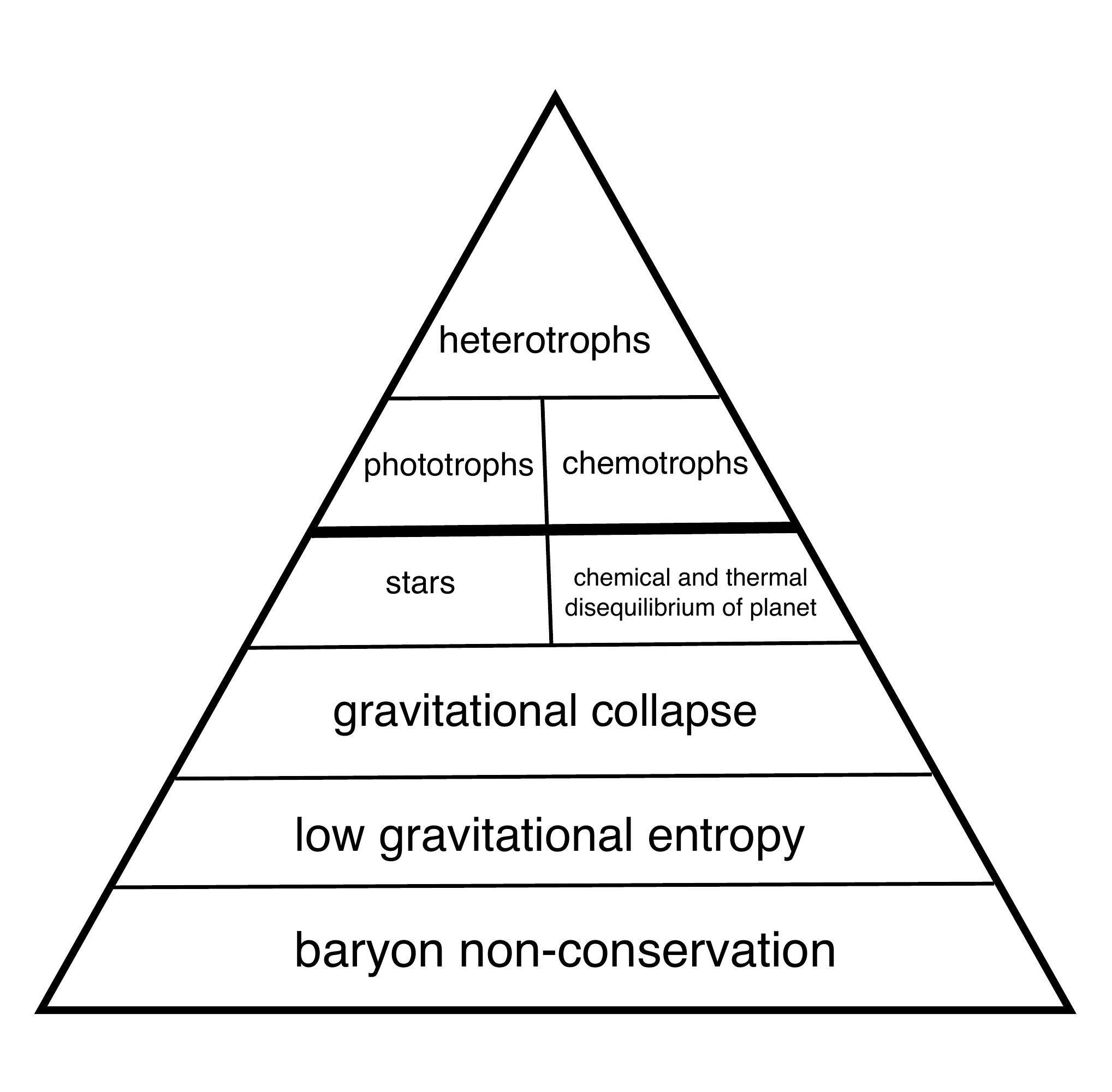}
\caption[The free energy pyramid.]{Pyramid of Free Energy Production.  
The free energy available at one level comes from the
level below it. The lower levels are prerequisites for the life above it.
The top two levels are traditionally classified as life forms in the
primary production pyramid. The pyramid shape represents the decreasing amount
of free energy available at higher trophic levels.
}
    \label{fig:A1pyramid}
  \end{center}
\end{figure*}

In our search for extraterrestrial life, we can use the most fundamental aspects of terrestrial life to guide us.  At the top
of the list is life's requirement for free energy. 
Despite uncertainties in the temperature limits of life ($ < 130^{\deg}$ C?), despite uncertainties in which solvent 
life can use (water?), despite uncertainties in its chemistry (carbon-based?)
 - extraterrestrial life, like terrestrial life, will need a source of free energy.
Free energy is a more basic requirement that all life anywhere must have.  Thus, instead of ``follow the water'',
our most fundamental life-detection strategy should be ``follow the free energy''.  To find chemistry-based life we should
look for the redox gradients between electron donors and acceptors.
These considerations  motivate us to quantify and understand the origin of free energy (Fig. \ref{fig:A1pyramid}).

In the beginning, 13.7 billion years ago, the universe was very hot. There was no life and there were no 
structures in the universe. The universe was a thermal heat bath of photons and a soup of nuclei (and later atoms) in chemical
equilibrium. Life is not possible in such an environment. In thermal 
equilibrium and chemical equilibrium, no free energy is available. As the universe expanded, the heat bath cooled
and life emerged. Life did not emerge simply because the universe cooled down to have the right temperature 
for $H_{2}O$ to be a liquid.  Life needed a source of free energy unavailable from an environment in chemical and
thermal equilibrium. In this paper we try to clarify the idea that the origin of all sources of free energy can be 
traced back to the initial low gravitational entropy of the unclumped matter in the universe (e.g. Penrose 2004). 
The gravitational collapse of this matter produced galaxies, stars and planets and is the source of all dissipative
structures and activities, including life in the universe. See Dyson (1979), Zotin (1984) and  Chaisson (2001) for discussion of how
life (unlike abiotic dissipative structures) seems to evolve toward more complexity. 

\subsection{The Pyramid of Free Energy Production}   

Usually bacteria are considered to be at the bottom of the food chain or at the base of the primary production pyramid,  but an interesting 
perspective comes when we add layers to the base of the pyramid.
At the top of Fig. \ref{fig:A1pyramid} are heterotrophs,  who eat (= extract free energy from )  organic compounds
(including other heterotrophs)  produced by the primary producers one level down. 
Heterotrophs include wolfs, humans, fish and mushrooms.
Supporting all heterotrophic life are the primary producers (phototrophs and chemotrophs). 
Although phototrophs and chemotrophs are usually considered to be primary producers, 
they get their free energy from solar photons and inorganic compounds, respectively.  
Phototrophs include plants and cyanobacteria and all photosynthesizers. 
Chemotrophs include iron and manganese oxidizing bacteria living off the non-equilibrium chemistry of igneous lava rock. 

The vertical line in Fig. \ref{fig:A1pyramid}  indicates that stars are the free energy sources for phototrophs while
the chemical and thermal disequilibrium of the Earth is the source of the free energy in the inorganic compounds 
used in the metabolisms of chemotrophs.
The source of both the free energy provided by stars and by planets comes from gravitational collapse in the level below
in the sense that 
the source of starlight is the fusion reactions taking place in the hot, dense center of the Sun that is
the result of gravitational collapse. 
The chemical and thermal disequilibrium of the Earth also has its source in the free energy of gravitational collapse.

Moving one level lower in the pyramid,
gravitational collapse is made possible by an initially very diffuse, almost unclumped distribution of baryons.
Unclumped baryons in the early universe provided the initial low entropy of the universe.
At the lowest level in the pyramid, the source of these almost unclumped baryons is baryon non-conservation (Sakharov 1967). 
The low initial gravitational entropy of the universe and baryon non-conservation are discussed further in Section \ref{sec:A1gravityentropy}. 


\begin{figure*}[!p]
  \begin{center}
    \includegraphics[scale=0.55]{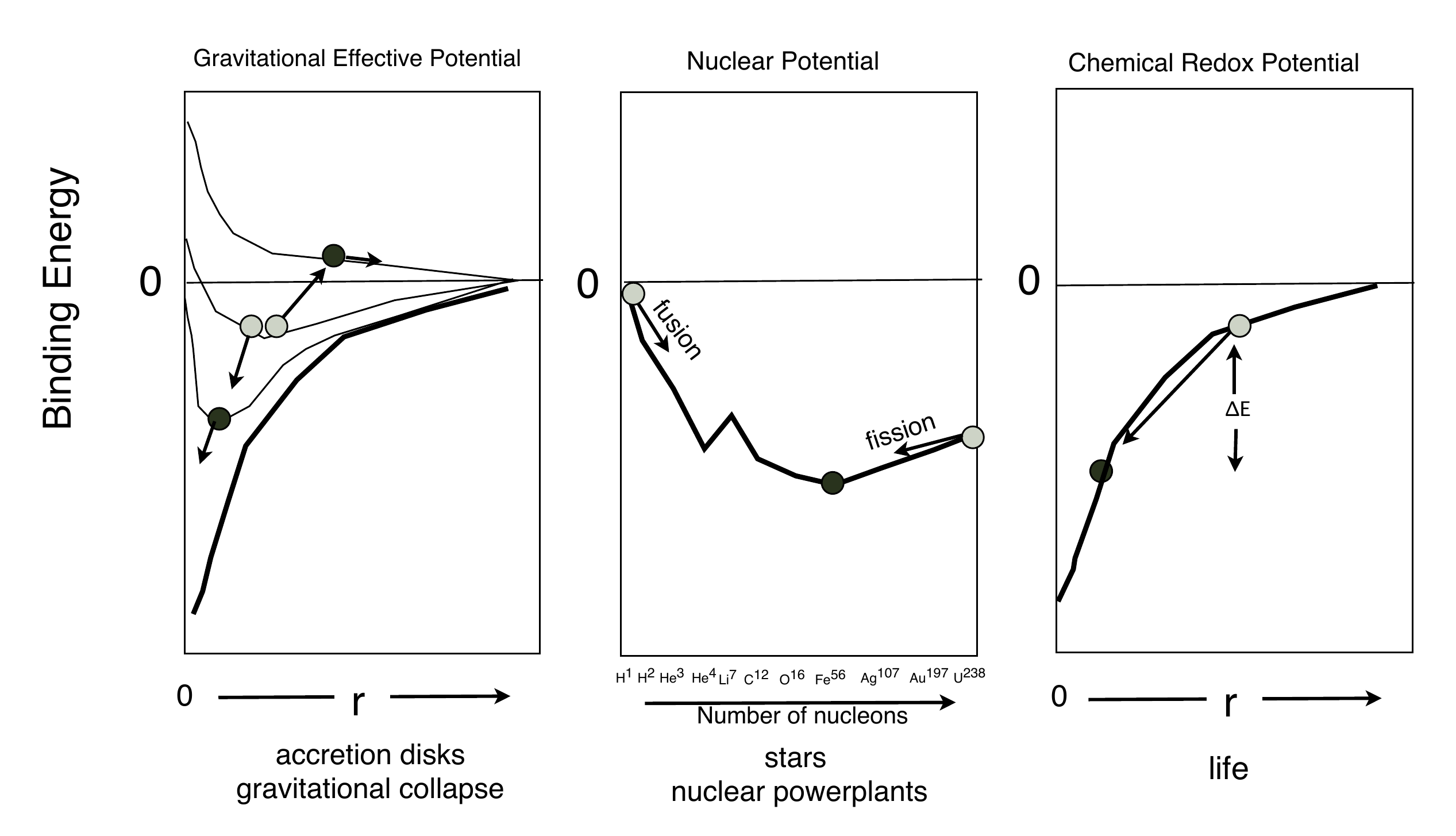}
\caption[Sources of free energy.]
{Sources of Free Energy in the Universe: Gravitational (left), Nuclear (middle) and Chemical (right).
Left panel: dissipation in an accretion disk leads to angular momentum exchange between two small masses (two light grey balls).
The mass that loses angular momentum falls in.  The one that gains momentum is expelled.
Middle panel: the binding energy per nucleon due to the strong nuclear force provides the gradient that makes 
fusion and fission drive nuclei towards iron.
Right panel: the energy that heterotrophic life extracts from organic compounds or that chemotrophic life extracts from inorganic compounds
can be understood as electrons sinking deeper into an electrostatic potential well $\phi(r) \propto 1/r$. 
In every redox pair, the electron starts out high in the electron
donor (light grey ball) and ends up (black ball) lower in the potential of the electron acceptor 
(cf. Nealson and Conrad 1999, their Fig. 3). 
}
    \label{fig:A1potentials}
  \end{center}
\end{figure*}

\begin{figure*}[!p]
  \begin{center}
   \includegraphics[scale=0.40]{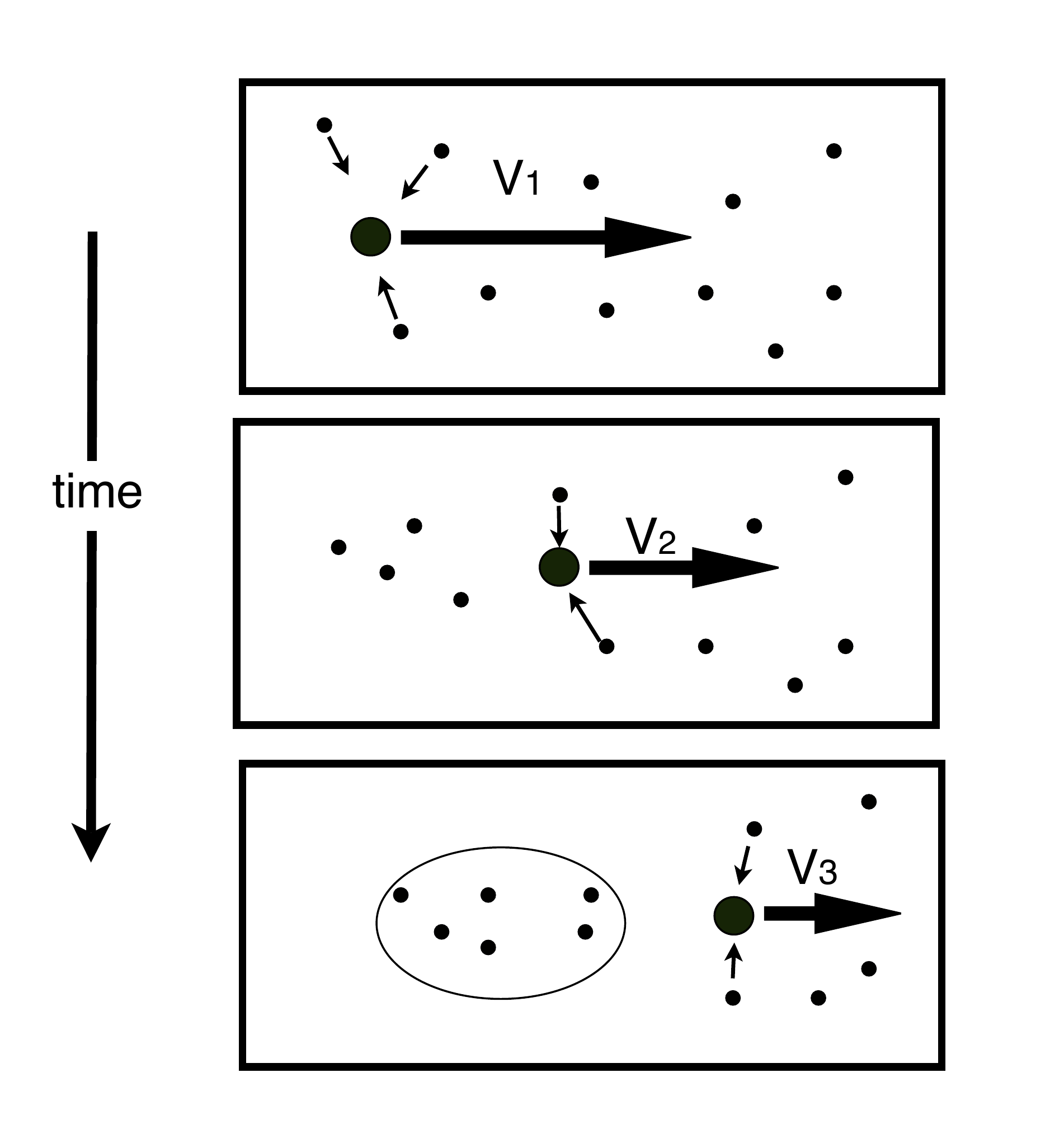}
\caption[Dynamical friction.]{Dynamical Friction.
Consider a massive particle with velocity $V_{1}$ moving through a cloud of less massive particles.
The less massive particles are attracted to the massive particle and end up clumped in the wake of the massive particle.
From there, the less massive particles will have a net gravitational force slowing down the massive particle.
This causes the most massive objects to fall into the center of the potential and is why clusters of galaxies have
massive cD galaxies at their cores.
(Binney and Tremaine 1987).}
    \label{fig:A1dynamicfriction}
  \end{center}
\end{figure*}

The sources of free energy in the universe are summarized in Fig. \ref{fig:A1potentials}.
In a gravitational system (left panel), such as a protoplanetary accretion disk,  
consider a small mass $m$ in orbit at distance $r$
from a large mass $M$ at $r=0$. The effective potential, including angular momentum $L$  
is  $\phi(L,r) = \frac{L^{2}}{2mr^{2}} - \frac{GmM}{r}$ (e.g. Goldstein 1980). 
Angular momentum $L$ must be reduced for gravitational collapse to happen.
Consider two small masses, originally in identical effective potentials (two light grey balls).  
They come close to each other and  exchange some angular momentum. 
The one that lost $L$, sinks into the well closer to $M$, the one that gained $L$ distances itself from $M$.
Since the $L$ of each mass has changed, their effective potentials have diverged.
One $m$ collapses, the other is expelled. 
Without the dissipation of energy, expulsion of matter and transfer of angular momentum that occurs in the 
turbulence and viscosity of an accretion disk (e.g. Balbus 2003), matter would not gravitationally collapse.
The efficiency of star formation in a molecular cloud is a few percent.
A substantial fraction of the infalling matter is scattered, or receives a large dose of angular momentum 
as it is processed through the accretion disk and then expelled (Balbus 2003).
Therefore accretion disks are also expulsion disks
(see Fig. \ref{fig:A1dynamicfriction} for the role of dynamical friction in the gravitational collapse of less-viscous non-accretionary systems). 
Gravitational collapse creates entropy by radiating away the $MG/r$ potential energy and expelling high velocity, high angular momentum material.

Fusion in the core of the Sun was made possible by the gravitational
collapse of $\sim 10^{31}$ kg of hydrogen
resulting in high densities 
and temperatures.  
Gravitational collapse also provides the conditions in the cores of stars
to make matter roll down the nuclear binding energy curve to the energy minimum  
(middle panel, Fig. \ref{fig:A1potentials}).

The right panel of Fig. \ref{fig:A1potentials} shows that the amount of energy extractable
from chemical bonds depends on the energy difference $\Delta E$ between the electron in the potential well of the
donor and that of the acceptor
(Nealson \& Conrad 1999).
Life takes in energy-rich atoms with electrons in high orbitals (electron donors) and excretes the same atoms 
with the electrons in the deeper atomic or molecular orbitals of electron acceptors.
Solar photons provide the energetic kick $\Delta E$ to lift the electrons back up during photosynthesis in phototrophs, who  provide 
the energy-rich materials for heterotrophs (e.g. Szent-Gyorgi 1961).

\begin{figure*}[!h!t!p]
  \begin{center}
   \includegraphics[width=0.9\linewidth]{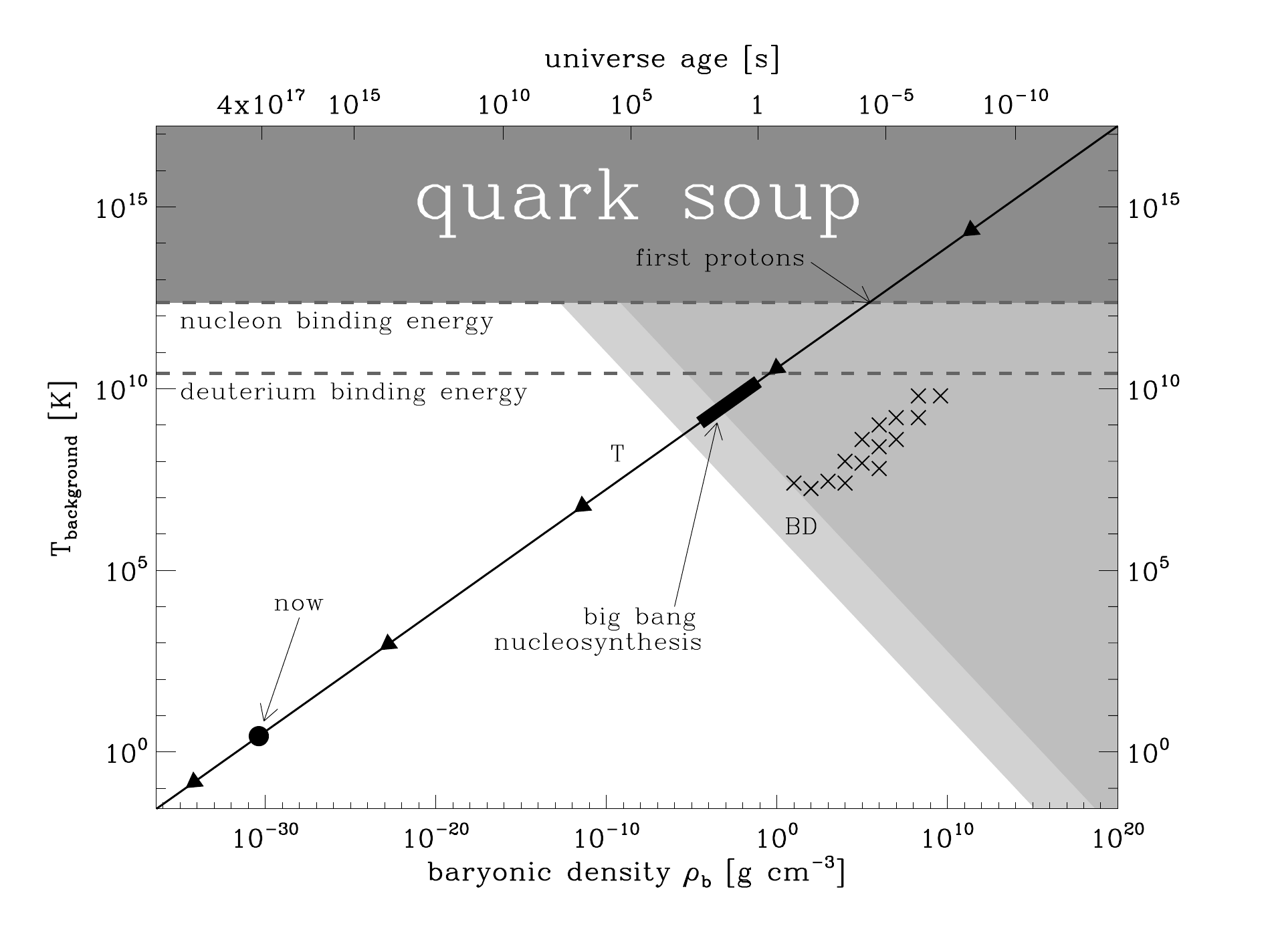}
\caption[Evolution of the Universe through the density-temperature plane.]
{Regions of the density-temperature plane where nuclear fusion reactions occur.
Our universe cooled along the line from top right to lower left. Between one second and three
minutes after the big bang, big bang nucleosynthesis (BBN) produced deuterium  $^{3}$He, $^{4}$He, and several other
light isotopes whose abundances we can measure in stellar atmospheres today (Weinberg 1977).
If the baryonic density of the universe were much larger and the expansion rate of the universe were slower (e.g. Peacock 2000),  
BBN would have produced many other elements and could have burned all the hydrogen into iron and precluded 
the production of starlight from stellar fusion.
After BBN, most of the baryons in the universe were in hydrogen and helium.  Thus the universe was in a state of low nuclear 
entropy.  This allowed stars to subsequently access
the free energy from nuclear fusion in their hot, dense cores. 
The cores of main sequence stars are labeled ``X''.
The cores of brown dwarfs, where deuterium (but not hydrogen) is fusing into helium, is labeled ``BD''.
The conditions inside a Tokamak reactor are labeled ``T''.
The diagonal shades indicate contours of constant reaction rate $\propto \rho^{2}T^{4}$: 
the light region indicates reaction rates similar to those
in BDs; the darker region indicates reaction rates $\gsim$ those in main sequence stars.
}
    \label{fig:A1rhoT}
  \end{center}
\end{figure*}

\subsection{Big Bang Nucleosynthesis and the Subsequent Low Entropy of Nuclei}

As the universe expands, the scale factor $R$ increases, the temperature decreases  ($T \propto R^{-1}$)
and the density decreases $\rho \propto R^{-3}$. Thus, 
$\frac{T}{T_{i}}  = \left( \frac{\rho}{\rho_{i}} \right ) ^{1/3}$.
This is the path the universe takes in Fig. \ref{fig:A1rhoT} starting at some initial temperature and density: $T_{i}, \rho_{i}$.
%
%
The early universe expanded and cooled too quickly for big bang nucleosynthesis to fuse hydrogen into iron and reach 
equilibrium at the lowest
nuclear binding energy per nucleon (middle panel, Fig. \ref{fig:A1potentials}).  
Thus, big bang nucleosynthesis 
left nuclei in a low entropy, high energy state.
Similarly, reheating after inflation (Kolb \& Turner 1990) left unclumped baryons in a state of low gravitational entropy since 
the baryons are not at the bottom of the gravitational potential wells.

Entropy is produced when free energy is extracted from the sources of potential energy shown in 
Fig. \ref{fig:A1potentials}.
For example,
dissipation of gravitational energy (left panel) happens when the turbulent viscosity and friction of an accretion disk transfers angular
momentum away from the central mass and makes some material fall onto the central mass while other material is expelled. 
Without such collisions, turbulence
and friction, 
angular momentum would not be transferred and material would not gravitationally collapse or be expelled.
Figure \ref{fig:A1globular} illustrates gravitational systems with minimal dissipation.

Dissipation happens and entropy is produced whenever a photon gets absorbed by a material at a temperature colder than the emission temperature.  
The photon energy gets reemitted and distributed among many photons.
This happens as a gamma ray produced by fusion at the center of the Sun makes its way to the photosphere where its energy
is distributed among millions of photons (Frautschi 1988). 
It also happens when the energy of solar photons ($T = 5760$ K) are harvested for photosynthesis by plants at temperatures below $T=5760$ K,
and when the Earth reemits solar energy at infrared wavelengths.

Energy cannot be created or destroyed.  Therefore, strictly speaking, we cannot ``use energy'' or ``waste energy''. Energy can however be
degraded.  Low-entropy, high-grade energy dissipates into high-entropy, low-grade energy. 
Life does not ``use'' energy since the same amount of energy that enters the biosphere, leaves the biosphere.
Life needs a source of free energy, and is unable to use high entropy energy. Life takes in energy at low entropy and excretes it at high entropy.
Any engine does the same thing. When coal burns, energy is conserved.  Electrons are high in the electric potentials of the fuel
and lower in the potentials of the ashes and exhaust gases. The difference
($\Delta E$ in the right panel of Fig. \ref{fig:A1potentials})  has been transfered into heat and work.

Two types of free energy are described in the literature: Gibbs free energy and Helmholtz free energy (e.g. Sears \& Salinger 1975).
For simplicity and convenience (cf. Appendix \ref{sec:A1which}) 
we focus on the Helmholtz free energy $F$ of a system: 
\be
 F = U - TS,
\label{eq:A1Helmholtz}
\ee

where $U$ is the internal energy of the system, $T$ is its temperature and $S$ is its entropy. 
The free energy $F$, is the amount of energy that can be extracted from the system to do any kind of useful work such
as climbing a tree or assembling fat molecules.
Equation \ref{eq:A1Helmholtz}  shows that all of the internal energy $U$ is not available to be extracted as free energy $F$.  
There is an entropic tax: $TS$.
$TS$ is the penalty one must pay for extracting the energy from the system and using it to do any useful work.
$U$ is how much money is in the bank and $TS$ are the bank fees you have to pay to get it out. 
The higher the temperature $T$ and the higher the entropy  $S$ of the
system, the higher the penalty and the lower will be the extractable, usable, life-supporting, life-giving free energy $F$.
Good engines produce minimal entropy and have $TS << U$ and thus $F \approx U$ (Bejan 2006, Eq. 3.7).


\begin{figure*}[!p]
  \begin{center}
    \includegraphics[scale=0.35]{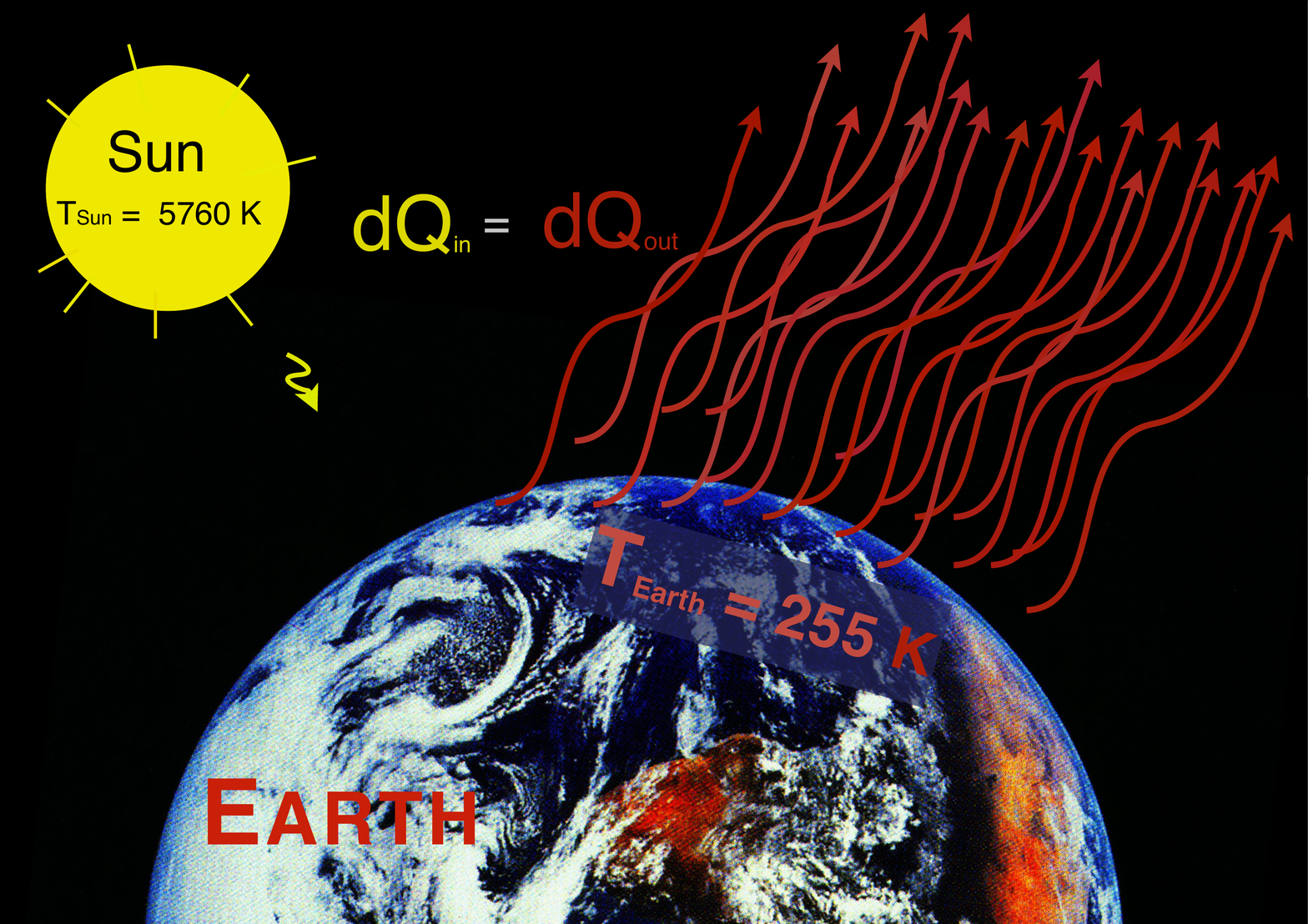}
\caption[Entropy production on Earth.]
{The Sun provides the Earth with a continuing source of free energy.
Energy that comes to Earth from the Sun  ``$dQ_{in}$'' is balanced
by the energy radiated by the Earth into outer space ``$dQ_{out}$''. The temperature of the incoming
photons is the temperature of the photosphere of the Sun: $T_{Sun} = 5760$ K.  The temperature
of the outgoing photons is the effective temperature of the Earth: $T_{Earth} = 255$ K.
When the Earth absorbs one solar photon (yellow squiggle), 
the Earth emits 20 photons (red squiggles, Eq. \ref{eq:A120}) with wavelengths 20 times longer.
The entropy of photons is proportional to the number of photons (Eq. \ref{eq:A1SN}).
Hence when the Earth absorbs a  high energy solar photon at low entropy and 
distributes that energy among 20 photons and radiates them back to space, the Earth
is exporting entropy; the waste entropy from the maintenance of the low entropy structures on Earth.}   
    \label{fig:A1SunEarth}
  \end{center}
\end{figure*}

\section{The Sun is the Source of Earth's Free Energy}
\label{sec:A1SunEarth}

Consider the amount of free energy that is delivered to the Earth in solar photons.
We make the reasonable assumption that the Earth is in a steady state (e.g. Kleidon 2008).  
It is not in equilibrium because there is an energy flow in to the system (sunlight) and out 
of the system (Earth radiates to space, Fig. \ref{fig:A1SunEarth}).

Steady state means that the average effective temperature
of the Earth is constant. 
It also means that the amount of energy delivered to the Earth in solar photons $dQ_{in}$, is the same
as the amount of energy radiated away by the Earth as infrared photons, $dQ_{out}$. Thus, 
$dQ_{in} = dQ_{out}$. 
If this were not so, the internal energy $U$ of the Earth would be increasing -- the Earth would be getting hotter or the speed of winds and the number of 
hurricanes would increase, or there would be a net increase in biomass.
However, the number of organisms that are born is about the same as the number that die.  The strength of the winds that dissipate the 
pole to equator temperature gradients are about the same and the number of hurricanes which equilibrate thermal, pressure and humidity
gradients, is about the same.
(We are ignoring variations in the temperature of the Earth
due to variations of the greenhouse gas content of the Earth or the Milankovich cyles or the secular increase in solar luminosity.)
Thus, in steady state, the Earth is at a constant temperature $T$, constant energy $U$, constant entropy $S$ and
constant free energy $F$. 

Let $\frac{dS_{D}}{dt}$ be all the entropy produced by all the dissipative structures on Earth 
(including  winds, hurricanes, ocean currents, life forms, and
the thermal dissipation when heat is transfered through the soil from hot sunny spots to cool shady spots). Then we have for the entropy of the Earth:

\be
\frac{dS}{dt}  = \frac{dS_{\gamma}}{dt} + \frac{dS_{D}}{dt} = 0,   
\label{eq:A1dS1}
\ee

or the net decrease in entropy from photons coming in and out ($\frac{dS_{\gamma}}{dt}$) is compensated for
by the increase in entropy from all the dissipative, low entropy structures on Earth. Thus,

\be
\frac{dS_{\gamma}}{dt} = -\frac{dS_{D}}{dt}.
\label{eq:A1dS2}
\ee

Since the energy in the photons arriving and leaving is equal, $dQ_{in} = dQ_{out}$, we have  $|dS_{in,\gamma}| < |dS_{out,\gamma}|$ since:

\bea
dS_{in,\gamma} &=& \frac{dQ_{in}}{T_{Sun}}\\
dS_{out,\gamma} &=& - \frac{dQ_{in}}{T_{Earth}}.
\eea

Thus, 
\be
\left | \frac{dS_{out,\gamma}}{dS_{in,\gamma}}\right | =\frac{T_{Sun}}{T_{Earth}} = \frac{5760}{255} \sim 20
\ee 
Thus, the Earth exports twenty times as much entropy as it receives. Equation (\ref{eq:A1SN})
then tells us that
the ratio of the number of emitted photons to the number of absorbed photons is:
\be
\frac{N_{out,\gamma}}{N_{in,\gamma}} \sim 20.
\label{eq:A120}
\ee

This is shown in Fig. \ref{fig:A1SunEarth} with its 1 incoming solar photon and 20 outgoing infrared photons.
The entropy flux to and from the Earth from the absorbtion of solar photons and the emission of infrared photons is:
\bea
\frac{dS_{\gamma}}{dt} &=&  \frac{dS_{in,\gamma}}{dt} + \frac{dS_{out,\gamma}}{dt} \nonumber\\
                       &=& \frac{dQ}{dt} \left( \frac{1}{T_{Sun}} - \frac{1}{T_{Earth}}\right)  \nonumber  \\
                       &=& - \frac{dQ}{dt} \frac{1}{T_{Earth}} \left( 1 - \frac{T_{Earth}}{T_{Sun}}\right) \nonumber\\
                      &=& - \frac{dQ}{dt} \frac{1}{T_{Earth}} (0.95).
\label{eq:A1dSdQ}
\eea

Since the amount of free energy is not building up in the Earth, we have  $\frac{dF}{dt} = 0$.
Let  $\frac{dF_{\gamma}}{dt}$ be the amount of free energy delivered to the Earth by solar photons and
 $\frac{dF_{D}}{dt}$ be the amount of free energy dissipated by all the dissipative structures
on Earth, then (cf. Eqs. \ref{eq:A1dS1} and \ref{eq:A1dS2}) we have, 

\be
\frac{dF}{dt}  = \frac{dF_{\gamma}}{dt}  +  \frac{dF_{D}}{dt} = 0  
\label{eq:A1dF1}
\ee

or 

\be
\frac{dF_{\gamma}}{dt} = -\frac{dF_{D}}{dt}
\label{eq:A1dF2}
\ee
where the minus sign indicates that
$\frac{dF_{D}}{dt}$ is the loss or dissipation of free energy.
Eqs. (\ref{eq:A1dF1}) and (\ref{eq:A1dF2}) are the key to understanding 
how the Earth can keep absorbing free energy from the Sun without the amount of free energy in the Earth going up.
The free energy in the food we eat goes to cell repair and movement, and is dissipated when we die or move.
Similarly, all of the free energy delivered by the Sun is dissipated in winds, hurricanes, ocean currents, life forms, or thermal conduction
through soil between sunny spots and shady spots.

The Earth is exporting much entropy but the entropy of the Earth is not decreasing. That is because the dissipative structures on
the Earth are producing the entropy that is exported.  They need the input of free energy to stay at low entropy -- just as a refridgerator needs
free energy to stay at a constant low temperature (= low entropy steady state).
Without a supply of free energy a fridge will heat up, a hot water tank will cool down, and life will die.
Things approach equilibrium.
It takes free energy to keep a fridge cool, the tank hot and the chemical order in life forms.
Free energy (or work) is needed to remove the heat and entropy that naturally leaks into the fridge.
The lower the temperature of the fridge and the more imperfect the insulation, the more free energy is needed to maintain
the low entropy steady state.

The export of entropy does not lower the entropy of the Earth. Rather it keeps the entropy of the Earth at a constant
low level.
In the absence of a flow of negentropy, the low entropy structures, such as hurricanes, dust devils, the hydrological cycle,
thermal gradients and life forms would run down and dissipate away.  The export of entropy compensates for this natural
dissipation and is the reason why low entropy structures endure.

\subsection{How much entropy is produced and how much free energy can be
extracted from a solar photon?}

We can compute the amount of free energy available on the Earth to drive the winds, 
hurricanes and all of life (Kleidon 2008).
Starting from Eq. \ref{eq:A1Helmholtz}, taking differentials and then dividing by $dt$ yields the rate of increase of 
free energy of the Earth:
\be
\frac{dF}{dt} = \frac{dU}{dt} - T\frac{dS}{dt} - S\frac{dT}{dt}.
\label{eq:A1FUTS}
\ee

Since we are assuming steady state, F, U, T and  S  are all constants and all of the terms
in Eq. \ref{eq:A1FUTS} are zero.
However, using Eqs. (\ref{eq:A1dS1}) and (\ref{eq:A1dF1}) we can write:
\be
\frac{dF_{\gamma}}{dt} + \frac{dF_{D}}{dt} =  - T_{Earth} \left( \frac{dS_{\gamma}}{dt} + \frac{dS_{D}}{dt}\right).
\ee
Separating terms to count only the contribution from photons we get:

\be
\frac{dF_{\gamma}}{dt}  =  - T_{Earth} \frac{dS_{\gamma}}{dt}.
\ee

With  Eq. \ref{eq:A1dSdQ} this yields,

\be
\frac{dF_{\gamma}}{dt}  =  \frac{dQ}{dt} 0.95
\label{eq:A1dF}
\ee

Thus, $95\%$ of the incoming solar energy can be used to do work, i.e. photovoltaics at the temperature of the
Earth have a maximum efficiency of $95\%$.  To get a numerical value for the free energy in Eq. (\ref{eq:A1dF}):  
the solar flux impinging on the disk of the Earth ($\pi R_{Earth}^{2}$) at 1 AU from the Sun is $1366 \; W m^{-2}$.  
Since $dQ_{in} = dQ_{out}$, the average flux $I_{o}$ from the Earth's surface ($4 \pi R_{Earth}^{2}$) balances the solar flux: 
\be
\pi R_{Earth}^{2} 1366 \;W m^{-2} = 4 \pi R_{Earth}^{2} I_{o}
\ee

where, $I_{o} = 342 \; W m^{-2}$.

Therefore, to get a numerical value for $\frac{dQ}{dt}$ (= the flux density of solar radiation through a unit area of $ 1\; m^{2}$) 
we have:

\bea
\frac{dQ}{dt} &=& \sigma \; T_{Earth}^{4}   =  I_{o} (1-A_{Earth})\\
              &=& 342 \; (0.7)  \;W m^{-2}  \nonumber\\
              &=& 238 \;W m^{-2}
\label{eq:A1Qt}
\eea
where $A_{Earth} \approx 0.3$ is the albedo of the Earth, and
the Stefan-Boltzmann constant is $\sigma = 5.67 \times 10^{-8} \; W m^{-2} K^{-4}$.
Thus the flux of free energy through unit area (Eq. \ref{eq:A1dF})  is
\be
\frac{dF_{\gamma}}{dt} =  238 \; W m^{-2} (0.95) =  228 \; W m^{-2}
\label{eq:A1Ft}
\ee

This flux of free energy maintains  all thermal gradients on the surface of the Earth, all winds and hurricanes and all life, and
is equal to the flux of free energy that is dissipated by all dissipative structures (Eq. \ref{eq:A1dF2}).

The total free energy available from sunlight is the flux per unit area times the area of the Earth:
$228 \; W m^{-2} \times 4 \pi R_{Earth}^{2}  \sim 1.2 \times 10^{17} $ W, 
which is about ten thousand times larger than the $1.3 \times 10^{13}$ W of global power consumption from burning fossil fuels.
Terrestrial life  (including humans) is a subdominant dissipator of the free energy delivered to Earth (Kleidon 2008).

There are no hurricanes or ocean currents on the Moon, so how does the free energy delivered by solar photons get dissipated there?
Performing the same computation for the Moon as we did for the Earth we have:
\be
I_{o} (1-A_{Moon}) = \sigma \; T_{Moon}^{4}
\ee
where the Moon's albedo is lower than the Earth's, ($A_{Moon} \approx 0.07$).
The Moon's effective temperature $T_{Moon} = 274 \; K$ is higher than the Earth's because of the Moon's lower albedo.
Instead of having hurricanes, winds, ocean currents and life forms, the free energy of the Moon is dissipated by heat flow due to the 
large temperature gradients (low entropy structures)  between regolith in the sunshine at $350$ K and the shadows at $150$ K.  
The input of low entropy solar radiation maintains the gradients.
The maximum temperature variation on the Moon is $\Delta T \sim 300$ K  between $\sim 390$ K at the equator in the early afternoon and $\sim 70$ K in the shade at the poles.
On Earth, this variation is only $\Delta T \sim 120$ K between $\sim 320$ K and $\sim 200$ K. 
If the Moon were the same temperature as the Sun,  $T_{Sun} \approx 5760$ K then the ``shadows'' would be the same temperature
as the Sun and there would be no export of entropy.
If the Moon were a smooth ball instead of having a bumpy surface then the large scale hemispheric temperature gradient
would be the only low entropy structure and a larger temperature gradient would be created to dissipate the same
constant amount of free energy from the low entropy photons.
A further refinement to the computation above would consider the low entropy associated with sunlight coming from a particular direction rather
than isotropically.

\section{The Entropy of the Cosmic Microwave Background Remains Constant as
the Universe Expands}
\label{sec:A1Sexpansion}

\begin{figure}[!p]
  \begin{center}
    \includegraphics[scale=0.4]{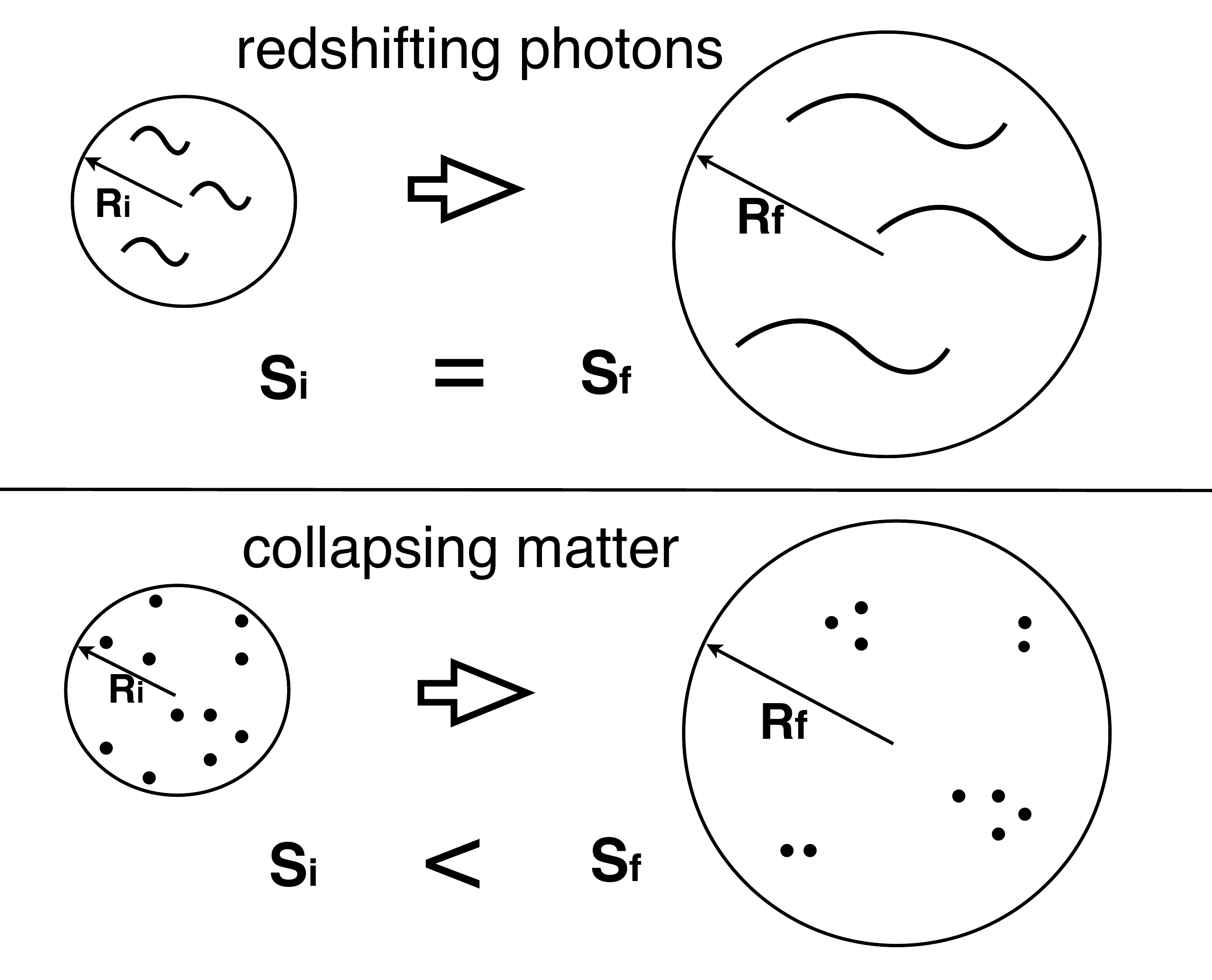}
\caption[Radiation and matter entropies in an expanding Universe.]
{The entropy of the universe changes as the universe expands.
Notice that the photons (squiggles) stay spread out while the baryons (dots) clump due to gravity. 
Top: as the universe expands, the entropy of cooling redshifted photons remains constant (Eq. \ref{eq:A1Sconstant})
while the entropy produced when material clumps into galaxies, stars and planets, increases the total 
entropy of the universe.
At a given initial time, the circles on the left represent an arbitrary volume of the universe with an initial
scale factor $R_{i}$. The small circle on top contains 3 cosmic microwave background photons (squiggles).
At a later time (right), the volume has expanded but contains the same number of photons and thus 
$S_{\gamma,i} = S_{\gamma,f}$ (see Eq. \ref{eq:A1SN}).
The entropy of the universe, however, includes contributions from photons and the net effect of gravitational collapse.
In the lower panel, the 11 baryons (dots) start out fairly unclumped then clump.
The net entropic effect of clumped matter and the heat given off to allow the clumping and the matter expelled to allow the clumping 
is: $S_{m, i} < S_{m, f}$.
Thus, the total entropy of the universe increases:  $S_{i} < S_{f}$.  
The photon to baryon ratio of our universe is about one billion, not the 3/11 shown here. 
}
    \label{fig:A1SexpansionConstant}
  \end{center}
\end{figure}

It is difficult to talk about the total entropy in the universe without knowing how big the 
universe is, so we talk about the entropy in a representative sample of the expanding universe.  Typically 
we put an imaginary sphere around a few thousand galaxies and consider the entropy in this expanding sphere -- the 
entropy per comoving volume.  We parameterize the expansion of the universe with a scale factor R.  
This means that when the universe increases in size by a given factor, R increases by the same 
factor (Fig. \ref{fig:A1SexpansionConstant}).

The expansion of the universe is adiabatic since the photons in any arbitrary volume of the universe have the same temperature
as the surrounding volume. There is no net flow of heat. 
The entropy of a photon gas does not increase under adiabatic expansion.
Specifically, the entropy $S$
of a gas of photons in a volume $V$ at temperature $T$ is
$S \propto VT^{3}$ (e.g. Eq. \ref{eq:A1gammaS} or Bejan 2006, eq. 9.20).  

The photon wavelengths $\lambda$, increase (are redshifted) with the scale factor: $\lambda \propto R$.
There is no absorbtion or reemission associated with the redshifting of cosmic microwave background photons.
These photons were last scattered at the surface of last scattering $\sim 480,000$ years after the big bang.  
%
Since the volume $V$ increases as $V \propto R^{3}$, and 
since the temperature of the microwave background goes down as the universe 
expands: $T \propto \frac{1}{R}$,  we have the result that the entropy of a given comoving volume of space is constant
(Kolb \& Turner 1990, Frautshi 1982, 1988):

\be
S \propto VT^{3} \propto R^{3} \left(\frac{1}{R^{3}}\right) = constant.
\label{eq:A1Sconstant}
\ee

The adiabatic expansion (or contraction) of a gas in equilibrium is reversible.
Thus the expansion of the universe by itself is not responsible for any 
entropy increase in the photons (Fig. \ref{fig:A1SexpansionConstant} top panel). 
Another way to understand that the entropy of the photons in the universe remains constant 
as the universe expands, is to realize that entropy is proportional to the number of photons
$S_{\gamma} \propto N_{\gamma}$  (Eq. \ref{eq:A1SN}). 
The number of photons $N_{\gamma}$ in the volume remains constant and therefore so does the entropy.
Thus, we obtain the result  indicated in the top panel of Fig. \ref{fig:A1SexpansionConstant}:
$S_{\gamma, i} = S_{\gamma, f}$.



\section{Gravity and Entropy }
\label{sec:A1gravityentropy}

In the big bang model, the early universe was in thermal and chemical equilibrium.  In the previous section
we showed how the expansion of the universe is not responsible for changing the entropy of the photons in the universe.
If the universe were in equilibrium, it should have stayed in equilibrium (top panel of Fig. \ref{fig:A1SexpansionConstant}).
Our existence shows that the universe could not have started from equilibrium.
The missing ingredient that solves this dilemma is gravity.  Matter, evenly distributed throughout 
the universe, has much potential energy and low entropy.  In the standard inflationary scenario 
describing the earliest moments after the big bang, matter originates (during a short period at 
the end of inflation called reheating) from the decay of the evenly distributed potential energy 
of a scalar field.  ‘False vacuum’ decays into our true vacuum.  Vacuum energy cannot clump.  
However, once the potential energy of the scalar field is dumped almost uniformly into the universe in the form 
of relativistic particles, these can cool and clump if they have mass (Fig. \ref{fig:A1SexpansionConstant}, lower panel).
Unclumped matter has a lower entropy than clumped matter:
\be
S_{unclumped} < < S_{clumped}.
\ee

By $S_{clumped}$ we mean the entropy of the phase space volume of the collapsed material as well as the phase
space volume of the  material expelled during the clumping, plus the entropy of the heat given off during the collapse and dumped into
the environment  which allowed the unclumped baryons to clump. That is a lot to include but ignoring the full picture has led to 
much confusion about the relationship between gravity and entropy.

The gravitational potential energy is enormous.
In this inflationary picture the potential energy of the false vacuum is the ultimate 
source of all energy and the matter/antimatter pairs which annihilate and create a bath of photons.  
Because of an intrinsic asymmetry (baryon non-conservation), the annihilation is incomplete and leaves one baryon for every 
billion photons.  The subsequent cooling (due to the expansion) and clumping of the residual baryons 
(due to gravity) is the source of all the free energy, dissipative structures and life in the universe
(bottom level in Fig. \ref{fig:A1pyramid}).

The relationship between entropy and gravity is similar to the relationship between energy and heat 200 years ago
when the concept of energy conservation in thermodynamics was being developed.
It took many decades for the different forms of energy to be
recognized.   Kinetic energy was different from potential energy,
``caloric'' became heat energy and Einstein showed us there was energy in mass and in the momentum of 
massless particles: $E^{2}=p^{2}c^{2} + m^{2}c^{4}$.
 It seems to be taking even longer to recognize and define the different
forms of entropy, including gravitational entropy and informational entropy (Brissard 2005, Shannon 1950).

There is some confusion about life being in violation of the second law.  If $dS > 0$ how can life be
so ordered? The answer is that the order and low entropy of life is maintained by the production and
export of entropy. 
Similarly, there is confusion about clumped material or gravitational structures being in violation 
of the second law. The resolution is the same: the entropy of the environment needs to be included in the calculation.

Life is trying to maintain its order,  while the second law is trying to decrease order. 
Superficially it seems that life and the second law are at cross purposes.
In fact, life and the second law are allies, since the maintainence of a highly ordered structure
increases the disorder of the universe more than would be the case without the structure.
Similarly,  maintaining the low entropy of the structures produced during gravitational collapse (e.g. bipolar outflows of
active galactic nuclei and accretion disks) exports entropy such that the net result is an increase of the entropy of
the universe, not a decrease.

\begin{figure}[!p]
  \begin{center}
    \includegraphics[scale=0.55]{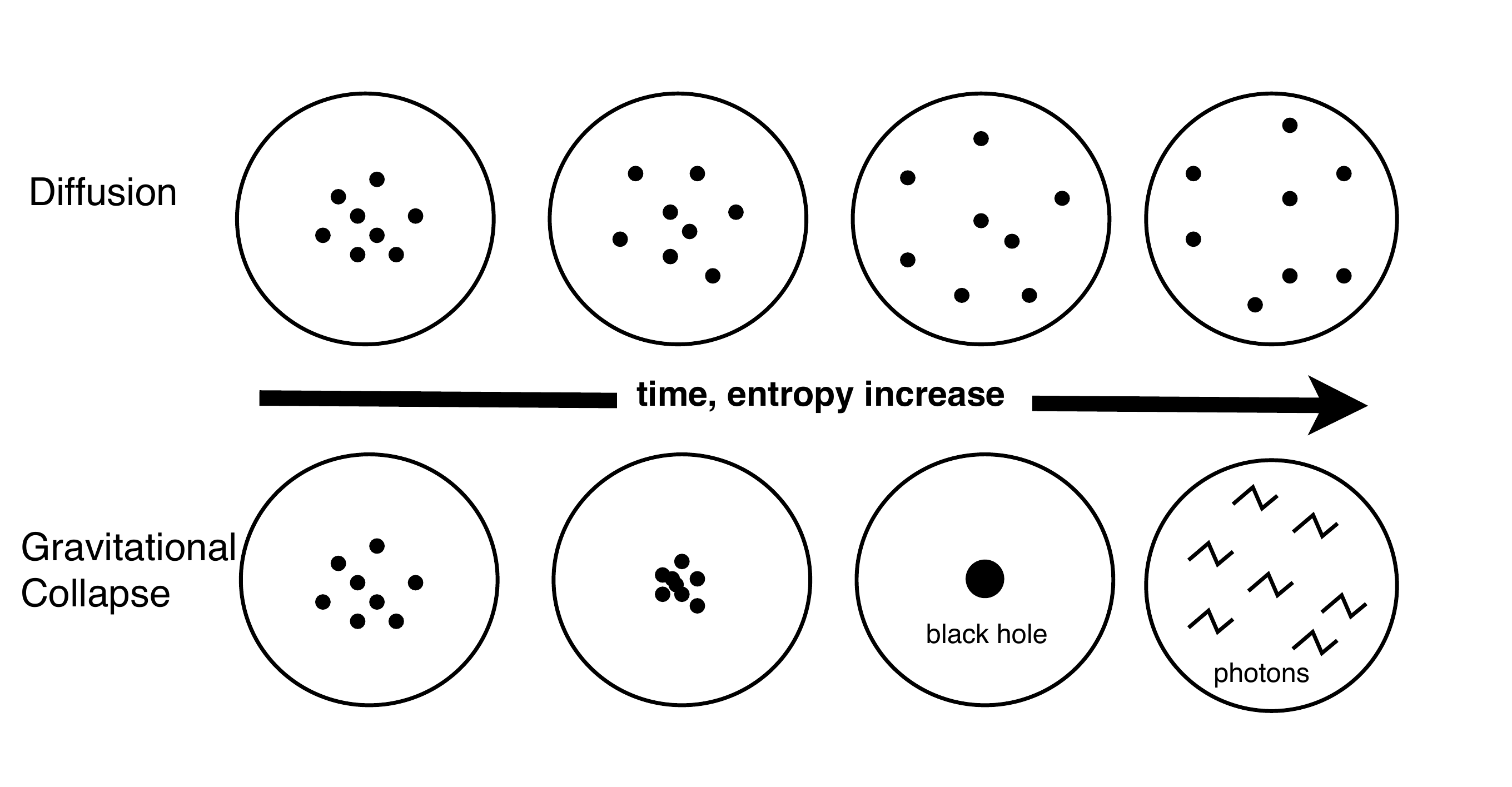}
\caption[Comparing diffusion to gravitational collapse and subsequent evaporation by the Hawking process.]
{Entropy increases during both diffusion (top) and gravitational clumping (bottom). 
If thermal energy dominates the gravitational binding energy (top), 
then entropy will increase as material diffuses and spreads out over the entire volume (think perfume diffusing in a room). 
If gravitational binding energy dominates thermal energy (bottom), then entropy will increase as some material
and angular momentum is expelled to allow other matter to have lower angular momentum and  gravitationally collapse into
galaxies and stars, which eventually collapse/accrete into a black hole.  If the temperature of the background photons is lower than
the temperature of the black hole, the black hole will evaporate to produce the maximum entropy state of photons spread 
out over the entire volume (last circle in lower panel). Compare this figure to Fig. 27.10 of Penrose (2004).  
}   
    \label{fig:A1diffusiongravity}
  \end{center}
\end{figure}

\subsection{Diffusion and Gravitational Collapse} 

 A misleading idea is that entropy makes things spread out while gravitational collapse makes things clump together, and
therefore gravity seems to work against or even violate the second law.  

\begin{quote}
``A recurring theme throughout the life of the universe is the continual struggle between the force of gravity and the tendency
for physical systems to evolve toward more disorganized conditions.  The amount of disorder in a physical system is measured by its entropy
content. In the broadest sense, gravity tends to pull things together and thereby organizes physical structures. Entropy production works 
in the opposite  direction and acts to make physical systems more disorganized and spread out. The interplay between these two competing 
tendencies provides much of the drama in astrophysics.''   Adams and Laughlin (1999).  
\end{quote}

The part of this quote that is easily misleading is
``In the broadest sense, gravity tends to pull things together and thereby organizes physical structures. Entropy production works 
in the opposite  direction ...''

See Fig 27.10 of Penrose (2004) for some clarity on this issue.
Gravity organizes physical structures but at the expense of disorganizing and expelling other material.
This supposed struggle between entropy and gravity is misleading  because lots of material is expelled (then ignored in the computation). The heat is ignored too.
Consideration of only the centralized accreted remains, does not encompass the full entropic effects of gravitational collapse
(Binney and Tremaine 1987).  

\begin{quote}
``If one part of the system becomes well ordered and loses entropy, the system as a whole must pay for it by increasing
its entropy somewhere else for compensation'' (Adams \& Laughlin 1999).  
\end{quote}

Gravity can only pull things together if angular momentum and energy are exported. If we ignore the entropy associated with
the angular momentum and energy export, it is easy to imagine that gravity pulling things together is acting in the opposite direction of the second
law, just as it is easy to believe that life is acting in the opposite direction of the second law. 
If one focuses on the collapsed object while ignoring the increased entropy of the surrounding distribution of stars (which puffs up when part of it collapses), 
one could believe that:

\begin{quote}
``The gravitational contribution to entropy is negative and the correlations of clustering decrease this entropy.
If we retain the notion that systems evolve in the direction of an entropy extreme (a maximum negative value in the gravitational
case), then we should expect infinite systems of galaxies to form tighter and tighter clumps over larger and larger scales....
Spherical systems of stars evolve toward maximum negative gravitational entropy.'' (Saslaw 1985 p. 65)
\end{quote}

When we ignore the entropy produced during the gravitational collapse of the ``spherical systems of stars'', and concentrate only on the collapsed system
itself, then Saslaw may be correct, but this seems to contradict the idea that the entropy of a black
hole is large and positive $S_{BH} > 0$.  The transition from a negative value to a positive value when an 
object collapses into its event horizon is problematic  (Frampton 2008, Hsu \& Reeb 2008).

Neither gravitational collapse nor life violate the second law when we include the increased
entropy of the environment.
Thus, the maintenance of a fridge or an accretion disk or life, increases the total entropy of the universe.

Thermal (random kinetic energy) can be written as $E_{kin} = \frac{p^{2}}{2m}$, while gravitational binding energy is $E_{grav} = \frac{GMm}{r}$.
When things are hot, $E_{kin} >> E_{grav}$ and diffusion dominates. The maximum entropy state is reached when the atoms, or molecules 
or stars or galaxies fill up the space randomly.  They occupy a larger volume of phase space.
This is labeled ``Diffusion'' in the top panel of Fig. \ref{fig:A1diffusiongravity}.
When things are cold, $E_{kin} < < E_{grav}$, gravitational collapse occurs and leads to a black hole.  
Fig. \ref{fig:A1diffusiongravity}  describes what happens in a universe that is not expanding. 
However, consider what happens to $E_{kin}$ and $E_{grav}$ when the universe expands (the scale factor $R$ increases 
as is shown in Fig. \ref{fig:A1SexpansionConstant}).
Since $p = \frac{p_{o}}{R}$ and  any distance scales as $r = r_{o} R$ we have
$E_{kin} = \frac{E_{kin,o}}{R^{2}}$ and $E_{grav} = \frac{E_{grav,o}}{R}$.
Thus, as the universe expands, $E_{kin}$ decreases faster than $E_{grav}$ and we will always eventually have $E_{kin} << E_{grav}$, which leads
to gravitational collapse, black holes and then their evaporation into a diffuse gas of photons -- the maximum entropy state of the universe, within which
no life can exist.
Thus, to depict our universe, the bottom panel of Fig. \ref{fig:A1diffusiongravity} 
should be moved to the right and tacked onto the top panel.

\subsection{Black Holes and Heat Death}

Bekenstein (1973) and Hawking (1974) showed that a black hole of mass $M$ has a temperature,
$T_{BH}  = \frac{\hbar c^{3}}{8\pi G k}\frac{1}{M}$
and evaporates predominantly as photons  when its temperature is hotter than the background temperature.
Thus, although the entropy of a black hole, 
$S = \frac{4\pi k G}{\hbar c}M^{2}= \frac{kc^{3}}{\hbar G}\frac{A}{4}$,
 is sometimes referred to as a maximum entropy state,
the sharp gravitational gradient at the event horizon leads to evaporation, photon emission and a higher entropy 
state of randomly distributed photons.
If the background temperature is larger than $T_{BH}$ then the black hole will increase in mass and cool down.
However in an expanding universe, $T_{CMB} \propto \frac{1}{R}$ and as the universe expands, $R$ increases, $T_{CMB}$ decreases and eventually we have
$T_{BH} > T_{CMB}$, which leads to the evaporation of the black hole and the diffusion of the photons produced.

\begin{figure}[!h!t!p]
  \begin{center}
    \includegraphics[scale=0.4]{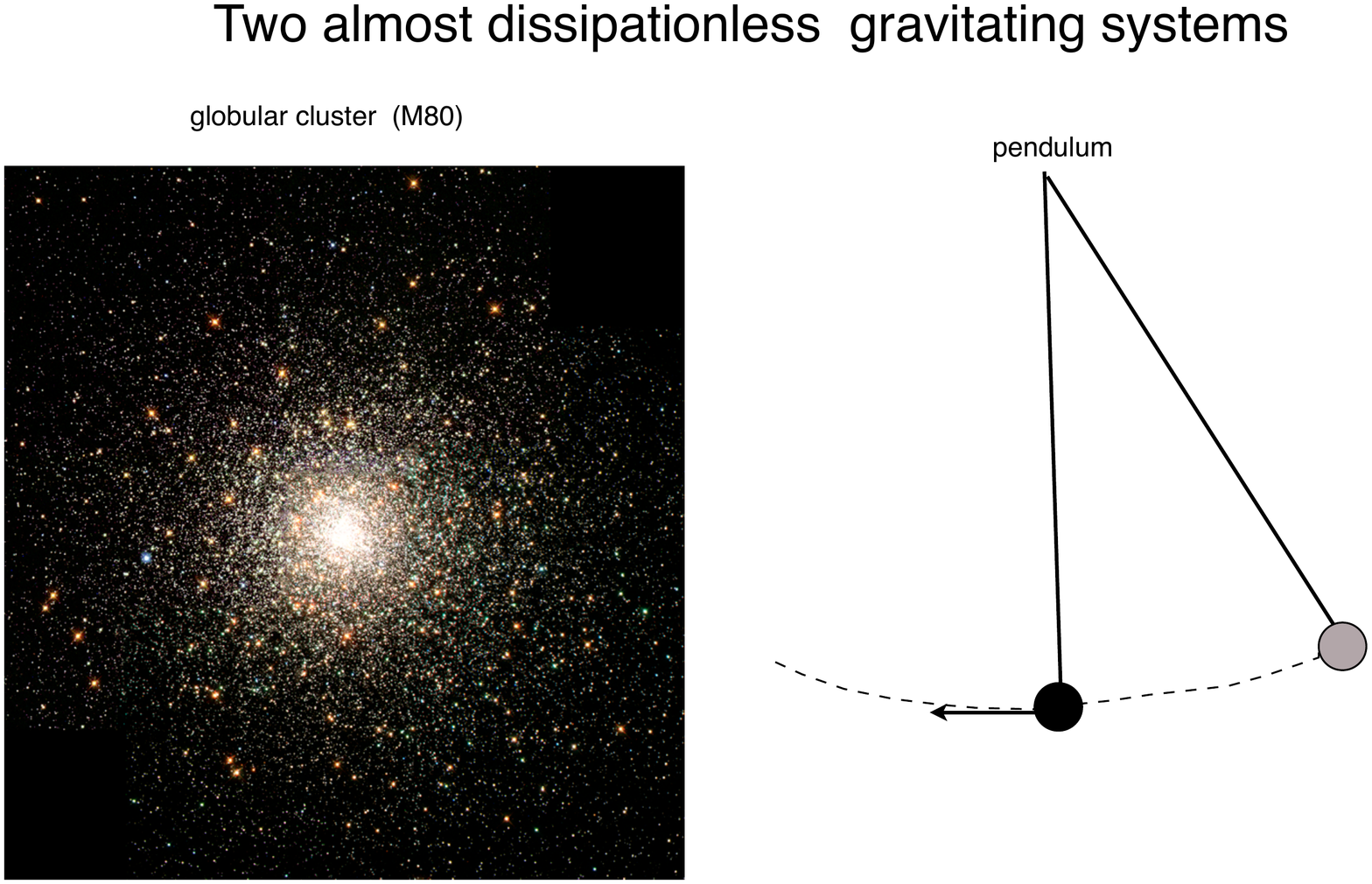}
\caption[Two almost dissipationless gravitating systems.]
{ Two almost dissipationless gravitating systems: a globular cluster and a pendulum.
Globular clusters are some of the oldest structures in the universe ($\sim 12$ billion years).
If there were no friction we would have a Hamiltonian system in which energy inside the system is conserved
as it sloshes back and forth between
kinetic and potential energy.  
Such a system cannot collapse further. The resulting isothermal sphere is the maximum entropy solution (Binney \& Tremaine 1987).
Even nominally Hamiltonian systems such as galaxies and globular clusters, emit gravitational waves and collapse.
These almost Hamiltonian systems should be contrasted with the large dissipation and entropy production of protoplanetary 
accretion disks that
allow stars to form and the much larger accretion disks in 
active galactic nuclei (AGN) which feed black holes in the center of galaxies.
(Image of M80 credit: Hubble Heritage Team, AURA/ STScI/ NASA)
}
    \label{fig:A1globular}
  \end{center}
\end{figure}

The second law establishes the arrow of time. Since we are far from equilibrium dissipative structures, we must
move through time in the direction in which entropy increases and in which free energy is available. 
Since all observers are dissipative, our existence depends on $dS > 0$.  The situation $dS = 0$ is unobservable.
This may be an anthropic explanation for the initial low entropy of the universe.
No other explanations are known to us.
Just as a universe, with a value of a cosmological
constant that is too big, is unobservable because stars never form (e.g. Weinberg 1987), so too, a universe that starts at maximum entropy
is unobservable. 
Since life (and any other dissipative structures) needs gradients to form and survive, the 
initial condition of any universe that contains life will be one of low entropy, not high entropy.  You can not 
start an observable universe from a heat death.

In  the multiverse scenario, we imagine universes with varying degrees of baryon non-conservation.
If baryon number were conserved, the early universe would have had the same amount of matter as anti-matter.
The universe would be filled with a diffuse gas of photons at maximum entropy.  There would be no matter homogeneously
distributed that would provide the low initial gravitational entropy.  Low energy photons, spread out evenly over the volume
of the universe, is a  maximum entropy state. Baryons spread out evenly, is a minimal gravitational entropy state.

In addition to baryon non-conservation, a requirement for life is that the baryons not be already clumped into black holes.
They can be very smoothly distributed, or clumped a bit, but
not too much.  In other words, non-clumped (but clumpable) matter is required to start the universe at low entropy.

Penrose (1979, 1987, 1989, 2004)  has been concerned with the relationship between entropy 
and gravity for more than three decades (see also Barrow \& Tipler 1986). 
He has stressed the amazingly unlikely initial low gravitational entropy 
of the universe that ensured that dissipative structures formed as gravity
clumped matter and produced gradients to drive dissipative structures.

This low initial entropy of the universe is quantified by the low amplitude of the power spectrum of density 
perturbations measurable in the cosmic microwave background 
and in the large scale structure of galaxies.  
According to the inflationary scenario, these low amplitude density fluctuations have their
origin in irreducible vacuum fluctuations that became real during inflation, in a manner analogous to the way electrons and positrons are
created out of the vacuum by ultra-strong electric fields between capacitor plates.
The initial low amplitude of fluctuations is measured as the amplitude $Q$ of CMB fluctuations or the amplitude  $A$ of 
the power spectrum $P(k)$ of large scale structure.
The lower the initial values of $Q$ or $A$, the lower the degree of clumpiness and the lower the initial gravitational entropy
of the universe.
Penrose (1979) describes these low entropy initial conditions in terms of small values for the Weyl curvature tensor.
We are uncertain how to explain the low values of Q or A or the Weyl curvature. In a multiverse scenario, perhaps there is some mother distribution
of values from which each universe gets its own initial entropy and ours is low because it has to be for us to 
evolve and observe it (Tegmark \& Rees 1998). 

\begin{figure*}[!p]
  \begin{center}
    \includegraphics[scale=0.4]{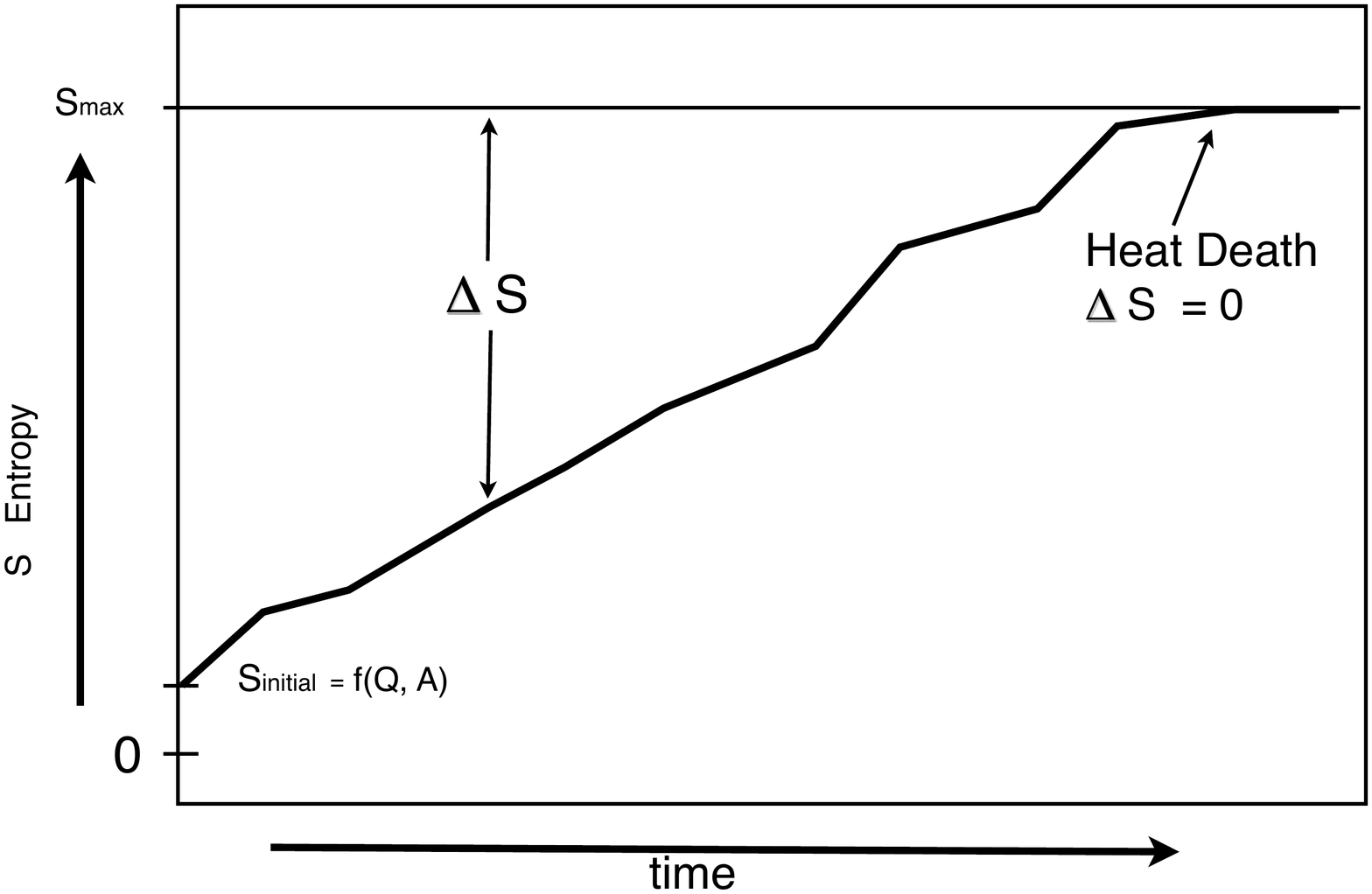}
\caption[The diminishing entropy gap.]{
The universe starts off at low entropy  (not zero) due to the low level of density perturbations in the early universe -- low $Q$ and low $A$ 
(e.g. Tegmark and Rees 1998) -- where ``low'' means less than the maximum value $S_{max}$.  At $S_{max}$  all the energy density
of the universe is in massless particles in equilibrium at a common temperature.  Thus the universe starts off with a large entropy gap  $\Delta S$. 
The parameters $Q$ and $A$ are the observable normalizations of the primordial density fluctuations and set the initial
gravitational entropy of the universe.
There is no general agreement on the curve shown here. See for example Fig. 7.3 of Davies (1994) and Fig. 1.2 of Frautschi (1988). 
}
\label{fig:A1DeltaS}
  \end{center}
\end{figure*}

\section{The Entropy Gap and the Heat Death of the Universe}
\label{sec:A1heatdeath}
Is the entropy of the universe getting closer or further from its maximum value  $S_{max}$? That is:
Is the entropy gap, $\Delta S(t) = S_{max} - S(t)$, increasing or decreasing? 
Through gravitational collapse, and the irreversible, dissipative
processes produced by the density and chemical gradients that result, the entropy of the universe increases, while
$S_{max}$ may be constant (Egan \& Lineweaver 2009).
Fig. \ref{fig:A1DeltaS} shows a monotonically decreasing entropy gap leading
to a heat death with no possibility for life thereafter.
The concept of a heat death was introduced
by Thomson (1851, 1862) and has dominated the discussion of the far future fate of the universe.

Tolman (1934) showed that if the universe could bounce back from a contraction into an expansion,
a cyclic universe could not be one that is infinitely old, since with each cycle, the entropy of the material would increase,
and the cycles would get longer and longer. Steinhardt and Turok (2007) have a model
which gets around this entropy problem by reducing entropy with the free energy of a semi-infinite gravitational potential.

\begin{figure*}[!p]
  \begin{center}
    \includegraphics[scale=0.4]{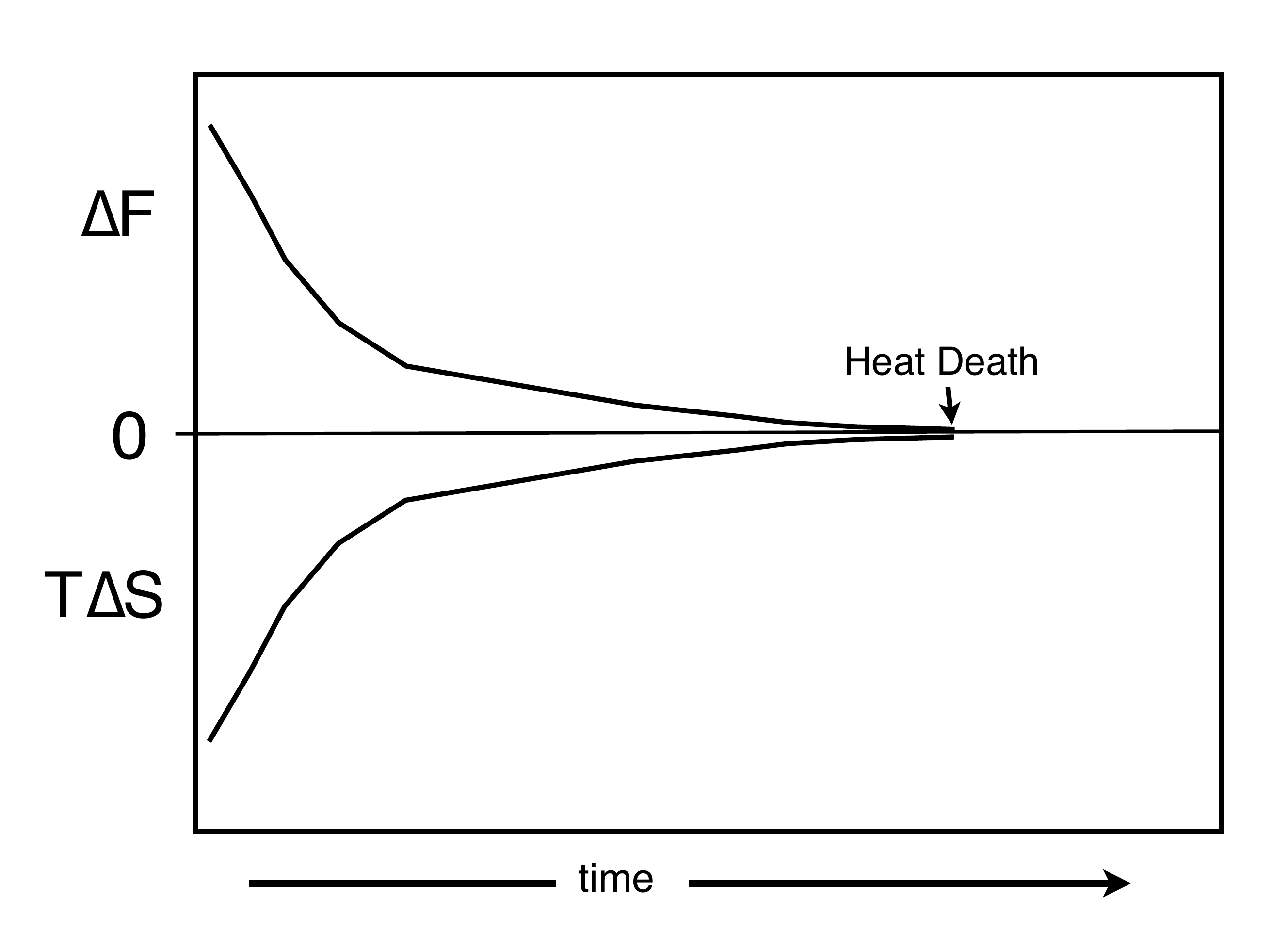}
\caption[Diminishing free energy.]{ The free energy in the Universe: $\Delta F = - T \Delta S$.
As long as there is an entropy gap in the universe, i.e., as long as $\Delta S > 0$ (Fig. \ref{fig:A1DeltaS}), 
there will be a flow of free energy to make life possible.  
As $\Delta S \rightarrow 0$ and $T \rightarrow 0$ then  $\Delta F \rightarrow 0$ and life can no longer survive.   
 }
    \label{fig:A1FTS}
  \end{center}
\end{figure*}

In a universe where the energy is conserved ($\Delta U = 0$), the free energy
available to do work (to maintain far from equilibrium structures) is $\Delta F = - T \Delta S$ and this is
plotted in Fig. \ref{fig:A1FTS}. 

Entropy is the unifying concept of life because the second law is universal; it applies to everything (Schneider \& Kay 1994). 
Man, machine, microbe or the entire cosmos -- there is no scale or material to which the second law does not apply.
However, the degree to which the equations of thermodynamics apply to near equilibrium situations, steady state
situations and far from equilibrium situations is still problematic (see however, Dewar 2003, 2005).

If Darwin had read Carnot (1824), Prigogine (1978) and Penrose (1979, 2004) rather than Newton, Malthus and Lyell, the last paragraph
of the Origin of Species would have read something like

\begin{quote}
``There is grandeur in this view of life, with its dissipative powers, having been originally induced into
many forms of far from equilibrium dissipative systems, and that, 
whilst irreversible processes on this planet have produced entropy according to the fixed second law of thermodynamics, from so 
simple a low gravitational entropic state, endless forms most beautiful and most wonderful have been and continue to increase the
entropy of the universe as they destroy the gradients which spawned them.''
\end{quote}


\section*{Appendix A. Entropy of Blackbody Photons}

Here we show that the entropy of a system of N particles is $S \sim N$.
For reference,  
Boltzmann's constant is $k = 1.38066 \times 10^{-23} J/K$.
The Stephn-Boltzmann constant
$\sigma = \frac{\pi^{2}}{15}\frac{k^{4}}{\hbar^{3}c^{3}}  = 7.565 \times 10^{-16} J m^{-3} K^{-4}$.
%
Consider a photon gas at temperature $T$ in a volume $V$  (e.g. Sears \& Salinger 1975).
The internal energy is,
\be
U = V \sigma T^{4}.
\label{eq:A1gammaU}
\ee
The entropy is
\be
S= \frac{4}{3}\sigma V T^{3}.
\label{eq:A1gammaS}
\ee
The pressure is
\be
p=\frac{1}{3}\sigma T^{4},
\ee
and the number of photons is
\bea
N_{\gamma} &=& \frac{36.06}{\pi^{4}k} \sigma V T^{3}\\
           &=&   \frac{27.045}{k\pi^{4}} \times \frac{4}{3}\sigma V T^{3}  \nonumber\\
           &=&   \frac{27.045}{k\pi^{4}} S.  
\label{eq:A1N}
\eea
Thus, 
\be
S= 0.2776 \; k \; N_{\gamma},
\label{eq:A1SN}
\ee

and to measure the entropy of the microwave background  we just need to count photons. If the number of photons in a given 
volume of the universe is N, then the entropy of photons in that volume is $S \sim kN$.   
The photonic entropy of the universe is in the cosmic microwave background.  
Starlight cannot change that.  If all the matter in the universe were transformed into 3 K blackbody 
radiation, the number of photons would add up to only $\sim$ 1 \% of the number of CMB photons.  
The entropy of the universe would increase by only $\sim 1\%$ (Frautschi 1982).

\section*{Appendix B. Which Free Energy is most useful F or G?}
\label{sec:A1which}

Thermodynamic equilibrium may be characterized by the minimization of the Helmholtz free energy (Eq. \ref{eq:A1Helmholtz})
$F = U - TS$  (e.g. Prigogine 1978).
When $U = TS$, no free energy can be extracted from the system, but this is not the same as equilibrium.
When there are pressure gradients that can do $pdV$ work and drive organization, i.e., hurricanes, $F$ is the most
relevant free energy. When pressure cannot be used, i.e., life on Earth, or photon pressure of the cosmic microwave background, 
then $G$ is more relevant.

Chemists are used to dealing with the Gibb's Free energy of a reaction $G$, where 
$G = U -TS - pV$ and G is the extractable energy, or free energy under constant pressure conditions (the usual conditions under our 
stable atmosphere and in the universe except at shock fronts and hurricanes). 
 $G$ does not include the $pdV$ work that could be done by a pressure gradient of the atmosphere, while $F$ does.
We use the Helmholtz free energy because we are interested in the most generic situations.  We want to know the extractable energy under any
conditions.

Since the free energy can never be more than the internal energy, $TS$ will always be positive,
or $(TdS + SdT) >  0$.
Using Eqs. Eq. \ref{eq:A1Helmholtz}, \ref{eq:A1gammaU}, \ref{eq:A1gammaS}, the free energy of a photon gas is,

\bea
F &=& U-TS\\
&=& V\sigma T^{4} - \frac{4}{3}\sigma V T^{4}  \nonumber\\
&=&- \frac{1}{3}\sigma V T^{4}.
\eea

This is the work that the photon gas could do if it were surrounded by zero pressure.
However the photon gas fills the universe, and can do no work on itself.

The Gibbs free energy of a photon gas in equilibrium does not include $pdV$ work and is equal to zero:
\bea
G &=& U-TS +pV\\
  &=& F + pV   \nonumber \\
&=& - \frac{1}{3}\sigma V T^{4} + \frac{1}{3} V \sigma T^{4}  \nonumber \\
&=& 0
\label{eq:A1zero}
\eea


No $pdV$ work is used to drive the chemistry-based metabolisms of terrestrial life forms, however 
some dissipative structures are driven by $pdV$ work and so here we use $F$ in our computations.
We interpret Eq. \ref{eq:A1zero} as ``no Gibbs free energy can be extracted from a photon gas at equilibrium''

\cleardoublepage

\fancypagestyle{plain}{									
\fancyfoot[LE]{\usefont{OT1}{qbk}{m}{n}\selectfont \thepage}		
\fancyfoot[RO]{\usefont{OT1}{qbk}{m}{n}\selectfont \thepage}		
\renewcommand{\headrulewidth}{0pt}						
\renewcommand{\footrulewidth}{0.5pt}						
}
%

\phantomsection\addcontentsline{toc}{chapter}{Bibliography}
\bibliographystyle{astron}
\pagestyle{plain}
\bibliography{thesis}

\begin{thebibliography}{}

\bibitem[\protect\astroncite{Abe et~al.}{2008}]{Abe2008}
Abe, S. et~al.: 2008,
\newblock {\em Phys. Rev. Lett.} {\bf 100}, 221803

\bibitem[\protect\astroncite{{Adams} and {Laughlin}}{1997}]{Adams1997}
{Adams}, F.~C. and {Laughlin}, G.: 1997,
\newblock {\em Reviews of Modern Physics} {\bf 69}, 337

\bibitem[\protect\astroncite{Adamson et~al.}{2008}]{Adamson2008}
Adamson, P. et~al.: 2008,
\newblock {\em Phys. Rev. Lett.} {\bf 101}, 131802

\bibitem[\protect\astroncite{{Aguirre} et~al.}{2007}]{Aguirre2006}
{Aguirre}, A., {Gratton}, S., and {Johnson}, M.~C.: 2007,
\newblock {\em \prd} {\bf 75(12)}, 123501

\bibitem[\protect\astroncite{{Aguirre} and {Tegmark}}{2005}]{Aguirre2005}
{Aguirre}, A. and {Tegmark}, M.: 2005,
\newblock {\em Journal of Cosmology and Astro-Particle Physics} {\bf 1}, 3

\bibitem[\protect\astroncite{{Ahmed} et~al.}{2004}]{Ahmed2004}
{Ahmed}, M., {Dodelson}, S., {Greene}, P.~B., and {Sorkin}, R.: 2004,
\newblock {\em \prd} {\bf 69(10)}, 103523

\bibitem[\protect\astroncite{{Albrecht} et~al.}{2006}]{Albrecht2006}
{Albrecht}, A., {Bernstein}, G., {Cahn}, R., {Freedman}, W.~L., {Hewitt}, J.,
  {Hu}, W., {Huth}, J., {Kamionkowski}, M., {Kolb}, E.~W., {Knox}, L.,
  {Mather}, J.~C., {Staggs}, S., and {Suntzeff}, N.~B.: 2006,
\newblock {\em ArXiv Astrophysics e-prints}

\bibitem[\protect\astroncite{{All{\`e}gre} et~al.}{1995}]{Allegre1995}
{All{\`e}gre}, C.~J., {Manh{\`e}s}, G., and {G{\"o}pel}, C.: 1995,
\newblock {\em \gca} {\bf 59}, 1445

\bibitem[\protect\astroncite{{Allende Prieto}}{2006}]{Allende2006}
{Allende Prieto}, C.: 2006,
\newblock {\em ArXiv Astrophysics e-prints}

\bibitem[\protect\astroncite{{Amendola}}{2000a}]{Amendola2000}
{Amendola}, L.: 2000a,
\newblock {\em \prd} {\bf 62(4)}, 043511

\bibitem[\protect\astroncite{{Amendola}}{2000b}]{Amendola2000b}
{Amendola}, L.: 2000b,
\newblock {\em \mnras} {\bf 312}, 521

\bibitem[\protect\astroncite{{Amendola} et~al.}{2007}]{Amendola2007}
{Amendola}, L., {Campos}, G.~C., and {Rosenfeld}, R.: 2007,
\newblock {\em \prd} {\bf 75(8)}, 083506

\bibitem[\protect\astroncite{{Amendola} et~al.}{2006}]{Amendola2006}
{Amendola}, L., {Quartin}, M., {Tsujikawa}, S., and {Waga}, I.: 2006,
\newblock {\em \prd} {\bf 74(2)}, 023525

\bibitem[\protect\astroncite{{Amendola} and {Quercellini}}{2003}]{Amendola2003}
{Amendola}, L. and {Quercellini}, C.: 2003,
\newblock {\em \prd} {\bf 68(2)}, 023514

\bibitem[\protect\astroncite{{Armendariz-Picon}
  et~al.}{2000}]{Armendariz-Picon2000}
{Armendariz-Picon}, C., {Mukhanov}, V., and {Steinhardt}, P.~J.: 2000,
\newblock {\em Physical Review Letters} {\bf 85}, 4438

\bibitem[\protect\astroncite{{Armendariz-Picon}
  et~al.}{2001}]{Armendariz-Picon2001}
{Armendariz-Picon}, C., {Mukhanov}, V., and {Steinhardt}, P.~J.: 2001,
\newblock {\em \prd} {\bf 63(10)}, 103510

\bibitem[\protect\astroncite{{Astier} et~al.}{2006}]{Astier2006}
{Astier}, P., {Guy}, J., {Regnault}, N., {Pain}, R., {Aubourg}, E., {Balam},
  D., {Basa}, S., {Carlberg}, R.~G., {Fabbro}, S., {Fouchez}, D., {Hook},
  I.~M., {Howell}, D.~A., {Lafoux}, H., {Neill}, J.~D.,
  {Palanque-Delabrouille}, N., {Perrett}, K., {Pritchet}, C.~J., {Rich}, J.,
  {Sullivan}, M., {Taillet}, R., {Aldering}, G., {Antilogus}, P.,
  {Arsenijevic}, V., {Balland}, C., {Baumont}, S., {Bronder}, J., {Courtois},
  H., {Ellis}, R.~S., {Filiol}, M., {Gon{\c c}alves}, A.~C., {Goobar}, A.,
  {Guide}, D., {Hardin}, D., {Lusset}, V., {Lidman}, C., {McMahon}, R.,
  {Mouchet}, M., {Mourao}, A., {Perlmutter}, S., {Ripoche}, P., {Tao}, C., and
  {Walton}, N.: 2006,
\newblock {\em \aap} {\bf 447}, 31

\bibitem[\protect\astroncite{{Bahcall} and {Soneira}}{1980}]{Bahcall1980}
{Bahcall}, J.~N. and {Soneira}, R.~M.: 1980,
\newblock {\em \apjs} {\bf 44}, 73

\bibitem[\protect\astroncite{{Balbus} and {Hawley}}{2002}]{Balbus2002}
{Balbus}, S.~A. and {Hawley}, J.~F.: 2002,
\newblock {\em \apj} {\bf 573}, 749

\bibitem[\protect\astroncite{{Barrow}}{1994}]{Barrow1994}
{Barrow}, J.~D.: 1994,
\newblock {\em {The origin of the universe}},
\newblock New York : BasicBooks, c1994.

\bibitem[\protect\astroncite{{Barrow} and {Tipler}}{1986}]{Barrow1986}
{Barrow}, J.~D. and {Tipler}, F.~J.: 1986,
\newblock {\em Atmospheric Chemistry \& Physics}

\bibitem[\protect\astroncite{{Basu} and {Lynden-Bell}}{1990}]{Basu1990}
{Basu}, B. and {Lynden-Bell}, D.: 1990,
\newblock {\em \qjras} {\bf 31}, 359

\bibitem[\protect\astroncite{{Bean} et~al.}{2001}]{Bean2001b}
{Bean}, R., {Hansen}, S.~H., and {Melchiorri}, A.: 2001,
\newblock {\em \prd} {\bf 64(10)}, 103508

\bibitem[\protect\astroncite{{Bekenstein}}{1973}]{Bekenstein1973}
{Bekenstein}, J.~D.: 1973,
\newblock {\em \prd} {\bf 7}, 2333

\bibitem[\protect\astroncite{{Bekenstein}}{1974}]{Bekenstein1974}
{Bekenstein}, J.~D.: 1974,
\newblock {\em \prd} {\bf 9}, 3292

\bibitem[\protect\astroncite{{Bekenstein}}{1981}]{Bekenstein1981}
{Bekenstein}, J.~D.: 1981,
\newblock {\em \prd} {\bf 23}, 287

\bibitem[\protect\astroncite{{Bekenstein}}{2005}]{Bekenstein2005}
{Bekenstein}, J.~D.: 2005,
\newblock {\em Foundations of Physics} {\bf 35}, 1805

\bibitem[\protect\astroncite{{Bell} and {de Jong}}{2001}]{Bell2001}
{Bell}, E.~F. and {de Jong}, R.~S.: 2001,
\newblock {\em \apj} {\bf 550}, 212

\bibitem[\protect\astroncite{{Bensby} and {Feltzing}}{2006}]{Bensby2006}
{Bensby}, T. and {Feltzing}, S.: 2006,
\newblock {\em \mnras} {\bf 367}, 1181

\bibitem[\protect\astroncite{{Bensby} et~al.}{2005}]{Bensby2005}
{Bensby}, T., {Feltzing}, S., {Lundstr{\"o}m}, I., and {Ilyin}, I.: 2005,
\newblock {\em \aap} {\bf 433}, 185

\bibitem[\protect\astroncite{{Bento} et~al.}{2002}]{Bento2002}
{Bento}, M.~C., {Bertolami}, O., and {Sen}, A.~A.: 2002,
\newblock {\em \prd} {\bf 66(4)}, 043507

\bibitem[\protect\astroncite{{Binney} and {Tremaine}}{2008}]{Binney2008}
{Binney}, J. and {Tremaine}, S.: 2008,
\newblock {\em {Galactic Dynamics: Second Edition}},
\newblock Princeton University Press

\bibitem[\protect\astroncite{{Bludman}}{2004}]{Bludman2004}
{Bludman}, S.: 2004,
\newblock {\em \prd} {\bf 69(12)}, 122002

\bibitem[\protect\astroncite{{Bludman} and {Roos}}{2001}]{Bludman2001}
{Bludman}, S.~A. and {Roos}, M.: 2001,
\newblock {\em \apj} {\bf 547}, 77

\bibitem[\protect\astroncite{{Bostrom}}{2002}]{Bostrom2002}
{Bostrom}, N.: 2002,
\newblock {\em {Anthropic Bias: Observation Selection Effects in Science and
  Philosophy}},
\newblock Routledge, London

\bibitem[\protect\astroncite{{Bousso}}{1999}]{Bousso1999}
{Bousso}, R.: 1999,
\newblock {\em Journal of High Energy Physics} {\bf 7}, 4

\bibitem[\protect\astroncite{{Bousso}}{2001}]{Bousso2001}
{Bousso}, R.: 2001,
\newblock {\em Journal of High Energy Physics} {\bf 4}, 35

\bibitem[\protect\astroncite{{Bousso}}{2002}]{Bousso2002}
{Bousso}, R.: 2002,
\newblock {\em Reviews of Modern Physics} {\bf 74}, 825

\bibitem[\protect\astroncite{{Bousso} et~al.}{2007}]{Bousso2007}
{Bousso}, R., {Harnik}, R., {Kribs}, G.~D., and {Perez}, G.: 2007,
\newblock {\em \prd} {\bf 76(4)}, 043513

\bibitem[\protect\astroncite{{Caldwell}}{2002}]{Caldwell2002}
{Caldwell}, R.~R.: 2002,
\newblock {\em Physics Letters B} {\bf 545}, 23

\bibitem[\protect\astroncite{{Caldwell} et~al.}{1998}]{Caldwell1998}
{Caldwell}, R.~R., {Dave}, R., and {Steinhardt}, P.~J.: 1998,
\newblock {\em Physical Review Letters} {\bf 80}, 1582

\bibitem[\protect\astroncite{{Caldwell} et~al.}{2003}]{Caldwell2003}
{Caldwell}, R.~R., {Kamionkowski}, M., and {Weinberg}, N.~N.: 2003,
\newblock {\em Physical Review Letters} {\bf 91(7)}, 071301

\bibitem[\protect\astroncite{{Carroll}}{2001a}]{Carroll2001b}
{Carroll}, S.~M.: 2001a,
\newblock {\em SNAP (SuperNova Acceleration Probe) Yellow Book
  astro-ph/0107571}

\bibitem[\protect\astroncite{{Carroll}}{2001b}]{Carroll2001a}
{Carroll}, S.~M.: 2001b,
\newblock {\em Living Reviews in Relativity} {\bf 4}, 1

\bibitem[\protect\astroncite{{Carroll}}{2004}]{Carroll2004book}
{Carroll}, S.~M.: 2004,
\newblock {\em {Spacetime and geometry. An introduction to general
  relativity}},
\newblock Spacetime and geometry / Sean Carroll.~San Francisco, CA, USA:
  Addison Wesley, ISBN 0-8053-8732-3, 2004, XIV + 513 pp.

\bibitem[\protect\astroncite{{Carter}}{1974}]{Carter1974}
{Carter}, B.: 1974,
\newblock in M.~S. {Longair} (ed.), {\em IAU Symp. 63: Confrontation of
  Cosmological Theories with Observational Data}, pp 291--298

\bibitem[\protect\astroncite{{Carter}}{1983}]{Carter1983}
{Carter}, B.: 1983,
\newblock {\em Philosophical Transactions of the Royal Society} {\bf A 310},
  347

\bibitem[\protect\astroncite{{Chiba} et~al.}{2000}]{Chiba2000}
{Chiba}, T., {Okabe}, T., and {Yamaguchi}, M.: 2000,
\newblock {\em \prd} {\bf 62(2)}, 023511

\bibitem[\protect\astroncite{{Chimento} et~al.}{2000}]{Chimento2000}
{Chimento}, L.~P., {Jakubi}, A.~S., and {Pav{\'o}n}, D.: 2000,
\newblock {\em \prd} {\bf 62(6)}, 063508

\bibitem[\protect\astroncite{{Chimento} et~al.}{2003}]{Chimento2003}
{Chimento}, L.~P., {Jakubi}, A.~S., and {Pav{\'o}n}, D.: 2003,
\newblock {\em \prd} {\bf 67(8)}, 087302

\bibitem[\protect\astroncite{{Chyba}}{1999}]{Chyba1999}
{Chyba}, C.~F.: 1999,
\newblock {\em AAS/Division for Planetary Sciences Meeting Abstracts} {\bf 31},
  66.09

\bibitem[\protect\astroncite{{Cleveland} et~al.}{1998}]{Cleveland1998}
{Cleveland}, B.~T., {Daily}, T., {Davis}, R.~J., {Distel}, J.~R., {Lande}, K.,
  {Lee}, C.~K., {Wildenhain}, P.~S., and {Ullman}, J.: 1998,
\newblock {\em \apj} {\bf 496}, 505

\bibitem[\protect\astroncite{{Cohn}}{1998}]{Cohn1998}
{Cohn}, J.~D.: 1998,
\newblock {\em \apss} {\bf 259}, 213

\bibitem[\protect\astroncite{{Coleman} and {Roos}}{2003}]{Coleman2003}
{Coleman}, T.~S. and {Roos}, M.: 2003,
\newblock {\em \prd} {\bf 68(2)}, 027702

\bibitem[\protect\astroncite{{Copeland} et~al.}{2006}]{Copeland2006}
{Copeland}, E.~J., {Sami}, M., and {Tsujikawa}, S.: 2006,
\newblock {\em ArXiv e-prints hep-th/0603057}

\bibitem[\protect\astroncite{{Courteau} and {van den
  Bergh}}{1999}]{Courteau1999}
{Courteau}, S. and {van den Bergh}, S.: 1999,
\newblock {\em \aj} {\bf 118}, 337

\bibitem[\protect\astroncite{{Dalal} et~al.}{2001}]{Dalal2001}
{Dalal}, N., {Abazajian}, K., {Jenkins}, E., and {Manohar}, A.~V.: 2001,
\newblock {\em Physical Review Letters} {\bf 87(14)}, 141302

\bibitem[\protect\astroncite{{Davies}}{1987}]{Davies1987}
{Davies}, P.~C.~W.: 1987,
\newblock {\em Classical and Quantum Gravity} {\bf 4}, L225

\bibitem[\protect\astroncite{{Davies}}{1994}]{Davies1994}
{Davies}, P.~C.~W.: 1994,
\newblock in J. {Halliwell}, J. {P{\'e}rez-Mercader}, and W. {Zurek} (eds.),
  {\em Physical Origins of Time Asymmetry}, pp 119--130

\bibitem[\protect\astroncite{{Davis} et~al.}{2003}]{Davis2003}
{Davis}, T.~M., {Davies}, P.~C.~W., and {Lineweaver}, C.~H.: 2003,
\newblock {\em Classical and Quantum Gravity} {\bf 20}, 2753

\bibitem[\protect\astroncite{{Davis} et~al.}{2007}]{Davis2007}
{Davis}, T.~M., {M{\"o}rtsell}, E., {Sollerman}, J., {Becker}, A.~C.,
  {Blondin}, S., {Challis}, P., {Clocchiatti}, A., {Filippenko}, A.~V.,
  {Foley}, R.~J., {Garnavich}, P.~M., {Jha}, S., {Krisciunas}, K., {Kirshner},
  R.~P., {Leibundgut}, B., {Li}, W., {Matheson}, T., {Miknaitis}, G.,
  {Pignata}, G., {Rest}, A., {Riess}, A.~G., {Schmidt}, B.~P., {Smith}, R.~C.,
  {Spyromilio}, J., {Stubbs}, C.~W., {Suntzeff}, N.~B., {Tonry}, J.~L.,
  {Wood-Vasey}, W.~M., and {Zenteno}, A.: 2007,
\newblock {\em \apj} {\bf 666}, 716

\bibitem[\protect\astroncite{{Dehnen} and {Binney}}{1998}]{Dehnen1998}
{Dehnen}, W. and {Binney}, J.~J.: 1998,
\newblock {\em \mnras} {\bf 298}, 387

\bibitem[\protect\astroncite{{del Campo} et~al.}{2006}]{delCampo2006}
{del Campo}, S., {Herrera}, R., {Olivares}, G., and {Pav{\'o}n}, D.: 2006,
\newblock {\em \prd} {\bf 74(2)}, 023501

\bibitem[\protect\astroncite{{del Campo} et~al.}{2005}]{delCampo2005}
{del Campo}, S., {Herrera}, R., and {Pav{\'o}n}, D.: 2005,
\newblock {\em \prd} {\bf 71(12)}, 123529

\bibitem[\protect\astroncite{{di Pietro} and {Claeskens}}{2003}]{diPietro2003}
{di Pietro}, E. and {Claeskens}, J.-F.: 2003,
\newblock {\em \mnras} {\bf 341}, 1299

\bibitem[\protect\astroncite{{Dicke}}{1961}]{Dicke1961}
{Dicke}, R.~H.: 1961,
\newblock {\em \nat} {\bf 192}, 440

\bibitem[\protect\astroncite{{Dirac}}{1937}]{Dirac1937}
{Dirac}, P.~A.~M.: 1937,
\newblock {\em \nat} {\bf 139}, 323

\bibitem[\protect\astroncite{{Dirac}}{1938}]{Dirac1938}
{Dirac}, P.~A.~M.: 1938,
\newblock {\em Royal Society of London Proceedings Series A} {\bf 165}, 199

\bibitem[\protect\astroncite{{Dodelson} et~al.}{2000}]{Dodelson2000}
{Dodelson}, S., {Kaplinghat}, M., and {Stewart}, E.: 2000,
\newblock {\em Physical Review Letters} {\bf 85}, 5276

\bibitem[\protect\astroncite{{Driver} et~al.}{1994}]{Driver1994}
{Driver}, S.~P., {Phillipps}, S., {Davies}, J.~I., {Morgan}, I., and {Disney},
  M.~J.: 1994,
\newblock {\em \mnras} {\bf 268}, 393

\bibitem[\protect\astroncite{{Durrer} and {Maartens}}{2007}]{Durrer2007}
{Durrer}, R. and {Maartens}, R.: 2007,
\newblock {\em ArXiv e-prints 0711.0077}

\bibitem[\protect\astroncite{{Eddington}}{1931}]{Eddington1931}
{Eddington}, A.~S.: 1931,
\newblock {\em \nat} {\bf 127}, 447

\bibitem[\protect\astroncite{Efron}{1979}]{Efron1979}
Efron, B.: 1979,
\newblock {\em Ann. Statist.} 7(1)

\bibitem[\protect\astroncite{{Efstathiou}}{1995}]{Efstathiou1995}
{Efstathiou}, G.: 1995,
\newblock {\em \mnras} {\bf 274}, L73

\bibitem[\protect\astroncite{{Efstathiou} and {Bond}}{1999}]{Efstathiou1999}
{Efstathiou}, G. and {Bond}, J.~R.: 1999,
\newblock {\em \mnras} {\bf 304}, 75

\bibitem[\protect\astroncite{{Egan} and
  {Lineweaver}}{2008}]{EganLineweaver2008}
{Egan}, C.~A. and {Lineweaver}, C.~H.: 2008,
\newblock {\em \prd} {\bf 78(8)}, 083528

\bibitem[\protect\astroncite{{Egan} and
  {Lineweaver}}{2010a}]{EganLineweaver2009}
{Egan}, C.~A. and {Lineweaver}, C.~H.: 2010a,
\newblock {\em \apj} {\bf 710}, 1825

\bibitem[\protect\astroncite{{Egan} and
  {Lineweaver}}{2010b}]{EganLineweaver2009b}
{Egan}, C.~A. and {Lineweaver}, C.~H.: 2010b,
\newblock {\em {How High Could the Entropy be and will the Universe end in a
  Heat Death?}},
\newblock in preparation

\bibitem[\protect\astroncite{{Eisenhauer} et~al.}{2005}]{Eisenhauer2005}
{Eisenhauer}, F., {Genzel}, R., {Alexander}, T., {Abuter}, R., {Paumard}, T.,
  {Ott}, T., {Gilbert}, A., {Gillessen}, S., {Horrobin}, M., {Trippe}, S.,
  {Bonnet}, H., {Dumas}, C., {Hubin}, N., {Kaufer}, A., {Kissler-Patig}, M.,
  {Monnet}, G., {Str{\"o}bele}, S., {Szeifert}, T., {Eckart}, A.,
  {Sch{\"o}del}, R., and {Zucker}, S.: 2005,
\newblock {\em \apj} {\bf 628}, 246

\bibitem[\protect\astroncite{{Eisenstein} and {Hu}}{1998}]{EisensteinHu1998}
{Eisenstein}, D.~J. and {Hu}, W.: 1998,
\newblock {\em \apj} {\bf 496}, 605

\bibitem[\protect\astroncite{{Eisenstein} et~al.}{2005}]{Eisenstein2005}
{Eisenstein}, D.~J., {Zehavi}, I., {Hogg}, D.~W., {Scoccimarro}, R., {Blanton},
  M.~R., {Nichol}, R.~C., {Scranton}, R., {Seo}, H.-J., {Tegmark}, M., {Zheng},
  Z., {Anderson}, S.~F., {Annis}, J., {Bahcall}, N., {Brinkmann}, J., {Burles},
  S., {Castander}, F.~J., {Connolly}, A., {Csabai}, I., {Doi}, M., {Fukugita},
  M., {Frieman}, J.~A., {Glazebrook}, K., {Gunn}, J.~E., {Hendry}, J.~S.,
  {Hennessy}, G., {Ivezi{\'c}}, Z., {Kent}, S., {Knapp}, G.~R., {Lin}, H.,
  {Loh}, Y.-S., {Lupton}, R.~H., {Margon}, B., {McKay}, T.~A., {Meiksin}, A.,
  {Munn}, J.~A., {Pope}, A., {Richmond}, M.~W., {Schlegel}, D., {Schneider},
  D.~P., {Shimasaku}, K., {Stoughton}, C., {Strauss}, M.~A., {SubbaRao}, M.,
  {Szalay}, A.~S., {Szapudi}, I., {Tucker}, D.~L., {Yanny}, B., and {York},
  D.~G.: 2005,
\newblock {\em \apj} {\bf 633}, 560

\bibitem[\protect\astroncite{{Eke} et~al.}{2004}]{Eke2004}
{Eke}, V.~R., {Frenk}, C.~S., {Baugh}, C.~M., {Cole}, S., {Norberg}, P.,
  {Peacock}, J.~A., {Baldry}, I.~K., {Bland-Hawthorn}, J., {Bridges}, T.,
  {Cannon}, R., {Colless}, M., {Collins}, C., {Couch}, W., {Dalton}, G., {de
  Propris}, R., {Driver}, S.~P., {Efstathiou}, G., {Ellis}, R.~S.,
  {Glazebrook}, K., {Jackson}, C.~A., {Lahav}, O., {Lewis}, I., {Lumsden}, S.,
  {Maddox}, S.~J., {Madgwick}, D., {Peterson}, B.~A., {Sutherland}, W., and
  {Taylor}, K.: 2004,
\newblock {\em \mnras} {\bf 355}, 769

\bibitem[\protect\astroncite{{Elgar{\o}y} and
  {Multam{\"a}ki}}{2007}]{Elgaroy2007}
{Elgar{\o}y}, O. and {Multam{\"a}ki}, T.: 2007,
\newblock {\em \aap} {\bf 471}, 65

\bibitem[\protect\astroncite{{Elmegreen}}{2007}]{Elmegreen2007}
{Elmegreen}, B.~G.: 2007,
\newblock in Y.~W. {Kang}, H.-W. {Lee}, K.-C. {Leung}, and K.-S. {Cheng}
  (eds.), {\em The Seventh Pacific Rim Conference on Stellar Astrophysics},
  Vol. 362 of {\em Astronomical Society of the Pacific Conference Series}, pp
  269--+

\bibitem[\protect\astroncite{{Feng} et~al.}{2006}]{Feng2006}
{Feng}, B., {Li}, M., {Piao}, Y.-S., and {Zhang}, X.: 2006,
\newblock {\em Physics Letters B} {\bf 634}, 101

\bibitem[\protect\astroncite{{Ferreira} and {Joyce}}{1998}]{Ferreira1998}
{Ferreira}, P.~G. and {Joyce}, M.: 1998,
\newblock {\em \prd} {\bf 58(2)}, 023503

\bibitem[\protect\astroncite{{Frampton} et~al.}{2008}]{Frampton2008}
{Frampton}, P., {Hsu}, S.~D.~H., {Kephart}, T.~W., and {Reeb}, D.: 2008,
\newblock {\em ArXiv e-prints 0801.1847}

\bibitem[\protect\astroncite{{Frampton}}{2009a}]{Frampton2009}
{Frampton}, P.~H.: 2009a,
\newblock {\em ArXiv e-prints 0904.2934}

\bibitem[\protect\astroncite{{Frampton}}{2009b}]{Frampton2009b}
{Frampton}, P.~H.: 2009b,
\newblock {\em Journal of Cosmology and Astro-Particle Physics} {\bf 10}, 16

\bibitem[\protect\astroncite{{Frampton} and {Kephart}}{2008}]{Frampton2008b}
{Frampton}, P.~H. and {Kephart}, T.~W.: 2008,
\newblock {\em Journal of Cosmology and Astro-Particle Physics} {\bf 6}, 8

\bibitem[\protect\astroncite{{Fran{\c c}a}}{2006}]{Franca2006}
{Fran{\c c}a}, U.: 2006,
\newblock {\em Physics Letters B} {\bf 641}, 351

\bibitem[\protect\astroncite{{Fran{\c c}a} and {Rosenfeld}}{2004}]{Franca2004}
{Fran{\c c}a}, U. and {Rosenfeld}, R.: 2004,
\newblock {\em \prd} {\bf 69(6)}, 063517

\bibitem[\protect\astroncite{{Frautschi}}{1982}]{Frautschi1982}
{Frautschi}, S.: 1982,
\newblock {\em Science} {\bf 217}, 593

\bibitem[\protect\astroncite{{Frautschi}}{1988}]{Frautschi1988}
{Frautschi}, S.: 1988,
\newblock in D.~J. {Depew}, B.~H. {Weber}, and J.~D. {Smith} (eds.), {\em
  Entropy, information, and evolution : new perspectives on physical and
  biological evolution}, MIT Press, Cambridge, Mass., California State
  University, Fullerton.

\bibitem[\protect\astroncite{{Fryer} and {Kalogera}}{2001}]{Fryer2001}
{Fryer}, C.~L. and {Kalogera}, V.: 2001,
\newblock {\em \apj} {\bf 554}, 548

\bibitem[\protect\astroncite{{Fukugita} and {Peebles}}{2004}]{Fukugita2004}
{Fukugita}, M. and {Peebles}, P.~J.~E.: 2004,
\newblock {\em \apj} {\bf 616}, 643

\bibitem[\protect\astroncite{{Garriga} et~al.}{1999}]{Garriga1999}
{Garriga}, J., {Livio}, M., and {Vilenkin}, A.: 1999,
\newblock {\em \prd} {\bf 61(2)}, 023503

\bibitem[\protect\astroncite{{Garriga} and {Vilenkin}}{2000}]{Garriga2000}
{Garriga}, J. and {Vilenkin}, A.: 2000,
\newblock {\em \prd} {\bf 61(8)}, 083502

\bibitem[\protect\astroncite{{Garriga} and {Vilenkin}}{2001}]{Garriga2001}
{Garriga}, J. and {Vilenkin}, A.: 2001,
\newblock {\em \prd} {\bf 64(2)}, 023517

\bibitem[\protect\astroncite{{Gibbons} and {Hawking}}{1977}]{Gibbons1977}
{Gibbons}, G.~W. and {Hawking}, S.~W.: 1977,
\newblock {\em \prd} {\bf 15}, 2738

\bibitem[\protect\astroncite{{Glazebrook} et~al.}{2007}]{Glazebrook2007}
{Glazebrook}, K., {Blake}, C., {Couch}, W., {Forbes}, D., {Drinkwater}, M.,
  {Jurek}, R., {Pimbblet}, K., {Madore}, B., {Martin}, C., {Small}, T.,
  {Forster}, K., {Colless}, M., {Sharp}, R., {Croom}, S., {Woods}, D., {Pracy},
  M., {Gilbank}, D., {Yee}, H., and {Gladders}, M.: 2007,
\newblock {\em ArXiv Astrophysics e-prints}

\bibitem[\protect\astroncite{{Gnedin} and {Gnedin}}{1998}]{Gnedin1998}
{Gnedin}, N.~Y. and {Gnedin}, O.~Y.: 1998,
\newblock {\em \apj} {\bf 509}, 11

\bibitem[\protect\astroncite{{Gonzalez}}{1999a}]{Gonzalez1999a}
{Gonzalez}, G.: 1999a,
\newblock {\em \mnras} {\bf 308}, 447

\bibitem[\protect\astroncite{{Gonzalez}}{1999b}]{Gonzalez1999b}
{Gonzalez}, G.: 1999b,
\newblock {\em Astronomy and Geophysics} {\bf 40}, 25

\bibitem[\protect\astroncite{{Gonzalez} et~al.}{2001}]{Gonzalez2001}
{Gonzalez}, G., {Brownlee}, D., and {Ward}, P.: 2001,
\newblock {\em Icarus} {\bf 152}, 185

\bibitem[\protect\astroncite{{Gould} et~al.}{1996}]{Gould1996}
{Gould}, A., {Bahcall}, J.~N., and {Flynn}, C.: 1996,
\newblock {\em \apj} {\bf 465}, 759

\bibitem[\protect\astroncite{{Graham} et~al.}{2007}]{Graham2007}
{Graham}, A.~W., {Driver}, S.~P., {Allen}, P.~D., and {Liske}, J.: 2007,
\newblock {\em \mnras} {\bf 378}, 198

\bibitem[\protect\astroncite{{Grether} and {Lineweaver}}{2006}]{Grether2006}
{Grether}, D. and {Lineweaver}, C.~H.: 2006,
\newblock {\em \apj} {\bf 640}, 1051

\bibitem[\protect\astroncite{{Grether} and {Lineweaver}}{2007}]{Grether2007}
{Grether}, D. and {Lineweaver}, C.~H.: 2007,
\newblock {\em \apj} {\bf 669}, 1220

\bibitem[\protect\astroncite{{Gunn} and {Gott}}{1972}]{Gunn1972}
{Gunn}, J.~E. and {Gott}, J.~R.~I.: 1972,
\newblock {\em \apj} {\bf 176}, 1

\bibitem[\protect\astroncite{{Guo} and {Zhang}}{2005}]{Guo2005}
{Guo}, Z.-K. and {Zhang}, Y.-Z.: 2005,
\newblock {\em \prd} {\bf 71(2)}, 023501

\bibitem[\protect\astroncite{{Gustafsson}}{1998}]{Gustafsson1998}
{Gustafsson}, B.: 1998,
\newblock {\em Space Science Reviews} {\bf 85}, 419

\bibitem[\protect\astroncite{{Gustafsson} et~al.}{1999}]{Gustafsson1999}
{Gustafsson}, B., {Karlsson}, T., {Olsson}, E., {Edvardsson}, B., and {Ryde},
  N.: 1999,
\newblock {\em \aap} {\bf 342}, 426

\bibitem[\protect\astroncite{{Guth}}{1981}]{Guth1981}
{Guth}, A.~H.: 1981,
\newblock {\em \prd} {\bf 23}, 347

\bibitem[\protect\astroncite{{Hansen} et~al.}{2004}]{Hansen2004}
{Hansen}, B.~M.~S., {Richer}, H.~B., {Fahlman}, G.~G., {Stetson}, P.~B.,
  {Brewer}, J., {Currie}, T., {Gibson}, B.~K., {Ibata}, R., {Rich}, R.~M., and
  {Shara}, M.~M.: 2004,
\newblock {\em \apjs} {\bf 155}, 551

\bibitem[\protect\astroncite{{Harrison}}{1995}]{Harrison1995}
{Harrison}, E.~R.: 1995,
\newblock {\em \apj} {\bf 446}, 63

\bibitem[\protect\astroncite{{Hawking}}{1976}]{Hawking1976}
{Hawking}, S.~W.: 1976,
\newblock {\em \prd} {\bf 13}, 191

\bibitem[\protect\astroncite{{Hebecker} and {Wetterich}}{2001}]{Hebecker2001}
{Hebecker}, A. and {Wetterich}, C.: 2001,
\newblock {\em Physics Letters B} {\bf 497}, 281

\bibitem[\protect\astroncite{{Heger} et~al.}{2005}]{Heger2005}
{Heger}, A., {Woosley}, S.~E., and {Baraffe}, I.: 2005,
\newblock in R. {Humphreys} and K. {Stanek} (eds.), {\em The Fate of the Most
  Massive Stars}, Vol. 332 of {\em Astronomical Society of the Pacific
  Conference Series}, pp 339--+

\bibitem[\protect\astroncite{{Henry}}{2006}]{Henry2006}
{Henry}, T.~J.: 2006,
\newblock {\em RECONS database}

\bibitem[\protect\astroncite{{Hinshaw}}{2006}]{Hinshaw2006}
{Hinshaw}, G.: 2006,
\newblock {\em {Legacy Archive for Microwave Background Data Analaysis
  (LAMBDA)}},
\newblock Online Archive: http://lambda.gsfc.nasa.gov/product/map/\\
  current/params/lcdm\_wmap\_sdss.cfm

\bibitem[\protect\astroncite{{Hopkins}}{2006}]{Hopkins2006}
{Hopkins}, A.~M.: 2006,
\newblock {\em ArXiv e-prints astro.ph/0611283}

\bibitem[\protect\astroncite{{Hulse} and {Taylor}}{1975}]{HulseTaylor1975}
{Hulse}, R.~A. and {Taylor}, J.~H.: 1975,
\newblock {\em \apjl} {\bf 195}, L51

\bibitem[\protect\astroncite{{Huterer} and {Cooray}}{2005}]{Huterer2005}
{Huterer}, D. and {Cooray}, A.: 2005,
\newblock {\em \prd} {\bf 71(2)}, 023506

\bibitem[\protect\astroncite{{Ida} and {Lin}}{2005}]{Ida2005}
{Ida}, S. and {Lin}, D.~N.~C.: 2005,
\newblock {\em \apj} {\bf 626}, 1045

\bibitem[\protect\astroncite{{Jarrett} et~al.}{2003}]{Jarrett2003}
{Jarrett}, T.~H., {Chester}, T., {Cutri}, R., {Schneider}, S.~E., and {Huchra},
  J.~P.: 2003,
\newblock {\em \aj} {\bf 125}, 525

\bibitem[\protect\astroncite{{Jaynes}}{1968}]{Jaynes1968}
{Jaynes}, E.~T.: 1968,
\newblock {\em IEEE Transactions on System Science and Cybernetics} {\bf
  SSC-4}, 227

\bibitem[\protect\astroncite{Jordan}{1955}]{Jordan1955}
Jordan, P.: 1955,
\newblock {\em Schwerkraft und Weltall},
\newblock Braunschweig, Germany

\bibitem[\protect\astroncite{{Kamenshchik} et~al.}{2001}]{Kamenshchik2001}
{Kamenshchik}, A., {Moschella}, U., and {Pasquier}, V.: 2001,
\newblock {\em Physics Letters B} {\bf 511}, 265

\bibitem[\protect\astroncite{{Kolb} and {Turner}}{1981}]{Kolb1981}
{Kolb}, E.~W. and {Turner}, M.~S.: 1981,
\newblock {\em \nat} {\bf 294}, 521

\bibitem[\protect\astroncite{{Kolb} and {Turner}}{1990}]{Kolb1990}
{Kolb}, E.~W. and {Turner}, M.~S.: 1990,
\newblock {\em {The early universe}},
\newblock Frontiers in Physics, Reading, MA: Addison-Wesley, 1988, 1990

\bibitem[\protect\astroncite{{Kroupa}}{2002}]{Kroupa2002}
{Kroupa}, P.: 2002,
\newblock {\em Science} {\bf 295}, 82

\bibitem[\protect\astroncite{{Kuchner} and {Seager}}{2005}]{Kuchner2005}
{Kuchner}, M.~J. and {Seager}, S.: 2005,
\newblock {\em ArXiv e-prints astro-ph/0504214}

\bibitem[\protect\astroncite{{Lanzetta} et~al.}{2002}]{Lanzetta2002}
{Lanzetta}, K.~M., {Yahata}, N., {Pascarelle}, S., {Chen}, H.-W., and
  {Fern{\'a}ndez-Soto}, A.: 2002,
\newblock {\em \apj} {\bf 570}, 492

\bibitem[\protect\astroncite{{Layzer}}{2009}]{Layzer2009}
{Layzer}, D.: 2009,
\newblock {\em www.informationphilosopher.com}

\bibitem[\protect\astroncite{{Linde}}{1982}]{Linde1982}
{Linde}, A.~D.: 1982,
\newblock {\em Physics Letters B} {\bf 108}, 389

\bibitem[\protect\astroncite{{Linde}}{2009}]{Linde2009}
{Linde}, A.~D.: 2009,
\newblock {\em private communication}

\bibitem[\protect\astroncite{{Linder}}{1997}]{Linder1997}
{Linder}, E.: 1997,
\newblock {\em {First Principles of Cosmology}},
\newblock First Principles of Cosmology, by E.V. Linder. Addison-Wesley,
  Harlow, England, 1997

\bibitem[\protect\astroncite{{Linder}}{2006a}]{Linder2006b}
{Linder}, E.~V.: 2006a,
\newblock {\em Astroparticle Physics} {\bf 26}, 102

\bibitem[\protect\astroncite{{Linder}}{2006b}]{Linder2006}
{Linder}, E.~V.: 2006b,
\newblock {\em \prd} {\bf 73(6)}, 063010

\bibitem[\protect\astroncite{{Linder} and {Huterer}}{2005}]{Linder2005}
{Linder}, E.~V. and {Huterer}, D.: 2005,
\newblock {\em \prd} {\bf 72(4)}, 043509

\bibitem[\protect\astroncite{{Lineweaver}}{1998}]{Lineweaver1998}
{Lineweaver}, C.~H.: 1998,
\newblock {\em \apjl} {\bf 505}, L69

\bibitem[\protect\astroncite{{Lineweaver}}{1999}]{Lineweaver1999}
{Lineweaver}, C.~H.: 1999,
\newblock {\em Science} {\bf 284}, 1503

\bibitem[\protect\astroncite{{Lineweaver}}{2001}]{Lineweaver2001}
{Lineweaver}, C.~H.: 2001,
\newblock {\em Icarus} {\bf 151}, 307

\bibitem[\protect\astroncite{{Lineweaver} and {Davis}}{2002}]{Lineweaver2002}
{Lineweaver}, C.~H. and {Davis}, T.~M.: 2002,
\newblock {\em Astrobiology} {\bf 2}, 293

\bibitem[\protect\astroncite{{Lineweaver} and {Davis}}{2003}]{Lineweaver2003a}
{Lineweaver}, C.~H. and {Davis}, T.~M.: 2003,
\newblock {\em Astrobiology} {\bf 3}, 241

\bibitem[\protect\astroncite{{Lineweaver} and {Egan}}{2007}]{Lineweaver2007}
{Lineweaver}, C.~H. and {Egan}, C.~A.: 2007,
\newblock {\em \apj} {\bf 671}, 853

\bibitem[\protect\astroncite{{Lineweaver} and
  {Egan}}{2008}]{LineweaverEgan2008}
{Lineweaver}, C.~H. and {Egan}, C.~A.: 2008,
\newblock {\em Physics of Life Reviews} {\bf 5}, 225

\bibitem[\protect\astroncite{{Lineweaver} and
  {Grether}}{2003}]{Lineweaver2003b}
{Lineweaver}, C.~H. and {Grether}, D.: 2003,
\newblock {\em \apj} {\bf 598}, 1350

\bibitem[\protect\astroncite{{Loveday}}{2000}]{Loveday2000}
{Loveday}, J.: 2000,
\newblock {\em \mnras} {\bf 312}, 557

\bibitem[\protect\astroncite{{Lynden-Bell}}{1967}]{LyndenBell1967}
{Lynden-Bell}, D.: 1967,
\newblock {\em \mnras} {\bf 136}, 101

\bibitem[\protect\astroncite{{Malquarti} et~al.}{2003}]{Malquarti2003}
{Malquarti}, M., {Copeland}, E.~J., and {Liddle}, A.~R.: 2003,
\newblock {\em \prd} {\bf 68(2)}, 023512

\bibitem[\protect\astroncite{{Martel} et~al.}{1998}]{Martel1998}
{Martel}, H., {Shapiro}, P.~R., and {Weinberg}, S.: 1998,
\newblock {\em \apj} {\bf 492}, 29

\bibitem[\protect\astroncite{{Mather} et~al.}{1994}]{Mather1994}
{Mather}, J.~C., {Cheng}, E.~S., {Cottingham}, D.~A., {Eplee}, Jr., R.~E.,
  {Fixsen}, D.~J., {Hewagama}, T., {Isaacman}, R.~B., {Jensen}, K.~A., {Meyer},
  S.~S., {Noerdlinger}, P.~D., {Read}, S.~M., {Rosen}, L.~P., {Shafer}, R.~A.,
  {Wright}, E.~L., {Bennett}, C.~L., {Boggess}, N.~W., {Hauser}, M.~G.,
  {Kelsall}, T., {Moseley}, Jr., S.~H., {Silverberg}, R.~F., {Smoot}, G.~F.,
  {Weiss}, R., and {Wilkinson}, D.~T.: 1994,
\newblock {\em \apj} {\bf 420}, 439

\bibitem[\protect\astroncite{{Mather} et~al.}{1999}]{Mather1999}
{Mather}, J.~C., {Fixsen}, D.~J., {Shafer}, R.~A., {Mosier}, C., and
  {Wilkinson}, D.~T.: 1999,
\newblock {\em \apj} {\bf 512}, 511

\bibitem[\protect\astroncite{{Mbonye}}{2004}]{Mbonye2004}
{Mbonye}, M.~R.: 2004,
\newblock {\em Modern Physics Letters A} {\bf 19}, 117

\bibitem[\protect\astroncite{{Mersini-Houghton} and
  {Adams}}{2008}]{Mersini2008}
{Mersini-Houghton}, L. and {Adams}, F.~C.: 2008,
\newblock {\em Classical and Quantum Gravity} {\bf 25(16)}, 165002

\bibitem[\protect\astroncite{Metropolis and Ulam}{1949}]{Metropolis1949}
Metropolis, N. and Ulam, S.: 1949,
\newblock {\em Journal of the American Statistical Association} {\bf 44(247)},
  335

\bibitem[\protect\astroncite{{Nagamine} et~al.}{2006}]{Nagamine2006}
{Nagamine}, K., {Ostriker}, J.~P., {Fukugita}, M., and {Cen}, R.: 2006,
\newblock {\em \apj} {\bf 653}, 881

\bibitem[\protect\astroncite{{Nojiri} and {Odintsov}}{2006}]{Nojiri2006}
{Nojiri}, S. and {Odintsov}, S.~D.: 2006,
\newblock {\em Physics Letters B} {\bf 637}, 139

\bibitem[\protect\astroncite{{Nordstr{\"o}m} et~al.}{2004}]{Nordstrom2004}
{Nordstr{\"o}m}, B., {Mayor}, M., {Andersen}, J., {Holmberg}, J., {Pont}, F.,
  {J{\o}rgensen}, B.~R., {Olsen}, E.~H., {Udry}, S., and {Mowlavi}, N.: 2004,
\newblock {\em \aap} {\bf 418}, 989

\bibitem[\protect\astroncite{{Olivares} et~al.}{2005}]{Olivares2005}
{Olivares}, G., {Atrio-Barandela}, F., and {Pav{\'o}n}, D.: 2005,
\newblock {\em \prd} {\bf 71(6)}, 063523

\bibitem[\protect\astroncite{{Olivares} et~al.}{2007}]{Olivares2007}
{Olivares}, G., {Atrio-Barandela}, F., and {Pavon}, D.: 2007,
\newblock {\em ArXiv e-prints 0706.3860} 706

\bibitem[\protect\astroncite{{{\"O}zer} and {Taha}}{1987}]{Ozer1987}
{{\"O}zer}, M. and {Taha}, M.~O.: 1987,
\newblock {\em Nuclear Physics B} {\bf 287}, 776

\bibitem[\protect\astroncite{{Page}}{1981}]{Page1981}
{Page}, D.~N.: 1981,
\newblock {\em General Relativity and Gravitation} {\bf 13}, 1117

\bibitem[\protect\astroncite{{Page} and {McKee}}{1981}]{Page1981a}
{Page}, D.~N. and {McKee}, M.~R.: 1981,
\newblock {\em \nat} {\bf 291}, 44

\bibitem[\protect\astroncite{{Pagel}}{1997}]{Pagel1997}
{Pagel}, B.~E.~J.: 1997,
\newblock {\em {Nucleosynthesis and Chemical Evolution of Galaxies}},
\newblock Nucleosynthesis and Chemical Evolution of Galaxies, by Bernard
  E.~J.~Pagel, pp.~392.~ISBN 0521550610.~Cambridge, UK: Cambridge University
  Press, October 1997.

\bibitem[\protect\astroncite{{Pav{\'o}n} and {Zimdahl}}{2005}]{Pavon2005}
{Pav{\'o}n}, D. and {Zimdahl}, W.: 2005,
\newblock {\em Physics Letters B} {\bf 628}, 206

\bibitem[\protect\astroncite{{Peacock}}{1999}]{Peacock1999}
{Peacock}, J.~A.: 1999,
\newblock {\em {Cosmological Physics}},
\newblock Cosmological Physics, by John A.~Peacock, pp.~704.~ISBN
  052141072X.~Cambridge, UK: Cambridge University Press, January 1999.

\bibitem[\protect\astroncite{{Peebles} and {Vilenkin}}{1999}]{Peebles1999}
{Peebles}, P.~J.~E. and {Vilenkin}, A.: 1999,
\newblock {\em \prd} {\bf 59(6)}, 063505

\bibitem[\protect\astroncite{{Penrose}}{1979}]{Penrose1979}
{Penrose}, R.: 1979,
\newblock in S.~W. {Hawking} and W. {Israel} (eds.), {\em General Relativity:
  An Einstein centenary survey}, pp 581--638

\bibitem[\protect\astroncite{{Penrose}}{1987}]{Penrose1987}
{Penrose}, R.: 1987,
\newblock {\em {Newton, quantum theory and reality.}}, pp 17--49,
\newblock Three hundred years of gravitation, p.~17 - 49

\bibitem[\protect\astroncite{{Penrose}}{2004}]{Penrose2004}
{Penrose}, R.: 2004,
\newblock {\em {The road to reality : a complete guide to the laws of the
  universe}},
\newblock The road to reality : a complete guide to the laws of the universe,
  by Roger Penrose.~ London: Jonathan Cape, 2004

\bibitem[\protect\astroncite{{Perlmutter} et~al.}{1999}]{Perlmutter1999}
{Perlmutter}, S., {Aldering}, G., {Goldhaber}, G., {Knop}, R.~A., {Nugent}, P.,
  {Castro}, P.~G., {Deustua}, S., {Fabbro}, S., {Goobar}, A., {Groom}, D.~E.,
  {Hook}, I.~M., {Kim}, A.~G., {Kim}, M.~Y., {Lee}, J.~C., {Nunes}, N.~J.,
  {Pain}, R., {Pennypacker}, C.~R., {Quimby}, R., {Lidman}, C., {Ellis}, R.~S.,
  {Irwin}, M., {McMahon}, R.~G., {Ruiz-Lapuente}, P., {Walton}, N., {Schaefer},
  B., {Boyle}, B.~J., {Filippenko}, A.~V., {Matheson}, T., {Fruchter}, A.~S.,
  {Panagia}, N., {Newberg}, H.~J.~M., {Couch}, W.~J., and {The Supernova
  Cosmology Project}: 1999,
\newblock {\em \apj} {\bf 517}, 565

\bibitem[\protect\astroncite{{Pogosian} and {Vilenkin}}{2007}]{Pogosian2007}
{Pogosian}, L. and {Vilenkin}, A.: 2007,
\newblock {\em Journal of Cosmology and Astro-Particle Physics} {\bf 1}, 25

\bibitem[\protect\astroncite{{Press} and {Schechter}}{1974}]{Press1974}
{Press}, W.~H. and {Schechter}, P.: 1974,
\newblock {\em \apj} {\bf 187}, 425

\bibitem[\protect\astroncite{{Ratra} and {Peebles}}{1988}]{Ratra1988}
{Ratra}, B. and {Peebles}, P.~J.~E.: 1988,
\newblock {\em \prd} {\bf 37}, 3406

\bibitem[\protect\astroncite{{Reddy} et~al.}{2003}]{Reddy2003}
{Reddy}, B.~E., {Tomkin}, J., {Lambert}, D.~L., and {Allende Prieto}, C.: 2003,
\newblock {\em \mnras} {\bf 340}, 304

\bibitem[\protect\astroncite{{Reid}}{2002}]{Reid2002}
{Reid}, I.~N.: 2002,
\newblock {\em \pasp} {\bf 114}, 306

\bibitem[\protect\astroncite{{Riess} et~al.}{1998}]{Riess1998}
{Riess}, A.~G., {Filippenko}, A.~V., {Challis}, P., {Clocchiatti}, A.,
  {Diercks}, A., {Garnavich}, P.~M., {Gilliland}, R.~L., {Hogan}, C.~J., {Jha},
  S., {Kirshner}, R.~P., {Leibundgut}, B., {Phillips}, M.~M., {Reiss}, D.,
  {Schmidt}, B.~P., {Schommer}, R.~A., {Smith}, R.~C., {Spyromilio}, J.,
  {Stubbs}, C., {Suntzeff}, N.~B., and {Tonry}, J.: 1998,
\newblock {\em \aj} {\bf 116}, 1009

\bibitem[\protect\astroncite{{Riess} et~al.}{2007}]{Riess2007}
{Riess}, A.~G., {Strolger}, L.-G., {Casertano}, S., {Ferguson}, H.~C.,
  {Mobasher}, B., {Gold}, B., {Challis}, P.~J., {Filippenko}, A.~V., {Jha}, S.,
  {Li}, W., {Tonry}, J., {Foley}, R., {Kirshner}, R.~P., {Dickinson}, M.,
  {MacDonald}, E., {Eisenstein}, D., {Livio}, M., {Younger}, J., {Xu}, C.,
  {Dahl{\'e}n}, T., and {Stern}, D.: 2007,
\newblock {\em \apj} {\bf 659}, 98

\bibitem[\protect\astroncite{{Riess} et~al.}{2004}]{Riess2004}
{Riess}, A.~G., {Strolger}, L.-G., {Tonry}, J., {Casertano}, S., {Ferguson},
  H.~C., {Mobasher}, B., {Challis}, P., {Filippenko}, A.~V., {Jha}, S., {Li},
  W., {Chornock}, R., {Kirshner}, R.~P., {Leibundgut}, B., {Dickinson}, M.,
  {Livio}, M., {Giavalisco}, M., {Steidel}, C.~C., {Ben{\'{\i}}tez}, T., and
  {Tsvetanov}, Z.: 2004,
\newblock {\em \apj} {\bf 607}, 665

\bibitem[\protect\astroncite{{Robles} et~al.}{2008a}]{Robles2008c}
{Robles}, J.~A., {Egan}, C.~A., and {Lineweaver}, C.~H.: 2008a,
\newblock in {\em Australian Space Science Conference Series: 8th Conference
  Proceedings}

\bibitem[\protect\astroncite{{Robles} et~al.}{2008b}]{Robles2008a}
{Robles}, J.~A., {Lineweaver}, C.~H., {Grether}, D., {Flynn}, C., {Egan},
  C.~A., {Pracy}, M.~B., {Holmberg}, J., and {Gardner}, E.: 2008b,
\newblock {\em \apj} {\bf 684}, 691

\bibitem[\protect\astroncite{{Robles} et~al.}{2008c}]{Robles2008b}
{Robles}, J.~A., {Lineweaver}, C.~H., {Grether}, D., {Flynn}, C., {Egan},
  C.~A., {Pracy}, M.~B., {Holmberg}, J., and {Gardner}, E.: 2008c,
\newblock {\em \apj} {\bf 689}, 1457

\bibitem[\protect\astroncite{{Rocha-Pinto} et~al.}{2000}]{RochaPinto2000}
{Rocha-Pinto}, H.~J., {Scalo}, J., {Maciel}, W.~J., and {Flynn}, C.: 2000,
\newblock {\em \aap} {\bf 358}, 869

\bibitem[\protect\astroncite{{Sahni}}{2002}]{Sahni2002}
{Sahni}, V.: 2002,
\newblock {\em Classical and Quantum Gravity} {\bf 19}, 3435

\bibitem[\protect\astroncite{{Sahni} and {Wang}}{2000}]{Sahni2000}
{Sahni}, V. and {Wang}, L.: 2000,
\newblock {\em \prd} {\bf 62(10)}, 103517

\bibitem[\protect\astroncite{{Sassi} and {Bonometto}}{2007}]{Sassi2007}
{Sassi}, G. and {Bonometto}, S.~A.: 2007,
\newblock {\em New Astronomy} {\bf 12}, 353

\bibitem[\protect\astroncite{{Scherrer}}{2005}]{Scherrer2005}
{Scherrer}, R.~J.: 2005,
\newblock {\em \prd} {\bf 71(6)}, 063519

\bibitem[\protect\astroncite{{Seljak} et~al.}{2006}]{Seljak2006}
{Seljak}, U., {Slosar}, A., and {McDonald}, P.: 2006,
\newblock {\em Journal of Cosmology and Astro-Particle Physics} {\bf 10}, 14

\bibitem[\protect\astroncite{{Soderblom}}{1983}]{Soderblom1983}
{Soderblom}, D.~R.: 1983,
\newblock {\em \apjs} {\bf 53}, 1

\bibitem[\protect\astroncite{{Spergel} et~al.}{2007}]{Spergel2007}
{Spergel}, D.~N., {Bean}, R., {Dor{\'e}}, O., {Nolta}, M.~R., {Bennett}, C.~L.,
  {Dunkley}, J., {Hinshaw}, G., {Jarosik}, N., {Komatsu}, E., {Page}, L.,
  {Peiris}, H.~V., {Verde}, L., {Halpern}, M., {Hill}, R.~S., {Kogut}, A.,
  {Limon}, M., {Meyer}, S.~S., {Odegard}, N., {Tucker}, G.~S., {Weiland},
  J.~L., {Wollack}, E., and {Wright}, E.~L.: 2007,
\newblock {\em \apjs} {\bf 170}, 377

\bibitem[\protect\astroncite{{Spergel} et~al.}{2006}]{Spergel2006}
{Spergel}, D.~N., {Bean}, R., {Dore'}, O., {Nolta}, M.~R., {Bennett}, C.~L.,
  {Hinshaw}, G., {Jarosik}, N., {Komatsu}, E., {Page}, L., {Peiris}, H.~V.,
  {Verde}, L., {Barnes}, C., {Halpern}, M., {Hill}, R.~S., {Kogut}, A.,
  {Limon}, M., {Meyer}, S.~S., {Odegard}, N., {Tucker}, G.~S., {Weiland},
  J.~L., {Wollack}, E., and {Wright}, E.~L.: 2006,
\newblock {\em astro-ph/0603449
  http://lambda.gsfc.nasa.gov/product/map/dr2/params/lcdm-all.cfm}

\bibitem[\protect\astroncite{{Steinhardt}}{2008}]{Steinhardt2009}
{Steinhardt}, P.: 2008,
\newblock {\em private communication}

\bibitem[\protect\astroncite{{Steinhardt}}{2003}]{Steinhardt2003}
{Steinhardt}, P.~J.: 2003,
\newblock {\em Royal Society of London Philosophical Transactions Series A}
  {\bf 361}, 2497

\bibitem[\protect\astroncite{{Steinhardt} et~al.}{1999}]{Steinhardt1999}
{Steinhardt}, P.~J., {Wang}, L., and {Zlatev}, I.: 1999,
\newblock {\em \prd} {\bf 59(12)}, 123504

\bibitem[\protect\astroncite{{Strominger} and {Vafa}}{1996}]{Strominger1996}
{Strominger}, A. and {Vafa}, C.: 1996,
\newblock {\em Physics Letters B} {\bf 379}, 99

\bibitem[\protect\astroncite{{Susskind}}{1995}]{Susskind1995}
{Susskind}, L.: 1995,
\newblock {\em Journal of Mathematical Physics} {\bf 36}, 6377

\bibitem[\protect\astroncite{{Szyd{\l}owski} et~al.}{2006}]{Szydlowski2006}
{Szyd{\l}owski}, M., {Kurek}, A., and {Krawiec}, A.: 2006,
\newblock {\em Physics Letters B} {\bf 642}, 171

\bibitem[\protect\astroncite{{'t Hooft}}{1993}]{tHooft1993}
{'t Hooft}, G.: 1993,
\newblock {\em ArXiv General Relativity and Quantum Cosmology e-prints}

\bibitem[\protect\astroncite{{Thomas}}{2009}]{Thomas2009}
{Thomas}, A.: 2009,
\newblock {\em www.ipod.org.uk}

\bibitem[\protect\astroncite{{Thompson} et~al.}{2006}]{Thompson2006}
{Thompson}, R.~I., {Eisenstein}, D., {Fan}, X., {Dickinson}, M., {Illingworth},
  G., and {Kennicutt}, Jr., R.~C.: 2006,
\newblock {\em \apj} {\bf 647}, 787

\bibitem[\protect\astroncite{{Thomson}}{1852}]{Thomson1852}
{Thomson}, W.: 1852,
\newblock {\em Philosophical Magazine} {\bf 4}, 304

\bibitem[\protect\astroncite{{Turner}}{2001}]{Turner2001}
{Turner}, M.~S.: 2001,
\newblock {\em Nuclear Physics B Proceedings Supplements} {\bf 91}, 405

\bibitem[\protect\astroncite{{Valenti} and {Fischer}}{2005}]{Valenti2005}
{Valenti}, J.~A. and {Fischer}, D.~A.: 2005,
\newblock {\em \apjs} {\bf 159}, 141

\bibitem[\protect\astroncite{{Vilenkin}}{1995a}]{Vilenkin1995b}
{Vilenkin}, A.: 1995a,
\newblock {\em \prd} {\bf 52}, 3365

\bibitem[\protect\astroncite{{Vilenkin}}{1995b}]{Vilenkin1995a}
{Vilenkin}, A.: 1995b,
\newblock {\em Physical Review Letters} {\bf 74}, 846

\bibitem[\protect\astroncite{{Vilenkin}}{1996a}]{Vilenkin1996a}
{Vilenkin}, A.: 1996a,
\newblock in N. {Sanchez} and A. {Zichichi} (eds.), {\em String Gravity and
  Physics at the Planck Energy Scale}, pp 345--367

\bibitem[\protect\astroncite{{Vilenkin}}{1996b}]{Vilenkin1996b}
{Vilenkin}, A.: 1996b,
\newblock in K. {Sato}, T. {Suginohara}, and N. {Sugiyama} (eds.), {\em
  Cosmological Constant and the Evolution of the Universe}, pp 161--+

\bibitem[\protect\astroncite{{Wang} et~al.}{2000}]{Wang2000}
{Wang}, L., {Caldwell}, R.~R., {Ostriker}, J.~P., and {Steinhardt}, P.~J.:
  2000,
\newblock {\em \apj} {\bf 530}, 17

\bibitem[\protect\astroncite{{Wang} and {Mukherjee}}{2006}]{Wang2006}
{Wang}, Y. and {Mukherjee}, P.: 2006,
\newblock {\em \apj} {\bf 650}, 1

\bibitem[\protect\astroncite{{Wang} and {Tegmark}}{2004}]{Wang2004b}
{Wang}, Y. and {Tegmark}, M.: 2004,
\newblock {\em Physical Review Letters} {\bf 92(24)}, 241302

\bibitem[\protect\astroncite{{Wang} and {Tegmark}}{2005}]{Wang2005}
{Wang}, Y. and {Tegmark}, M.: 2005,
\newblock {\em \prd} {\bf 71(10)}, 103513

\bibitem[\protect\astroncite{{Weinberg}}{1987}]{Weinberg1987}
{Weinberg}, S.: 1987,
\newblock {\em Physical Review Letters} {\bf 59}, 2607

\bibitem[\protect\astroncite{{Weinberg}}{1989}]{Weinberg1989}
{Weinberg}, S.: 1989,
\newblock {\em Reviews of Modern Physics} {\bf 61}, 1

\bibitem[\protect\astroncite{{Weinberg}}{2000a}]{Weinberg2000}
{Weinberg}, S.: 2000a,
\newblock {\em \prd} {\bf 61(10)}, 103505

\bibitem[\protect\astroncite{{Weinberg}}{2000b}]{Weinberg2000b}
{Weinberg}, S.: 2000b,
\newblock {\em ArXiv e-prints astro-ph/0005265}

\bibitem[\protect\astroncite{{Weisberg} and {Taylor}}{2005}]{Weisberg2005}
{Weisberg}, J.~M. and {Taylor}, J.~H.: 2005,
\newblock in F.~A. {Rasio} and I.~H. {Stairs} (eds.), {\em Binary Radio
  Pulsars}, Vol. 328 of {\em Astronomical Society of the Pacific Conference
  Series}, pp 25--+

\bibitem[\protect\astroncite{{Wetherill}}{1996}]{Wetherill1996a}
{Wetherill}, G.~W.: 1996,
\newblock in L.~R. {Doyle} (ed.), {\em Circumstellar Habitable Zones}, p. 193

\bibitem[\protect\astroncite{{Wood-Vasey} et~al.}{2007}]{Wood-Vasey2007}
{Wood-Vasey}, W.~M., {Miknaitis}, G., {Stubbs}, C.~W., {Jha}, S., {Riess},
  A.~G., {Garnavich}, P.~M., {Kirshner}, R.~P., {Aguilera}, C., {Becker},
  A.~C., {Blackman}, J.~W., {Blondin}, S., {Challis}, P., {Clocchiatti}, A.,
  {Conley}, A., {Covarrubias}, R., {Davis}, T.~M., {Filippenko}, A.~V.,
  {Foley}, R.~J., {Garg}, A., {Hicken}, M., {Krisciunas}, K., {Leibundgut}, B.,
  {Li}, W., {Matheson}, T., {Miceli}, A., {Narayan}, G., {Pignata}, G.,
  {Prieto}, J.~L., {Rest}, A., {Salvo}, M.~E., {Schmidt}, B.~P., {Smith},
  R.~C., {Sollerman}, J., {Spyromilio}, J., {Tonry}, J.~L., {Suntzeff}, N.~B.,
  and {Zenteno}, A.: 2007,
\newblock {\em ArXiv Astrophysics e-prints}

\bibitem[\protect\astroncite{{Wright} et~al.}{2004}]{Wright2004}
{Wright}, J.~T., {Marcy}, G.~W., {Butler}, R.~P., and {Vogt}, S.~S.: 2004,
\newblock {\em \apjs} {\bf 152}, 261

\bibitem[\protect\astroncite{{Yang} and {Wang}}{2005}]{Yang2005}
{Yang}, G. and {Wang}, A.: 2005,
\newblock {\em General Relativity and Gravitation} {\bf 37}, 2201

\bibitem[\protect\astroncite{{Zel'Dovich}}{1967}]{ZelDovich1967}
{Zel'Dovich}, Y.~B.: 1967,
\newblock {\em Soviet Journal of Experimental and Theoretical Physics Letters}
  {\bf 6}, 316

\bibitem[\protect\astroncite{{Zhang}}{2005}]{Zhang2005}
{Zhang}, X.: 2005,
\newblock {\em Modern Physics Letters A} {\bf 20}, 2575

\bibitem[\protect\astroncite{{Zimdahl} et~al.}{2001}]{Zimdahl2001}
{Zimdahl}, W., {Pav{\'o}n}, D., and {Chimento}, L.~P.: 2001,
\newblock {\em Physics Letters B} {\bf 521}, 133

\bibitem[\protect\astroncite{{Zlatev} et~al.}{1999}]{Zlatev1999}
{Zlatev}, I., {Wang}, L., and {Steinhardt}, P.~J.: 1999,
\newblock {\em Physical Review Letters} {\bf 82}, 896

\end{thebibliography}

\end{document}